\documentclass[a4paper,11pt]{report}

\newcommand{\linespacing}{1.5}
\renewcommand{\baselinestretch}{\linespacing}

\usepackage{cite}

\usepackage{graphicx,color}
\usepackage{epstopdf}
\usepackage[british]{babel}
\usepackage{amsmath}
\usepackage{amssymb}
\usepackage[T1]{fontenc}
\usepackage[a4paper,top=2.5cm,bottom=2.5cm,left=4cm,right=2cm,headsep=10pt]{geometry}
\usepackage{fancyhdr}
\fancypagestyle{plain}{
	\fancyhf{}
	\chead{\thepage}
	
}
\usepackage[colorlinks,pagebackref,pdfusetitle,urlcolor=blue,citecolor=blue,linkcolor=blue,bookmarksnumbered,plainpages=false]{hyperref}
\newcommand{\titletext}{{Constraining the physics of the early Universe}}

\newcommand{\newform}[2]{#1}
\begin{document}

\pagenumbering{roman}

\thispagestyle{empty}
\begin{flushright}
\includegraphics[width=6cm]{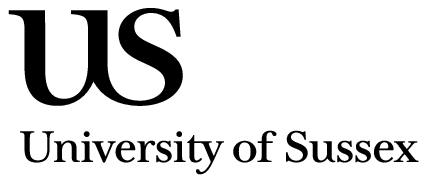}
\end{flushright}	
\vskip40mm
\begin{center}
\huge\textbf{\titletext}
\vskip2mm
\vskip5mm
\Large\textbf{Jos\'{e} Pedro Pinto Vieira}
\normalsize
\end{center}
\vfill
\begin{flushleft}
\large
Submitted for the degree of Doctor of Philosophy \\
University of Sussex	\\
March 2018
\end{flushleft}		

\chapter*{Declaration}
I hereby declare that this thesis has not been and will not be submitted in whole or in part to another University for the award of any other degree.
\vskip2mm
\noindent The work in this thesis was done in collaboration with Christian Byrnes, Djuna Croon, Antony Lewis, Carlos Martins, Sonali Mohapatra, and Paul Shellard. This thesis incorporates the following papers:

\begin{itemize}

\item J.P.P. Vieira, Christian T. Byrnes, and Antony Lewis. ``Cosmology with Negative Absolute Temperatures'', published in JCAP 1608 (2016) no.08, 060. DOI: 10.1088/1475-7516/2016/08/060. arXiv:1604.05099.

I was responsible for the original idea for this paper.
The paper and calculations were completed by myself under the supervision of Christian Byrnes and Antony Lewis,
who also made some minor adjustments and additions.

\item J.P.P. Vieira, C.J.A.P. Martins, and E.P.S. Shellard. ``Models for small-scale structure on cosmic strings. II. Scaling and its stability'', published in Phys. Rev. D94 (2016) no.9, 096005. DOI: 10.1103/PhysRevD.94.099907. arXiv: arXiv:1611.06103.

This paper follows naturally from a previous one by the same authors \cite{PAP1}.
Most calculations were completed by myself under the supervision of Carlos Martins.
Most of the theoretical portions of the paper were written by myself with important contributions and adjustments by Carlos Martins.
I made a few minor contributions to the writing of the remainder of the paper.
The main results in \ref{findscale} and \ref{scasol} were found during my MSc at the University of Porto and are also in my MSc thesis.

\item J.P.P. Vieira, Christian T. Byrnes, and Antony Lewis. ``	
Can power spectrum observations rule out slow-roll inflation?'', published in JCAP 1801 (2018) no.01, 019. DOI: 10.1088/1475-7516/2018/01/019. arXiv:1710.08408. 

The concept for this paper came from my adaptation of an original idea by Christian Byrnes.
The paper and calculations were completed by myself under the supervision of Christian Byrnes and Antony Lewis,
who also made some minor adjustments and additions.

\end{itemize}

\vskip5mm
Signature:
\vskip20mm
Jos\'{e} Pedro Pinto Vieira

\thispagestyle{empty}
\newpage
\null\vskip10mm
\begin{center}
\large
\underline{UNIVERSITY OF SUSSEX}
\vskip20mm
\textsc{Jos\'{e} Pedro Pinto Vieira, Doctor of Philosophy}
\vskip20mm
\underline{\textsc{\titletext}}
\vskip0mm
\vskip20mm
\underline{\textsc{Summary}}
\vskip2mm
\end{center}
\renewcommand{\baselinestretch}{1.0}
\small\normalsize

The established cosmological theory which describes the history of the Universe since shortly after the ``Big Bang'' until today is remarkably successful. Thanks to the increasing precision of available observational data, we are now able to considerably constrain the geometry and composition of the Universe --- and to glimpse how these will evolve in the near future. However, this success comes at a price: one must assume the Universe ``started'' in a highly fine-tuned initial condition. Understanding what came before this is therefore one of the main goals of modern cosmology. 

This thesis attempts to further our understanding of the epoch before this initial condition in three different ways.

Firstly, the concept of negative absolute temperatures (NAT) is introduced and its potential relevance for cosmology is investigated. In particular, it is shown that a Universe at a NAT should undergo a period of inflation --- although it is unclear whether this would be consistent with current observations. 

Secondly, work is done on the topic of the evolution of networks of cosmic strings --- topological defects which are expected to form in a broad class of phase transitions the Universe may have gone through. A model which takes into account the presence of small-scale structure in strings is used to address questions concerning the existence and stability of scaling regimes for these networks.

Finally, it is investigated how future experiments might try to falsify a simple class of canonical single-field slow-roll inflation models by measuring the running and the running of the running of the spectral index of scalar perturbations.

\chapter*{Acknowledgements}
\renewcommand{\baselinestretch}{\linespacing}
\small\normalsize
First of all, my most sincere thanks to my supervisors: Christian Byrnes and Antony Lewis. I could not have asked for more supportive and patient  guidance, and I am truly grateful for all that I have learnt from them\footnote{Like not to write too many unnecessary footnotes.}. I must also thank the other collaborators without whose hard work and invaluable insights this thesis would be undoubtedly poorer: Djuna Croon, Carlos Martins, Sonali Mohapatra, and Paul Shellard.

Throughout these (almost) four years, I have been blessed with a great many companions who have made my life fuller. I am especially thankful to my fellow PhD students and former housemates, Ridwan Barbhuiyan, Luc\'{i}a Fonseca, and Michaela Lawrence - they have put up with me more than anyone should have to and I will always cherish their friendship. To my office mates, Pippa Cole, Kiattisak Devasurya, Luc\'{i}a Fonseca (look at you, being mentioned twice!), Mateja Gosenca, Mark Mirmelstein, and Sam Young, I am thankful for the best working environment in MPS and for the ability to beat British people at Scrabble. Special thanks also to D\'{a}niel Moln\'{a}r, who introduced me to the world of boardgaming and taught me there is no reason one cannot study the cosmos while conquering it. To Ridwan Barbhuiyan, Christian Byrnes (wow, two second mentions in a row!), Beno\^{i}t Fournier, and Alex Eggemeier (a.k.a. the glorious lunchmeister), I am thankful for having fairly regularly squashed me. To the remainder of the PhD students and staff in the department, especially Sunayana Bhargava, Scott Clay, Steven Duivenvoorden, Ciaran Fairhurst, Azizah Hosein, Hannah Ross, and David Sullivan, I am thankful for countless things ranging from stimulating lunchtime discussions (you know what I'm talking about) to Game of Thrones viewing sessions and an appreciation of rugby. I am also thankful to everyone at the Chess Society of the University of Sussex and the Catholic Chaplaincy of the Universities of Brighton and Sussex for having greatly enriched my evenings and weekends during this period.

Finally, a very special word of thanks to those who are always behind the scenes but without whose love and support I would have never got here: my family. To my grandparents, whom my success gladdens so much more than me; to my parents, who have always been there and who have made me into who I am; to my sister, who has taught me the real meaning of strength; and to my wife, who has shared this fantastic journey with me and who kept me sane through the writing of this thesis: thank you all so much; I am truly fortunate to have you all in my life!


\newpage
\pdfbookmark[0]{Contents}{contents_bookmark}
\tableofcontents
\newpage
\phantomsection
\addcontentsline{toc}{chapter}{List of Tables}
\listoftables
\newpage
\phantomsection
\addcontentsline{toc}{chapter}{List of Figures}
\listoffigures
\newpage
\begin{flushright}

``$\ll$Faz-te merc\^{e}, bar\~{a}o, a Sapi\^{e}ncia\\
Suprema de, cos olhos corporais,\\
Veres o que n\~{a}o pode a v\~{a} Ci\^{e}ncia\\
Dos errados e m\'{\i}seros mortais.\\
Sigue-me firme e forte, com prud\^{e}ncia,\\
Por este monte espesso, tu cos mais.$\gg$\\
Assi lhe diz e o guia por um mato\\
\'{A}rduo, dif\'{\i}cil, duro a humano trato.''\\
\ \\
Lu\'{\i}s Vaz de Cam\~{o}es, in \textit{Os Lus\'{\i}adas}
\\
\ \ \ \\
\ \ \ \\
``M\'{a}quina do Mundo\\
\ \ \ \\
O Universo \'{e} feito essencialmente de coisa nenhuma.\\
Intervalos, dist\^{a}ncias, buracos, porosidade et\'{e}rea.\\
Espa\c{c}o vazio, em suma.\\
O resto, \'{e} a mat\'{e}ria.\\
Da\'{\i}, que este arrepio,\\
este cham\'{a}-lo e t\^{e}-lo, ergu\^{e}-lo e defront\'{a}-lo,\\
esta fresta de nada aberta no vazio,\\
deve ser um intervalo.''
\\
\ \\
Ant\'{o}nio Gede\~{a}o, in \textit{M\'{a}quina de Fogo}
\end {flushright}
\thispagestyle{empty}
\newpage

\newpage
\begin{flushright}
``$\ll$To thee supremest wisdom guerdon gave,\\
Baron ! who hast beheld with fleshly eyne\\
what things the Future hath the pow'er to save\\
from Mortals' petty pride and science vain.\\
Follow me firmly, prudent as thou'rt brave,\\
to yonder craggy brake with all thy train !$\gg$\\
Thus she, and straightway through a long wood led\\
arduous, gloomy, fere for foot to tread.''\\
\ \\
Lu\'{\i}s Vaz de Cam\~{o}es, in \textit{Os Lus\'{\i}adas} [translation by Richard Francis Burton]
\\
\ \ \ \\
\ \ \ \\
``Engine of the World\\
\ \ \ \\
The Universe is made essentially of no thing.\\
Intervals, distances, holes, aetherial porosity.\\
Empty space, in sum.\\
The rest, is matter.\\
Thus, this shiver,\\
this calling it and having it, holding it and facing it,\\
this crack of nothing open in the void,\\
must also be an interval.''
\\
\ \\
Ant\'{o}nio Gede\~{a}o, in \textit{M\'{a}quina de Fogo}
\end {flushright}
\thispagestyle{empty}
\newpage

\newpage
\pagenumbering{arabic}




\chapter{Introduction}
\label{chap:intro}

The study of the early Universe is one of the most important endeavours of modern physics --- not only because of its implications for some of the oldest ``big questions'' about our origin and place in the Universe, but also because the primordial Universe is an ideal system to test theories of particle physics at the highest energies. At a time when ever more precise cosmological observations are becoming available, a solid understanding of the implications of new physics in the early Universe is of the utmost importance to gain knowledge of cosmic history and physics as a whole. This thesis seeks to contribute to this understanding by way of three different projects making up three separate chapters.

The current chapter provides an introduction to standard cosmology and useful concepts which will be needed ahead. Afterwards, the cosmological consequences of negative absolute temperatures in the early Universe are investigated in chapter \ref{chap:NAT}. Chapter \ref{chap:wiggly} explores the problem of modelling the evolution of networks of cosmic strings with small-scale structure. Chapter \ref{chap:SFSRI} studies how the scale-dependence of the inferred spectrum of primordial perturbations may be used to test a simple class of slow-roll inflation. Finally, chapter \ref{chap:conc} summarises the main conclusions of each project and outlines possible directions of future work.

For the sake of simplicity, natural units with $c=\hbar=k_{B}=\left(8\pi G\right)^{-1}=1$ are used thoughout this thesis.

\section{Standard cosmology}

\subsection{Short-sighted cosmology}

In a sense, cosmology (defined as the study of the structure, dynamics, origin, and ultimate fate of the Universe) is one of the most ancient fields of human enquiry. Notwithstanding, it was one of the last to mature into a fully-fledged (or even just a fledgling) science - which is not surprising, given that for most of human history we have lacked the basic theoretical and observational tools necessary for this pursuit.

While astronomy has been with us pretty much since the dawn of civilization, its initial reliance on naked-eye observations meant that, before telescopes became available in the early 17th century, there was a limited number of observations of objects beyond our immediate cosmic vicinity. To make matters worse, it took millenia of observations before we even realised just how distant these objects are\footnote{The visible part of the Milky Way, for example, was believed by Aristotle to be due to fires in the upper atmosphere \cite{sep-ibn-bajja}, and it wasn't until Ibn al-Haytham's observations in the 11th century that it became clear that it had to be much more distant from us \cite{AlHaythamWorkshop}.}. Not to mention how hard it was for us to grasp the basic dynamics of just the easily observable bits of our solar system.

And yet progress was made. Granted, it was slow progress, but little by little our knowledge of our surroundings did improve. By the beginning of the 20th century, this progress may not have seemed like much compared to what was about to come; however, it was a radical enough departure from the sort of cosmology that was the norm until Copernicus and Galileo. Radical enough for Neptune's existence to have been inferred from its effect on Uranus' orbit. Radical enough for there being a debate about whether our galaxy might not constitute the bulk of the Universe. Radical enough that cosmology was ready to play a role in the revolution that marked physics in the first few decades of the century...

\subsection{The Big Bang: story of an idea}

The advent of modern cosmology began with the introduction of general relativity by Albert Einstein, in 1915. For the first time in history, the scientific paradigm which was about to be established provided mathematical tools which enabled a consistent and rigorous study of the geometry and dynamics of the Universe as a whole\footnote{Interestingly, in \emph{``Cosmological considerations in the general theory of relativity''} \cite{1917SPAW.......142E}, Einstein does briefly discuss an attempt to apply Newtonian gravity to the entire Universe. This approach, however, was marred by the difficulty of maintaining the stable finite Universe it required. It is perhaps ironic that similar considerations ended up leading to the abandonment of the Einstein Universe model put forward in this work \cite{ORaifeartaigh:2017uct}. Failure to consider these issues, it has been argued, is the real reason Einstein alledgedly made his famous ``biggest blunder'' comment \cite{ORaifeartaigh2017}.}. When, in 1917, Einstein wrote his \emph{``Cosmological considerations in the general theory of relativity''} \cite{1917SPAW.......142E}, he was still arguably more interested in using cosmological considerations as added constraints on general relativity (leading to the infamous ``blunder'' of the introduction of a cosmological constant) than the other way round\footnote{Not that there is anything ``uncosmological'' about that sort of mindset. In fact, using the Universe as a ``laboratory'' for constraining fundamental physics is a most common motivation. Nevertheless, the converse is still an important component of cosmology --- and Einstein's dismissal of that component may explain why something like the Friedmann equations did not feature in this paper.}. Nonetheless, the door was finally open for a fruitful symbiosis between cosmology and relativity.

The first to take full advantage of this new relationship was the theoretical meteorologist Alexander Friedmann, who in 1922 put forth the possibility of an expanding (or contracting) Universe allowed by general relativity \cite{1922ZPhy...10..377F,0038-5670-6-4-E01}. Unfortunately, Einstein didn't appreciate the physical relevance of Friedmann's contribution (even after retracting an initial attack on his calculations \cite{Einstein1923,ORaifeartaigh2017}) and, despite Friedmann's own international connections, his work seems to not have reached that many outside of the USSR.

In 1927, two years after Friedmann's death, Georges Lema\^{\i}tre independently rederived Friedmann's expanding solution and predicted a linear relation between distance and velocity for extragalactic nebulae \cite{1927ASSB...47...49L}. This relation was observationally demonstrated two years later by Edwin Hubble \cite{1929PNAS...15..168H}, and thus became known as Hubble's law\footnote{Lema\^{\i}tre himself is actually responsible for the first estimate of the numerical value of the proportionality constant (now known as Hubble's constant), but he did not have access to enough data to prove there was a linear relation in the first place. Hubble and his assistant, Milton Humason, did so by combining their own distance measurements with earlier velocity measurements by Vesto Slipher \cite{1917PAPhS..56..403S}. Sadly, Hubble would only give due credit to Slipher two years later, once the discovery had already become associated with his own name \cite{1931ApJ....74...43H}.}. This development crucially contributed to the growing acceptance of Lema\^itre's approach, notably by Einstein (who had initially opposed it as he had Friedmann's).

Not long after, in 1931, Lema\^itre once again espoused controversy by suggesting that the Universe may have had a beginning ``a little before the beginning of space and time''\footnote{Nowadays, it is commonplace to justify the idea of a Universe with a beginning in a mysterious singularity by applying the sorts of models Friedmann and Lema\^itre introduced to look at the past instead of the future. Curiously, Lema\^itre's original argument relied not on relativity but on a quantum formulation of thermodynamical principles. This argument is where the term ``primeval atom'' comes from: in his original theory, Lema\^itre supposed that, in its initial state, the Universe would be a literal atom ``the atomic weigh of which is the total mass of the universe'' which would then decay into ``smaller and smaller atoms''.} \cite{1931Natur.127..706L}. At first, this was not a popular assumption, with even Eddington, the most influential proponent of Lema\^itre's earlier work on expanding cosmologies \cite{1930MNRAS..90..668E} (and under whom Lema\^itre had worked in Cambridge), finding the basic notion ``repugnant'' \cite{IBSUSJ} --- despite his own reflections on the nature of time and the ``end of the world'' having provided the seed for Lema\^itre's original argument\cite{1931Natur.127..447E}.

This hypothesis, which came to be known as the Big Bang theory, would remain an important bone of contention for decades. Although initially viewed with cautious suspicion (which Lema\^itre's status as a Catholic priest did little to dispel), the idea survived through the 1930's proliferation of alternative cosmologies \cite{1937RSPSA.158..324M,1935rgws.book.....M,DIRAC_CONST,1934PNAS...20..169T,1934rtc..book.....T,1929PNAS...15..773Z} and eventually found its most persistent contender in Fred Hoyle's post-war (1948) steady state theory \cite{1948MNRAS.108..372H} (which assumed particle creation made up for the then-established expansion so that the Universe would look the same at any given time). 

It was also around that time that Friedmann's former student, George Gamow, started working on cosmology on the side of the Big Bang theory \cite{GamowEncyclopaediaBritannica2017}. Drawing from his background in nuclear physics, Gamow investigated how chemical elements could be produced in a hot expanding Universe. On the 1st of April of 1948, Gamow and his student Ralph Alpher (famously, Hans Bethe's name was added as an author to the paper for its comic potential) showed how light elements could be produced by neutron capture in the early Universe \cite{abc}, effectively introducing the field of Big Bang nucleosynthesis. Although Gamow's and Alpher's original vision of all elements being produced in the hot early Universe \cite{1948PhRv...74.1577A} ended up being refuted by Hoyle's (later in collaboration with William Fowler and Margaret and George Burbidge) work on stellar nucleosynthesis \cite{1946MNRAS.106..343H,1954ApJS....1..121H,1957RvMP...29..547B}, the success of this approach at explaining the observed overabundance of helium (compared to what would be expected from just stellar nucleosynthesis) became an important argument in favour of the Big Bang.

The last conclusive piece of evidence that definitely established the Big Bang as the dominant paradigm was the discovery of the cosmic microwave background (CMB) --- just two years before Lema\^itre's death. The CMB is a thermal black body spectrum left over from the time of recombination (when the first hydrogen atoms formed), which was initially predicted by Ralph Alpher and Robert Herman in 1948 \cite{1948Natur.162..774A}. This prediction was however only briefly mentioned in Alpher and Herman's paper and was thus easily overlooked\footnote{The main focus of the paper was to correct the calculations in a previous Gamow paper on Big Bang nucleosynthesis \cite{1946PhRv...70..572G}. The reference to the CMB is simply a one-sentence statement of the ``temperature of the gas'' and no suggestion of its relevance as a potentially observable signal is made.}. Having been independently repredicted by Robert Dicke and Jim Peebles in the early 1960s, the CMB was first presented as a detectable relic by Andrei Doroshkevich and Igor Novikov in 1964 \cite{1964SPhD....9..111D}. Later that same year, Robert Dicke, David Wilkinson, and Peter Roll set out to measure the CMB; just late enough for Arno Penzias and Robert Wilson to inadvertedly do it first\footnote{Interestingly, some have argued that this was actually only the second ``accidental'' discovery of the CMB. As early as 1940, Andrew McKellar proposed that then-unidentified spectral lines of interstellar origin could indicate the presence of cyanogen (CN) and methyne (CH); and from the analysis of those spectral lines concluded that the effective temperature of interstellar space ``must be extremely low'', going as far as giving an estimate of $2.7\mathrm{K}$ as an upper limit for this temperature. However, it seems unlikely that at the time he thought of this as related to anything like the CMB, having then remarked that ``the concept of such a temperature in a region with so low a density of both matter and radiation'' may have no meaning \cite{1940PASP...52..187M}.} (for which they would receive part of the 1978 Nobel Prize in Physics) \cite{Penzias_NL}. While alternative theories for a ``more local'' origin of the CMB \cite{1967Natur.216...43N} meant this new detection did not change opinions on the issue of the Big Bang as quickly as Hubble's observations did on the issue of expansion, the following decade was marked by a steady shift in opinion as new measurements confirmed the black body nature of the observed spectrum \cite{1991Natur.352..769P}.

Nowadays, Big Bang cosmology is almost universally accepted as the standard framework for dealing with the early Universe --- with the important caveat that the ``initial state'' is taken to be slightly later than the actual singularity\footnote{Lest general relativity (and all known physics) not apply.}, thus for the time being keeping the question of whether the Universe had a beginning outside of the realm of scientific enquiry.

\subsection{Growing a simple Universe}

The first and most basic postulate on which standard cosmology is based is the so-called cosmological principle: the assumption that the Universe is statistically homogeneous and isotropic over its spatial dimensions. It follows naturally from the notion that Earth-dwellers are not privileged observers and thus, were we to be at any other point in the Universe, we should expect our observations to be statistically indistinguishable from the ones we have access to\footnote{Of course, although both Occam's razor and historical caution suggest this is a reasonable assumption to make, the fact that we are mostly confined to the Earth means that we are extremely limited in our ability to test this hypothesis. Therefore, it can be argued that this assumption originally arose more from necessity of mathematical simplicity than from any sort of real understanding of the cosmos \cite{pittphilsci9062} --- although later tests of violations of this principle in the context of general relativity have provided no conclusive evidence against it \cite{Saadeh:2016sak,Bengaly:2016amk,Clarkson:1999zq}.}.

At large scales, the most general metric which is consistent with the cosmological principle is the Friedmann-Lema\^itre-Robertson-Walker (FLRW) metric \cite{Hitchin413}, defined by the line element
\begin{equation}
ds^2=dt^2-a^2\left(t\right)\left[\frac{d r^2}{1-Ka^2\left(t\right) r^2}+ r^2 d\theta^2+ r^2 \sin^2\left(\theta\right)d\phi^2\right],\label{FLRW}
\end{equation}
where $t$ is cosmic time; $r$ is a radial coordinate; $\theta$ and $\phi$ are angular coordinates; $a\left(t\right)$ is (for now) a free function called the scale factor; and $K$ can be negative, null, or positive, depending on whether the Universe is closed, flat, or open, respectively. A correspondingly general energy-momentum tensor for a homogenous and isotropic perfect fluid takes the simple form $T^{\mu}_{\nu}=\mathrm{diag}\left(\rho\left(t\right),P\left(t\right),P\left(t\right),P\left(t\right)\right)$, where $\rho$ is the fluid energy density and $P$ is its pressure. The Einstein field equations for this metric and energy-momentum tensor are called the Friedmann equations and can be expressed as
\begin{equation}
H^2\equiv\left(\frac{\dot{a}}{a}\right)^2=\frac{\rho+\Lambda}{3}-\frac{K}{a^2}\label{Friedmann1}
\end{equation}
and
\begin{equation}
\dot{\rho}=-3H\left(\rho+P\right),\label{Friedmann_conservation}
\end{equation}
where $H$ is the Hubble parameter, $\Lambda$ is the cosmological constant, and a dot denotes differentiation with respect to cosmic time. Eq.~\eqref{Friedmann_conservation} is equivalent to the requirement that the energy-momentum tensor be covariantly conserved. Eq.~\eqref{Friedmann1} is often written using the alternative notation
\begin{equation}
1=\Omega+\Omega_k+\Omega_\Lambda\label{Omegas}
\end{equation}
where, making the formal identification between the terms with $\Lambda$ and $K$ and energy densities (so that $\rho_\Lambda\equiv\Lambda$ and $\rho_K\equiv -3Ka^{-2}$), $\Omega_i$ are density parameters defined so that $\Omega_i=\rho_i/\rho_c$ ($\rho_c\equiv 3H^2$ being the critical density, which would correspond to an exactly flat Universe if $\Lambda=0$).

Of particular interest is the case of a barotropic fluid with a constant equation of state $w\equiv P/\rho$, as the contents of the cosmic fluid can usually be taken to be in the form of either radiation (with $w=1/3$) or matter (often also called dust; with $w=0$\footnote{Naturally, realistically most dust will not have exactly null pressure. Regardless, as long as the peculiar velocities of its constitutent particles are non-relativistic, this should be a good approximation \cite{Lyth:2009zz}.}) and the cosmological constant can be interpreted as a component with $w=-1$. For this important special case, Eq.~\eqref{Friedmann_conservation} imposes a simple scaling of the energy density with the scale factor,
\begin{equation}
\rho\propto a^{-3\left(1+w\right)}.\label{w_rho_a}
\end{equation}
Note that for matter ($w=0$) this gives the result one would expect simply from the assumption that the total energy in a comoving volume remains constant as said volume increases by $a^3$. Conversely, the result for radiation ($w=1/3$) corresponds to an added loss of $a^{-1}$ due to radiation being redshifted as $a$ increases, on top of the aforementioned dilution effect.

Writing $\rho=\rho_m+\rho_r$ (with $\rho_m$ corresponding to the energy density in matter and $\rho_r$ to the energy density in radiation) and assuming that the two components do not exchange energy\footnote{Which is usually a fair approximation, although realistically there will be times when it will fail; i.e., when a component of initially ultra-relativistic massive particles (which behave like radiation to a good approximation) slows down sufficiently (behaving like pressureless matter in the non-relativistic limit).}, Eq.~\eqref{Friedmann1} becomes
\begin{equation}
H^2=\frac{\rho_{r0}}{3}a^{-4}+\frac{\rho_{m0}}{3}a^{-3}-Ka^{-2}+\frac{\Lambda}{3},\label{Friedmann_scale}
\end{equation}
where the subscript $0$ indicates a quantity evaluated at the time when $a=1$. Equivalently, Eq.~\eqref{Omegas} can be written as
\begin{equation}
\left(\frac{H}{H_0}\right)^2=\Omega_{r0}a^{-4}+\Omega_{m0}a^{-3}+\Omega_{k0}a^{-2}+\Omega_{\Lambda 0},\label{Friedmann_omega}
\end{equation}
where the scale factor is normalised so that $a=1$ today.

Some important epochs in cosmic history can be well described by the simple limit in which the right-hand sides of Eq.~\eqref{Friedmann_scale} and Eq.~\eqref{Friedmann_omega} are dominated by just one term. When radiation is the dominant component
\begin{equation}
a\left(t\right)\propto t^{1/2},\label{arad}
\end{equation}
whereas if the matter term prevails then
\begin{equation}
a\left(t\right)\propto t^{2/3}.\label{amat}
\end{equation}
If instead the curvature term eclipses the rest\footnote{This can only happen if the curvature is non-positive since $H$ must always be real. If the curvature is positive and the Universe is expanding then once $H=0$ the Universe will start contracting.\label{fn:kbounce}} then
\begin{equation}
a\left(t\right)\propto t.\label{ak}
\end{equation}
Finally, if the $\Lambda/3$ term takes over (as must eventually happen in a strictly expanding\footnote{If $K=1$ and $\Lambda$ is sufficiently small, it is possible for $H$ to be forced to change to a negative sign before the other terms become smaller.} Universe), then
\begin{equation}
a\left(t\right)\propto \exp{\left(\pm\sqrt{\frac{\Lambda}{3}}t\right)},\label{aLambda}
\end{equation}
where the sign will depend on whether the Universe is ``initially'' contracting or expanding and $\Lambda$ has to be positive if it dominates (see footnote \ref{fn:kbounce}).

\subsection{Recombination, the CMB, and the primordial perturbation}

When it was very young, the Universe was too energetic for atoms to exist; instead, there was a hot plasma in which light atomic nuclei (mostly hydrogen and helium) and electrons were tightly coupled: with each other due to electromagnetic interactions, and with photons due to the high cross section of Compton scattering in such an electron-rich medium. As a result, the photon mean free path was then too short for there to be an appreciable chance of light from that time to ever reach us. Only after the epoch of recombination, when the Universe became cool enough for charged particles to form atoms, did photons decouple from the rest of the cosmic fluid (then transitioning to the form of a neutral gas) and become able to propagate freely so that they could reach us today \cite{Lyth:2009zz}. These are thus the oldest photons we can see in the Universe, and constitute what is known as the cosmic microwave background (CMB) --- originally, they would be at a much higher frequency, consistent with a correspondingly hot thermal spectrum, but have since been severely redshifted by cosmic expansion.

An interesting use of the CMB is the study of its anisotropies. If the Universe were exactly homogeneous and isotropic over all scales (which there is no good reason to expect) then the CMB would also be an isotropic thermal spectrum, corresponding to the same temperature in every direction. Since we do observe small angular variations in the CMB, we can trace back their evolution using cosmological perturbation theory. Thus the observed CMB anisotropies can be related to a primordial perturbation which had to be present in the ``initial state'' of Big Bang cosmology. Moreover, this calculation assumes knowledge of the evolution of the scale factor between this ``initial state'' and today, making the CMB an important probe of the consistency of the accepted cosmological model.

This primordial perturbation is most conveniently characterised in terms of the departure of the actual metric from (flat) FLRW in a physically meaningful gauge. In the uniform density gauge (in which $\rho$ is merely a function of time), we consider linearly perturbed metrics whose line element can be written as
\begin{equation}
ds^2=dt^2-a^2\left(t\right)\left(1+2\zeta\right)\left(\delta_{ij}+2h_{ij}\right)d\mathrm{x}^i d\mathrm{x}^j,\label{perturbg}
\end{equation}
where $\zeta$ is the scalar perturbation and the tensor perturbation, $h_{ij}$, is transverse ($\partial_i h_{ij}=0$) and trace-free ($h^i_i=0$). In general, there could also be a vector perturbation, but we neglect such a component because it is expected to decay in an expanding Universe \cite{EU_lectnotes}.

Usually, these primordial perturbations are assumed to be well described by small Gaussian stochastic fluctuations defined by their $2$-point correlators,
\begin{equation}
\langle \zeta_\mathbf{k}\zeta_\mathbf{k^\prime} \rangle \equiv \frac{2\pi^2}{k^3}\mathcal{P}\left(k\right)\delta^3_{\mathbf{k}+\mathbf{k^\prime}},\label{power_spectrum}
\end{equation}
\begin{equation}
16\langle h_{\left(+\right)\mathbf{k}}h_{\left(+\right)\mathbf{k^\prime}} \rangle = 16\langle h_{\left(-\right)\mathbf{k}}h_{\left(-\right)\mathbf{k^\prime}} \rangle \equiv \frac{2\pi^2}{k^3}\mathcal{P}_T\left(k\right)\delta^3_{\mathbf{k}+\mathbf{k^\prime}};\label{power_spectrum_T}
\end{equation}
where $\langle Q \rangle$ is the expected value of the stochastic quantity $Q$; the subscripts $\mathbf{k}$ and $\mathbf{k^\prime}$ indicate a Fourier transform (defined ``symmetrically'', with a $\left(2\pi\right)^{-3/2}$ factor); $\delta^3_{\mathbf{k}+\mathbf{k^\prime}}$ is a 3-dimensional Dirac delta ``function'' in Fourier space; $h_{\left(+\right)}$ and $h_{\left(-\right)}$ are the two independent components of $h_{ij}$; $\mathcal{P}$ is called the scalar power spectrum; and $\mathcal{P}_T$ is called the tensor power spectrum. The assumption of Gaussianity means that higher-order correlators can be straightforwardly computed if the power spectra are known (all odd ones just vanishing). Often constraints on $\mathcal{P}_T$ are expressed as constraints on the tensor-to-scalar ratio,
\begin{equation}
r\equiv\frac{\mathcal{P}_T}{\mathcal{P}}.\label{rdef}
\end{equation}

\subsection{Observing the primordial perturbation today}

The simplest way to learn about the primordial perturbation is via the study of CMB anisotropies. Thanks to the scales probed by CMB observations being so large, the evolution of the perturbation since its generation until recombination (when the CMB is formed) can be accurately modelled using linear perturbation theory at these scales. In recent decades, the characterisation and study of these anisotropies has been a major goal of observational cosmology. Notably, NASA's COBE (1989-1993) and WMAP (2001-2010), as well as ESA's \emph{Planck} satellite (2009-2013), have collected remarkably precise CMB data which has crucially contributed to progress in this area. 

Scalar perturbations are seen in the CMB as temperature fluctuations. When seen from the Earth, these fluctuations are most conveniently decomposed in spherical harmonics according to
\begin{equation}
\frac{\delta T}{T}\left(\mathbf{e}\right)=\sum_{\ell m}a_{\ell m}Y_{\ell m}\left(\mathbf{e}\right),\label{alm}
\end{equation}
where $\delta T/T$ is the observed relative temperature fluctuation, $Y_{\ell m}$ are spherical harmonics, $\mathbf{e}$ is the unit vector corresponding to the direction of an incident CMB photon, and $a_{\ell m}$ are the observed CMB multipoles. Note that the Earth's motion relative to the CMB rest frame affects the values of the multipoles, but only noticeably for $\ell=1$ (for which it dominates) and $\ell=2$ (for which it contributes at about 10\% of the observed value) \cite{Lyth:2009zz}.

The stochastic properties of these fluctuations are usually described in terms of $2$-point correlators, which due to (stochastic) rotation invariance must be of the form
\begin{equation}
\langle a_{\ell m} a^*_{\ell^\prime m^\prime}\rangle=\delta_{\ell\ell^\prime}\delta_{m m^\prime} C_\ell,\label{Cl}
\end{equation}
where $C_\ell$ is called the spectrum of the CMB anisotropy and is completely determined if the three-dimensional scalar power spectrum (which evolved from the primordial perturbation) is known.

More information can be gathered if one has access to the polarisation of CMB photons, as we do since WMAP and \emph{Planck} \cite{Adam:2015rua,2013ApJS..208...20B}. Of particular interest to us is the possibility of B-mode polarisation due to the primordial tensor power spectrum being detected --- although current B-mode observations only allow upper limits to be set on the scale of this spectrum\footnote{$r<0.12$ being the current $2\sigma$ limit at the \emph{Planck} pivot scale of $k=0.05\mathrm{Mpc^{-1}}$ \cite{BKPanalysis}.}.

Even more information (on different scales and different parameters) can be obtained by a number of alternative methods. In this thesis we are interested in two promising examples in particular: spectral distortions and 21cm observations.

Spectral distortions are deviations in the CMB from a perfect black-body spectrum. Despite no such distortions having been measured to date (making the CMB spectrum the most perfect black-body spectrum ever to be measured in Nature \cite{White:1999nh}), they are expected to be caused by several out-of-equilibrium processes in the early Universe, including a few, like recombination, which are assumed in the current standard model of cosmology \cite{Chluba:2014sma}. The most well-studied types of spectral distortions are: $\mu-$distortions, characterised by the introduction of a chemical potential (typically generated at very early epochs when Compton scattering by free electrons is still common); and $y-$distortions, characterised by a non-constant ``temperature'' which increases with frequency (typically generated at later epochs when Compton scattering is inefficient) \cite{Chluba:2011hw}. A detection of these types of distortions would constrain an integrated version of the power spectrum for scales in the range $1{\rm Mpc}^{-1}\lesssim k\lesssim 10^4{\rm Mpc}^{-1}$\footnote{For comparison, CMB observations constrain the power spectrum most strongly for scales in the range $10^{-3}{\rm Mpc}^{-1}\lesssim k\lesssim 0.3{\rm Mpc}^{-1}$ \cite{Planck_inf2015}.}, but a more complicated type of distortion could provide even more valuable scale-dependent information \cite{Chluba:2013vsa}.

21cm observations provide an observational window into the so-called dark ages, the time between recombination and the formation of the first stars. During this time, the Universe is expected to have been filled with neutral hydrogen which must have resonantly absorbed CMB photons of wavelength $\sim$21cm, corresponding to the hydrogen atom's spin-flip transition \cite{Loeb:2003ya}. Looking for the resulting absorption signal in the CMB thus allows the inhomogeneities in the hydrogen gas distribution during this time to be probed, providing information about the power spectrum down to scales as small as $k\sim 100\mathrm{Mpc^{-1}}$.

\subsection{$\Lambda$CDM}

The $\Lambda$CDM model is the simplest known model which accounts for currently available cosmological observations \cite{Ade:2015xua,Lyth:2009zz}. It assumes a flat\footnote{Current constraints by the \emph{Planck} collaboration give $\left|\Omega_k\right|<0.005$ to a $2\sigma$ accuracy \cite{Ade:2015xua}.} FLRW cosmology with a non-null cosmological constant in which the Universe is filled with radiation as well as non-relativistic matter in two forms: ordinary (baryonic) matter and cold dark matter (CDM). Dark matter is a hypothetical type of matter which interacts with known particles primarily (possibly exclusively) through gravity on large scales. Its existence is inferred from a number of observed gravitational phenomena\footnote{Notably, discrepancies in galaxy rotation curves relative to what would be expected if all the mass in galaxies were currently visible to us \cite{2012arXiv1201.3942P}.} which, in the framework of Einstein gravity, cannot be explained without it\footnote{Naturally there are those who worry about the potentially unsurmountable obstacles in the way of detection of dark matter particles \cite{2012arXiv1201.3942P,Undagoitia:2015gya} and find it more scientific to attribute such phenomena to manifestations of a more general theory of gravity \cite{Nojiri:2017ncd}. At the time of writing, however, there is no known theory of gravity which can fit the available data as well as the assumption of dark matter.}. Its ``coldness''  refers to this dark matter being assumed to be non-relativistic (i.e., behaving like pressureless dust).

Additionally, a small nearly-Gaussian scalar perturbation is assumed and its power spectrum\footnote{In general, a tensor perturbation with a corresponding tensor power spectrum is also expected. Here we shall omit it as its observational consequences are yet to be detected --- thus ``power spectrum'' will henceforth be taken to mean the scalar one unless explicitly stated otherwise.} is parameterised by
\begin{equation}
\mathcal{P}\left(k\right)\equiv \mathcal{A}_{s} \left(\frac{k}{k_0}\right)^{n_s\left(k\right)-1},\label{Asns}
\end{equation}
where $n_s$ is called the spectral index and $\mathcal{A}_{s}$ is the spectral amplitude at the pivot scale $k_0$. Current observations are consistent with a constant spectral index, therefore $\Lambda$CDM need only assume two numbers to characterise primordial fluctuations: $\mathcal{A}_s$ and $n_s$ at a pivot scale. In general, however, the spectral index is not necessarily constant and its variations with scale are still allowed to be relatively large (a possibility which provides the motivation for chapter \ref{chap:SFSRI}) \cite{Planck_inf2015}.

In the end, the $\Lambda$CDM model requires the specification of six parameters to fully determine the evolution of the Universe, for example: the Hubble parameter today, $H_0$ (often specified by $h\equiv H_0/100\mathrm{km\, s^{-1}Mpc^{-1}}$); the present values of the density parameters for baryonic matter ($\Omega_{B0}$), cold dark matter ($\Omega_{c0}$), and the cosmological constant ($\Omega_{\Lambda 0}$); as well the spectral amplitude ($\mathcal{A}_s$) and the spectral index ($n_s$) of the primordial perturbation. Note that the present value of the density parameter for radiation ($\Omega_{r0}$) can be computed from Eq.~\eqref{Omegas} if the rest are known. However, this parameter can be more accurately determined to be $\Omega_{r0}=2.47\times 10^{-5}h^{-2}$ directly from the observed CMB temperature $T=2.7255\pm 0.0006\ \mathrm{K}$ \cite{PDG_review} (neglecting contributions from ultra-relativistic neutrinos). Given this remarkable precision, it is often not varied when fitting other parameters, effectively bringing the number of independent parameters in the model down by one --- i.e., making it so that only two of the density parameters are needed.

In practice, an additional parameter is needed to make up for our lack of understanding of the processes that lead to the reionization of the cosmic medium following the formation of early objects: the optical depth, $\tau$, defined so that $e^{-\tau}$ is the probability that a photon emitted between recombination and reionization is scattered \cite{Lyth:2009zz}.
Current observational constraints from the \emph{Planck} collaboration for these parameters indicate $h=0.6774\pm 0.0046$\footnote{There seems to be a tension with more direct estimations of the Hubble parameter which prefer slightly higher values (with $h\approx 0.70$) \cite{Ade:2015xua}.}, $\Omega_{B0}h^2=0.02230\pm 0.00014$, $\Omega_{c0}h^2=0.1188\pm 0.0010$, $\Omega_{\Lambda 0}=0.6911\pm 0.0062$, $\ln\left(10^{10}\mathcal{A}_s\right)=3.064\pm 0.023$ (for the \emph{Planck} pivot scale of $0.05\mathrm{Mpc^{-1}}$), $n_s=0.9667\pm 0.0040$, and $\tau=0.066\pm 0.012$.

According to this picture, the Universe was initially dominated by hot radiation which was quickly overcome by the matter component due to its slower scaling with the scale factor. The matter component then dominated the Universe during most of its history, until recent times when the cosmological constant is starting to become the main driver of cosmic expansion.

\begin{figure}[h]
\includegraphics[scale=0.6]{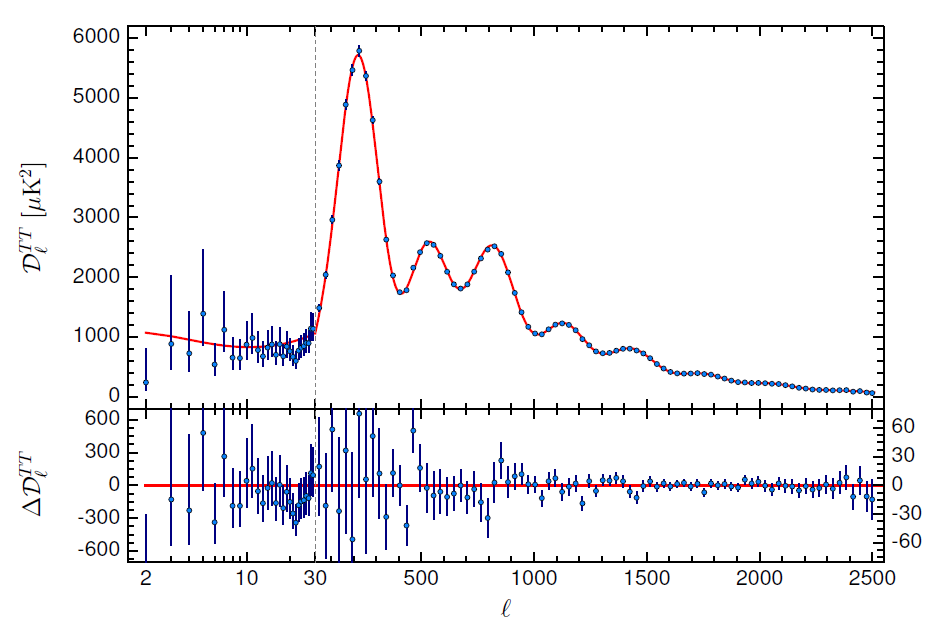}
\caption{\emph{Planck} 2015 temperature power spectrum rescaled by $\mathcal{D}_\ell^{TT}\equiv\ell\left(\ell+1\right)C_\ell/\left(2\pi\right)$ (blue with $1\sigma$ error bars) against the best-fit base $\Lambda$CDM theoretical spectrum (red line) --- image from Ade \emph{et al} \cite{Ade:2015xua}. The bottom plot shows residuals relative to $\Lambda$CDM and changes scale (to the one on the right-hand side) for $\ell>30$.\label{fig:cmbtt}}
\end{figure}

One of the most iconic successes of $\Lambda$CDM is the remarkable accuracy to which the CMB temperature spectrum predicted by the model's best fit to data approximates observations (see figure \ref{fig:cmbtt}). Although this relatively simple model is not free of problems, this kind of observational success makes it the most widely accepted model of our Universe at the largest scales.

\subsection{Trouble on the horizon}
\label{problems}

One of the main shortcomings of the standard picture we have presented is arguably the lack of justification for our assumed ``initial state''. While there are other well-known problems with it, particularly when small scales are considered \cite{DelPopolo:2016emo}, those can usually be blamed on computational difficulties when modelling complex phenomena, unknown features of more fundamental physical theories (e.g., quantum gravity), or fairly minor modifications of $\Lambda$CDM (e.g., adding a new component, but keeping the basic framework). This, however, raises critical questions about the theoretical and epistemic foundations of Big Bang cosmology. To illustrate these, we focus on three specific problems: the flatness problem, the horizon problem, and the relic problem.

\subsubsection{The flatness problem}

The Universe we observe is spatially flat as far as any current observation can tell. An immediate consequence of this fact is that the density of the Universe must be very close to the critical value $\rho_c=3H^2$. However, $\rho=\rho_c$ is known to be an unstable equilibrium point for the Friedmann equations under normal circumstances. This can be easily seen by neglecting $\Lambda$ (which in $\Lambda$CDM is a fair approximation during most of cosmic history) and rewriting Eq.~\eqref{Friedmann1} as
\begin{equation}
\left(1-\Omega^{-1}\right)\rho a^2=3K=const.\label{flatproblem}
\end{equation}
Since, as we have seen, $\rho$ is usually expected to decrease much faster than $a^{-2}$, $\Omega$ must evolve away from its critical value ($\Omega=1$) correspondingly fast as long as $\Omega\neq 1$. In particular, the condition that $\left|1-\Omega\right|=\left|\Omega_k\right|<0.005$ today (imposed by \emph{Planck} constraints \cite{Ade:2015xua}) leads to the expectation that $\left|\Omega_k\right|\lesssim 10^{-58}$ at the Planck time (a popular time to choose for the ``initial condition'' due to quantum gravity becoming important then) \cite{VSH}. The flatness problem refers to the lack of a natural explanation for such stringent fine-tuning in the ``initial state'' of Big Bang cosmology.

\subsubsection{The horizon problem}

A well-known lesson of relativistic physics is that no information can travel faster than the speed of light in vacuum. A consequence of this in the context of a finite-age infinite Universe is that, at any given time, sufficiently distant points will be causally disconnected (i.e., unable to have influenced each other as light has not had time to travel between them). The maximum distance that light can have travelled since the ``Big Bang'', when $a=0$, is called the particle horizon and can be written as
\begin{equation}
d_H\left(t\right)=a\left(t\right)\intop_0^t\frac{dt^\prime}{a\left(t^\prime\right)}.\label{phor}
\end{equation}

If $a\left(t\right)\propto t^\lambda$, as is the case when the Universe is dominated by either matter ($\lambda=2/3$) or radiation ($\lambda=1/2$), then (for $\lambda\neq 1$)
\begin{equation}
d_H\left(t\right)=\frac{t}{1-\lambda}=\frac{\lambda}{1-\lambda}H^{-1},\label{phorl}
\end{equation}
meaning that for most of cosmic history $d_H$ is well approximated (up to a multiplicative factor of order unity) by $H^{-1}$, which is simply called the horizon (and is usually more useful to estimate the distance over which two points may interact at a given time before being ``cut off'' by cosmic expansion).

The horizon problem is the realization that the observable Universe contains regions which appear to be correlated (due to the observed homogeneity/isotropy) even though they should be causally disconnected. For example, the smallness of CMB anisotropies indicates that the entire observable Universe was at the same temperature at the time of recombination; yet CMB photons coming from regions separated by more than a few degrees in the sky should have been outside of each other's horizons when they were emitted \cite{Lyth:2009zz}.

\subsubsection{The relic problem}

Tracing back the evolution of the Universe requires knowledge of particle physics at high energies and, conversely, cosmological observations probing the early Universe provide a wealth of information about particle physics. A classic example of the latter is neutrino physics, with CMB observations being sufficient to strongly constrain the effective number of neutrino species ($N_\mathrm{eff}=2.94\pm 0.38$) and the sum of neutrino masses ($\Sigma m_\nu<0.589$ to $2\sigma$ accuracy) \cite{Ade:2015xua}.

If the ``initial state'' of Big Bang cosmology were taken to be shortly after the Planck time (corresponding to temperatures around $T\sim 10^{19}\mathrm{GeV}$), then the Universe must have gone through a number of important phase transitions as it cooled. Some of these transitions are well understood from the points of view of both cosmology and particle physics: like recombination ($T\sim 1\mathrm{eV}$) and nucleosynthesis ($T\sim 1\mathrm{MeV}$).
At high enough temperatures, however,
knowledge of particle physics beyond the standard model is required and thus we don't even know what transitions to expect.

Broad classes of possible phase transitions at very high energies predict the formation of so-called exotic relics: exotic particles; small black-holes; and topological defects, which correspond to non-trivial field solutions arising due to topological considerations in symmetry-breaking transitions (they are typically classified as monopoles, strings\footnote{Chapter \ref{chap:wiggly} is specially dedicated to the study of cosmic strings.}, walls, and textures, depending on whether they are effectively point-like, line-like, membrane-like, or three-dimensional, respectively). In particular, grand unified theories (GUTs) seem to very generally lead to the overproduction of monopoles.

The relic problem is that it seems to be difficult to write down a well-motivated theory of particle physics at high energies which does not lead to the production of large quantities of these relics which should have persisted until today despite none having ever been observed \cite{VSH}.

\section{Cosmic inflation}

The problems with the standard Big Bang cosmology described in subsection \ref{problems} can be thought of as essentially fine-tuning problems. Some might (and do) argue that such fine-tuning problems are not so much an indication of something wrong with the theory as a sign of theorists struggling to make physical sense of unphysical models. A typical argument in this sort of epistemological tradition might say that a physical model is successful merely if it can reproduce available observations, and that it is thus moot to ask questions about what it would have to say about alternative imaginary realities\footnote{An interesting review of alternative arguments and counter-arguments regarding the validity of these sorts of fine-tuning concerns can be found in \cite{sep-fine-tuning}.}.

In the end, the importance of these problems is undeniable because of the fact that Big Bang cosmology is intrinsically incomplete, in the sense that its foundations must break down at sufficiently early times. In order to understand what happened before this time, a new and more fundamental theory of high energy physics will be needed. From this point of view, these problems with the ``initial state'' of Big Bang cosmology are of the utmost importance not because they raise doubts about the successes of the theory in the regimes where its validity is well-established but because they provide important constraints for any theory seeking to explain cosmology before this point.

We now briefly review the paradigm of cosmic inflation, which is currently the most popular solution to these problems.

\subsection{A natural solution}

Let us first focus on the flatness problem as stated in terms of the dynamical repeller nature of the $\Omega=1$ solution to Eq.~\eqref{flatproblem}. If we momentarily abstract from the physical meaning of $\rho$, we are allowed to ask under which conditions the critical solution could become an attractor instead. As it turns out, this is when
\begin{equation}
\frac{d}{dt}\left(\rho a^2\right)>0\Rightarrow \frac{P}{\rho}\equiv w<-\frac{1}{3},\label{solveflat}
\end{equation}
where the second inequality is derived assuming only Eq.~\eqref{Friedmann_conservation} and $H>0$ (in a contracting Universe there would be no flatness problem).

Interestingly, if we assume that the flatness problem is solved by the Universe being dominated by some exotic component obeying Eq.~\eqref{solveflat} during early times, we find that while this component dominates and the Universe can be assumed to be flat we have $\ddot{a}>0$ or, equivalently,
\begin{equation}
\frac{d}{dt}\left(aH\right)^{-1}<0;\label{horinf}
\end{equation}
i.e., the comoving radius of the horizon shrinks, meaning that the horizon expands more slowly than the cosmic fluid. In practice, this will make it so that the particle horizon during (and, consequently, after) this epoch is larger than it appears --- and thus this tentative solution to the flatness problem can potentially solve the horizon problem as well.

The requirement that the scale factor be accelerating (or, equivalently, Eq.~\eqref{horinf}) is usually taken as defining an epoch of inflation\footnote{So technically an epoch of $\Lambda$ domination is also an inflationary phase, although many authors reserve the term for hypothetical eras taking place before the ``initial state'' of Big Bang cosmology.}. Whether it can really solve the flatness and horizon problems essentially boils down to whether it can last long enough for $\Omega_k$ to be naturally negligible today and for today's observable Universe to have been inside the horizon before inflation. Typical inflation models can, usually lasting long enough for the scale factor to increase by at least around $60$ e-folds (i.e., by a multiplicative factor of $\gtrsim e^{60}$) \cite{Lyth:2009zz}. Such a dramatic amount of expansion also provides a natural solution to the relic problem: even if copious amounts of relics were produced before inflation, this rapid expansion would have diluted them so much that the likelihood of coming across one is very low.

It should be noted that inflation is not the only known way to solve these problems of the standard model. An equally natural solution (at least before specific realisations are considered) is given by ekpyrosis, inflation's old rival theory \cite{Buchbinder:2007ad,2001PhRvD..64l3522K}. Assuming that before the ``Big Bang'' the Universe was contracting and contained an exotic component with $w\gg 1$, this component should come to dominate the Universe (due to Eq.~\eqref{w_rho_a}) --- solving the flatness and relic problems due to the contributions of curvature and relics to cosmic evolution becoming negligible\footnote{Note that our corresponding arguments for inflation could have been phrased in this manner too, as we can see from Eq.~\eqref{w_rho_a} that $w<-1/3$ corresponds to the condition that the energy density of the component sourcing inflation decays with the scale factor slower than $\Omega_k$ (and thus than any matter-like component).}. Given enough time to contract (before the ``bounce'' which must eventually happen if the standard picture is to be recovered), the horizon problem is solved in the same way as in inflation, since during this epoch
\begin{equation}
\frac{d}{dt}\left|aH\right|^{-1}<0.\label{ekpyrhor}
\end{equation}

\subsection{Implementing inflation}

Currently, there are myriads of different known models of inflation \cite{Martin:2013tda,Martin2014}. Without modifying general relativity \cite{Bamba:2015uma}, the most natural way to implement inflation is by postulating that at early times the energy density of the Universe is dominated by some exotic component which obeys Eq.~\eqref{solveflat}. If, for the sake of simplicity, this component is assumed to be made up of a single uniform scalar field\footnote{Inflation models which can be written down in this way are known as (canonical) single-field inflation models, as opposed to multi-field inflation models in which more than one field is involved. Note that, while inflaton fields are usually assumed to be scalar (due to difficulties in obtaining a homogeneous and isotropic Universe if they are not \cite{Golovnev:2008cf}), in chapter \ref{chap:NAT} we focus on a model in which inflation is driven by fermions.} (usually called the inflaton) with the canonical Lagrangian density
\begin{equation}
\mathcal{L}=-\frac{1}{2}\partial^\mu\phi\partial_\mu\phi-V\left(\phi\right),\label{SFI}
\end{equation}
then its energy density and pressure are given by
\begin{equation}
\rho=\frac{1}{2}\dot{\phi}^2+V\left(\phi\right),\ P=\frac{1}{2}\dot{\phi}^2-V\left(\phi\right);\label{SFI_rhoP}
\end{equation}
the Friedmann equations (Eqs.~\eqref{Friedmann1} and \eqref{Friedmann_conservation}, respectively) now being expressable (neglecting $\Lambda$) as
\begin{equation}
3H^2=\frac{1}{2}\dot{\phi}^2+V\left(\phi\right)\label{F1phi}
\end{equation}
and
\begin{equation}
\ddot{\phi}+3H\dot{\phi}+V^\prime\left(\phi\right)=0,\label{F2phi}
\end{equation}
where prime denotes differentiation with respect to $\phi$ (and Eq.~\eqref{F2phi} is just the equation of motion associated with the Lagrangian in Eq.~\eqref{SFI}).
Imposing Eq.~\eqref{solveflat} thus leads to the condition that inflation will take place if and only if
\begin{equation}
\epsilon\equiv -\frac{\dot{H}}{H^2}\equiv\frac{1}{2}\frac{\dot{\phi}^2}{H^2}<1,\label{epsinf}
\end{equation}
where the two ratios on the left-hand side can be seen to coincide using the relation 
\begin{equation}
\dot{H}=-\frac{1}{2}\dot{\phi}^2,\label{F2phiH}
\end{equation}
which follows from Eqs.~\eqref{F1phi} and \eqref{F2phi}.

Of special interest is the case of slow-roll inflation, defined by the homonymous approximation\footnote{With a few exceptions, most inflation models require the slow-roll approximation to hold most of the time in order to inflate for long enough.}
\begin{equation}
\epsilon\ll 1,\ \left|\delta_1\right|\equiv\left|\frac{\ddot{\phi}}{H\dot{\phi}}\right|\ll 1;\label{slowroll}
\end{equation}
due to which $\epsilon$ and $\delta_1$ are known as the slow-roll parameters\footnote{$\delta_1$, in particular, has a few mostly equivalent definitions, which are often called $\eta$ in the literature.}.

With this approximation, the system of equations to be solved is significantly simplified. Eqs.~\eqref{F1phi} and \eqref{F2phi} can now be written as the slow-roll equations:
\begin{equation}
3H^2\simeq V\left(\phi\right)\label{SR1}
\end{equation}
and
\begin{equation}
3H\dot{\phi}\simeq -V^\prime\left(\phi\right);\label{SR2}
\end{equation}
which can be solved in a more straightforward manner (even if not necessarily easily for all choices of $V\left(\phi\right)$).

\subsection{The seeds of structure}

Although inflation does provide a natural explanation to some of the main problems of standard cosmology, the reason it is currently the most popular paradigm of the very early Universe is its ability to also explain the generation of small primordial perturbations consistent with CMB observations. The basic mechanism is qualitatively easy to understand: while pre-existing large-scale inhomogeneities are expected to be smoothed out by inflation, the Universe expands so fast that quantum fluctuations will at some point find themselves at cosmological scales.

If we do not assume that the inflaton field is uniform, but still assume the FLRW metric (i.e., if we work in the so-called flat gauge\footnote{Note that, in this gauge, the field perturbation is related to the curvature perturbation in the uniform field gauge by $\zeta=-H\frac{\delta\phi}{\dot{\phi}}$ \cite{Lyth:2009zz}.}), the equation of motion dictated by Eq.~\eqref{SFI} is
\begin{equation}
\ddot{\phi}+3H\dot{\phi}-\frac{\nabla^2\phi}{a^2}+V^\prime=0.\label{geom}
\end{equation}
Making the change of variables $\phi\left(t,\mathbf{x}\right)\equiv \phi\left(t\right)+\delta\phi\left(t,\mathbf{x}\right)$ (where the background quantity $\phi\left(t\right)$ still obeys Eq.~\eqref{F2phi}) and taking a Fourier transform we find the equation of motion for the field perturbation in Fourier space (neglecting second-order terms),
\begin{equation}
\ddot{\delta\phi}_\mathbf{k}+3H\dot{\delta\phi}_\mathbf{k}+\left(\frac{k}{a}\right)^2\delta\phi_\mathbf{k}+V^{\prime\prime}\delta\phi_\mathbf{k}=0.\label{keom}
\end{equation}

In order to quantise these fluctuations it is convenient to rewrite Eq.~\eqref{keom} in conformal time (given by $d\eta=a^{-1}dt$ and varying from $\eta=-\infty$ in the past to $\eta=0$ in the future) and using $\varphi\equiv a\delta\phi$, so that it can be made to look like the equation of motion of a harmonic oscillator:
\begin{equation}
\frac{d^2\varphi_\mathbf{k}\left(\eta\right)}{d\eta^2}+\omega^2_k\left(\eta\right)\varphi_\mathbf{k}\left(\eta\right)=0,\label{eomHO}
\end{equation}
where
\begin{equation}
\omega^2_k\equiv k^2-\frac{1}{z}\frac{d^2 z}{d\eta^2}\label{omegak}
\end{equation}
and $z\equiv \frac{a\dot{\phi}}{H}$.

We now quantise these fluctuations by defining the field operator
\begin{equation}
\hat{\varphi}\left(\eta,\mathbf{x}\right)\equiv \intop \frac{d^3\mathbf{k}}{\left(2\pi\right)^{3/2}}\left[\varphi_k\left(\eta\right)e^{i\mathbf{k\cdot x}}\hat{a}_\mathbf{k}+\varphi_k^*\left(\eta\right)e^{-i\mathbf{k\cdot x}}\hat{a}_{\mathbf{k}}^\dagger\right],\label{quantphi}
\end{equation}
where $\hat{a}_\mathbf{k}^\dagger$ and $\hat{a}_\mathbf{k}$ are creation and annihilation operators, respectively, and the mode functions $\varphi_k\left(\eta\right)$ obey Eq.~\eqref{eomHO} as well as
\begin{equation}
\varphi_k\left(\eta\right)\frac{d\varphi_k^*\left(\eta\right)}{d\eta}-\varphi_k^*\left(\eta\right)\frac{d\varphi_k\left(\eta\right)}{d\eta}=i,\label{Wronskian}
\end{equation}
so that $\left[\hat{\varphi}\left(\eta,\mathbf{x}\right),\frac{\hat{\varphi}\left(\eta,\tilde{\mathbf{x}}\right)}{d\eta}\right]=i\delta^3_{\mathbf{x}-\tilde{\mathbf{x}}}$, as required by canonical commutation relations\footnote{That $\frac{\hat{\varphi}\left(\eta,\tilde{\mathbf{x}}\right)}{d\eta}$ is the canonical conjugate of $\hat{\varphi}\left(\eta,\mathbf{x}\right)$ can be seen, for example, from the action that generates Eq.~\eqref{eomHO}: $S=\intop\frac{1}{2}\left[\left(\frac{d\varphi}{d\eta}\right)^2-\left(\nabla\varphi\right)^2+\left(\frac{1}{z}\frac{d^2 z}{d\eta^2}\right)\varphi^2\right]d\eta d^3\mathbf{x}$.}. 

Additionally, it can be shown that $\varphi=-z\zeta$ and it is known that, in single-field inflation, $\zeta_\mathbf{k}$ is usually\footnote{Ultra slow-roll inflation (for which $\epsilon\ll 1$ and $\delta_1=-3$) is a notable exception. Since in this thesis we only deal with slow-roll inflation, we need not worry about such cases \cite{Romano:2015vxz,Kinney:2005vj}.} conserved when scales $\sim k^{-1}$ are well outside the horizon (i.e., when $k\ll aH$) \cite{WANDS_SEPARATE}, therefore Eq.~\eqref{eomHO} is to be solved with the boundary conditions
\begin{equation}
\varphi_k\left(\eta\right)\longrightarrow
\begin{cases}
\frac{1}{\sqrt{2k}}e^{-ik\eta}, & -k\eta\rightarrow \infty\\
A_k z, & -k\eta\rightarrow 0
\end{cases}
,
\label{bound}
\end{equation}
where $A_k$ is a constant and the $-k\eta\rightarrow \infty$ limit corresponds to the usual flat space vacuum for scales well inside the horizon.

Finally, we can use the equivalent form of Eq.~\eqref{power_spectrum},
\begin{equation}
\langle\zeta\left(\mathbf{x}\right)\zeta\left(\tilde{\mathbf{x}}\right)\rangle=\intop \frac{d^3\mathbf{k}}{4\pi k^3}\mathcal{P}\left(k\right)e^{i\mathbf{k\cdot\left(x-\tilde{x}\right)}},\label{xcorr}
\end{equation}
together with Eq.~\eqref{quantphi} (making the correspondence $\langle Q\rangle =\left\langle 0\right|\hat{Q}\left| 0\right\rangle$, where $\left| 0\right\rangle$ is the vacuum state of the Fock space generated by $\hat{a}_\mathbf{k}$) to find that the power spectrum is given by
\begin{equation}
\mathcal{P}\left(k\right)=\frac{k^3}{2\pi^2}\lim_{-k\eta\rightarrow 0}\left|\frac{\varphi_k\left(\eta\right)}{z}\right|^2=\frac{k^3}{2\pi^2}\left|A_k\right|^2.\label{infPs}
\end{equation}

Eq.~\eqref{infPs} shows us how, if the background evolution of the inflaton is known, the power spectrum can in principle be found by solving Eq.~\eqref{eomHO}. In general, the resulting system of differential equations can be arbitrarily difficult to solve, requiring numerical approaches. The assumption of slow-roll, however, simplifies this task enough that remarkably accurate spectra can usually be computed from relatively simple formulas.

Under the slow-roll approximation (Eq.~\eqref{slowroll}), the Hubble parameter varies slowly (as $\dot{\left(H^{-1}\right)}=\epsilon\ll 1$). When $H$ is almost constant $\eta\simeq -1/aH$ and thus Eq.~\eqref{omegak} becomes
\begin{equation}
\omega_k^2\simeq k^2-\frac{2}{\eta^2},\label{omegaHconst}
\end{equation}
so that the solution to Eq.~\eqref{eomHO} with the appropriate boundary conditions is
\begin{equation}
\varphi_k\left(\eta\right)=\frac{\left(k\eta-i\right)}{\sqrt{2k^3}\eta}e^{-ik\eta}\xrightarrow[\eta\rightarrow 0]{}-\frac{i}{\sqrt{2k^3}}\frac{1}{\eta}\simeq\frac{iz}{\sqrt{2k^3}}\frac{H^2}{\dot{\phi}},\label{y0phi}
\end{equation}
leading to the well-known result (which now follows from Eq.~\eqref{infPs})
\begin{equation}
\mathcal{P}\left(k\right)\simeq\frac{H^4}{\left(2\pi\dot{\phi}\right)^2},\label{Ps4eva}
\end{equation}
where in practice the right-hand side is to be evaluated around the time of horizon crossing (when $k\sim aH$) because that is when cosmological dynamics have a sizeable impact on perturbations\footnote{In other words, since $\varphi_k$ well inside the horizon is set by the flat space vacuum condition and $\zeta$ is ``frozen out'' on superhorizon scales, we are allowed to use our result derived with the constant $H$ approximation provided that for each $k$ we use the value of $H$ that corresponds to the transition period between these limits.}.

Crucially, we can differentiate Eq.~\eqref{Ps4eva} to find
\begin{equation}
n_s\left(k\right)-1=\frac{d\ln\mathcal{P}}{d\ln k}=-4\epsilon-2\delta_1,\label{ns4eva}
\end{equation}
where the right-hand side is still to be evaluated at horizon crossing.

Interestingly, this calculation can be adapted to yield the tensor power spectrum. Working in the uniform field gauge, the perturbed metric can be written as
\begin{equation}
ds^2=dt^2-a^2\left(t\right)\left(\delta_{ij}+2h_{ij}\right)d\mathrm{x}^i d\mathrm{x}^j,\label{ds2infT}
\end{equation}
for which the Einstein equations give \cite{Lyth:2009zz}
\begin{equation}
\ddot{h}+3H\dot{h}+\left(\frac{k}{a}\right)^2 h=0,\label{heominf}
\end{equation}
where $h$ can be taken to be any component of $\left(h_{ij}\right)_\mathbf{k}$.

The formal equivalence between Eq.~\eqref{heominf} here and Eq.~\eqref{keom} for a massless field is because the perturbed version of the full action (i.e., with the Lagrangian density from Eq.~\eqref{SFI} as well as the Einstein-Hilbert term) for $h$ in this gauge is the same as that for $\delta\phi/\sqrt{2}$ with $V^{\prime\prime}=0$ in the flat gauge. Therefore, comparing with Eq.~\eqref{power_spectrum_T}, the tensor power spectrum can be found from the scalar result to be
\begin{equation}
\mathcal{P}_T\left(k\right)=\frac{16}{2}\times\frac{\dot{\phi}^2}{H^2}\times\mathcal{P}\left(k\right)=8\left(\frac{H}{2\pi}\right)^2,\label{Tfromzeta}
\end{equation}
where the right-hand side is still to be evaluated at horizon crossing. The two spectra are usually related by the tensor-to-scalar ratio, defined by
\begin{equation}
r\equiv\frac{\mathcal{P}_T}{\mathcal{P}}=16\epsilon.\label{rinf}
\end{equation}

\begin{figure}[h]
\includegraphics[scale=0.6]{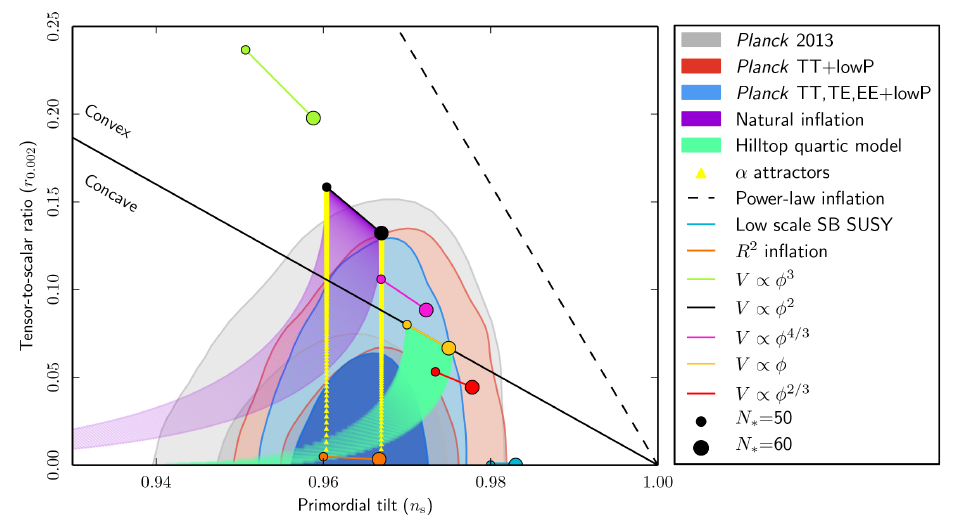}
\caption{Marginalised 68\% and 95\% limits for $n_s$ at $k=0.05\mathrm{Mpc^{-1}}$ and $r$ at $k=0.002\mathrm{Mpc^{-1}}$ from \emph{Planck} in combination with other data sets, compared to the theoretical predictions of selected inflationary models ($N_*$ being the number of e-folds to the end of inflation) --- image from Ade \emph{et al} \cite{Planck_inf2015}.\label{fig:nsr}}
\end{figure}

Despite relatively accurate observational limits on $n_s$ and $r$, it is fairly easy to write down inflation models whose spectrum is compatible with current observational constraints (see, e.g., figure \ref{fig:nsr}). In fact, it is so easy to do so that critics of inflation often present the huge number of inflation models compatible with current observations \cite{Martin:2013tda,Martin2014} as evidence that the inflationary paradigm is practically unfalsifiable \cite{infchall} --- which is why the search for observational tests able to rule out classes of inflation models, and particularly the simple class of canonical single-field slow-roll inflation models, is a major goal of modern cosmology\footnote{In chapter \ref{chap:SFSRI} we discuss a specific attempt to do this with parameters which quantify the scale dependence of $n_s\left(k\right)$.}.

\subsection{Reheating and the end of inflation}

Useful as inflation is, in order to effectively solve the problems of standard Big Bang cosmology it is meant to solve, it must eventually end to give way to the hot radiation-dominated Universe required by nucleosynthesis. Reheating is the name given to the intermediate epoch between the end of inflation and the establishment of this radiation-dominated initial state.

Ending inflation is not usually hard from a model-building perspective. In single-field inflation, for example, it is usually relatively simple to make sure that $\epsilon\geq 1$ at some point by making the potential steep enough. Alternatively, there are even models for which inflation never really ends classically but in which parts of the Universe still spontaneously stop inflating due to stochastic quantum processes \cite{Guth:2007ng}.

Reheating is a different matter. Since at the end of inflation the Universe is practically empty of ``normal'' matter, it requires the dominant component of the cosmic fluid to almost completely decay into standard model particles which need to be able to thermalise at some temperature $T\gtrsim 1\mathrm{MeV}$ (needed at the onset of nucleosynthesis). In the simplest models for single-field scenarios this is achieved by having the inflaton field lose a fraction of its energy as it oscillates around a minimum of its potential, but in general it can be a lot more complicated, possibly with more than one reheating phase taking place before the Universe becomes radiation-dominated \cite{Lyth:2009zz}. To make matters worse, in inflation models with more than one degree of freedom there is no guarantee that the curvature perturbation is conserved after horizon exit, leading to the necessity of an understanding of reheating mechanisms in order to make reliable observational predictions \cite{WANDS_SEPARATE,Martin:2013tda,Martin2014}.

\section{Cosmic strings}

Cosmic strings are line-like topological defects whose production in phase transitions in the early Universe is predicted in several particle physics scenarios (particularly in grand unification theories) \cite{Copeland:2009ga}. If cosmic strings are formed before inflation, there should be almost no chance of coming across one today. However, it is possible for them to be produced at the end of inflation, in particular in popular models in good agreement with observations such as brane inflation and hybrid inflation \cite{Achucarro:2008fn,Sarangi:2002yt}. If this is the case, an eventual cosmic string detection may be able to teach us a great deal about particle physics at very high energies in general and inflation in particular --- but only if we are able to properly model the evolution of the properties of realistic cosmic string networks through cosmic history. Chapter \ref{chap:wiggly} is dedicated to an effort to improve the way in which the evolution of the simplest cosmic string networks is modelled. In preparation for that, this section briefly reviews some basic concepts of cosmic string formation and dynamics.

\subsection{Topological defects from the Kibble mechanism}

The Kibble mechanism (also known as the Kibble-Zurek mechanism\footnote{The mechanism revolving around the topology of broken symmetry groups was first proposed by Tom Kibble when he introduced the study of topological defects to cosmology \cite{1976JPhA....9.1387K}. Wojciech Zurek later had his name added in recognition for his work in condensed matter tests of this scenario \cite{1985Natur.317..505Z}. Here we will favour the designation of ``Kibble mechanism'' because we will merely be referring to Kibble's original argument without any of the insights introduced by Zurek.}) is the main process by which topological defects can be formed in cosmological symmetry-breaking phase transitions.

The basic idea of this mechanism can be grasped in a situation in which a field is initially in a symmetric state (relative to some symmetry of its action) and is forced to settle in a broken-symmetry ground state (of which there must be more than one as a symmetry of the action must produce a ground state when acting on a ground state). Since all ground states are energetically equivalent, the specific ground state the field will ``choose'' will be determined by random fluctuations which must be uncorrelated over large enough distances (e.g. distances larger than the horizon). Depending on the specific topology of the set of ground states, it is possible that the requirement that the field be continuous (which is generally imposed by a kinetic term in the action) in the face of these fluctuations will force some regions of space not to be in any ground state --- and these regions will correspond to topological defects (the exact type of defect depending on the topology of the set of ground states\footnote{More specifically, on its homotopy group \cite{VSH}.}).

\begin{figure}[h]
\begin{center}
\includegraphics[scale=0.6]{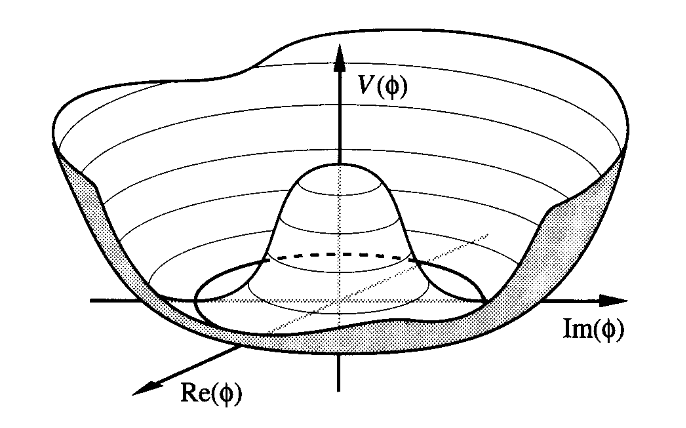}
\end{center}
\caption{Illustration of the ``mexican hat'' potential of Eq.~\eqref{mexican} for a complex field $\phi$ --- image from Vilenkin and Shellard \cite{VSH}.\label{fig:mexican}}
\end{figure}

As an illustrative example one can think of a scalar field with a ``mexican hat'' potential whose Lagrangian density is given by
\begin{equation}
\mathcal{L}=\partial_\mu\phi\partial^\mu\phi^*-\frac{\lambda}{4}\left(\left|\phi\right|^2-\alpha^2\right)^2,\label{mexican}
\end{equation}
where $\lambda$ and $\alpha$ are positive constants. If $\phi$ is allowed to be a complex field, this theory is invariant under global transformations of the form $\phi\left(x\right)\rightarrow e^{i\theta}\phi\left(x\right)$, where $\theta$ is any real constant; whereas if $\phi$ is forced to be a real field this symmetry only relates to sign changes $\phi\left(x\right)\rightarrow -\phi\left(x\right)$. Note that the potential term in Eq.~\eqref{mexican} (illustrated in figure \ref{fig:mexican}) has a local maximum at $\phi=0$ (where the field tends to settle at high temperatures) and a set of global minima at $\left|\phi\right|=\alpha$ (where the field settles at low temperatures). Therefore, if this field cools down from a high enough temperature there should be a transition at which $\phi$ will be forced to choose a phase in different regions of space.

Depending on whether the field is real or complex, this should lead to the formation of different topological defects. If the field is real, there are only two ground states ($\phi=\pm\alpha$) and it is easy to see that distant regions of space will ``choose'' different ones leading to there being relatively large three-dimensional domains of positive and negative $\phi$, at the border of which the field must quickly vary between them; effectively defining ``false vacuum'' membranes which are examples of domain walls. If, instead, the field is complex, the set of ground states corresponds to the circle defined by $\phi=\alpha e^{i\theta}$ and distant regions will have to settle for different values of $\theta$. It is then likely that there will be closed paths in space along which the field will wind around this circle (i.e., $\theta$ will continuously vary from $0$ to a multiple of $2\pi$). For this to be possible in a situation in which $\phi$ is continuous everywhere, in every two-dimensional surface with support in this closed path there must be at least one point at which $\theta$ is not defined. In other words, if there is at least one infinite line-like region of false vacuum going through the closed path in which the field is not in a ground state\footnote{Note that, although this is a little harder to visualise than in the domain wall case, there must be an actual one-dimensional region where $\phi=0$ (like in the domain wall example there must be a two-dimensional one) so that all such paths may be continuously ``glued together''.}: an example of a cosmic string.

\subsection{Goto-Nambu strings}

There are many different field theories which may give rise to many different types of cosmic strings: with different interaction terms, different (linear) energy densities, different associated charges, etc. Here we are interested in the simplest class of cosmic strings, which are free of long-range interactions and whose large-scale dynamics (when typical distances are considerably larger than the string thickness) depend only on their configuration. These are well described by the Goto-Nambu action
\begin{equation}
S=-\mu_0\intop\sqrt{-\gamma }d^2\sigma,\label{SGN}
\end{equation}	
where $\mu_0$ is a constant related to the symmetry breaking scale and $\gamma$ is the determinant of the metric
\begin{equation}
\gamma_{ab}=g_{\mu\nu}x^\mu_{,a}x^\nu_{,b},\label{gammametric}
\end{equation}
induced by the metric $g_{\mu\nu}$ on the $\left(1+1\right)$-dimensional string worldsheet parameterised by $x^\mu\left(\sigma\right)\equiv x^\mu\left(\sigma^0,\sigma^1\right)$, where $\sigma^0$ is a timelike coordinate and $\sigma^1$ is spacelike.

The equations of motion which follow from Eq.~\eqref{SGN} are given by
\begin{equation}
\nabla^2 x^\mu+\Gamma^\mu_{\nu\lambda}\gamma^{ab}x^\nu_{,a}x^\lambda_{,b}=0,\label{GNeom}
\end{equation}
where $\Gamma^\mu_{\nu\lambda}$ are the Christoffel symbols associated with the background metric $g_{\mu\nu}$ and the Laplacian operator acts on $x^\mu$ as
\begin{equation}
\nabla^2 x^\mu=\frac{1}{\sqrt{-\gamma}}\partial_a\left(\sqrt{-\gamma}\gamma^{ab}x^\mu_{,b}\right).\label{laplgamma}
\end{equation}

For a flat FLRW background metric and in the transverse temporal gauge, defined by $\sigma^0=\eta$ and $\mathbf{\dot{x}\cdot x^\prime}=0$ (where $\dot{Q}\equiv \frac{dQ}{d\eta}$ and $Q^\prime\equiv\frac{dQ}{d\sigma^1}$), Eq.~\eqref{GNeom} yields
\begin{equation}
\dot{\epsilon}+2\epsilon\frac{\dot{a}}{a}\mathbf{\dot{x}}^2=0,\label{GNFLRW0}
\end{equation}
\begin{equation}
\mathbf{\ddot{x}}+2\frac{\dot{a}}{a}\mathbf{\dot{x}}\left(1-\mathbf{\dot{x}}^2\right)=\frac{1}{\epsilon}\left(\frac{\mathbf{x^\prime}}{\epsilon}\right)^\prime,\label{GNFLRWi}
\end{equation}
where we have defined
\begin{equation}
\epsilon\equiv -\frac{x^{\prime 2}}{\sqrt{-\gamma}}=\sqrt{\frac{\mathbf{x^\prime}^2}{1-\mathbf{\dot{x}}^2}},\label{epsdefstrings}
\end{equation}
so that the energy in a string segment is given by \cite{VSH}
\begin{equation}
E=\mu_0 a\intop\epsilon d\sigma^1.\label{EGNstring}
\end{equation}

\begin{figure}[h]
\includegraphics[scale=0.6]{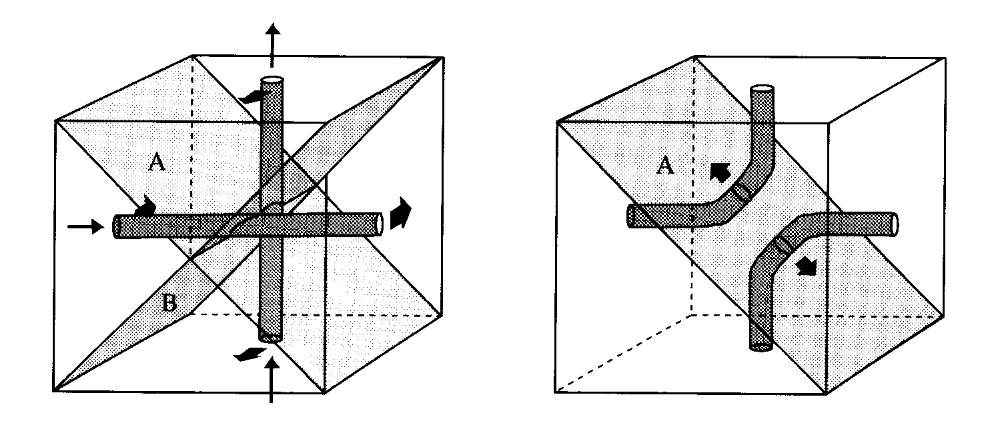}
\caption{Outcome of an intercommutation event between two strings (the bisecting planes A and B serving the purpose of evidencing the two-dimensional nature of the interaction) --- image from Vilenkin and Shellard \cite{VSH}.\label{fig:intercommute}}
\end{figure}

Note that often the Nambu-Goto equations of motion will predict that two segments of string will intersect each other. When this happens, the zero-width approximation behind Eq.~\eqref{SGN} breaks down and microscopic interactions may have to be taken into account. In the simplest models (and in the simplest Goto-Nambu simulations which will be considered in chapter \ref{chap:wiggly}) these intersections result in intercommutation, meaning that the two segments ``break'' at the point of intersection and the resulting ends reattach to the other segment's corresponding ends (see figure \ref{fig:intercommute}). Intercommutation leads to the appearances of kinks in strings and often leads to the production of string loops. In more complicated models, in addition to intercommutation, there may be a probability of the two segments becoming connected by a ``bridge''\footnote{Called a junction if it is point-like and a zipper if it is string-like.} or of no interaction taking place and the segments just going through each other.

\subsection{The Velocity-dependent One-Scale model}


On large scales, cosmic string networks are expected to resemble a collection of trajectories of Brownian random walks. This expectation is what leads to the popular simplifying assumption (known as the one-scale approximation) that the large-scale properties of these networks should be determined by a single characteristic length scale (meaning that it is assumed that quantities like the string correlation length and the string curvature radius can be assumed to be roughly the same) \cite{FRAC,KIB,MS2,ABELIAN}. This characteristic length, $L$, is usually defined in terms of the energy density in long strings (i.e., excluding string loops) as
\begin{equation}
\rho\equiv\frac{\mu_0}{L^2},\label{Ldefintro}
\end{equation}
where it is assumed that there is about one string segment of length $L$ per volume $L^3$ (since $L$ is also the typical separation between segments).
 
A less trivial consequence of this one-scale approximation is that the rate of energy loss in the form of loops inside a correlation volume $L^3$ must be just proportional to the characteristic velocity of string segments, $v$\footnote{This is because the number of loops of size $l$ being formed per intercommutation should only depend on $l/L$, but the rate at which they form is proportional to the probability of a segment of size $l$ encountering one of the other segments within a time $\delta t$, which is of order $vl\delta t/L^2$. In other words, the typical size of loops being formed in the correlation volume $L^3$ is proportional to $L$, but the probability of them forming is proportional to $v/L$.}. As a result, we can write
\begin{equation}
\left(\frac{d\rho}{dt}\right)_\mathrm{to\ loops}=cv\frac{\rho}{L},\label{elossloops}
\end{equation}
where $c$ is known as the loop chopping efficiency parameter (which depends on the specific model in which strings arise) and for the characteristic velocity $v$ we consider the root mean square velocity
\begin{equation}
v^2\equiv\frac{\intop\epsilon\mathbf{\dot{x}}^2 d\sigma^1}{\intop\epsilon d\sigma^1}.\label{vrmsintro}
\end{equation}

Finally, integrating Eqs.~\eqref{GNFLRW0} and \eqref{GNFLRWi} over the whole network (introducing the definitions from Eqs.~\eqref{EGNstring}, \eqref{Ldefintro}, and \eqref{vrmsintro}, and adding Eq.~\eqref{elossloops} by hand), we arrive at the system of equations which defines the Velocity-dependent One-Scale (VOS) model for the evolution of a network of Goto-Nambu cosmic strings which only loses energy by redshifting and by loop production,
\begin{equation}
2\frac{dL}{dt}=2HL\left(1+v^2\right)+cv,\label{VOSL}
\end{equation}
\begin{equation}
\frac{dv}{dt}=\left(1-v^2\right)\left[\frac{k}{L}-2Hv\right],\label{VOSv}
\end{equation}
where $k$ is called the momentum parameter and is defined by
\begin{equation}
k\equiv\frac{\langle\left(1-\mathbf{\dot{x}}^2\right)\left(\mathbf{\dot{x}\cdot u}\right)\rangle}{v\left(1-v^2\right)},\label{kmomintro}
\end{equation}
where $\mathbf{u}$ is a unit vector parallel to the curvature vector, and in most relevant limits $k$ can just be written as \cite{MS3}
\begin{equation}
k\left(v\right)=\frac{2\sqrt{2}}{\pi}\frac{1-8v^6}{1+8v^6}.\label{kvintro}
\end{equation}

An important property of this model is that when $a\propto t^\lambda$ it has the attractor scaling solution
\begin{equation}
\frac{L}{t}=\sqrt{\frac{k\left(k+c\right)}{4\lambda\left(1-\lambda\right)}},\ v^2=\frac{k\left(1-\lambda\right)}{\lambda\left(k+c\right)},\label{scalingintro}
\end{equation}
which is useful for comparison with simulations assuming matter or radiation domination.

\subsection{Small-scale structure on cosmic strings: a mathematical formalism}

The VOS model has become a mainstream tool in the field of cosmic strings \cite{Planck} owing to its relative simplicity and quantitative agreement with simulations (particularly in scaling regimes) \cite{ABELIAN,FRAC}. Moreover, even though the VOS model as we have described it applies only to networks of simple strings (defined by the Goto-Nambu action), its basic framework can be naturally extended to describe networks of more complex strings - for example, superconducting strings \cite{SUPERC1,SUPERC2}.

One of the most serious shortcomings of the VOS model is its inability to account for the presence and effect of small-scale structure (on scales much below $L$) in a cosmic string network throughout its evolution. The one-scale approximation makes the model intrinsically limited at such small scales; however, realistic networks should develop a non-negligible amount of such small-scale struture (chiefly due to kinks created by intercommutation) \cite{FRAC,1990PhRvL..65.1705A}.

One way of generalising the VOS model in order to take this small-scale structure into account \cite{PAP1} involves substituting the Goto-Nambu action (Eq.~\eqref{SGN}) with the elastic string action given by
\begin{equation}
S=-\mu_0\intop\sqrt{1-\gamma^{ab}\phi_{,a}\phi_{,b}}\sqrt{-\gamma }d^2\sigma,\label{elasticS}
\end{equation}
where the additional scalar field $\phi$ is to be thought of as a stream function (defined on the string worldsheet) that is constant along the flow lines of an associated current --- which in this case we interpret as a mass current due to wiggles propagating along the string.

The main idea is that, in a course-grained sense (i.e., as long as we are only interested in quantities defined as averages over large string segments), a smooth elastic string obeying the action in Eq.~\eqref{elasticS} should behave in the same way as a Goto-Nambu string with small-scale structure \cite{WIG1}.

The resulting equations of motion, analogous to Eqs.~\eqref{GNFLRW0} and \eqref{GNFLRWi}, can now be written as
\begin{equation}
\left(\frac{\epsilon}{w}\right)^{\bf \dot{}}+\left(\frac{\epsilon}{w}\right)\frac{\dot{a}}{a}\left[2w^{2}\mathbf{\dot{x}}^2+\left(1+\mathbf{\dot{x}}^2\right)\left(1-w^2\right)\right]=0,\label{weom1}
\end{equation}
\begin{equation}
\mathbf{\ddot{x}}+\frac{\dot{a}}{a}\mathbf{\dot{x}}\left(1-\mathbf{\dot{x}}^2\right)\left(1+w^2\right)=\frac{w^2}{\epsilon}\left(\frac{\mathbf{x^\prime}}{\epsilon}\right)^\prime,\label{weom2}
\end{equation}
\begin{equation}
\frac{\dot{w}}{w}=\left(1-w^2\right)\left(\frac{\dot{a}}{a}+\frac{\mathbf{x^\prime}\cdot\mathbf{\dot{x}^\prime}}{\mathbf{x^\prime}^2}\right),\label{weom3}
\end{equation}
where
\begin{equation}
w\equiv\sqrt{1-\gamma^{ab}\phi_{,a}\phi_{,b}}.\label{wdef}
\end{equation}

The formula for the total energy in a piece of string is now
\begin{equation}
E=\mu_0 a\intop\frac{\epsilon}{w} d\sigma^1,\label{Ewstring}
\end{equation}
whereas the corresponding result neglecting small-scale structure is called the bare energy and is written as (according to Eq.~\eqref{EGNstring})
\begin{equation}
E_0=\mu_0 a\intop\epsilon d\sigma^1.\label{E0string}
\end{equation}
Each of these energies will be associated with a different characteristic length (via Eq.~\eqref{Ldefintro}). We shall henceforth refer to the one associated with $E$ as $L$ and to the one associated with $E_0$ as $\xi$.

The greater the difference between these two energies/lengths the more small-scale structure there is in a given network. Therefore, a natural measure of the wiggliness in a cosmic string network is the renormalised string mass per unit length factor given by
\begin{equation}
\mu=\frac{E}{E_0}=\frac{\xi^2}{L^2}=\frac{\intop\frac{\epsilon}{w} d\sigma^1}{\intop\epsilon d\sigma^1},\label{mudefintro}
\end{equation}
which is unity in the absence of small-scale structure and increases in its presence. However, it must be stressed that this quantity is not an intrinsic property of the network: rather, it depends on the characteristic scale of the coarse-graining procedure from which the distinction between small and large-scale structure on the strings arises\footnote{Much like the relation between the total length of a segment of coastline and its length on a map depends on the scale of the map.}. If we designate this coarse-graining scale (also sometimes called a renormalisation scale) by $\ell$, the scale dependence of $\mu$ must follow
\begin{equation}
\frac{\partial\ln\mu}{\partial\ln\ell}\sim d_m\left(\ell\right)-1,\label{fracintro}
\end{equation}
where $d_m\left(\ell\right)$ is the multifractal dimension of the network at the scale $\ell$.

In order to obtain VOS-like equations from these results, it will be necessary to take averages of Eqs.~\eqref{weom1}-\eqref{weom3}. The most natural way to define averages in this context is according to
\begin{equation}
\langle Q\rangle\equiv\frac{\intop Q\frac{\epsilon}{w} d\sigma^1}{\intop\frac{\epsilon}{w} d\sigma^1}.\label{wavgintro}
\end{equation}
Consistently, we also define the root mean square velocity as
\begin{equation}
v^2\equiv\langle \mathbf{\dot{x}}^2\rangle=\frac{\intop\frac{\epsilon}{w}\mathbf{\dot{x}}^2 d\sigma^1}{\intop\frac{\epsilon}{w} d\sigma^1}.\label{vrmsw}
\end{equation}

At last, Eqs.~\eqref{weom1}-\eqref{weom3} can be integrated to yield (identifying the typical curvature of the network with $\xi$, as in the VOS model)
\begin{equation}
\frac{\dot E}{E}=\frac{\dot\rho}{\rho}+3\frac{\dot a}{a}=\frac{\dot E_0}{E_0}+\frac{\dot \mu}{\mu}=
\left[\langle w^2\rangle-v^2-\langle w^2{\dot {\bf x}}^2\rangle\right]\frac{\dot a}{a}, \label{wigint_1}
\end{equation}
\begin{equation}
\frac{\dot\mu}{\mu}=\frac{a\mu}{\xi}\langle w(1-w^2)({\dot{\bf x}}\cdot {\bf u})\rangle+\frac{\dot a}{a}\left[\langle w^2\rangle-1+\langle (\mu w-1)(1+w^2){\dot {\bf x}}^2\rangle\right] + [d_m(\ell)-1]\frac{\dot\ell}{\ell}, \label{wigint_2}
\end{equation}
\begin{equation}
\dot{\left(v^2\right)}=\frac{2a}{\xi}\langle w^2(1-{\dot {\bf x}}^2)({\dot{\bf x}}\cdot {\bf u})\rangle -\frac{\dot a}{a}\langle (v^2+{\dot {\bf x}}^2)(1+w^2)(1-{\dot {\bf x}}^2)\rangle+\frac{1-v^{2}}{1+\left\langle w^{2}\right\rangle }\frac{\partial\left\langle w^{2}\right\rangle }{\partial \ell}\dot{\ell}, \label{wigint_3}
\end{equation}
where the final terms in the last two equations are called scale drift terms and take into account the possible temporal dependence of $\ell$\footnote{The scale drift term in Eq.~\eqref{wigint_2} following trivially from Eq.~\eqref{fracintro}, while the one in Eq.~\eqref{wigint_3} results from the realisation that Eq.~\eqref{wigint_1} must not depend on $\ell$ (and using also the assumption that the microscopic string velocity is locally independent of $w$).}.

At this point, the only thing needed for these equations to constitute a full VOS-like model of a wiggly string network is the addition of energy loss terms analogous to the one in Eq.~\eqref{elossloops}. This, as well as an in-depth exploration of the resulting model, is done in chapter \ref{chap:wiggly}.

\section{Thermodynamics and Negative Absolute Temperatures}

Chapter \ref{chap:NAT} is dedicated to the study of possible cosmological consequences of the existence of negative absolute temperatures (NAT). The present section provides a preliminary introduction to this work by reviewing some basic results associated with this exotic concept.

\subsection{What is a temperature?}

Temperature is a rather unique physical concept. Despite being practically ubiquitous in quotidian scenarios, its actual physical meaning remains elusive to the intuition of most who use it.

What the vast majority of people really have in mind when they use the word ``temperature'' in informal settings is the related concept of hotness. When they say something is at a higher temperature than something else, what they really mean is that it feels hotter; i.e., touching it will cause heat to be transferred to their skin. In fact, if two objects at different temperatures are put in thermal contact, most intuitions will unquestioningly expect heat to flow from the higher temperature one to the other --- in line with the definition of ``hotness'' implied in the Clausius formulation of the second law of thermodynamics\footnote{\emph{``No process is possible whose sole result is the transfer of heat from a colder to a hotter body.''} \cite{blundell}}.

In reality, temperature is a little more complicated than that (and even than more sophisticated versions of that which use temperature as merely a complicated measure of the internal energy in a system). Naturally, such views endure because in daily life, and indeed in most applications, temperature can be thought of in this way. However, those expectations do break down in rare but achievable conditions, and thus should not be mistaken for fundamental properties.

In the usual formulation of the first law of thermodynamics \cite{blundell}, the temperature of an isolated system is defined by
\begin{equation}
\frac{1}{T}=\left(\frac{\partial S}{\partial U}\right)_{V,\, N},\label{eq:thermo_Tintro}
\end{equation}
where $T$ is the temperature, $U$ the internal energy, $V$ the volume, $N$ the number of particles, and $S$ is the Boltzmann entropy defined as
\begin{equation}
S=k_{B}\ln W,\label{eq:Boltzmann_entropyintro}
\end{equation}
where $k_{B}$ is the Boltzmann constant and $W$ is the number of microstates corresponding to the macrostate the system is in.

Alternatively, in a canonical ensemble, temperature can be given in terms of the occupation numbers of one-particle states given by the Fermi-Dirac and the Bose-Einstein distributions:
\begin{equation}
\langle n_i\rangle=\frac{1}{e^{\frac{\left(\epsilon_i -\mu\right)}{k_B T}}\pm 1}, \label{eq:fermi_dirac_bose_einstein}
\end{equation}
where $\langle n_i\rangle$ denotes the average occupancy of a single-particle state of energy $\epsilon_i$, $\mu$ is the chemical potential, and the $\pm$ sign is positive for fermions (Fermi-Dirac distribution) and negative for bosons (Bose-Einstein distribution). Mathematically, in this picture $1/k_B T$ is a Lagrange multiplier associated with energy conservation and $-\mu/k_B T$ is a Lagrange multiplier associated with number conservation. Physically, the temperature indicates how much more likely a particle is to be in a lower-energy state than in a higher-energy one.

\subsection{Negative absolute temperatures}

In order to illustrate how common intuitions regarding the behaviour of temperature may fail, let us consider the example of an isolated system with a finite number of distinguishable particles which can each be in one of two states with different energies. A notable example of such a system is given by an Ising model of a ferromagnet subject to an external magnetic field \cite{RevModPhys.39.883}. In the following analysis we call the lower-energy one the ground state and the higher-energy one the excited state.

In this system, macrostates are characterised only by their energy --- or, equivalently, by the number of particles in the excited state. Therefore, the number of microstates associated with a given macrostate is just the number of ways in which the particles can be distributed with the corresponding number in the excited state:
\begin{equation}
W=\frac{N!}{n!\left(N-n\right)!},\label{W2level}
\end{equation}
where $N$ is the total number of particles and $n$ is the number of particles in the excited state.

Crucially, $W$ given by Eq.~\eqref{W2level} remains unchanged if $n$ is redefined as the number of particles in the ground state (as it should, since $W$ is a purely combinatorial quantity). As a result, the entropy as a function of $n$ must be symmetric around $n=N/2$ (corresponding to equal numbers of particles in both states and the maximum of $W$) and is a decreasing function for $n>N/2$.  In fact, inserting this result into Eq.~\eqref{eq:Boltzmann_entropyintro} and then into Eq.~\eqref{eq:thermo_Tintro} leads to (using the Stirling approximation for large $N$)
\begin{equation}
\frac{1}{T}=\frac{k_B}{\Delta \epsilon}\left[\ln\left(\frac{N}{n}-1\right)\right],\label{T_NATintro}
\end{equation}
where $\Delta \epsilon$ is the energy difference between the two states.

In Eq.~\eqref{T_NATintro} we find a classical example of a temperature defying the usual expectations whenever $n>N/2$, i.e., when there are more particles in the excited state than in the ground state and the temperature becomes non-positive (when all particles are excited both the temperature and the entropy vanish, as in the symmetric configuration\footnote{In fact, Eq.~\eqref{T_NATintro} is odd with respect to all transformations which swap all the particles in both states.}; otherwise the temperature is negative). Interestingly, the temperature here can also be seen to be singular at $n=N/2$; a sign that the physically meaningful quantity is the inverse of the temperature rather than the temperature itself.

The role of these negative absolute temperatures (NAT) is easier to see in the canonical ensembles assumed in Eq.~\eqref{eq:fermi_dirac_bose_einstein}. Regardless of the spectrum of available one-particle states, positive-temperature distributions favour the occupation of lower-energy states whereas NAT lead to higher-energy states being preferred (and if all states are uniformly occupied that corresponds to a singularity in the temperature). This also makes it easier to see that the key feature of the two-level quantum system that makes NAT possible is the existence of a maximum possible energy of the system: without that, NAT distributions following Eq.~\eqref{eq:fermi_dirac_bose_einstein} would not be normalisable.


\subsection{Negative temperatures and negative pressures}

In the end, NAT are a consequence of the physically non-obvious (although mathematically rigorous) manner in which temperature is canonically definded. At first glance, it appears that one could simply think in terms of occupation numbers and everything else would work similarly as it does for $T>0$. However, the importance of temperature in thermodynamics does lead to some exotic results in the calculation of specific observables. In this work, we are especially interested in the consequences for the pressure: which will be naturally negative when $T<0$.

One way of showing this result \cite{BraunNAT} uses the fact that the maximum entropy principle generally requires
\begin{equation}
\left( \frac{\partial S}{\partial V}\right)_U\geq 0,\label{maxentintro}
\end{equation}
as otherwise the system would spontaneously contract to increase its entropy.

Using the total differential of the internal energy,
\begin{equation}
dU=TdS-PdV,\label{dUintro}
\end{equation}
where $P$ is the pressure, we can then find that
\begin{equation}
\left( \frac{\partial S}{\partial V}\right)_U=\frac{P}{T},\label{NAPintro}
\end{equation}
and thus the pressure must have the same sign as the temperature\footnote{An independent derivation which takes into account how chemical potentials may preserve the positivity of $P$ even when $T<0$ can be found in subsection \ref{subsec:NAP}.}. The potential relevance of this result for cosmology is discussed in the following chapter.

\chapter{Cosmology with Negative Absolute Temperatures}
\label{chap:NAT}

\begin{center}

J.P.P. Vieira,$^{1}$ Christian T. Byrnes,$^{1}$ and Antony Lewis$^{1}$\\[0.5cm]
$^{1}$Department of Physics \& Astronomy, University of Sussex, Brighton BN1 9QH, UK

\end{center}

\ \\

Negative absolute temperatures (NAT) are an exotic thermodynamical
consequence of quantum physics which has been known since the 1950's
(having been achieved in the lab on a number of occasions). Recently,
the work of Braun \emph{et al} \cite{BraunNAT} has rekindled interest in negative temperatures
and hinted at a possibility of using NAT systems in the lab as dark
energy analogues. This paper goes one step further, looking into the cosmological consequences
of the existence of a NAT component in the Universe. NAT-dominated expanding Universes experience a borderline phantom expansion ($w<-1$) with no Big Rip, and their contracting counterparts are
forced to bounce after the energy density becomes sufficiently large.
Both scenarios might be used to solve horizon and
flatness problems analogously to standard inflation and bouncing cosmologies.
We discuss the difficulties in obtaining and ending a NAT-dominated epoch, and
possible ways of obtaining density perturbations with an acceptable spectrum.

\newpage

\section{Introduction}

\subsection{How can temperature be negative?}

Say the words ``negative absolute temperatures'' (NAT) to anyone
who hasn't heard of them before, and your remark will most likely be
met with a look of bewilderment (and perhaps the question in the title).
Even more than sixty years after negative temperatures were achieved in the lab, this
is by no means an unexpected reaction. In informal parlance we all get used to
perceiving temperature as a measure of the energy in a macroscopic system, and thus necessarily a positive
quantity.
In fact, temperature is canonically defined in terms of the rate of change
of entropy with internal energy in thermal equilibrium, which can be negative. Specifically
\begin{equation}
\frac{1}{T}=\left(\frac{\partial S}{\partial U}\right)_{V,\, N,\, X_{i}}\label{eq:thermo_T}
\end{equation}
where $T$ is the absolute temperature, $U$ the internal energy,
$S$ the entropy, $V$ the volume, $N$ the number of particles and
$X_{i}$ represents any other (eventually) relevant extensive property
of the system. In this work, $S$ is defined as%
\footnote{
\label{fn:Boltazmann-Gibbs_controversy}
There has recently been some controversy \cite{inconsist_NAT,comment_consist_NAT,Gibbs_Boltzmann_Teq,
Repy_comment_inconsist_NAT,Boltz_vs_Gibbs,thermo_isolated,inform_gibbshertz,dispute_boltzgibbs}
regarding whether this quantity, known as the Boltzmann entropy, is correct;
the alternative being the Gibbs-Hertz entropy, brought under the spotlight
by \cite{inconsist_NAT} (in the original reference, they just call
it the ``Gibbs entropy'' since Gibbs was apparently the first to
propose this entropy formula despite it traditionally being credited
to Hertz). While this debate is an important one (especially for
anyone interested in NAT, which are impossible in the Gibbs-Hertz
formalism), it is not completely clear in which situations the disagreement
actually affects obervables in the thermodynamic limit \cite{Boltz_vs_Gibbs}.
Moreover, it has recently been shown \cite{dispute_boltzgibbs}
that the Boltzmann formula is the appropriate one for systems
with equivalence of statistical ensembles.
}
\begin{equation}
S=k_{B}\ln W\label{eq:Boltzmann_entropy}
\end{equation}
where $k_{B}$ is the Boltzmann constant and $W$ is the number
of microstates corresponding to the macrostate the system is in.

The reason we do not expect NAT in classical scenarios is that for
those we generally expect the number of states with energy $U$ to
increase with $U$. In quantum mechanical systems, however, it is
fairly easy to construct situations in which the energy is bounded
from above as well as from below. When that happens, if the entropy
is a continuous function of the energy, $S$ must
have a maximum somewhere between the upper and the lower energy bounds
(where $S$ is zero). By Eq.~\eqref{eq:thermo_T}, $T$ must then
allow negative values.

The simplest example is a two-level quantum
system which can be populated by a fixed number of distinguishable
particles. As the energy of the system is increased, more particles will populate the higher-energy level.
At infinite temperature the number of particles is the same in both energy levels
(corresponding to maximum entropy), but it
is quite possible to give the system more energy than that, so that there are then more particles in the
higher-energy state, corresponding to a negative temperature. Note that the system at a negative temperature
has more energy, and is therefore ``hotter'', than at a positive temperature.

In practice, negative temperatures can be realized in a number of ways. As an illustration,
consider a lattice of localized spin-$1/2$ particles interacting with an external
magnetic field. There are two one-particle energy levels, corresponding
to the two possible spin orientations relative to the magnetic field.
At low temperatures, we expect most spins to be in the lowest-energy
state. However, if the sign of the external magnetic field is switched
at very low temperatures, then suddenly the most populated state will
become the highest-energy state and if the system can then be isolated
(so that energy cannot be lost and most particles are forced to be
in the highest-energy state) then we are left with a state corresponding
to $T<0$. This was essentially the set-up used by Purcell and Pound
in 1951 \cite{NuclearNAT}, in the first experiment in which it is
claimed that NAT were measured (the magnetic material they used was
crystal of Lithium fluoride, which was known to have very long magnetic relaxation
times).

\subsection{From the lab to the sky}

The first thorough theoretical study of the conditions under which
NAT occur is due to Ramsay \cite{RamseyNAT}, five years after the
experiment by Purcell and Pound \cite{NuclearNAT} (although the first
known appearance of the concept of NAT seems to have been two years earlier,
when Onsager used them to explain the formation of large-scale
persistent vortices in turbulent flows \cite{NAT_turbul_vortex}). Even today, most
discussions revolving around NAT take this treatise as a starting
point.

After Ramsay (1956), there was not much big news regarding NAT until
2012, when Braun et al.~\cite{BraunNAT} reported the first experimental
realization of NAT in a system with motional degrees of freedom (an ultra-cold
boson gas). Important as this may be as an experimental landmark,
one of its main consequences was arguably the revival of theoretical
interest in NAT which led to the debate about Boltzmann vs Gibbs-Hertz
entropies we have already mentioned (see footnote \ref{fn:Boltazmann-Gibbs_controversy}).
Interestingly, Braun et al. also noticed that an (almost) inevitable consequence
of negative temperatures, negative pressures, are \emph{``of fundamental
interest to the description of dark energy in cosmology, where negative
pressure is required to account for the accelerating expansion of
the universe''}. Apparently, this remark was mostly interpreted as a
suggestion that known NAT systems could be useful as laboratory dark
energy analogues. Some people, however, seem to have read this hint
differently, meaning that some analogous mechanism could be responsible
for the observed accelerated expansion of the Universe. This interpretation
seems to have inspired Brevik and Gr\o n \cite{NEGVISC} to come up
with a class of models where, while not using NAT directly, an analogous
effect is achieved by means of a negative bulk viscosity. Nevertheless,
as far as we are aware, nobody has proposed a model where this is
done with actual negative temperatures, possibly due to not having
found a well-motivated physical assumption that could lead to NAT
at cosmological scales%
\footnote{A connection between NAT and phantom inflation seems to have
been first independently suggested in Ref.~\cite{Phantom_thermo}. However,
the word "temperature" there is really referring to an out-of-equilibrium
generalisation of temperature and none of their examples can
correspond to NAT as defined here. Those following the ensuing
discussion \cite{thermo_phantom2,thermo_phantom_mu1,thermo_phantom_mu2,thermo_phantom_mu3,thermo_phantom_mu4,scalarNATmail}
might be interested in the questions we raise regarding the introduction of a non-null
chemical potential in this context (see Appendix~\ref{sec:The-problem-of-mu}).%
}.

\subsection{A natural cut-off?\label{sub:A-natural-cut-off?}}

The key requirement for a NAT is an upper bound to the energy of the system.
This could either be an absolute upper bound, or there could be an energy gap allowing a metastable
population inversion. As long as the interaction time for particles below the energy gap is much shorter
than the typical time scale for thermal equilibrium to be reached,
an effective NAT can develop (as in the experimental realizations).

In the context of cosmology, where we are mainly interested in the properties of the total density,
a NAT could be obtained if there is a fundamental energy cut-off. This could
 be related to a minimum length scale, for example as discussed
in the context of quantum gravity (see for example \cite{min_length_review,Garay_min_length}
and references therein). For the purpose of this paper we are not assuming any particular model,
 and simply consider the possibility that the fundamental model features a cut-off and investigate the consequences. Nevertheless, it must be kept in mind that there is a number of non-trivial requirements that must be met by any well-motivated realization of this idea.

One major such difficulty is raised by the existence of particles which can never thermalize at NAT, like photons\footnote{Since they are not subject to number conservation, there can never be a maximum energy for a photon fluid, even if individual photons have their energy bounded from above.}. Even if a physically well-motivated scenario in which certain particles can reach thermal equilibrium at NAT can be found, for them to do so at cosmological scales it is necessary that they be able to do so despite the presence of photons. The problem here is that, if there is exchange of energy between the NAT fluid and a photon fluid, then thermal equilibrium must occur at the same temperature for both fluids --- and since the photon fluid cannot be at NAT then equilibrium at NAT becomes impossible. Moreover, even if there is no direct coupling between these two fluids, they must both couple to gravity and thus some measure of thermal contact is unavoidable. Therefore, in order for any eventual theory of NAT at cosmological scales to be successful, it must be possible to show that the characteristic timescale of this energy transfer is long enough compared to other relevant timescales that there can be an effective thermal equilibrium at NAT over cosmological time scales.

Another important requirement concerns the interaction time for dominant particles with energies up to the cut-off. This must be short relative to other relevant timescales, particularly the Hubble time. Otherwise, any population inversion could rapidly go out of equilibrium as the particles decouple, rendering the very concept of temperature meaningless.

In the end, we focus on the possibility that equilibrium is maintained and see what a phenomenological NAT description would imply. The relevant quantity that needs to
be extracted from an eventual fundamental theory is the number density
of states at a given energy $\epsilon$, $g\left(\epsilon\right)$,
which at low energies is constrained to take a standard form.
Given the lack of an actual complete fundamental theory to
work with, we shall express all results in the most general form possible.
Any time we want to illustrate a calculation for a specific model we consider
a simple ansatz for a gas of independent particles with a cut-off at $\epsilon=\Lambda$
and the right behaviour at low (i.e., currently observed) energies,
\begin{equation}
g\left(\epsilon\right)=\begin{cases}
\frac{g}{2\pi^{2}}\epsilon\sqrt{\epsilon^{2}-m^{2}} & \text{if}\ m<\epsilon<\Lambda\\
0 & \text{otherwise},
\end{cases}\label{eq:DOS_Lambda}
\end{equation}
where $g$ 
is the usual degeneracy factor and $m$ is the particle
mass (note we are using units in which $c=\hbar=M_{P}=k_{B}=1$).
Interestingly, it turns out that our most important results
in section \ref{sec:NATive-Cosmology} will be essentially independent
of the specific form of $g\left(\epsilon\right)$ as long as it behaves as it should at low energies.

In the remainder of this paper, we shall focus on the cosmological
implications of NAT. The discussion is organised as follows.
In section \ref{sec:Thermodynamical-functions} we show how to calculate
thermodynamical functions as model-independently as possible. In section
\ref{sec:NATive-Cosmology} we use the results from section \ref{sec:Thermodynamical-functions}
to model the evolution of generic expanding and contracting NATive
Universes. In particular, we show that exactly exponential inflation
is an attractor regime in these models and address the problems
associated with ending it. Finally, the main successes and problems
of this approach are summarised in section \ref{sec:Conclusions}.
Additionally, appendix \ref{sec:Thermal-Perturbation-Generation}
deals with the challenges of thermal perturbation generation at NAT.

\section{Thermodynamical functions\label{sec:Thermodynamical-functions}}

The main goal of this section is to investigate the temperature dependence
of the most relevant thermodynamical quantities (which we will later
need to substitute into the Friedmann equations in order to do cosmology).
In particular, we are interested in finding model-independent relations
between results at very low positive temperatures (the kind that has
been extensively studied) and results at negative temperatures very
close to $T=0^{-}$ (which we shall see generally corresponds to the
highest possible energy scales, which have never been probed).

\subsection{Negative pressure}
\label{subsec:NAP}

Our main motivation for studying NAT is that they naturally
give rise to negative pressures. Let us start by seeing why this is so.
One of the most straightforward ways of calculating the pressure of
a system is by making use of the grand potential, defined as
\begin{equation}
\Phi=U-ST-\mu N, \label{eq:Helmholtz}
\end{equation}
and whose gradient can be written as
\begin{equation}
d\Phi=-SdT-PdV-N d\mu+x^{i}dX_{i}, \label{eq:dF}
\end{equation}
where $\mu$ is the chemical potential and $x^{i}$ represent the
thermodynamic potentials corresponding to the quantities $X_{i}$.
Assuming there is no relevant $X_{i}$, we get the well-known Euler
relation:
\begin{equation}
P=-\left(\frac{\partial \Phi}{\partial V}\right)_{T,\, \mu}=-\rho+sT+\mu n.\label{eq:P_general}
\end{equation}

Note that when $T<0$ the only term in Eq.~\eqref{eq:P_general} which
is not necessarily negative is $\mu n$, and the pressure will be
very negative unless this term is significant in comparison to the
others. In particular, if $\mu=0$ (as must be the case in regimes
where the total number is not conserved) we recover the better-known
result
\begin{equation}
P=-\rho+sT,\label{eq:P_mu0}
\end{equation}
which corresponds to an equation of state with $w<-1$ (leading to
what is known as \emph{phantom inflation}) for any $T<0$.

\subsection{Fermions and holes}

For now we deal only with fermions (in appendix
\ref{sec:The-problem-of-mu} we discuss why we do not want to work
with bosons). We will therefore use the
Fermi-Dirac distribution,
\begin{equation}
{\cal N} \left(\epsilon;T,\mu\right)=\frac{1}{e^{\beta\left(\epsilon-\mu\right)}+1}, \label{eq:fermi_dirac}
\end{equation}
which should still be valid for $\beta=\left(k_{B}T\right)^{-1}<0$
since microstate probabilities are still associated with the Boltzmann
factor $e^{-\beta\left(E-\mu N\right)}$ (where $E$ is the total
energy associated with a specific microstate, so that $U=\left\langle E\right\rangle $).

We can now use standard thermostatistics to find the relevant quantities
as a function of temperature and chemical potential.
The energy and the number density are trivial,
\begin{equation}
\rho\left(T,\mu\right)=\intop_{m}^{\Lambda}\epsilon g\left(\epsilon\right){\cal N}\left(\epsilon;T,\mu\right)d\epsilon, \label{eq:rho_general}
\end{equation}
\begin{equation}
n\left(T,\mu\right)=\intop_{m}^{\Lambda}g\left(\epsilon\right){\cal N}\left(\epsilon;T,\mu\right)d\epsilon, \label{eq:n_general}
\end{equation}
as are their maximum possible values,
\begin{equation}
\rho_{\mathrm{max}}\equiv\intop_{m}^{\Lambda}\epsilon g\left(\epsilon\right)d\epsilon, \label{eq:rhomax}
\end{equation}
\begin{equation}
n_{\mathrm{max}}\equiv\intop_{m}^{\Lambda}g\left(\epsilon\right)d\epsilon.\label{eq:nmax}
\end{equation}
Note that these maximum values correspond only to the NAT fermion gas, so
in situations in which there is more than one component the total $\rho$ and $n$
can exceed these values.

The pressure is less simple, but can be found from the grand potential given by \cite{blundell}
\begin{equation}
\Phi=-\frac{1}{\beta}\ln Z, \label{eq:F_Z}
\end{equation}
where $Z$ is the grand canonical partition function. For fermions this
is just given by
\begin{equation}
Z=\sum_{s}e^{-\beta\left(E_{s}-\mu N_{s}\right)}=\sum_{\left\{ N_{i}\right\} }\prod_{i}e^{-\beta\left(\epsilon_{i}-\mu\right)N_{i}}
=\prod_{i}\left(1+e^{-\beta\left(\epsilon_{i}-\mu\right)}\right), \label{eq:Zg_fermions}
\end{equation}
where $s$ are the states of the whole system and we used $i$ to
label different one-particle states, $\epsilon_{i}$ and $N_{i}$
representing their energy and occupation number ($0$ or $1$) respectively,
and $\left\{ N_{i}\right\} $ represents a sum over all possible combinations
of $N_{i}$. Inserting Eq.~\eqref{eq:Zg_fermions} into Eq.~\eqref{eq:F_Z}
and then taking the continuous limit before applying Eq.~\eqref{eq:P_general} we finally find
\begin{equation}
P\left(T,\mu\right)=\frac{1}{\beta}\intop_{m}^{\Lambda}g\left(\epsilon\right)\ln\left[1+e^{-\beta\left(\epsilon-\mu\right)}\right]d\epsilon.\label{eq:f(T,mu)}
\end{equation}

So far, it looks as though all these results should be highly dependent
on the specific form of $g\left(\epsilon\right)$. The reason this
is not true is because we can relate results at positive and negative
temperatures using the well-known symmetry of the Fermi-Dirac distribution:
\begin{equation}
{\cal N}\left(\epsilon;T,\mu\right)=\frac{1}{e^{\beta\left(\epsilon-\mu\right)}+1}=1-\frac{1}{e^{-\beta\left(\epsilon-\mu\right)}+1}
=1-{\cal N}\left(\epsilon;-T,\mu\right).\label{eq:hole_sym}
\end{equation}
This allows us to borrow the concept of \emph{holes} from solid state
physics.
A hole here is just a way to conceptualize the absence of
a particle in a given state as a quasi-particle of negative energy
in a positive energy ``vacuum''. This just means that it is as valid
to describe our system in terms of which states are occupied by particles
as in terms of which states are unoccupied. For us it is particularly
useful in the limit where most particles are occupying the highest-energy
states (which correspond to $T<0$), since this can be thought of
as the limit where holes are populating the lower-energy states (corresponding
to $T>0$). Note that there exists a similar identity for the kind
of logarithmic term in the integral in Eq.~\eqref{eq:f(T,mu)},
\begin{equation}
\ln\left[1+e^{-\beta\left(\epsilon-\mu\right)}\right]=-\beta\left(\epsilon-\mu\right)+\ln\left[1+e^{\beta\left(\epsilon-\mu\right)}\right].\label{eq:log_property}
\end{equation}

It is now easy to combine Eqs.~\eqref{eq:hole_sym} and \eqref{eq:log_property}
with Eqs.~\eqref{eq:rho_general}, \eqref{eq:n_general}, and \eqref{eq:f(T,mu)}, yielding
\begin{equation}
\rho\left(T,\mu\right)=\rho_{\mathrm{max}}-\rho\left(-T,\mu\right)\label{eq:rho_holes}
\end{equation}
\begin{equation}
n\left(T,\mu\right)=n_{\mathrm{max}}-n\left(-T,\mu\right)\label{eq:n_holes}
\end{equation}
\begin{equation}
P\left(T,\mu\right)=-\rho_{\mathrm{max}}+\mu n_{\mathrm{max}}-P\left(-T,\mu\right).\label{eq:P_holes}
\end{equation}
These functions depend on very few
parameters from the fundamental theory as long as holes are at ``low''
temperatures (which here just means low enough that we know how physics
works at those temperatures). If $\mu=0$, as will be the case in
most relevant scenarios in this paper, the pressure has an even simpler
form%
\footnote{It is interesting to notice that this seemingly surprising relation still makes sense physically. Since (if $\mu=0$) $P=-\left(\frac{\partial U}{\partial V}\right)_{S}$, in a situation in which all single-particle states are filled the entropy is zero, and keeping the entropy constant as $V$ varies corresponds to always keeping all states filled, yielding $U=\rho_{\mathrm{max}}V$ and
thus $P=-\rho_{\mathrm{max}}$. If not all states are filled, then it makes sense to think of holes as negative momentum particles that contribute negatively to the total pressure as in Eq.~\eqref{eq:P_holes_mu0}.}:
\begin{equation}
P\left(T,\mu=0\right)=-\rho_{\mathrm{max}}-P\left(-T,\mu=0\right)\label{eq:P_holes_mu0}
\end{equation}
(note that only in this case can we be sure that a barotropic fluid
at $T>0$ will correspond to a barotropic fluid at $T<0$). Note also
the useful symmetry
\begin{equation}
\rho\left(T,\mu=0\right)+P\left(T,\mu=0\right)=
-\rho\left(-T,\mu=0\right)-P\left(-T,\mu=0\right).\label{eq:rho+P_sym}
\end{equation}

If, in addition to having $\mu=0$ and $T<0$, we have holes behaving
like cold matter (corresponding to $m\gg-T$), the quantity $\rho + P$ and the equation of state parameter $w\equiv P/\rho$
are given by
\begin{equation}
\rho + P= \rho -\rho_{\rm{max}}<0,\ \ \ \ \ \ w=-\frac{\rho_{\mathrm{max}}}{\rho}<-1,\label{eq:w_mu0_matter}
\end{equation}
whereas if they behave like radiation (the opposite limit)
\begin{equation}
\rho + P= \frac{4}{3}\left(\rho -\rho_{\rm{max}}\right)<0,\ \ \ \ \ \ w=-\frac{1}{3}\left(4\frac{\rho_{\mathrm{max}}}{\rho}-1\right)<-1.\label{eq:w_mu0_radiation}
\end{equation}

Alternatively, it can be interesting to consider the high $|T|$ region separating $T<0$ and $T>0$,
where $\left|\beta\Lambda\right|\ll1$.
Then, just looking at the limit when $\beta\rightarrow0$ yields
(from Eqs.~\eqref{eq:rho_general} and \eqref{eq:f(T,mu)}),
to leading order in $\beta$ and still assuming $\mu=0$,
\begin{equation}
\begin{cases}
\rho=\frac{1}{2}\rho_{\mathrm{max}}-\frac{\left\langle \epsilon^{2}\right\rangle _{0}}{4}\beta\\
P=\frac{\ln2}{\beta}n_{\mathrm{max}}-\frac{1}{2}\rho_{\mathrm{max}}+\frac{\left\langle \epsilon^{2}\right\rangle _{0}}{8}\beta
\end{cases},\label{eq:mu_T_inf}
\end{equation}
where
\begin{equation}
\left\langle \epsilon^{n}\right\rangle _{0}\equiv\intop_{m}^{\Lambda}\epsilon^{n}g\left(\epsilon\right)d\epsilon.\label{eq:avg_eps}
\end{equation}

\begin{figure}
\includegraphics[scale=0.30]{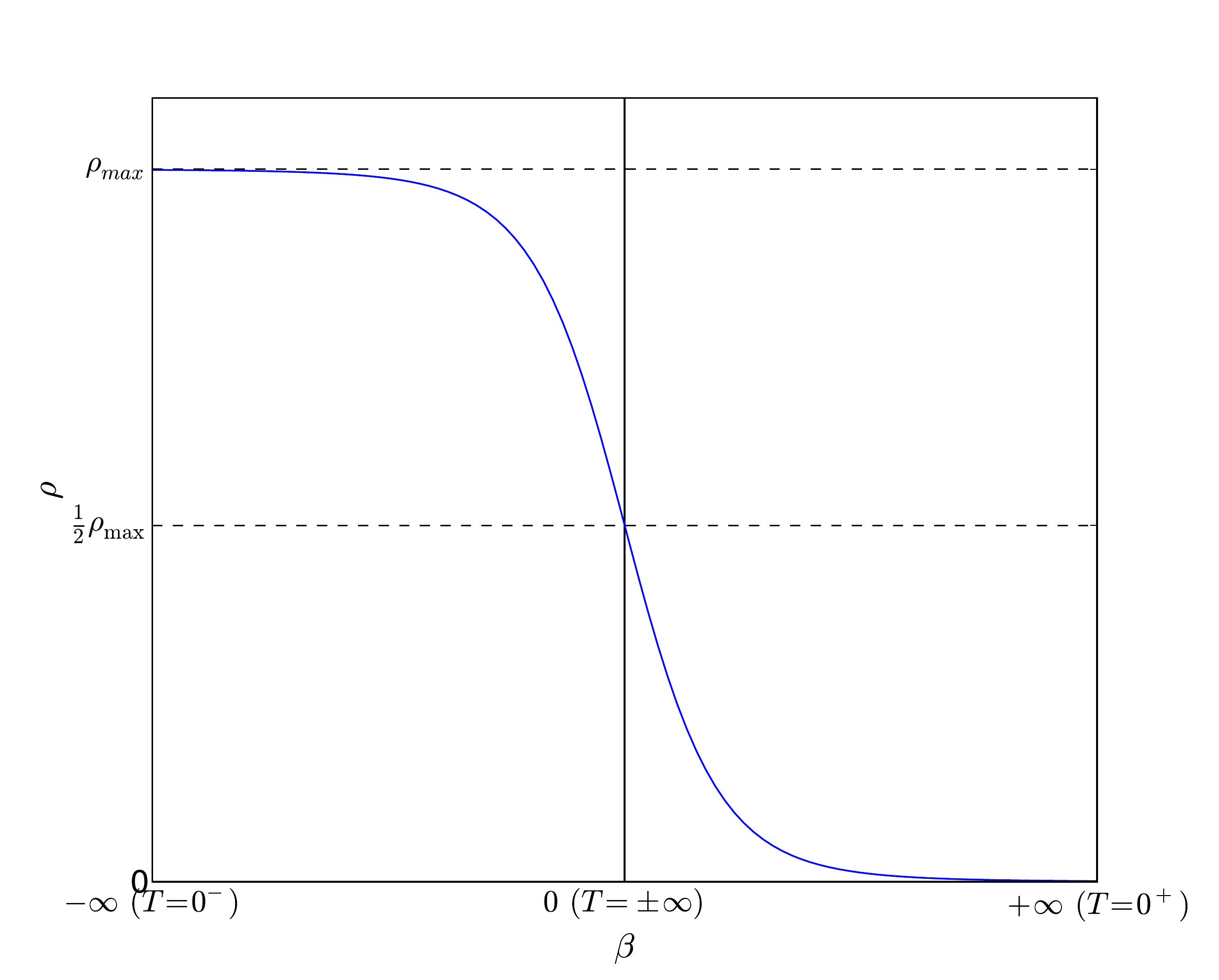}\includegraphics[scale=0.30]{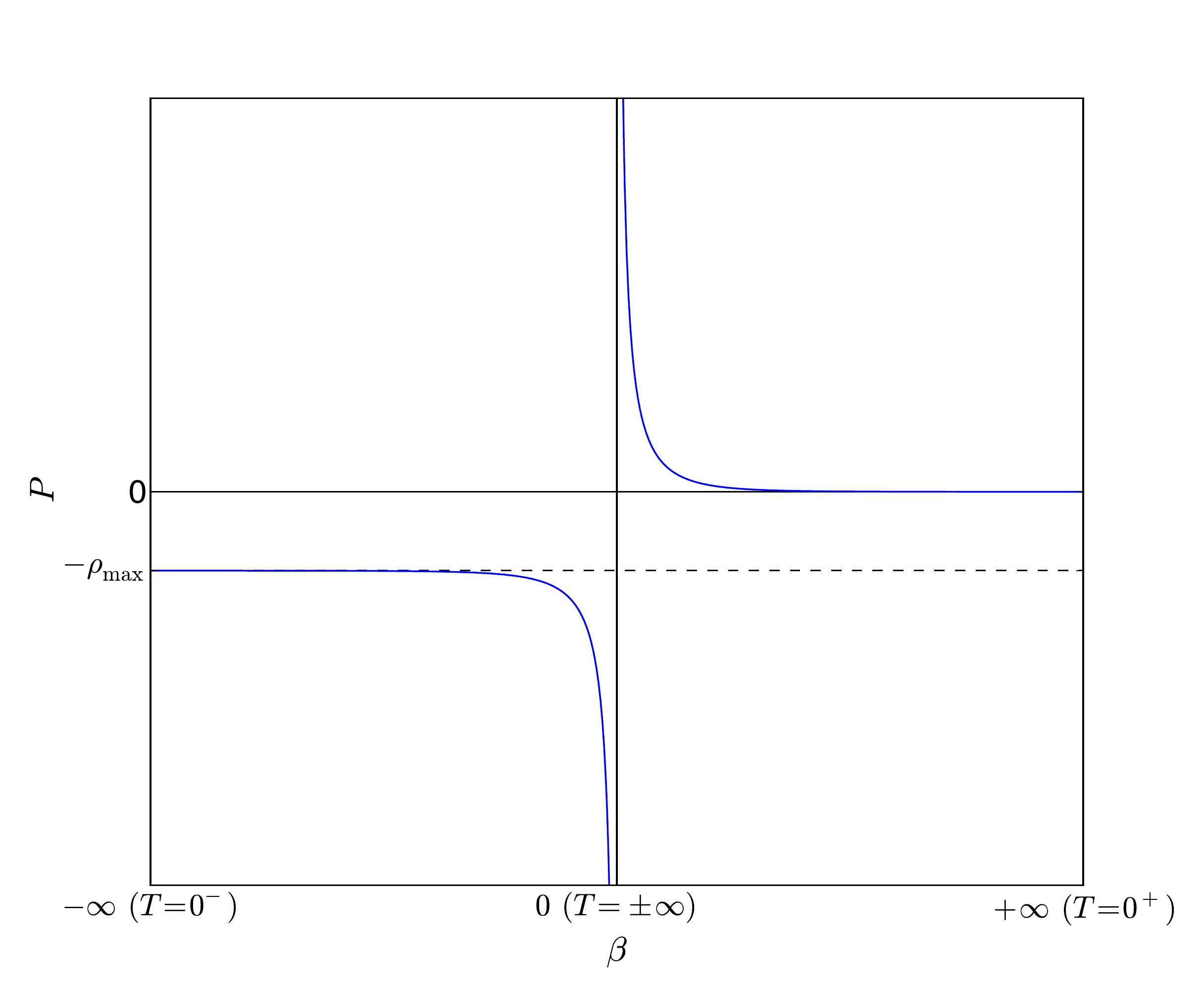}

\caption{\label{fig:Energy-density-and}Energy density and pressure as functions
of $\beta$ for a massless fermion with $\mu=0$ and $g\left(\epsilon\right)$
given by Eq.~\eqref{eq:DOS_Lambda}.
}
\end{figure}

Notice that thanks to this we can know that the energy density and pressure profiles
have to look like those in Fig.~\ref{fig:Energy-density-and} (except for intermediate
values of $\beta$).

%
%

\section{Cosmology\label{sec:NATive-Cosmology}}

We are finally ready to investigate the cosmological consequences
of NAT. In this section we answer the question ``How does a Universe
at negative absolute temperature behave?''. In order to answer
this, and motivated by our analysis so far, we first assume that an FLRW
Universe is filled by a single perfect fluid in thermal equilibrium at NAT,
 and that this fluid is made up of fermions not subject to number
conservation (which we shall refer to as \emph{temperons}).
The requirement of thermal equilibrium can probably be translated
into a requirement for temperon-producing interactions to operate quickly compared to the Hubble time.
We do not consider scenarios with number conservation and/or bosons
because those entail additional (model-dependent) problems discussed in Appendix \ref{sec:The-problem-of-mu}\footnote{Note that, even if those problems can be overcome, the cosmological relevance of temperons subject to number conservation is reduced by the fact that they cannot play an important role in the dynamics of an expanding Universe for more than a few e-foldings due to their quick dilution (although they might play a role in a contracting or bouncing scenario).}. We further assume that at ``low'' energy scales
these temperons should behave like all other known particles; i.e.,
like matter or radiation, depending on their mass.

New physics giving rise to the maximum energy cut-off required for NAT could produce new dynamics when many particles have energies close to the cut-off.
However, to make progress, here we simply assume that general relativity still holds at the relevant macroscopic scales so that
the dynamics of the NATive Universe will then be governed by the usual Friedmann equations
\begin{equation}
\begin{cases}
3H^{2}=\rho\\
\dot{H}=-\frac{1}{2}\left(\rho+P\right)
\end{cases},\label{eq:Friedmann_eq_Lambda0_k0}
\end{equation}
where $\rho$ and $P$ will be calculated according to the process
outlined in section \ref{sec:Thermodynamical-functions}. The energy conservation equation,
\begin{equation}
\dot{\rho}=-3H\left(\rho + P\right),\label{eq:rhocons}
\end{equation}
can also be integrated to give a useful relation between the number of e-foldings the Universe has
expanded (or contracted) and its initial and final energy densities:
\begin{equation}
N=-\frac{1}{3}\intop_{\rho_{i}}^{\rho_{f}}\frac{d\rho}{\rho+P}=-\frac{1}{3}\intop_{\rho_{i}}^{\rho_{f}}\frac{d\rho/\rho}{1+w},\label{eq:DN_rho}
\end{equation}
where, as usual, $N\equiv\ln\frac{a_{f}}{a_{i}}$ and subscripts $i$
and $f$ denote ``initial'' and ``final'', respectively.

In the first two subsections of this section we shall focus on analysing
the background dynamics of two qualitatively different scenarios:
NAT in expanding cosmologies (subsection \ref{sub:NATive-inflation}),
and NAT in contracting cosmologies (subsection \ref{sub:NATive-bounces}).
The remainder of this section will then be dedicated
to discussing perturbation generation and the transition to a normal positive-temperature
Universe.

\subsection{NATive inflation\label{sub:NATive-inflation}}

It can be easily seen from Eq.~\eqref{eq:rhocons} that expanding cosmological solutions with negative temperature ($H>0$, $T<0$) have an attractor fixed point
at $T=0^{-}$, corresponding to de Sitter expansion with $\rho=\rho_{\mathrm{max}}$ and $w=-1$.
This has the interesting consequence that
all expanding NATive Universes should tend towards a phase of exactly
exponential inflation (although not necessarily reaching it)%
\footnote{This property suggests it might be possible to explain the
accelerated expansion we measure today with a \emph{dark temperon}
component. Unfortunately, any such mechanism would have to rely on
a very low energy cut-off, and one would have to explain why this
dark temperon behaves so differently from every other particle at
that energy (we would expect
$\rho_{\mathrm{max}}\sim\rho_{\Lambda0}\simeq10.5h^{2}\Omega_{\Lambda0}\mathrm{GeVm^{-3}}$).%
} --- therefore, if this limit is reached, we should expect $\rho_{\rm{max}}\lesssim\rm{10^{111}GeVm^{-3}}$
just from the fact that we have not seen primordial tensor modes.

Interestingly, unlike with most phantom inflation models (recall that
Eq.~\eqref{eq:P_mu0} implies our expansion must either be phantom or
exactly exponential), we do not have to worry about a \emph{Big Rip}
--- a divergence of the scale factor in a finite interval of time \cite{BigRip}.
This is simply because the energy density (and therefore $H$) here is
bounded, so the evolution asymptotes to exponential expansion with constant density sufficiently quickly
that the impact of the transient phantom period is small.

We start our quantitative analysis by showing that even if we begin
very close to $T=-\infty$ we should expect to evolve towards the
vicinity of $T=0^{-}$ extremely rapidly. If we are in the high $|T|$ regime where $|\beta\Lambda|\ll 1$ then, from
Eqs.~\eqref{eq:mu_T_inf} and \eqref{eq:DN_rho}, the number of e-foldings
between two densities while in this regime is
\begin{multline}
N\approx\frac{4/3}{\left(\ln 2\right)  n_{\mathrm{max}}\left\langle \epsilon^{2}\right\rangle _{0}}\intop_{\rho_{i}}^{\rho_{f}}\left(\rho-\frac{1}{2}\rho_{\mathrm{max}}\right)d\rho
\\
=\frac{2/3}{\left(\ln 2\right) n_{\mathrm{max}}\left\langle \epsilon^{2}\right\rangle _{0}}\left[\left(\rho_{f}-\frac{1}{2}\rho_{\mathrm{max}}\right)^{2}-\left(\rho_{i}-\frac{1}{2}\rho_{\mathrm{max}}\right)^{2}\right]
\\
\approx \frac{\left\langle \epsilon^{2}\right\rangle _{0}}{n_{\mathrm{max}}}\frac{\beta_{f}^{2}-\beta_{i}^{2}}{24\ln 2},\label{eq:DN_T_inf}
\end{multline}
where we have used
\begin{equation}
\beta\simeq\frac{4}{\left\langle \epsilon^{2}\right\rangle _{0}}\left(\frac{1}{2}\rho_{\mathrm{max}}-\rho\right).\label{eq:beta_zero_rho}
\end{equation}

In order to get some intuition regarding the order of magnitude we
should expect from this $N$, we can assume the simple ansatz from
Eq.~\eqref{eq:DOS_Lambda} with $m=0$ (the order of magnitude should
not change significantly as long as $m\ll\Lambda$) and find
\begin{equation}
\begin{array}{c}
n_{\mathrm{max}}=\frac{g}{6\pi^{2}}\Lambda^{3}\\
\rho_{\mathrm{max}}=\frac{g}{8\pi^{2}}\Lambda^{4}\\
\left\langle \epsilon^{2}\right\rangle _{0}=\frac{g}{10\pi^{2}}\Lambda^{5}
\end{array}\label{eq:n_rho_eps2_max}
\end{equation}
leading to
\begin{equation}
N=\mathcal{O}\left\{ \frac{\left\langle \epsilon^{2}\right\rangle _{0}\beta_{f}^{2}}{n_{\mathrm{max}}}\right\} =\mathcal{O}\left\{ \left(\beta_{f}\Lambda\right)^{2}\right\}, \label{eq:order_N}
\end{equation}
which is small by definition. Therefore, we should not expect to remain in this low-$\left|\beta\right|$
regime long enough for this epoch to significantly contribute to the total number of e-foldings.

Once $-\beta$ becomes comparable to $\Lambda$ it is harder to make predictions as the specific shape of $g\left(\epsilon\right)$
we are working with starts to make a difference. In other words, as $-\beta$ increases,
we start needing more and more higher-order terms in the expansion in Eq.~\eqref{eq:mu_T_inf}
which makes model-independent predictions impossible. Nevertheless, we know
$\beta$ will have to keep evolving towards $-\infty$ and, sooner or later,
we will be in the opposite limit where $-T\ll\Lambda$ and we can make use of the fact that
holes should behave like either matter or radiation.

If holes behave like matter then
\begin{equation}
N=\frac{1}{3}\intop_{\rho_{i}}^{\rho_{f}}\frac{d\rho}{\rho_{\mathrm{max}}-\rho}
=-\frac{1}{3}\intop_{\rho_{\mathrm{max}}-\rho_{i}}^{\rho_{\mathrm{max}}-\rho_{f}}\frac{dx}{x}
=\frac{1}{3}\ln\left[\frac{\rho_{\mathrm{max}}-\rho_{i}}{\rho_{\mathrm{max}}-\rho_{f}}\right]=\frac{1}{3}\ln\left[\frac{1+w_{i}^{-1}}{1+w_{f}^{-1}}\right]\label{eq:DN_matterhole}
\end{equation}
where $w_{i}$ and $w_{f}$ are the initial and final $w$, respectively.

If instead holes behave like radiation then
\begin{equation}
N=\frac{1}{4}\intop_{\rho_{i}}^{\rho_{f}}\frac{d\rho}{\rho_{\mathrm{max}}-\rho}
=\frac{1}{4}\ln\left[\frac{\rho_{\mathrm{max}}-\rho_{i}}{\rho_{\mathrm{max}}-\rho_{f}}\right]
=\frac{1}{4}\ln\left[\frac{\left(1+w_{i}\right)\left(1-3w_{f}\right)}{\left(1+w_{f}\right)\left(1-3w_{i}\right)}\right]\label{eq:DN_radiationhole}
\end{equation}
with essentially the same type of behaviour.

Notice that the density becomes exponentially close to $\rho_{\rm{max}}$ in just a few e-foldings, since Eq.~\eqref{eq:DN_radiationhole} implies that
\begin{equation}
\rho_f  = \rho_{\mathrm{max}} - \left(\rho_{\mathrm{max}}-\rho_{i}\right)e^{-4N}
\end{equation}
and $\rho_i = \mathcal{O}(\rho_{\mathrm{max}}/2)$. An analogous result holds for Eq.~\eqref{eq:DN_matterhole}).

In addition, note that if we compute the adiabatic sound speed
\begin{equation}
c_{s}^{2}\equiv \frac{\dot{P}}{\dot{\rho}},\label{eq:sound_speed}
\end{equation}
we have
\begin{equation}
c_{s}^{2}=\begin{cases}
1+4\frac{\ln2}{\beta^{2}}\frac{n_{\mathrm{max}}}{\left\langle \epsilon^{2}\right\rangle _{0}}=\mathcal{O}\left\{ \frac{1}{\left(\beta\Lambda\right)^{2}}\right\} \gg1 & \text{if}\ \left|\beta\Lambda\right|\ll1\\
0  \qquad\text{if holes behave like matter}\\
\frac{1}{3}  \qquad\text{if holes behave like radiation}
\end{cases}\label{eq:sound_speed_limits}
\end{equation}
which shows that the sound speed only seems to be problematically
large in the very high (negative) temperature regime which should only
be valid at most during a very short time interval.

\subsection{NATive bouncing Universe\label{sub:NATive-bounces}}

Let us now turn our attention to a scenario where the Universe is
contracting (i.e. $H<0$) and, normally, there would be a Big Crunch.
For simplicity, we shall assume a spatially flat Universe (in the end
we should expect the same type of qualitative evolution).

With an energy cut-off, a fermion component cannot be indefinitely
compressed due to the Pauli exclusion principle. So either the fermions have to be destroyed
as the universe collapses, or the contraction has to stop, preventing a Big Crunch (or there is new physics).
If $w=-1$ exactly,
so that $\rho=\rho_{\mathrm{max}}$ and contraction does not change the temperon energy density,
we have the situation where fermions are destroyed at just the right rate for exponential contraction to continue indefinitely.
However, in other cases we can hope for a bounce.

An expanding Universe tends towards $T=0^{\pm}$ (depending
on the initial sign of $T$), but in the contracting case it should tend
towards $\beta=0$%
\footnote{Note that an interesting consequence of this fact is that the mere
existence of the energy cut-off will lead to exotic cosmological dynamics
due to ``excess'' positive pressure (in particular, as we shall
see, possibly preventing a Big Crunch) even if the Universe
is at a positive temperature all the time.%
}.
This is because the energy conservation equation forces $\dot{\rho}$
to have the same sign as $T$ and to be proportional to $-\frac{H}{\beta}$ once $\left|\beta\right|$
becomes sufficiently small. This causes $\rho$
to approach $\frac{1}{2}\rho_{\mathrm{max}}$, corresponding to $\beta=0$
(recall that $\rho+P$ must change sign at that point).
At some point, then, the small $\beta$ approximation must become
valid and we can follow the evolution of $H$ analytically%
\footnote{If one simply wishes to verify it is not possible to contract forever
in this regime it suffices to take a look at Eq.~\eqref{eq:DN_T_inf}
(for which the sign of $H$ makes no difference) and confirm that
$N$ is bounded.%
}.
Note also that the dynamics of this system should not
change appreciably even in the presence of other (normal)
types of matter. This is because the NATive pressure singularity (which occurs for finite $a$) should dominate
the Friedmann equations even if the energy density of temperons is subdominant
(as for "normal" matter $\rho + P$ can only diverge when $a=0$).

Combining Eqs.~\eqref{eq:Friedmann_eq_Lambda0_k0} and \eqref{eq:beta_zero_rho},
we can find a relation for the temperature as a function of $H^{2}$
\begin{equation}
\beta=\frac{2\rho_{\mathrm{max}}-12H^{2}}{\left\langle \epsilon^{2}\right\rangle _{0}}.\label{eq:beta_H_smallbeta}
\end{equation}
Using this we can write
\begin{equation}
\frac{dH}{dt}=-\frac{\ln2}{2}n_{\mathrm{max}}\frac{\left\langle \epsilon^{2}\right\rangle _{0}}{2\rho_{\mathrm{max}}-12H^{2}},\label{eq:dH_smallbeta}
\end{equation}
which can be integrated to yield
\begin{equation}
-2\left(H^{3}-H_{i}^{3}\right)+\rho_{\mathrm{max}}\left(H-H_{i}\right)+\frac{\ln2}{4}n_{\mathrm{max}}\left\langle \epsilon^{2}\right\rangle _{0}\left(t-t_{i}\right)
=0.\label{eq:3rdorder_smallbeta}
\end{equation}
This encodes the evolution of $H\left(t\right)$ in a cubic equation; it has a well-known set of solutions, however it is easier
to understand what happens next graphically.

\begin{figure}[h]
\begin{center}
\includegraphics[width=10cm]{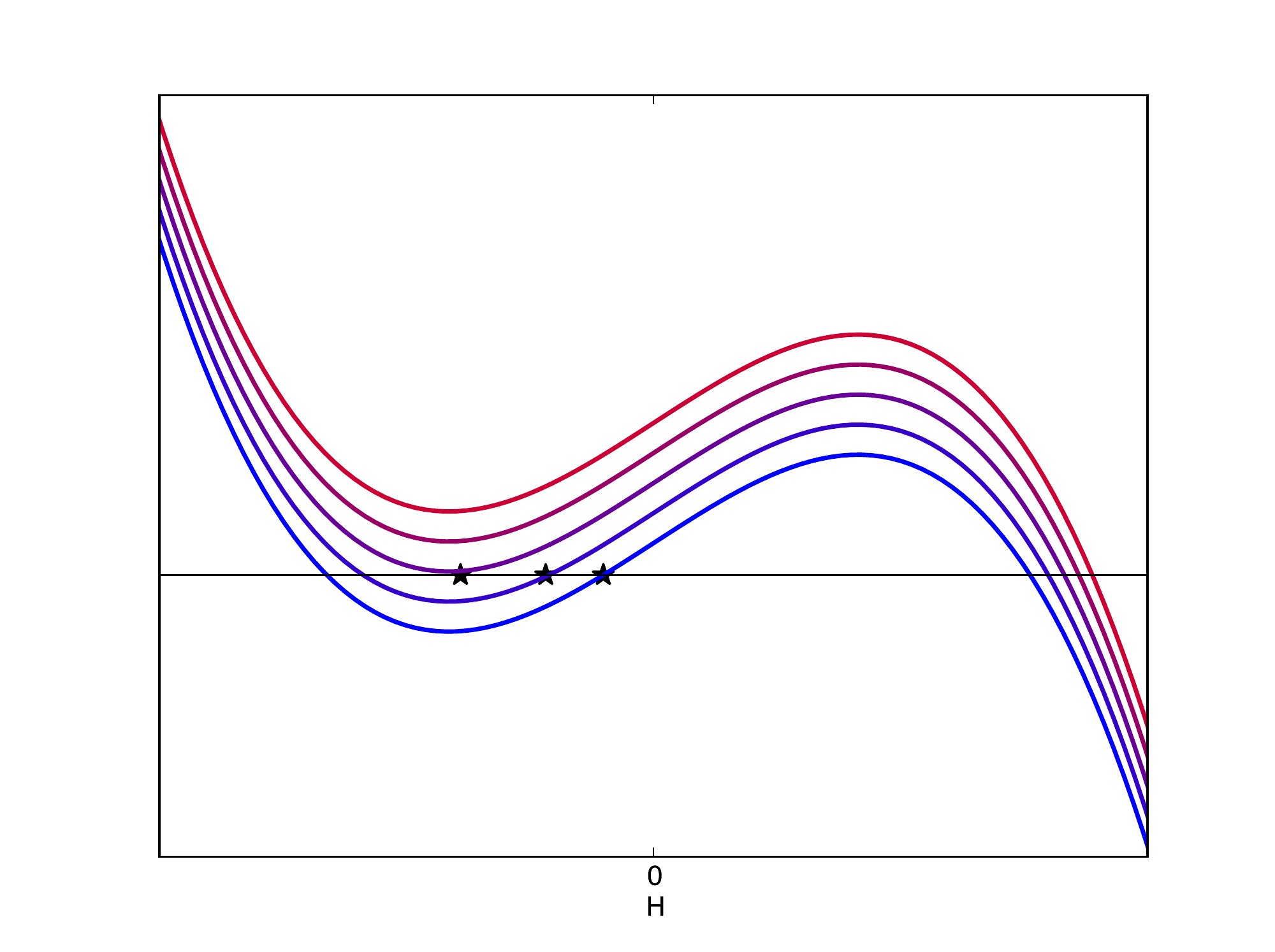}
\end{center}
\caption{A graphic representation of Eq.~\eqref{eq:3rdorder_smallbeta}
for increasing values of $t$ starting from $t_{i}$ (increasing from
blue to red and bottom to top). The physical value of $H$ (when it exists)
is indicated by a black star. The proportions between $\left\langle \epsilon^{2}\right\rangle _{0}$,
$n_{\mathrm{max}}$, and $\rho_{\mathrm{max}}$ here correspond to those in Eqs.~\eqref{eq:n_rho_eps2_max},
however, it can be shown that a different scenario would look qualitatively
the same.
\label{fig:polynomial_H}}
\end{figure}

From Eq.~\eqref{eq:3rdorder_smallbeta} we can see that $H$ at
a given time is given by a root of a third order polynomial whose
zeroth order coefficient is proportional to $t-t_{i}$ (see Fig.~\ref{fig:polynomial_H}).
At time $t=t_{i}$ there are three such roots, the physical solution
corresponding to the middle one ($H=H_{i}$), which must be followed
by continuity until the moment the temperature (and thus $\dot{H}$)
becomes infinite (when $H^{2}=\frac{1}{6}\rho_{\mathrm{max}}$, meaning $\beta=0$).
At that point, the root we were following disappears and there is
no physically meaningful solution to Eq.~\eqref{eq:3rdorder_smallbeta}%
\footnote{Note that we are not entitled to then follow the remaining root, as
it always corresponds (at this time) to $\rho=2\rho_{\mathrm{max}}$, which
is clearly physically impossible.%
} --- which is not surprising since our formula for the pressure yields a division by zero at this point.
Given that our equations are clearly invalid, we have to resort to physical arguments
in order to know what must happen next.
If we impose that the energy density is continuous and the thermal
equilibrium assumption remains valid then $H$ must
change sign discontinuously causing a bounce.
However, since the pressure is discontinuous
at that point, this is still not enough to determine the subsequent
cosmological evolution. Both a scenario with $\dot{\beta}<0$
 leading to the kind of NATive inflation discussed in subsection \ref{sub:NATive-inflation}
and a scenario with $\dot{\beta}>0$ leading immediately to a ``normal''
expanding Universe seem possible.
The discontinuity in $H$ is likely to be an indication that
our approach is not valid at the moment of the bounce.
Nevertheless, it is not unreasonable
to assume that thermal equilibrium should be restored relatively quickly
after the bounce, leading to one of these two options.

As in the previous section, a contracting Universe can solve the horizon problem.
 In this case, the mechanism would be essentially the same as in most other bouncing Universe models:
 homogeneity would be brought about by a large positive pressure
acting during a cosmological contraction.
In order to solve this
problem, bouncing cosmology models need to allow the quantity
\begin{equation}
\mathcal{N}_{H}\equiv\ln\left|aH\right|\label{eq:ekpyrefolds}
\end{equation}
to grow by a factor of order $60$ \cite{run_bounce}. This seems
to be achieved as long as the contraction starts at sufficiently small
$H$. For example, assuming a matter or radiation dominated Universe
at the beginning of the contraction,
\begin{equation}
\mathcal{N}_{H}\left(t\right)-\mathcal{N}_{H}\left(t_{i}\right)=\left(\frac{2}{3\left(1+w_{0}\right)}-1\right)\ln\left|\frac{H_{i}}{H\left(t\right)}\right|\label{eq:efolds_w_ekpyrosis}
\end{equation}
where $w_{0}=1/3$ if radiation dominates.
Since the left-hand side of Eq.~\eqref{eq:efolds_w_ekpyrosis}
is always negative and $H^{2}$ is increasing during the contraction,
it is always possible to get the right amount of contraction as long
as the initial energy density is low enough%
\footnote{Actually, one might raise the question of whether we are demanding
the initial energy density to be too low. Assuming that $\mathcal{N}\left(t_{\star}\right)-\mathcal{N}\left(t_{i}\right)\sim60$
and that $H\left(t_{\star}\right)$ is low enough that we can still
treat temperons as radiation, as is implicit in Eq.~\eqref{eq:efolds_w_ekpyrosis},
then $H_{i}/H_{\star}\sim e^{-120}$. In a flat Universe
this would correspond to $\rho_{i}/\rho_{\star}\sim10^{-104}$,
which is not a particularly small number if we keep in mind that if
$\rho_{\mathrm{max}}$ is of order $10^{111}\mathrm{GeVm^{-3}}$ (the maximum
order of magnitude for $\rho$ during inflation from tensor modes
constraints) then the ratio between the critical energy density today
and $\rho_{\mathrm{max}}$ is $\sim10^{-110}$. Moreover, Eq.~\eqref{eq:efolds_w_ekpyrosis}
should underestimate ${\cal N}$ since at very late times
a correct formula should account for the diverging increase in positive
pressure.%
}.

If, instead, the Universe is initially at a very low negative temperature
(let us assume, for simplicity, that holes behave like radiation), then
\begin{equation}
\mathcal{N}_{H}\left(t\right)-\mathcal{N}_{H}\left(t_{i}\right)=\frac{1}{4}\ln\left|\frac{\rho^{2}\left(t\right)\left(\rho_{\rm{max}}-\rho\left(t\right)\right)}{\rho^{2}_{i}\left(\rho_{\rm{max}}-\rho_{i}\right)}\right|\label{eq:efolds_nat_ekpyrosis}
\end{equation}
which can also be as large as needed provided that the initial hole density is small enough.

This would mean the NAT themselves would not really contribute to
solving the horizon problem (though the extra positive
pressure close to the bounce would help). In fact, NAT might not even
occur in this scenario --- it is enough for temperons to force a bounce
in a model that would otherwise still solve these problems but end
in a Big Crunch.

\subsection{Perturbations\label{sub:A-NATive-Curvaton}}

If NATive models are to be taken as realistic candidates to realise
inflation or bouncing cosmologies, then a complete study of perturbation
generation will be necessary. One of the main successes of standard
inflation is how easy it is to write down a model which yields a nearly
scale-invariant spectrum of scalar perturbations (which is in excellent
agreement with CMB observations).
One might think that the nearly-exponential expansion of a NAT fluid also would give a
scale invariant spectrum of thermal fluctuations. However an exactly de Sitter phase produces no density fluctuations,
and this limit is rapidly approached. Moreover, $\rho+P$ goes to zero sufficiently fast that
curvature perturbations rapidly increase with time, leading to a blue spectrum
until the de Sitter limit is saturated (see Appendix \ref{sec:Thermal-Perturbation-Generation} for details).
Thus pure NATive inflation cannot be a realistic model for the early Universe.

Instead we can consider a simple scenario with a spectator field that has negligible effect on the background evolution.
Suppose that besides the temperons
there exists a canonical scalar field $\sigma$ whose potential $V\left(\sigma\right)$
is much smaller than the temperon energy density (and does not interact with temperons).
Since the background evolution is almost unchanged, the temperon density will still tend
towards its maximum possible value with  $P_{t}=-\rho_{t}=-\rho_{\mathrm{max}}$,
and its contributions to the Friedmann equations will quickly become
constant. The evolution is then the same as we would have for
just a canonical scalar field with potential $V_{\rm eff}\left(\sigma\right)=V\left(\sigma\right)+\rho_{\mathrm{max}}$.

We can also look at the curvature perturbation produced in the same limit.
Using the fact that the density
perturbation due to temperons should tend to zero (since $\delta\rho_{t}=-\delta\rho_{\rm holes}$
and there are no holes at $T=0$), we have the interesting result that in the flat slicing
\begin{equation}
\zeta=-H\frac{\delta\rho}{\dot{\rho}}\rightarrow -H\frac{\delta\rho_{\sigma}}{\dot{\rho}_{\sigma}}\equiv\zeta_{\sigma},\label{eq:zeta_def-1}
\end{equation}
where $\zeta_{\sigma}$ is the curvature perturbation we would get
from the same scalar field (with potential $V_{\rm eff}\left(\sigma\right)=V\left(\sigma\right)+\rho_{\mathrm{max}}$).
However, since the spectator field has (by construction) negligible density, this would not significantly contribute
to an observable curvature perturbation if the dominant uniform temperon density somehow decays to give a radiation dominated universe. Instead,  the spectator field fluctuation would either have to become dynamically important after temperon decay or
somehow modulate the decay process. We discuss this further at the end of the next section.

\subsection{Ending NATive inflation\label{sub:The-end-of-NAT}}

The analysis so far has focused mostly on the basic cosmological implications
of the possibility of domination by a temperon gas. Since an inflationary
and a bouncing Universe both seem to be naturally realized in this
sort of scenario, it is worth considering whether the transition from NATive inflation to
a normal positive-temperature Universe --- which we may call \emph{recooling},
by analogy with reheating --- can also happen naturally, giving rise
to the standard Hot Big Bang cosmology.

The main difficulty in an expanding universe is that we have shown that a NAT fluid rapidly tends to the stable attractor solution with constant density $\rho=\rho_{\mathrm{max}}$, so on its own there is no dynamical evolution that could naturally set a timescale for recooling. However, as with reheating, the process of recooling to a universe dominated by familiar content must require some level of interaction with normal particles, however indirect, so it is possible that additional degrees of freedom could be responsible for ending NATive inflation.

Note that for
$\rho > \frac{1}{2}\rho_{\mathrm{max}}$,
the energy conservation equation for the temperons has
$\dot{\rho_t} = -3H(\rho_t+P_t)  > 0$
(with singular negative pressure term at the
$\rho=\frac{1}{2}\rho_{\mathrm{max}}$
threshold between positive and negative temperatures), which prevents the temperon fluid from dynamically evolving to normal temperatures even if other components modify the background. The temperons also cannot be in equilibrium with normal matter (involving particles which do not admit negative temperatures), since equilibrium would be reached with both systems at a positive temperature,
regardless of how small the additional positive-temperature system might be \cite{NO_NEGT}. Any end to the NATive epoch must therefore involve an out of equilibrium process.

With this in mind, if we naively postulate that above a critical energy density
temperons can interact with bosons slightly and even decay into bosons
with some low probability, we should expect to recover a positive-temperature
Universe some time after that critical energy density is reached.
This whole process would necessarily take us away from
equilibrium, so the formalism we have been using is no longer
valid and it is not possible to
make model-independent predictions. It seems plausible that it should
be possible to get more e-foldings of inflation by forcing the
temperon-photon interaction to be weaker, at the possible expense of fine-tuning the interaction timescale to be
close to the Hubble time.
However we can see from
Eqs.~\eqref{eq:DN_matterhole} and \eqref{eq:DN_radiationhole} that we would not get
more than a few e-foldings of expansion in equilibrium
unless the critical energy density is also fine-tuned to be extremely close
to $\rho_{\mathrm{max}}$: if we wanted about $60$ e-foldings in
this regime we would need $1-\frac{\rho_{c}}{\rho_{\mathrm{max}}}\lesssim10^{-6}$.
Note also that out of equilibrium
the perturbation calculations from Appendix \ref{sec:Thermal-Perturbation-Generation}
would also not be applicable.

An added difficulty is how to calculate the effective pressure in a non-equilibrium setting.
Unfortunately, this requires calculating the pressure from
first principles, which is non-trivial and model-dependent - even in equilibrium.
The mechanical pressure is usually given by the standard formula
\begin{equation}
P_{\rm{mech}}=\int_m^\Lambda\left(\frac{\epsilon^{2}-m^{2}}{3\epsilon}\right)g\left(\epsilon\right){\cal{N\left(\epsilon\right)}} d\epsilon.
\label{eq:P_standard}
\end{equation}
However, we have been using the result of Eq.~\eqref{eq:f(T,mu)} (which assumes thermal equilibrium).
These are not equivalent in the presence of a cut-off, and are only equivalent in the limit
where $\Lambda\rightarrow\infty$ if $\beta>0$. This can be seen
using integration by parts and assuming the ansatz in Eq.~\eqref{eq:DOS_Lambda} as well as ${\cal{N}}=1/\left(e^{\beta \epsilon}+1\right)$, which yields
\begin{equation}
P=P_{{\rm mech}}+\frac{g}{2\pi^{2}}\frac{\left(\Lambda^{2}-m^{2}\right)^{3/2}}{3\beta}\ln\left(1+e^{-\beta\Lambda}\right)\label{eq:P_mech_th}
\end{equation}
for $\beta>0$. For negative temperatures the result instead follows Eq.~\eqref{eq:P_holes}.

It may seem a problem that Eq.~\eqref{eq:P_standard} does not work for NAT (and in particular fails to even allow
$P<0$). However, it was originally written down for ideal
gases of classical particles and, at these high energy (and momentum)
scales, close to the cut-off, there is no reason why that picture
should still be valid. It is interesting to note that the pressure in the Friedmann equations
should always coincide with that given by Eq.~\eqref{eq:f(T,mu)}.
This can be seen by noting that the first Friedmann equation is equivalent to the First Law of
Thermodynamics in the case of adiabatic expansion/contraction.

Although we do not know how to calculate these "microscopic" pressures, some hints are given by the work of \cite{phantomGUP},
who did something similar for the case of a scalar field, finding that there
was a negative correction to the pressure that made some normal inflation
models become phantom.
The main idea is to make use of the known fact \cite{discrete_heisenberg}
that in these theories there is a significant deviation from the canonical
commutation relation between the usual position and momentum operators,
implying that the usual momentum operator is no longer the conjugate
momentum of the position operator and invalidating the standard result.
In principle, it should be possible to rewrite the Lagrangian in terms
of the correct momentum operator and from that compute the corrections
to the standard energy-momentum tensor due to this deformed algebra.
This would then have the effect of adding corrections to
both the pressure and the energy density%
\footnote{The changes to the energy density being interpretable as differences
in the function $g\left(\epsilon\right)$ due to in one case it being
related to the deformed momentum operator and in another to the actual
eigenvalues of the correct Hamiltonian%
}, essentially solving our problem and enabling the accurate calculation
of non-equilibrium pressures. The pursuit of this approach is left for future work.
In principle, if it succeeds, it may help us understand what is required for
a complete microphysical description of these fluids (at the Lagrangian level). 

An alternative way to end NATive inflation would be to make use of a spectator field as
described in Sec.~\ref{sub:A-NATive-Curvaton}.
A natural way to do this might be for the scalar field to precipitate the end of
inflation, for example by having it decay into bosons
which then interact with the temperons, ending inflation by
full thermalisation. However, since the energy density of the spectator field should be subdominant,
this would probably require a sharp feature in the
potential to compensate the large Hubble damping from the background.

The curvature perturbation from the spectator field (Eq.~\eqref{eq:zeta_def-1}) will have a nearly
scale-invariant spectrum during temperon domination provided that it is light compared to the Hubble scale,
but it does not automatically give rise to a significant amplitude of the curvature perturbation after recooling. Note that with two independent components
$\zeta$ can change in time on superhorizon scales \cite{WANDS_SEPARATE}.
However, the temperon fluid should be very homogeneous, so the
recooling surface would be determined by the scalar
field perturbations $\delta\rho_{\sigma}$. The quasi-scale-invariant $\delta\rho_\sigma$ fluctuations can therefore convert into
local variation in the recooling time, and hence a total curvature perturbation (i.e.~essentially the same mechanism as the modulated
reheating mechanism for multi-field inflation \cite{inhom_reheating_dvali}).
Non-Gaussianities could also be introduced at this stage
analogously to similar scenarios in the context of inflation \cite{reviewNG_multif,local_NG_inflation}.
A specific model would be required to make quantitative predictions, in particular the perturbation amplitude is model dependent even if the fluctuation scale dependence is more generally preserved.

\section{Conclusions\label{sec:Conclusions}}

A fluid with negative temperature is an interesting effective macroscopic model for a component of the Universe, even without any compelling microphysical motivation.
Regardless of the microphysics, the evolution of a NAT fluid dominated cosmology will be qualitatively the
same and depend only on the initial value of the Hubble parameter
(as summarised in table \ref{tab:Fate-of-a-NATive-uni}). It might be an attractive way to realise both inflation and bouncing cosmologies.

\begin{table}[h]
\begin{centering}
\begin{tabular}{|c|c|c|}
\hline
 & $H<0$ & $H>0$\tabularnewline
\hline
\hline
$H^{2}<\frac{1}{6}\rho_{\mathrm{max}}$ & NATive bouncing & standard cosmology\tabularnewline
\hline
$H^{2}>\frac{1}{6}\rho_{\mathrm{max}}$ & NATive bouncing & NATive inflation\tabularnewline
\hline
\end{tabular}
\par\end{centering}

\caption{Fate of a temperon-dominated Universe depending on its initial conditions.\label{tab:Fate-of-a-NATive-uni}}
\end{table}

However, there are a number of significant problems with a naive application to cosmology:
\begin{itemize}
\item The physical plausibility of obtaining a maximum energy cut-off is unclear.
\item For a NAT description to apply, the system must remain in equilibrium. In an expanding universe the NAT fluid has to be able to produce more particles rapidly as the universe expands (but no bosons), and any microphysical model would have to explain why this happens rather than simply rapidly decoupling and going out of equilibrium.
\item In an expanding universe, the NAT component rapidly becomes indistinguishable from a cosmological constant at the background level, and it is therefore of limited interest for obtaining realistic dynamics that lead to the end of inflation.
\item An additional component, such as a light scalar spectator field, would be required to produce an acceptable fluctuation spectrum.
\item Any end of NATive inflation, or resolution of a bounce, requires non-equilibrium evolution that cannot be modelled in a model-independent way.\newline
\end{itemize}

Nevertheless, it must be noted that, as long as the first assumption (about the existence
of the energy cut off) holds, there will be interesting consequences even if the following problems cannot be overcome
and our formalism cannot be used most of the time. Even if thermalization at a NAT turns out to
be impossible, a universe with a cut-off would likely still lead to interesting dynamics (for example due to the pressure discontinuity at $T=+\infty$).
Moreover, even if this component turns out to be unable to provide an acceptable explanation to
horizon and flatness problems, it may still have interesting consequences in systems where it might be found
if it exists --- as in the interior of black holes.

The discussion of Appendix \ref{sec:Thermal-Perturbation-Generation} suggests some ways in which acceptable
fluctuations could be produced, though with additional ingredients and fine-tuning such models have limited appeal.

Above all, we have shown that there are interesting cosmological consequences
of NAT, and that it is possible that popular paradigms like inflation
and bouncing cosmologies may be successfully realised in scenarios
which are fundamentally different from the usual domination by simple
scalar fields.

\section{Acknowledgements}

We thank Mariusz D\k{a}browski for bringing the work of \cite{Phantom_thermo}
to our attention and Yi Wang for helpful comments on \cite{thermo_perturb}.
We also thank Carlos Martins, Daniel Passos, Daniela Saadeh, Frank K\"{o}nnig, Henryk
Nersisyan, Jonathan Braden, and Sam Young for fruitful discussions, insightful
questions, and other useful references.

JV is supported by an STFC studentship, CB is supported by a Royal
Society University Research Fellowship, and AL acknowledges support
from the Science and Technology Facilities Council {[}grant number
ST/L000652/1{]} and European Research Council under the European Union's Seventh Framework Programme (FP/2007-2013) / ERC Grant Agreement No. [616170].

\chapter{Models for Small-Scale Structure on Cosmic Strings: II. Scaling and its stability}
\label{chap:wiggly}

\begin{center}

J.P.P. Vieira,$^{1,2}$ C.J.A.P. Martins,$^{1,3}$ and E.P.S. Shellard$^{4}$\\[0.5cm]
$^{1}$Centro de Astrof\'{\i}sica, Universidade do Porto, 
4150-762 Porto, Portugal\\
$^{2}$Department of Physics \& Astronomy, University of Sussex, Brighton BN1 9QH, UK\\
$^{3}$Instituto de Astrof\'{\i}sica e Ci\^encias do Espa\c co, CAUP, 
4150-762 Porto, Portugal\\
$^{4}$Department of Applied Mathematics and Theoretical Physics, Centre for Mathematical Sciences, University of Cambridge, Wilberforce Road, Cambridge CB3 0WA, UK

\end{center}

\ \\

We make use of the formalism described in a previous paper\cite{PAP1} to address general features of wiggly cosmic string evolution. In particular, we highlight the important role played by poorly understood energy loss mechanisms and propose a simple ansatz which tackles this problem in the context of an extended velocity-dependent one-scale model. We find a general procedure to determine all the scaling solutions admitted by a specific string model and study their stability, enabling a detailed comparison with future numerical simulations. A simpler comparison with previous Goto-Nambu simulations supports earlier evidence that scaling is easier to achieve in the matter era than in the radiation era. In addition, we also find that the requirement that a scaling regime be stable seems to notably constrain the allowed range of energy loss parameters.

\newpage

\section{\label{sint}Introduction}

Vortex-lines or topological strings are ubiquitous in physical contexts, with perhaps the most interesting and well-studied examples being cosmic strings in the early universe and vortex-lines in superfluid helium. (For extensive reviews on the subject see \cite{VSH,COND1,COND2,GEYER}.) Their nonlinear nature and interactions imply that the detailed quantitative understanding of their properties and experimental or observational consequences is a significant challenge, which is compounded by the the complexity of evolving a full network. This is particularly topical given the recent availability of high-quality data which one may use to constrain these models, such as that of the Planck satellite \cite{Planck}. In the future, gravitational waves should become an additional observational window \cite{DamourVilenkin}.

A significant part of this effort must therefore be based on numerical simulations, but these are both technically difficult and very computationally costly \cite{BB,AS,FRAC,RSB,VVO,Stuckey,Blanco,Hiramatsu}. This is among the motivations for developing complementary analytic approaches, essentially abandoning the detailed \textit{statistical physics} of the string network to concentrate on its \textit{thermodynamics}. For the simplest Goto-Nambu string networks, which have been the subject of most studies so far, the velocity-dependent one-scale (VOS) model \cite{MS1,MS2,MS3,MS4} has been exhaustively studied, and its quantitative success has been extensively demonstrated by direct comparison with both field theory and Goto-Nambu numerical simulations \cite{FRAC,ABELIAN}. The model allows one to describe the scaling laws and large-scale properties of string networks in both cosmological and condensed matter settings with a minimal number of free parameters. More elaborate approaches have also been adopted \cite{ACK,POLR}, though usually at a cost of a larger number of free phenomenological parameters and/or (arguably) loss of intuitive clarity.

However, cosmologically realistic string networks are not expected to be of Goto-Nambu type. In particular, the previously mentioned simulations of cosmic strings in expanding universes have established beyond doubt the existence of a significant amount of short-wavelength propagation modes (commonly called \textit{wiggles}) on the strings, on scales that can be several orders of magnitude smaller than the correlation length. In a previous paper \cite{PAP1} we introduced a mathematical formalism suitable for the description of the evolution of both large-scale and small-scale properties of a cosmic string network in expanding space. In particular, we arrived at a complete set of equations which allows us to model the evolution of such important quantities as the characteristic length of the network, a characteristic velocity, and both the multifractal dimension and the effective energy per unit length of the strings. There the focus of the applications was on two simplified limits of physical relevance: the tensionless and the linear limit (the latter being especially appropriate for comparison with Abelian-Higgs network simulations). 

This paper continues the exploration of this formalism. After a brief overview of the main results of the first paper, we focus our attention on a general study of the scaling regimes allowed by this model, including their attractor behavior. These results will be illustrated for the case of a simple ansatz which naturally generalizes the energy loss mechanisms considered in the simpler one-scale-type models. Finally, we use our results to make a first comparison with previously existing numerical simulations. A more detailed comparison will require new simulations (both because additional diagnostics should be output and because a higher resolution would be desirable) and is left for subsequent work.

\section{\label{elast}Elastic String Evolution}

The VOS model \cite{PHD,MS2,MS1} is the simplest and most reliable method for calculating the evolution of the large-scale properties of a network of Goto-Nambu cosmic strings obeying the action
\begin{equation}
S=-\mu_{0}\int\sqrt{-\gamma}d^{2}\sigma\label{eq:GN_action}
\end{equation}
where $\sigma^{a}$ are the string worldsheet coordinates, $\gamma$ is the determinant of $\gamma^{ab}$, the pullback metric on the worldsheet, and $\mu_{0}$ is the string mass per unit length (equal to the local string tension) which is generally expected to be of the order of the square of the symmetry breaking scale associated with the formation of the strings. At the expense of assuming there is only one relevant length scale $L$ in the network (as in Kibble's one-scale model \cite{KIB}), this model allows us to make quantitative predictions about the evolution of the energy in the network $E$ as well as a RMS velocity $v$ defined by
\begin{equation}
E=\mu_{0}a\int\epsilon d\sigma\propto\frac{\mu_{0}a^{3}}{L^{2}},\,\quad v^{2}=\frac{\int\mathbf{\dot{x}}^{2}\epsilon d\sigma}{\int\epsilon d\sigma}\label{eq:L_v}
\end{equation}
where $a$ is the scale factor of an FLRW metric
\begin{equation}
ds^{2}=a^{2}\left(d\tau^{2}-d\mathbf{x}^{2}\right).\label{FRW}
\end{equation}
In particular, it is found that if the scale factor behaves as a power law of the form
\begin{equation}
a\propto t^{\lambda}\label{a_power}
\end{equation}
where $\lambda$ is a constant between $0$ and $1$, then there is an attractor \textit{scaling} regime defined by $L/t=const.$ and $v=const$.

Throughout this discussion, our aim is to emulate the success of the VOS model whilst taking into account the presence and evolution of small-scale structure (i.e., wiggles) in the network - to which we are 'blind' in the standard VOS model due to the one-scale approximation. This is achieved by considering that the dynamics of a wiggly Goto-Nambu string can be approximated by that of a smoother (i.e., with no significant structure at scales below $L$) elastic string which obeys the generalised action \cite{CARTERA}
\begin{equation}
S=-\mu_{0}\int\sqrt{-\gamma}\sqrt{1-\gamma^{ab}\phi_{,a}\phi_{,b}}d^{2}\sigma\label{wiggly_S}
\end{equation}
where $\phi$ is a scalar field whose associated current is regarded as a mass current resulting from the propagation of wiggles on the string.
 
Note that $\phi$ is an effective quantity which is related to an undefined renormalization procedure by which structure below some length scale $\ell$ is smoothed. Naturally, $\ell$ should be no greater than the string correlation length, but still large enough for the effective string energy per unit length (and $\phi$) to depend solely on the worldsheet time, at least in regions large enough for an eventual spatial dependence to be negligible in the local equations of motion.

\subsection{\label{quant}Basic properties}

Besides affecting the evolution of the string configuration, the presence of this mass current also changes the way some relevant quantities are defined on the string. 

Given the mesoscopic nature of $\phi$ we can simplify our equations by introducing the dimensionless quantity
\begin{equation}
w=\sqrt{1-\gamma^{ab}\phi_{,a}\phi_{,b}}\label{w}
\end{equation}
in terms of which the local string tension and energy density can be simply written as
\begin{equation}
T=\mu_{0}w,\,\quad U=\mu_{0}w^{-1}.\label{T_U_w}
\end{equation}

As in the VOS case, the coordinate energy per unit length along the string is given by
\begin{equation}
\epsilon=\sqrt{\frac{\mathbf{x^{\prime}}^2}{1-\mathbf{\dot{x}^{2}}}}.\label{epsilon}
\end{equation}
However, there are now two relevant independent energies which can be defined: the total energy in a piece of string
\begin{equation}
E=\mu_{0}a\int{\frac{\epsilon}{w}d\sigma}\label{E}
\end{equation}
and the energy in a Goto-Nambu string with the same configuration as our smoothed elastic string, called the bare energy,
\begin{equation}
E_{0}=\mu_{0}a\int{\epsilon d\sigma}.\label{E0}
\end{equation}

Since it is generally assumed that the basic VOS assumptions apply to the smoothed string, it is the bare energy that should be associated with the network correlation length via
\begin{equation}
\rho_{0}=\frac{\mu_{0}}{\xi^{2}}.\label{xi}
\end{equation}

Analogously, there are now two natural averaging procedures defined for a generic quantity $Q$ by
\begin{equation}
\left\langle Q\right\rangle =\frac{\int Q\frac{\epsilon}{w}d\sigma}{\int\frac{\epsilon}{w}d\sigma}\label{average}
\end{equation}
and
\begin{equation}
\left\langle Q\right\rangle _{0}=\frac{\int Q\epsilon d\sigma}{\int\epsilon d\sigma}\label{average0}
\end{equation}
the former appearing more naturally in our equations but the latter possibly being more convenient to use in applications when the wiggliness of a string is not well known. Note that, in an infinite string, the two procedures are equivalent if and only if $Q$ is independent of $w$ (i.e., $\left<Qw\right>=\left<Q\right>\left<w\right>$).

Finally, these concepts can be combined in the definition of the renormalized string mass per unit length factor
\begin{equation}
\mu \equiv \frac{E}{E_{0}}\equiv\frac{\xi^{2}}{L^{2}}=\left\langle w\right\rangle ^{-1}=\left\langle w^{-1}\right\rangle _{0}\label{mu}
\end{equation}
which is trivially at least unity ($\mu=1$ corresponding to the Goto-Nambu limit, when there is no small-scale structure) and quantifies the wiggliness of a network.

\subsection{\label{aver}Averaged evolution}

The system of equations which define the model introduced in the previous paper \cite{PAP1} can be found by using the equations of motion obtainable from the action given by Eq.~(\ref{wiggly_S}) together with the following phenomenological terms that model energy loss to loops as well as energy transfer from the bare to the wiggly component due to kink formation by intercommutation
\begin{equation}
\left(\frac{1}{\rho}\frac{d\rho}{dt}\right)_{loops}=-cf\left(\mu\right)\frac{v}{\xi}\label{f}
\end{equation}
\begin{equation}
\left(\frac{1}{\rho_{0}}\frac{d\rho_{0}}{dt}\right)_{loops}=-cf_{0}\left(\mu\right)\frac{v}{\xi}\label{f0}
\end{equation}
\begin{equation}
\left(\frac{1}{\rho_{0}}\frac{d\rho_{0}}{dt}\right)_{wiggles}=-cs\left(\mu\right)\frac{v}{\xi}\label{s}
\end{equation}
where $v\equiv \left\langle\mathbf{\dot{x}}^2\right\rangle$, $c$ is a constant of order unity which corresponds to the loop-chopping parameter of the VOS model, and $f$, $f_{0}$, and $s$ are functions of $\mu$ which are unity (in the case of $f$ and $f_{0}$) and zero (in the case of $s$) if $\mu=1$, lest we not recover the VOS model in the Goto-Nambu limit.

Apart from these energy loss mechanisms, it is important to take into account that varying the renormalization scale $\ell$ is tantamount to redefining what small-scale structure is, and thus must have an effect on the value of $E_{0}$ (as well as $v$ since $w$ is also changed). This can be done by introducing the following scale-drift terms
\begin{equation}
\frac{1}{\mu}\frac{\partial\mu}{\partial\ell}\frac{d\ell}{dt}\sim\frac{d_{m}-1}{\ell}\frac{d\ell}{dt}\label{dm}
\end{equation}
\begin{equation}
\frac{\partial v^{2}}{\partial\ell}\frac{d\ell}{dt}=\frac{1-v^{2}}{1+\left\langle w^{2}\right\rangle}\frac{\partial\left\langle w^{2}\right\rangle}{\partial\ell}\frac{d\ell}{dt}\label{v_dm}
\end{equation}
where $d_{m}\left(\ell\right)$ is the multifractal dimension of a string segment at scale $\ell$ \cite{TAKAYASU}. Note that Eq.~(\ref{dm}) is essentially just a geometric identity whereas Eq.~(\ref{v_dm}) comes from imposing total energy conservation across different scales.

If we further assume uniform wiggliness (i.e., $w$ to be just a function of time) then the system of equations we are looking for is just
\begin{equation}
2\frac{d\xi}{dt}=H\xi\left[2+\left(1+\frac{1}{\mu^{2}}\right)v^{2}\right]+v\left[k\left(1-\frac{1}{\mu^{2}}\right)+c\left(f_{0}+s\right)\right]
+\left[d_{m}\left(\ell\right)-1\right]\frac{\xi}{\ell}\frac{d\ell}{dt}\label{qui_full}
\end{equation}
\begin{equation}
\frac{dv}{dt}=\left(1-v^{2}\right)\left[\frac{k}{\xi\mu^{2}}-Hv\left(1+\frac{1}{\mu^{2}}\right)-\frac{1}{1+\mu^{2}}\frac{\left[d_{m}\left(\ell\right)-1\right]}{v\ell}\frac{d\ell}{dt}\right]\label{v_full}
\end{equation}
\begin{equation}
\frac{1}{\mu}\frac{d\mu}{dt}=\frac{v}{\xi}\left[k\left(1-\frac{1}{\mu^{2}}\right)-c\left(f-f_{0}-s\right)\right]-H\left(1-\frac{1}{\mu^{2}}\right)
+\frac{\left[d_{m}\left(\ell\right)-1\right]}{\ell}\frac{d\ell}{dt}\label{mu_full}
\end{equation}
where $H\equiv\dot{a}/a$ is the Hubble parameter and $k$, called the momentum parameter, is defined as
\begin{equation}
k=\frac{\left\langle \left(1-\mathbf{\dot{x}}^{2}\right)\left(\mathbf{\dot{x}\cdot\hat{u}}\right)\right\rangle }{v\left(1-v^{2}\right)} \sim \frac{\left\langle \mathbf{\dot{x}\cdot\hat{u}}\right\rangle}{v}\label{momentum_parameter}
\end{equation}
and in the relevant relativistic regime it can be written as (see \cite{MS3}) 
\begin{equation}
k\left(v\right)=\frac{2\sqrt{2}}{\pi}\frac{1-8v^{6}}{1+8v^{6}}\,. \label{momentum_parameter_ansatz}
\end{equation}

Note that in order for this formalism to be consistent it is already necessary that the uniform wiggliness condition be locally true, even though it can still not be so over cosmological length scales (i.e., $w^{\prime}$ can be very small but non-zero). 

Some interesting considerations can be drawn from the fact that Eqs.~(\ref{dm}--\ref{v_dm}) can be integrated. The former trivially yields
\begin{equation}
\log{[\mu(\ell)]}=\int_0^\ell[d_m(\ell')-1]d\ln{\ell'}\label{solvdm}
\end{equation}
while for the latter, assuming uniform wiggliness and defining the convenient parameter
\begin{equation}
X\equiv\frac{1}{\mu^{2}}\label{X}
\end{equation}
we have
\begin{equation}
v^{2}\left(\ell\right)=1-2\frac{1-v^{2}\left(\ell=0\right)}{1+X\left(\ell\right)} \label{vrms}
\end{equation}
which is an important equation linking a 'microscopic' velocity to wiggliness, and which forces us to face a non-trivial crossroads.

The most natural way to proceed is clearly to keep to the spirit of the VOS model and just interpret the velocity for $\ell=0$ as the RMS velocity that was seen in that model. 
\begin{equation}
v^{2}\left(\ell\right)=1-\frac{2\left(1-v^{2}_{RMS}\right)}{1+X\left(\ell\right)}. \label{vrms1}
\end{equation}
That interpretation, however, necessarily entails an unexpected limitation to the application of the formalism: since this scale-dependent $v^2$ must still be positive, we have to be beyond our domain of applicability whenever $X\left(\ell\right)<1-2v^{2}_{RMS}$. In other words, we should expect our wiggly models to break down in the non-relativistic regime. In particular, this means that our formalism cannot make trustworthy predictions in the tensionless limit. If so, the calculations in this limit in the previous paper worked only because $v$ was artificially fixed at $v=0$ (although the calculations for a fixed $\ell$ should still hold). Even though there is in principle no reason why our formalism should be valid all the time (including in regimes in which the VOS model has not been properly tested) this should at least serve as motivation to entertain a possible alternative.

A perhaps more serious motivation for questioning the validity of Eq.~\eqref{vrms1} is related to a certain tension between different types of simulations regarding what this microscopic velocity should be. The RMS velocity measured in expanding universe Goto-Nambu simulations is close to, but slightly below $1/\sqrt{2}$ (highlighting the presence of small-scale wiggles), whereas in Minkowski space Goto-Nambu simulations or field theory simulations the measured velocities are consistent with $1/\sqrt{2}$. This might motivate an even simpler form for the scale dependence of the characteristic velocity,
\begin{equation}
v^{2}(\ell)=\frac{1}{1+\mu^2(\ell)}; \label{vvrms}
\end{equation}
which as we shall see is qualitatively (though not quantitatively) in agreement with numeric simulations if we interpret $v$ as the coherent velocity.

In the end, it seems that which formula is correct is related to whether Goto-Nambu or field theory simulations are more accurate at the relevant scales---see for example the comparison between both types of simulations in \cite{ABELIAN}. Naturally, Goto-Nambu simulations should never be expected to favour Eq.~\eqref{vvrms} over Eq.~\eqref{vrms1}, but one should keep in mind that ultimately we want to model realistic networks rather than simply fit the output of any type of simulation.

Moreover, there is even no guarantee that either formula has to be correct. The same way we have already mentioned there is no a priori reason why our formalism should have to be valid in the tensionless limit, there is no reason that it has to be valid down to arbitrarily small scales; especially if we keep in mind this formalism is based on a 'string renormalization' procedure, connecting wiggly and elastic strings, which we do not fully comprehend (especially when it comes to transforming velocity vectors). All we really need in order to use our evolution equations is that it be valid over a range of scales that includes our choice for $\ell$.

Nevertheless, it should be noted that this dilemma can have a non-trivial effect in the complexity of our equations. If Eq.~\eqref{vvrms} is true then we can reduce the number of equations in our system since $v$ and $\mu$ are now completely correlated and thus Eqs.~(\ref{v_full}--\ref{mu_full}) cannot be independent. This realisation allows us to relate the loop-chopping terms to the momentum parameter and the Hubble parameter via
\begin{equation}
\frac{v}{\xi} \left[2k - c \left( f - f_{0} - s \right)\right] = 2H \label{cons_muv}
\end{equation}
which in particular implies, since $k\left(v=\frac{1}{\sqrt{2}}\right)=0$, that
\begin{equation}
\xi\left(\ell=0\right)=-c\frac{f\left(1\right)-f_{0}\left(1\right)-s\left(1\right)}{2\sqrt{2}H} \label{cons_cor}
\end{equation}
and the numerator, usually assumed to be null in this limit, now has to be non-zero. This is not wholly unexpected since the null case corresponds to an attempt to recover the VOS model exactly as $\ell$ goes to zero, which this approach must necessarily contradict.

Finally, note that Eqs.~(\ref{vrms}--\ref{vvrms}) are all very useful tools since they provide us with a way to test whether a scale-dependent velocity is the characteristic velocity in our model (independently of the multifractal dimension), which may further our physical understanding of this formalism. Nonetheless, most of the following calculations will only assume Eq.~\eqref{vrms} simply  because most simulations available to us are Goto-Nambu and using Eq.~\eqref{vvrms} would require knowing more about energy-loss mechanisms (i.e., more freedom  in parametrizing $f$, $f_{0}$, and $s$). Regardless, it would be straightforward to carry out the analogous calculations, which would actually be simpler to solve, as they would typically involve systems of two equations instead of three, with Eq.~\eqref{cons_muv} working as a consistency relation among the parameters of the model.

\section{\label{sca}The Scaling Regime}

The prediction of an attractor scaling regime when the scale factor is a power law (as in Eq.~\eqref{a_power}) is one of the main predictions of the VOS model which is in quantitative agreement with numerical simulations. This regime is characterised by a constant velocity and a characteristic length proportional to time (or, equivalently, to the cosmological horizon length). Specifically, the VOS model predicts \cite{MS3}
\begin{equation}
\left(\frac{L}{t}\right)^{2}\equiv\gamma^{2}=\frac{k\left(k+c\right)}{4\lambda\left(1-\lambda\right)}\label{GN_gamma}
\end{equation}
\begin{equation}
v^{2}=\frac{k\left(1-\lambda\right)}{\lambda\left(k+c\right)}\label{GN_v}
\end{equation}
and since this result is confirmed by Abelian-Higgs simulations (for $c=0.23$) our corresponding prediction should not significantly deviate from this.

An important open question in cosmic string evolution is whether the small-scale component also scales, i.e., whether we should also expect $\mu$ to evolve towards a constant value. Despite current simulations not answering this question definitely \cite{FRAC}, they suggest that such a small-scale scaling is reached at least in a matter era (when $\lambda=2/3$). In the radiation era simulations show a more complex behavior, which could reflect the fact that the approach to scaling is slower in this case (since there is less Hubble damping) or could be due to the existence of more than one scaling solution.

\subsection{Finding wiggly scaling\label{findscale}}

Scaling solutions can be straightforwardly sought by making the appropriate substitutions on the left-hand side of Eqs.~(\ref{qui_full}--\ref{mu_full}) and assuming that $\ell$ is also scaling. At this point we need to specify a specific behavior for the fractal dimension $d_m$ as a function of the other parameters. (A mathematically simpler but physically less realistic alternative would be to consider it a constant phenomenological parameter at the scale $\xi$ that we'll be interested in.) This turns out to be a more subtle question than it may appear, and a full derivation is left for subsequent work, but we can nevertheless provide an approximate derivation here.

It is obvious that the fractal dimension will be scale-dependent, ranging from $d_m=1$ on very small scales to $d_m=2$ (Brownian) on super-horizon scales, and interpolating between the two limits on scales around the correlation length. Such a behavior has been explicitly shown to occur in Goto-Nambu simulations \cite{FRAC}. We can therefore construct a fairly generic phenomenological function that reproduces this behavior
\begin{equation}
d_m(\ell)=2-\left[1+B\left(\frac{\ell}{\xi}\right)^b\right]^{-1} \,. \label{fractal1}
\end{equation}
This allows freedom both in the characteristic scale at which the transition occurs an in how fast it occurs as one changes scale. Now, the fractal dimension and $\mu$
are related by Eq.~\eqref{solvdm} and in this case this yields
\begin{equation}
\mu(\ell)=\left[1+B\left(\frac{\ell}{\xi}\right)^b\right]^{1/b} \,. \label{fractal3}
\end{equation}
By simple substitution we can now remove the $\ell$ dependence and obtain an explicit relation between $d_m$ and $\mu$
\begin{equation}
d_m(\mu)=2-\frac{1}{\mu^b}\,. \label{fractal4}
\end{equation}
Notice that this depends only on the parameter $b$, not on $B$.

All that remains to be done is to fix the free parameter $b$. Comparing to expanding universe numerical simulations \cite{FRAC} we find that $b=2$ provides a fairly reasonable approximation. Thus in what follows we will use
\begin{equation}
d_{m}=2-\frac{1}{\mu^{2}}\label{dm_mu}\,.
\end{equation}
Note that combining this with Eq.~\eqref{vrms1} we can also write
\begin{equation}
v^{2}(\ell)=1-\frac{2\left(1-v^{2}_{RMS}\right)}{3-d_m(\ell)}=\frac{1-d_m(\ell)+2v^{2}_{RMS}}{3-d_m(\ell)}\,, \label{vrms2}
\end{equation}
or equivalently
\begin{equation}
d_m(\ell)=3-\frac{2\left(1-v^{2}_{RMS}\right)}{1-v^2(\ell)}=\frac{1+2v^{2}_{RMS}-3v^2(\ell)}{1-v^2(\ell)}\,; \label{vrms3}
\end{equation}
naturally the analogous expressions for the ansatz of Eq.~\eqref{vvrms} ensue by taking the particular case $v_{RMS}=1/\sqrt{2}$.

With these assumptions we can now reduce our problem to solving the algebraic system
\begin{equation}
v^{2}=\frac{\left[4X^{2}-2\lambda X\left(1+X\right)\right]\left(k/c\right)-X(1-X)\left(f_{0}+s\right)}{\lambda\left(1+X\right)^{2}\left[\left(k/c\right)+f_{0}+s\right]}\label{full_scaling_system1}
\end{equation}
\begin{equation}
\gamma_{\xi}=v\frac{k\left(1-X\right)+c\left(f_{0}+s\right)}{1+X-\lambda\left[2+\left(1+X\right)v^{2}\right]}\label{full_scaling_system2}
\end{equation}
\begin{equation}
\frac{v}{\gamma_{\xi}}\left[k\left(1-X\right)-c\left(f-f_{0}-s\right)\right]+\left(1-\lambda\right)\left(1-X\right)=0\label{full_scaling_system3}
\end{equation}
which interestingly has at most two solutions with the same fixed value of $X\neq 1$ (assuming that the shape of the energy loss functions is fixed). In other words, for any given $X$ there are at most two values of $c$ such that there is a scaling solution with that constant value of $X$; in what follows we will denote these by $c_{X}$. These solutions, if they exist, can be found by the following algorithm: first just compute
\begin{equation}
v_{X}^{2}=\frac{\left[4X^{2}-2\lambda X\left(1+X\right)\right]\varphi_{X}-X(1-X)\left(f_{0}+s\right)}{\lambda\left(1+X\right)^{2}\left[\varphi_{X}+f_{0}+s\right]}\label{eq:scaling_cons_V_func_X}
\end{equation}
where $\varphi_{X}$ is a real solution of the quadratic equation
\begin{equation}
A\varphi_{X}^{2}+B\varphi_{X}+C=0\label{eq:varphi_X}
\end{equation}
whose coefficients are
\begin{equation}
A=\left(1-\lambda\right)\left(1-X\right)\left(1-X^{2}\right) 
-\left(1-X\right)\left[4X^{2}-2\lambda\left(1+X\right)X\right]+\left(1-X^{2}\right)\left[1+X-2\lambda\right] \label{eq:scalingA}
\end{equation}
\begin{eqnarray}
B &=& \left(1-\lambda\right)\left(1-X^{2}\right)\left(2-X\right)\left(f_{0}+s\right)+\left(f-f_{0}-s\right)\left(4X^{2}-2\lambda\left(1+X\right)X\right)\\ \nonumber
 &+& \left(f_{0}+s\right)X\left(1-X\right)^{2} +\left[\left(f_{0}+s\right)\left(1-X\right)-f+f_{0}+s\right]\left[\left(1+X\right)^{2}-2\lambda\left(1+X\right)\right]\label{eq:scalingB}
\end{eqnarray}
\begin{equation}
C=\left(f_{0}+s\right)^{2}\left(1-\lambda\right)\left(1-X^{2}\right) 
-\left(f_{0}+s\right)\left(f-f_{0}-s\right)\left[X\left(1-X\right)+\left(1+X\right)^{2}-2\lambda\left(1+X\right)\right]\label{eq:scalingC}
\end{equation}
(of course, if there are no real solutions to Eq.~\eqref{eq:varphi_X} that
just means that scaling is impossible for that $X$), then compute
$k\left(v_{X}\right)$ using Eq.~\eqref{momentum_parameter_ansatz} and
the $c_{X}$ we are after is simply 
\begin{equation}
c_{X}=\frac{k\left(v_{X}\right)}{\varphi_{X}}\label{eq:scaling_c}
\end{equation}
if it is positive and less than $1$ - otherwise there is no scaling. Obviously, there is also no scaling if the velocity $v$ and the correlation coefficient $\gamma_{\xi}$ calculated in this way have non-physical values.

Interestingly, one can see by setting $X=1$ that the VOS solutions are also solutions of our model provided that $f_{0}\left(X=1\right)=f\left(X=1\right)=1$ and $s\left(X=1\right)=0$. That is by no means unexpected, since when building this model we required that the VOS equations be recovered whenever $X=1$, $f_{0}=f=1$, and $s=0$. This is not to be regarded as a problem since $s\left(X=1\right)=0$ is an approximation which is to some extent motivated by the success of the VOS predictions. In a way, we are just saying that $s\left(X=1\right)$ gives a contribution which is much weaker than those of competing energy loss mechanisms.

\subsection{Wiggly scaling stability\label{stable}}

Ultimately, the feature that made scaling regimes in the VOS model interesting was their attractor nature - which, in particular, enables us to use them to calibrate the loop-chopping efficiency $c$ by comparison with simulations. Therefore, a study of the stability of the non-trivial (here meaning those with $X\neq 1$) scaling solutions found above is needed.

With this in mind, it is straightforward to linearize our equations around these solutions
\begin{equation}
\left[\begin{array}{c}
\gamma_{\xi,}\\
v_{\,}\\
X_{\,}
\end{array}\right]\sim\left[\begin{array}{c}
\gamma_{ s}\\
v_{s}\\
X_{s}
\end{array}\right]+\left[\begin{array}{c}
\overline{\gamma_{\xi}}\\
\overline{v}\\
\bar{X}
\end{array}\right]\label{linearize}
\end{equation}
and write them in matrix form
\begin{equation}
t\frac{d}{dt}\left[\begin{array}{c}
\overline{\gamma_{\xi}}\\
\overline{v}\\
\bar{X}
\end{array}\right]=\left[\begin{array}{ccc}
\\
&M_{j}^{i}\\
\\
\end{array}\right]\left[\begin{array}{c}
\overline{\gamma_{\xi}}\\
\overline{v}\\
\bar{X}
\end{array}\right]\label{stability_M}
\end{equation}
where $\gamma_{s}$, $v_{s}$, and $X_{s}$ are the scaling values of $\gamma_{\xi}$, $v$, and $X$, respectively. The components of $M$ can be shown to be
\begin{equation}
M_{1}^{1}=-1+\lambda\left(\frac{2}{1+X_{s}}+v_{s}^{2}\right)\label{M11}
\end{equation}
\begin{equation}
M^{1}_{2}=2\lambda\gamma_{s}v_{s}+\frac{B_{s}+\left(k_{s}+v_{s}k_{\star}\right)\left(1-X_{s}\right)}{1+X_{s}}\label{M12}
\end{equation}
\begin{equation}
M^{1}_{3}=\frac{v_{s}\left(\lambda\gamma_{s}v_{s}-k_{s}+B_{\star}\right)}{1+X_{s}}\label{M13}
\end{equation}
\begin{equation}
M^{2}_{1}=\left(1-v_{s}^{2}\right)\left(-\frac{\lambda v_{s}}{\gamma_{s}}\left[1+X_{s}\right]-\frac{X_{s}\left[1-X_{s}\right]}{\gamma_{s}v_{s}\left[1+X_{s}\right]}\left(1+M_{1}^{1}\right)\right)\label{M21}
\end{equation}
\begin{multline}
M^{2}_{2}=\left(1-v_{s}^{2}\right)\left(\frac{X_{s}k_{s}}{\gamma_{s}v_{s}}+\frac{k_{\star}}{\gamma_{s}}-2\lambda\left[1+X_{s}\right]-\frac{X_{s}\left[1-X_{s}\right]}{\gamma_{s}v_{s}\left[1+X_{s}\right]}M_{2}^{1}\right)\\-\frac{2k_{s}v_{s}X_{s}}{\gamma_{s}}+2\lambda v_{s}^{2}\left(1+X_{s}\right)+\frac{2X_{s}\left(\-X_{s}\right)}{1+X_{s}}\label{M22}
\end{multline}
\begin{equation}
M^{2}_{3}=\left(1-v_{s}^{2}\right)\left(\frac{k_{s}\left[1+2X_{s}\right]}{\gamma_{s}\left[1+X_{s}\right]}-2\lambda v_{s}-\frac{\left(1-2X_{s}\right)}{v_{s}\left(1+X_{s}\right)}-\frac{X_{s}\left[1-X_{s}\right]}{\gamma_{s}v_{s}\left[1+X_{s}\right]}M_{3}^{1}\right)\label{M23}
\end{equation}
\begin{equation}
M^{3}_{1}=\frac{2\lambda X_{s}\left(1-X_{s}\right)}{\gamma_{s}}-\frac{2X_{s}\left(1-X_{s}\right)}{\gamma_{s}}\left[1+M_{1}^{1}\right]\label{M31}
\end{equation}
\begin{equation}
M^{3}_{2}=-\frac{2X_{s}\left[\left(1-X_{s}\right)v_{s}k_{\star}+\left(k_{s}\left[1-X_{s}\right]-M_{s}\right)\right]}{\gamma_{s}}-\frac{2X_{s}\left(1-X_{s}\right)}{\gamma_{s}}M_{2}^{1}\label{M32}
\end{equation}
\begin{equation}
M^{3}_{3}=2\frac{X_{s}}{\gamma_{s}}v_{s}\left(k_{s}+M_{\star}\right)+2X_{s}\left(1-\lambda\right)-\frac{2X_{s}\left(1-X_{s}\right)}{\gamma_{s}}M_{3}^{1}\label{M33}
\end{equation}
where
\begin{equation}
k\equiv k_{s}+k_{\star}\bar{v}\,,
\end{equation}
meaning that
\begin{equation}
k_{\star}=-k_{s}\frac{96v_{s}^{5}}{1-64v_{s}^{12}}\,,
\end{equation}
\begin{equation}
c\left(f_{0}+s\right)\equiv B_{s}+B_{\star}\bar{X}\,,
\end{equation}
and
\begin{equation}
c\left(f-f_{0}-s\right)\equiv M_{s}+M_{\star}\bar{X}\,.
\end{equation}
In writing these formulas for the components of $M$ one also has to assume the natural relation for the mesoscopic scale $\ell$
\begin{equation}
\ell\propto\xi\label{ell_prop}
\end{equation}
which is logical given that $\xi$ is the most important scale governing loop production (not to mention that it scales), but similar expressions for the components of $M$ could be found by assuming any alternative of similar form.

If $u_{k}$ (with $k=1,2,3$) are the eigenvectors of $M^{i}_{j}$ with eigenvalues $\alpha_{k}$ then it is easily seen that $u_{k}\propto t^{\alpha_{k}}$ and, in particular, a scaling solution is stable in this linearized limit if and only if the real parts of all three eigenvalues of $M^{i}_{j}$ are negative. Therefore, whether and how fast our three independent variables approach their scaling values is completely determined by the values of the three eigenvalues (and respective eigenvectors) of the matrix $M$. 

\subsection{Exploring Scaling\label{scasol}}

Let us now illustrate the procedure described above, starting by introducing a particular ansatz for the energy loss terms. As has been noted, the dependence of $f$ and $f_{0}$ on $\mu$ can in principle be investigated using high-resolution network simulations. In the absence of such information, however, when forced to consider a specific type of dependence, we shall resort to a more ad-hoc argument.

Recall that when the loop-chopping parameter $c$ is introduced in one-scale-type models it is usually as a result of the appearance of a loop-production function, $g$, which only depends on the ratio between the size of loops being produced and the correlation length of the network. This is typically defined \cite{VSH} so that
\begin{equation}
\left.\frac{d\rho_{0}}{dt}\right|_{loops}=-\frac{\mu_{0}v}{\xi^{3}}\intop_{0}^{\infty}g\left(l/\xi\right)\frac{dl}{\xi}\equiv-cv\frac{\rho_{0}}{\xi}\label{OS_loop_g}\,.
\end{equation}
Since we generally assume that the bare string is one for which the VOS assumptions apply, it makes sense to not change this relation and simply use
\begin{equation}
f_{0}=1\label{f0_ansatz}\,.
\end{equation}

Bearing in mind that deviations in the total energy lost to loops should be due to a second loop-production mechanism operating on a scale significantly smaller than the correlation length, it makes sense to expect that $f>1$. Furthermore we conjecture that, in the context of this formalism, the typical length of these smaller loops can be related to a combination of $L$ and $\xi$ that vanishes in the Goto-Nambu limit, when $L=\xi$. Clearly, the simplest such scale is just $\xi_{\star}=\xi-L$. We are then justified to write
\begin{equation}
\left.\frac{d\rho}{dt}\right|_{loops}=-\frac{\mu_{0}\mu v}{\xi^{3}}\intop_{0}^{\infty}g\left(l/\xi\right)\frac{dl}{\xi}-\frac{\mu_{0}\mu v}{\xi^{3}}\intop_{0}^{\infty}g_{\star}\left(l/\xi_{\star}\right)\frac{dl}{\xi}\label{wiggly_loop_g}
\end{equation}
 which corresponds to 
\begin{equation}
f\left(\mu\right)=1+\eta\left(1-\frac{1}{\sqrt{\mu}}\right)\label{f_ansatz}
\end{equation}
where we have defined $\eta=c^{-1}\int{g_{\star}\left(x\right)dx}$, which is a positive parameter quantifying how much energy is lost to small-scale loops. For the sake of simplicity, let us further assume that $s$ can be approximated by
\begin{equation}
s\left(\mu\right)\simeq D\left(1-\frac{1}{\mu^{2}}\right)\label{D_def}
\end{equation}
which we expect to be the case as long as $\mu$ is not too large.

To begin with, let us look for non-trivial scaling solutions without worrying about stability; we address the latter issue in the following section. We start by focusing on the matter era ( $\lambda=2/3$), which is when simulations suggest that it is the easiest to achieve scaling \cite{FRAC}. 

\begin{figure*}[!h]
\begin{center}
\includegraphics[scale=0.41]{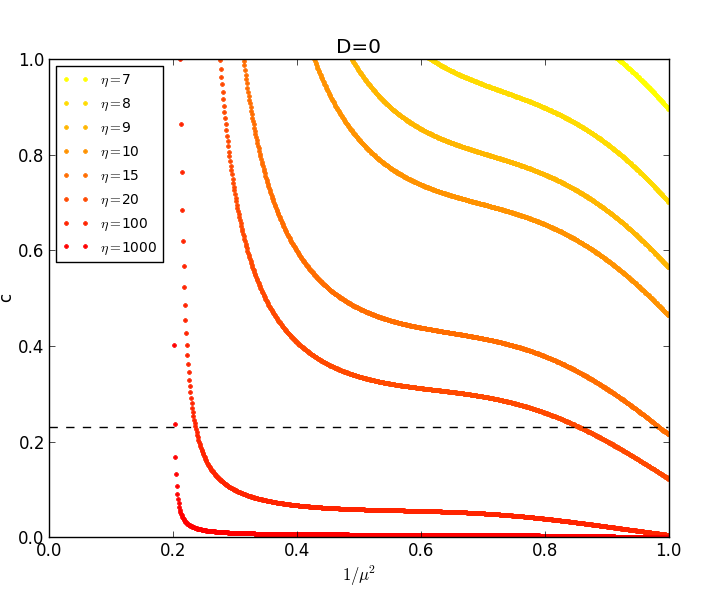}\includegraphics[scale=0.41]{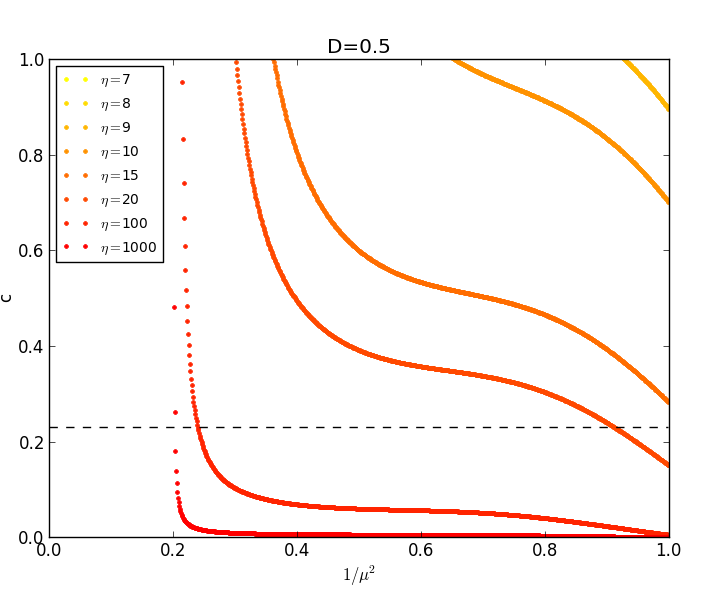}
\includegraphics[scale=0.41]{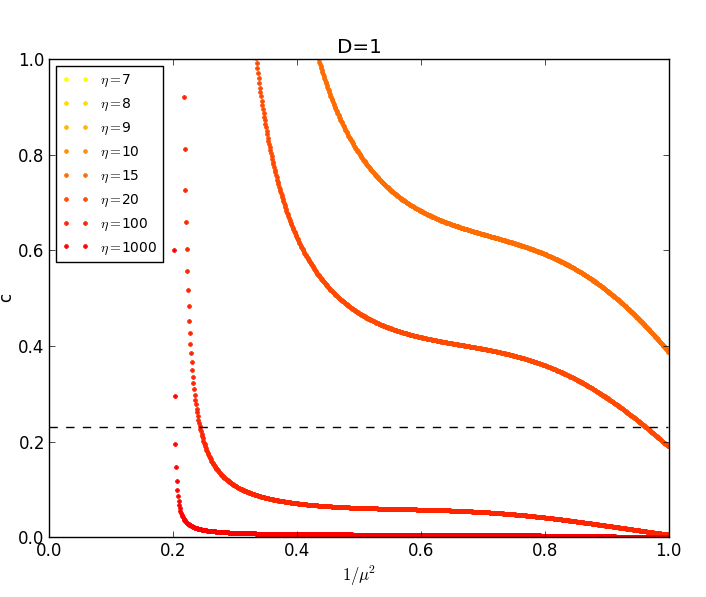}\includegraphics[scale=0.41]{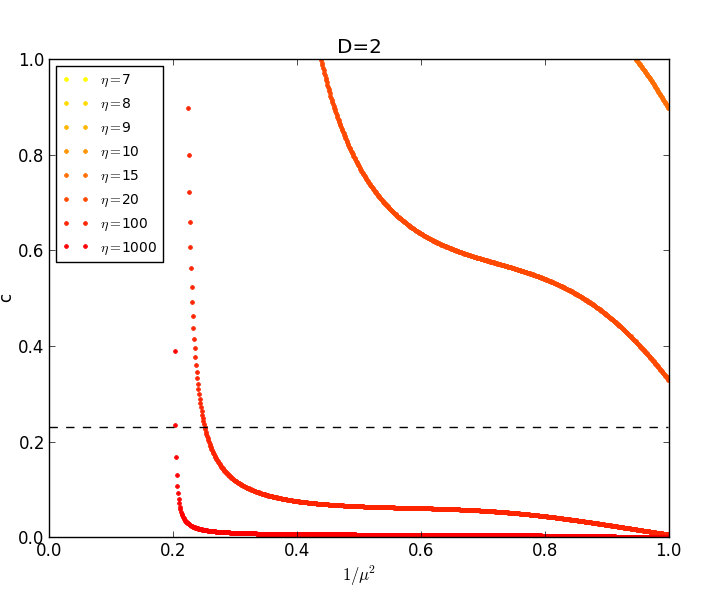}
\end{center}
\caption{\label{fig_matter} Values of the loop-chopping parameter $c$ for which there can be non-trivial scaling, as a function of wiggliness during scaling and for different values of $D$, calculated in the matter era. The dashed line is $c=0.23$, the best fit for the VOS model (the best fit for our model does not have to be the same, but we expect it to be close). We only show the physically meaningful values that stem from Eq.~\protect\ref{eq:varphi_X}---the complementary solution would lead to non-physical (negative) values of $c$.}
\end{figure*}

Applying the procedure described in subsection \ref{findscale} to find $c_{X}$, we get the results summarized in Fig. \ref{fig_matter}. In accordance with our simplistic interpretation of $\eta$ is the observation that increasing $\eta$ leads to a decrease in the $c_{X}$ necessary to maintain scaling with a fixed wiggliness value (essentially, since more energy is lost per collision, we need not be so efficient at colliding). More counterintuitive is the realization that an increase in $\eta$ for a fixed $c$ leads to a higher scaling wiggliness - one would naively expect the opposite behaviour, that more small-scale energy loss led to a lower scaling wiggliness.

Instead, our results indicate that the network needs a higher wiggliness in order to survive the more violent energy loss in equilibrium. In fact, this behaviour hints at something we will notice when we study the stability of these models: that the wiggly component of our equations leads to instabilities in the scaling regime of these simple models. In other words, the reason our intuition fails us in this analysis is because when we deviate the network from a non-trivial scaling regime it does not generically tend to go back to equilibrium on its own; these scaling regimes are not usually attractors. Also of particular interest is that for these small values of $D$ there appears to be a maximum allowed value of $\mu$ in scaling, $\mu\lesssim 2.2$. This feature disappears if we allow much larger values of this parameter, which however does not seem desirable when we study the stability of the model. Notice also how a slight increase in $D$ seems to dramatically decrease the amount of small-scale structure in any given model (with fixed $\eta$ and $c$) - or, conversely, how it seems to increase the value of $c$ necessary to maintain fixed values of $\mu$ and $\eta$.

\begin{figure*}[!h]
\begin{center}
\includegraphics[scale=0.41]{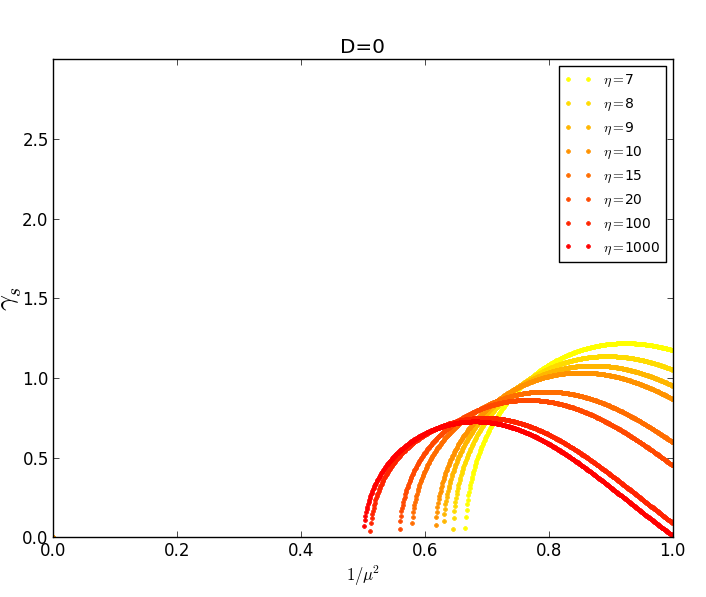}\includegraphics[scale=0.41]{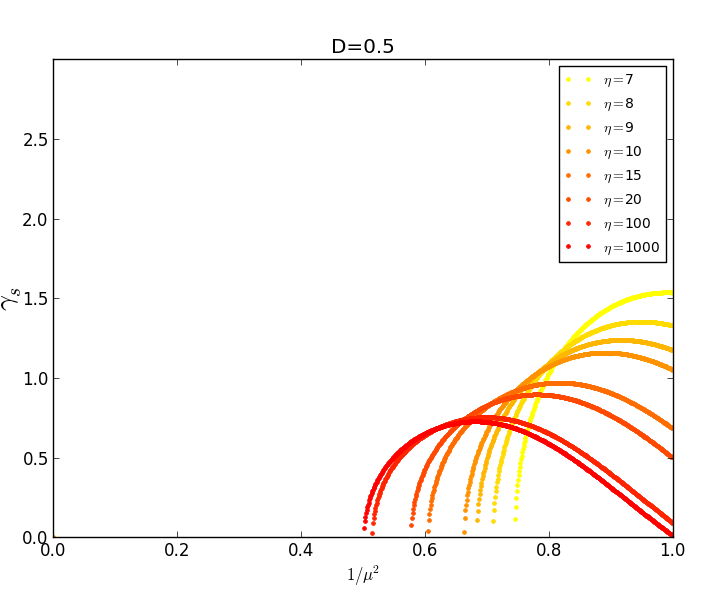}
\includegraphics[scale=0.41]{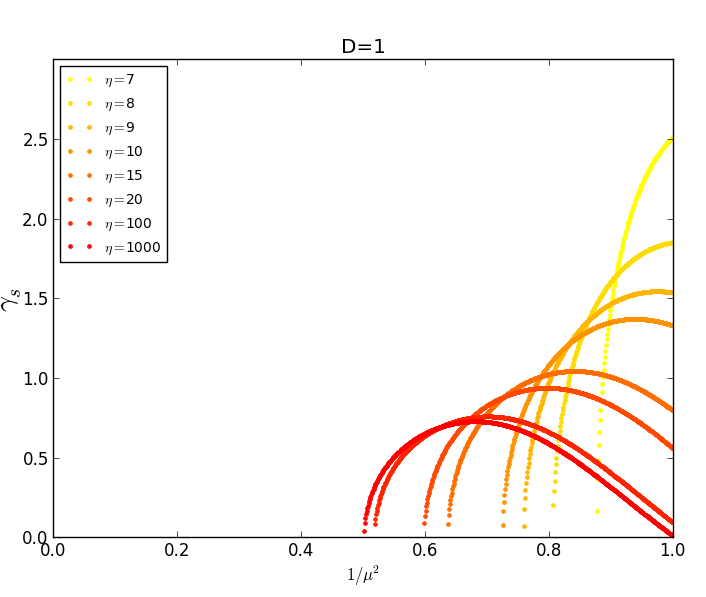}\includegraphics[scale=0.41]{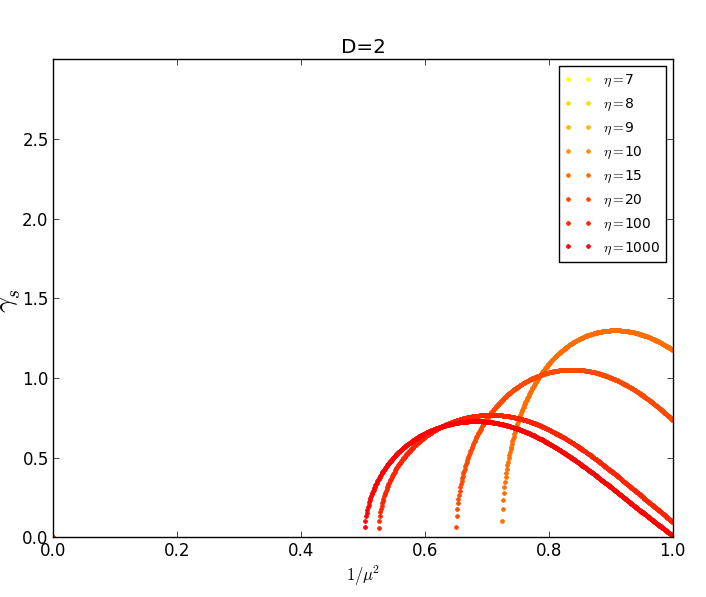}
\end{center}
\caption{\label{fig_matterg} Values of the correlation scaling parameter for which there can be non-trivial scaling, as a function of wiggliness during scaling and for different values of $D$ in the matter era. As before, only physically meaningful values are shown(in this case, $0<\gamma_{s}<\frac{1}{1-\lambda}$).}
\end{figure*}

\begin{figure*}[!h]
\begin{center}
\includegraphics[scale=0.41]{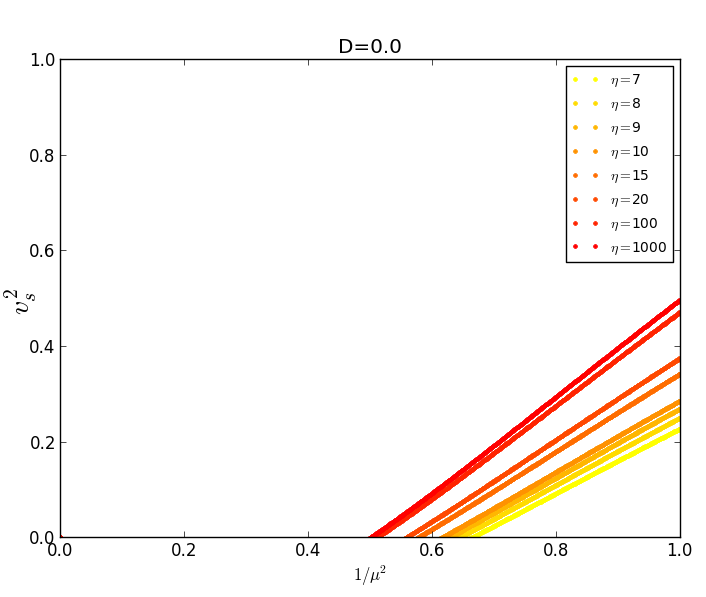}\includegraphics[scale=0.41]{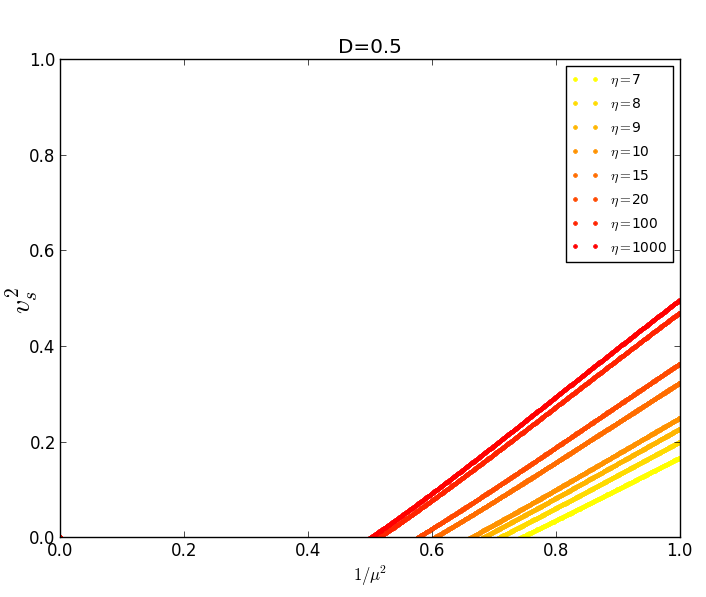}
\includegraphics[scale=0.41]{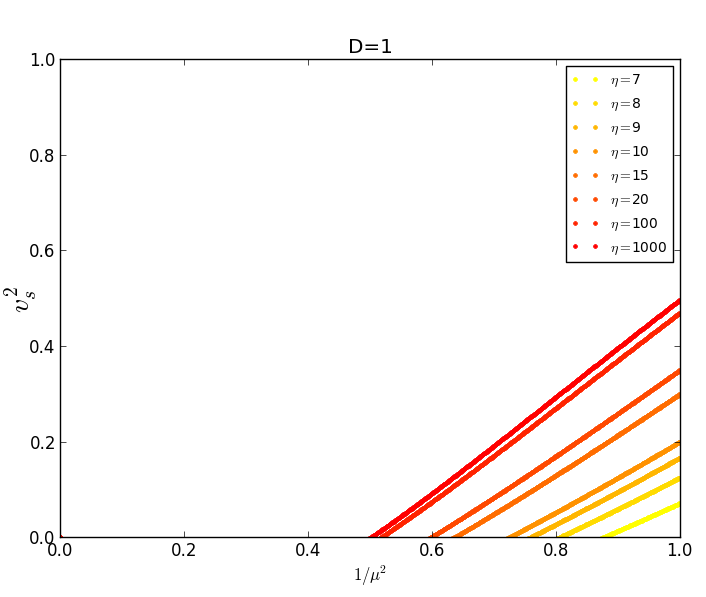}\includegraphics[scale=0.41]{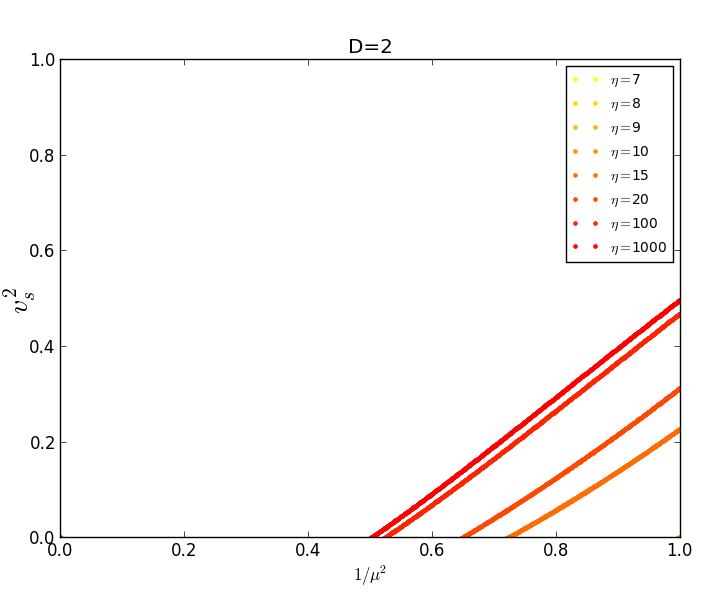}
\end{center}
\caption{\label{fig_matterv} Values of the velocity for which there can be non-trivial scaling, as a function of wiggliness during scaling and for different values of $D$ in the matter era. As before, only physically meaningful values are shown(in this case, $0<v^{2}_{s} <1$).}
\end{figure*}

The analogous results for $\gamma_{s}$ and $v^{2}_{s}$ can be found in Figs. \ref{fig_matterg} and \ref{fig_matterv}. Naturally, scaling is only allowed for a certain model if it is allowed in all three figures.

\begin{figure*}[!h]
\begin{center}
\includegraphics[scale=0.41]{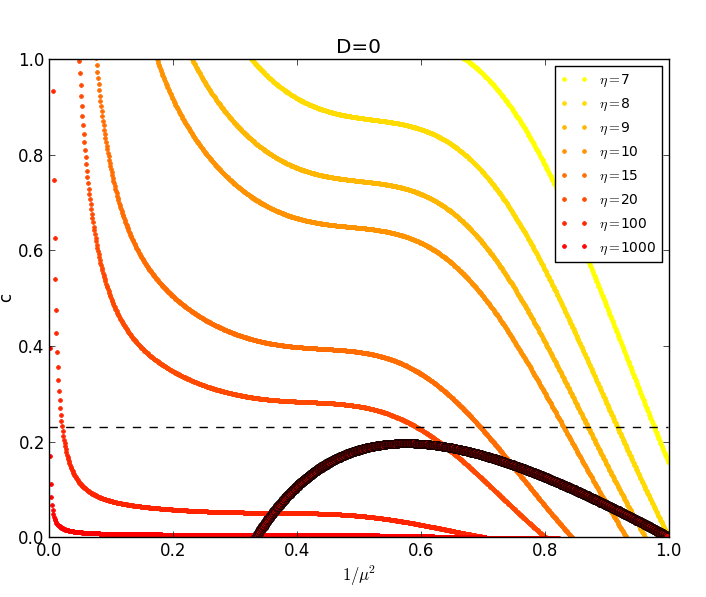}\includegraphics[scale=0.41]{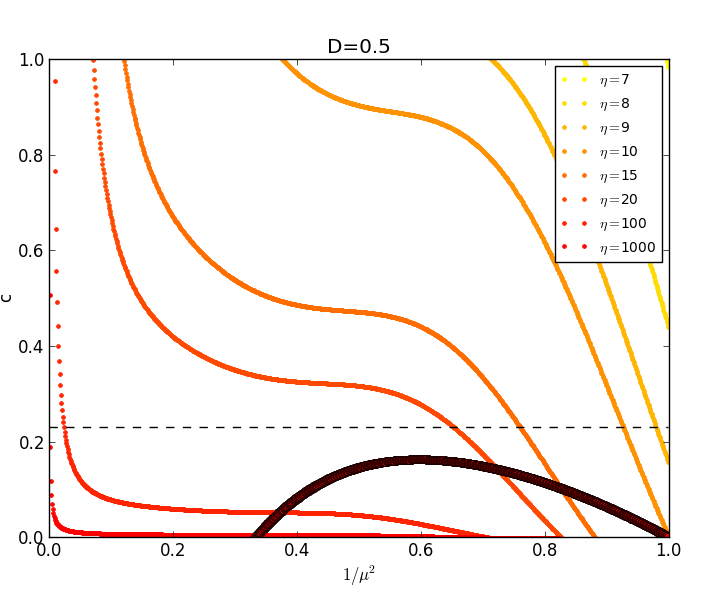}
\includegraphics[scale=0.41]{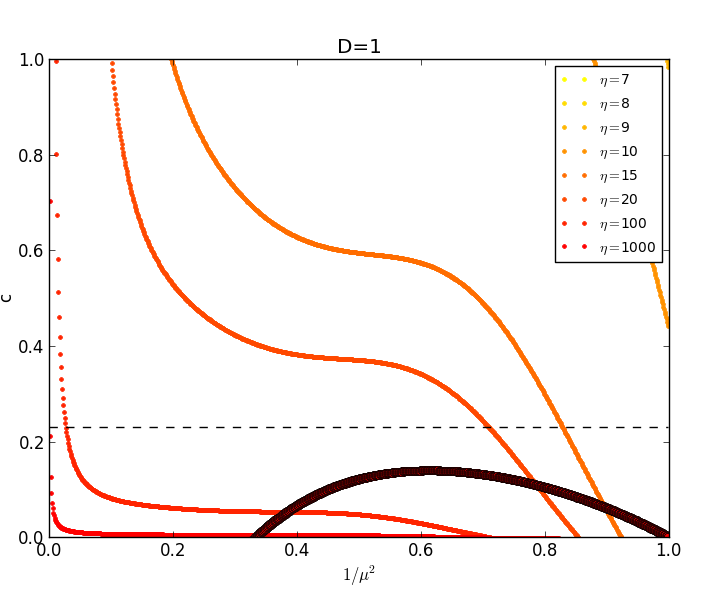}\includegraphics[scale=0.41]{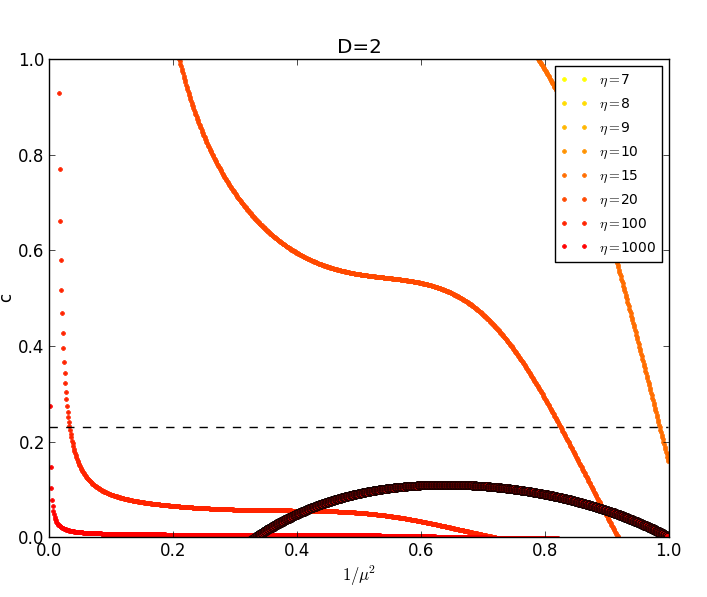}
\end{center}
\caption{\label{fig_rad} Values of the loop-chopping parameter $c$ for which there can be non-trivial scaling, as a function of wiggliness during scaling and for different values of $D$, calculated in the radiation era. The dashed line is $c=0.23$, the best fit for the VOS model (the best fit for our model does not have to be the same, but we expect it to be close). The darker line is there essentially because points of all colours are being plotted on top of each other. Notice that we are only showing the physically meaningful values (in this case, $0<c<1$).}
\end{figure*}

\begin{figure*}[!h]
\begin{center}
\includegraphics[scale=0.41]{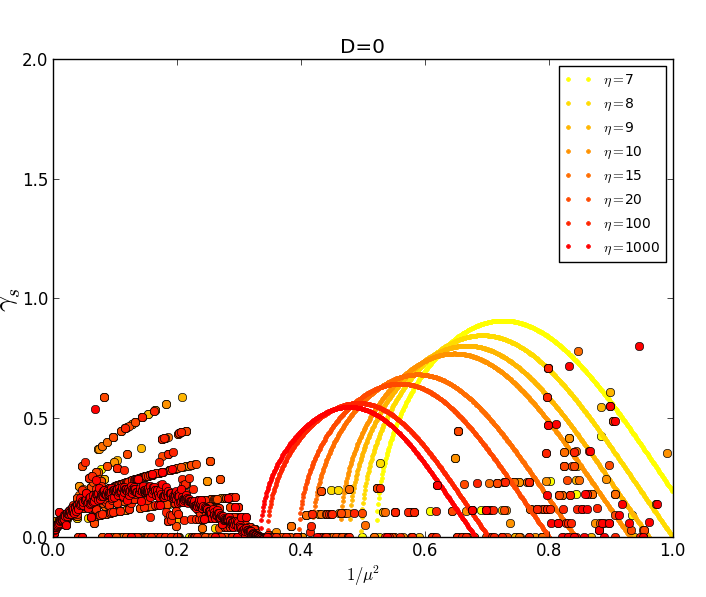}\includegraphics[scale=0.41]{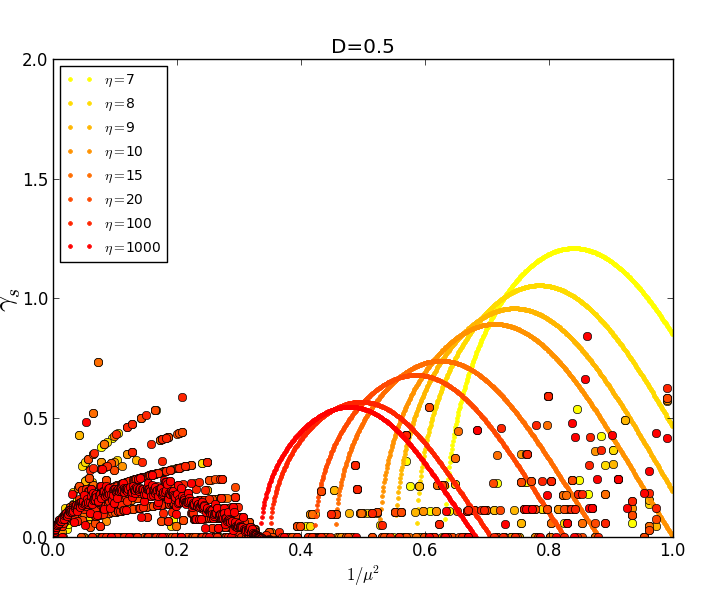}
\includegraphics[scale=0.41]{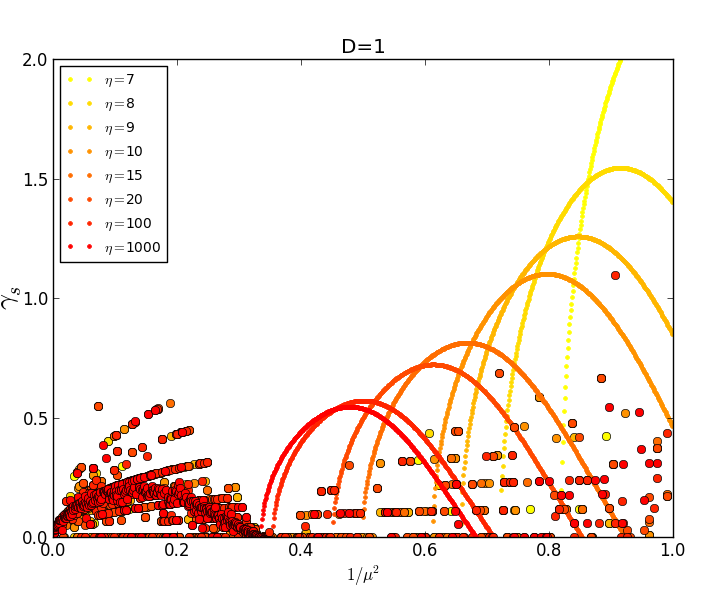}\includegraphics[scale=0.41]{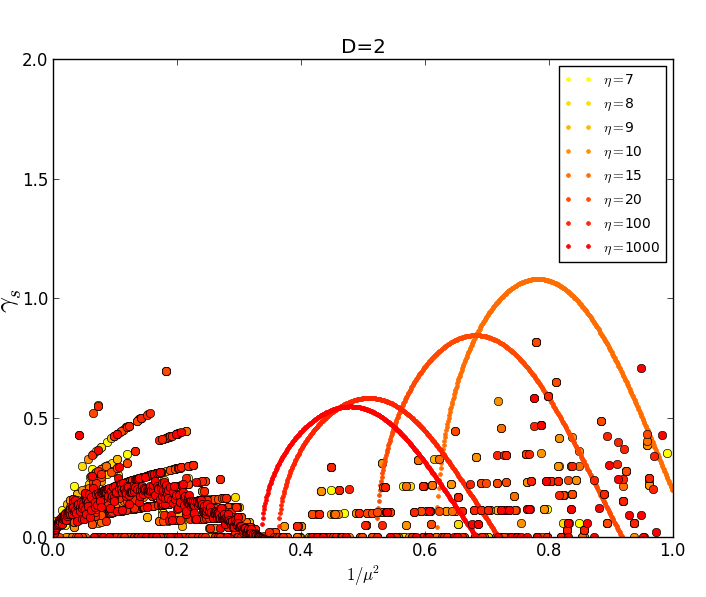}
\end{center}
\caption{\label{fig_radg} Values of the correlation scaling parameter for which there can be non-trivial scaling, as a function of wiggliness during scaling and for different values of $D$ in the radiation era. As before, only physically meaningful values are shown (in this case, $0<\gamma_{s}<\frac{1}{1-\lambda}$). Note that these graphs are fairly contaminated by "noise" generated by computational errors.}
\end{figure*}

\begin{figure*}[!h]
\begin{center}
\includegraphics[scale=0.41]{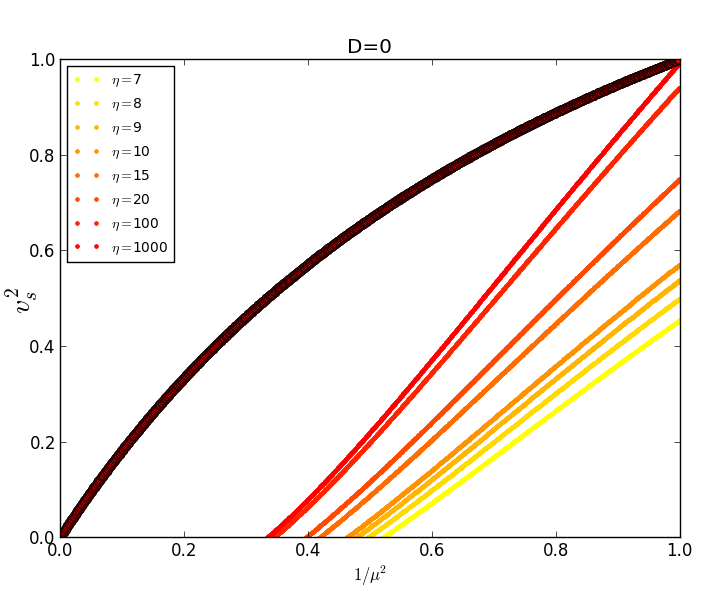}\includegraphics[scale=0.41]{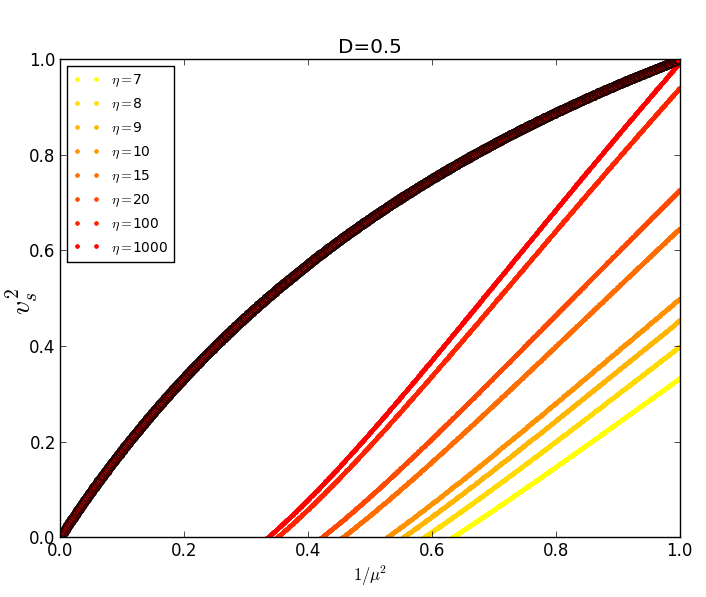}
\includegraphics[scale=0.41]{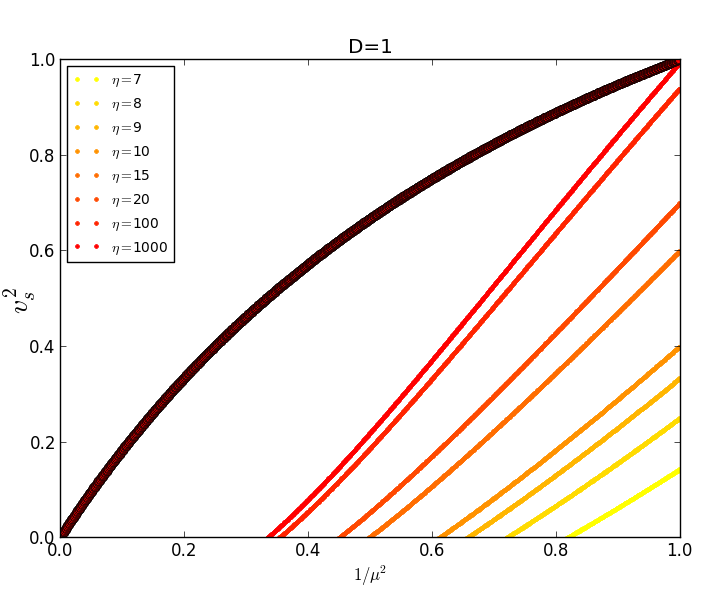}\includegraphics[scale=0.41]{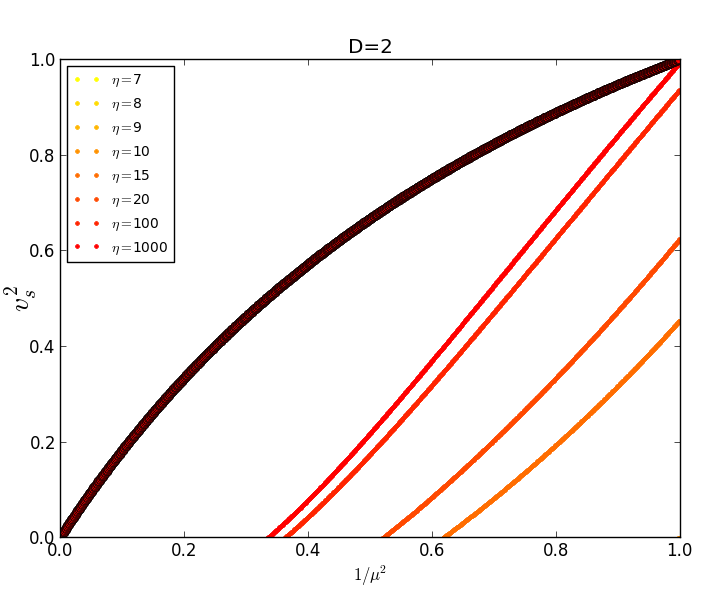}
\end{center}
\caption{\label{fig_radv} Values of the velocity for which there can be non-trivial scaling, as a function of wiggliness during scaling and for different values of $D$ in the radiation era.  The darker line is there essentially because points of all colours are being plotted on top of each other. As before, only physically meaningful values are shown (in this case, $0<v^{2}_{s} <1$).}
\end{figure*}

We can also carry out a similar analysis for the radiation era ($\lambda=1/2$), whose results for the solutions that come from using the greater roots of Eq.~\eqref{eq:varphi_X} are analogous to the ones we have just seen. The results from the other solution, however, are of a much less straightforward interpretation (and are probably of reduced physical significance). If we take a look at the analog of Fig. \ref{fig_matter}, which is Fig. \ref{fig_rad}, this difference is stark: not only is the line corresponding to this new solution of a much different shape and size, but it seems to be extremely insensitive to large variations of $\eta$ while being very sensitive to $D$ (which appears to consistently suppress it).

If we focus instead on the radiation epoch results for $\gamma_{s}$, shown in Fig. \ref{fig_radg}, the situation is even slightly worse: because at some point during our calculations we need to divide very small numbers, our graphs are vulnerable to computational uncertainties. Nevertheless, we are still able to discern a difference in the behaviour from the previous case, as well as a robust independence of $\eta$. Without this numerical 'noise', the same kind of differences can be seen in the velocity, which can be found in Fig. \ref{fig_radv}.

\subsection{Exploring (in)stability\label{explore_stable}}

Now that we have found a large family of non-trivial scaling solutions, the time has come to test their stability. We have already mentioned that the shapes we see in Figs. \ref{fig_matter} and \ref{fig_rad} suggest that the introduction of $\mu$ in our equations has spoiled the attractor feature of non-trivial scaling regimes.

\begin{figure*}[!h]
\begin{center}
\includegraphics[scale=0.4]{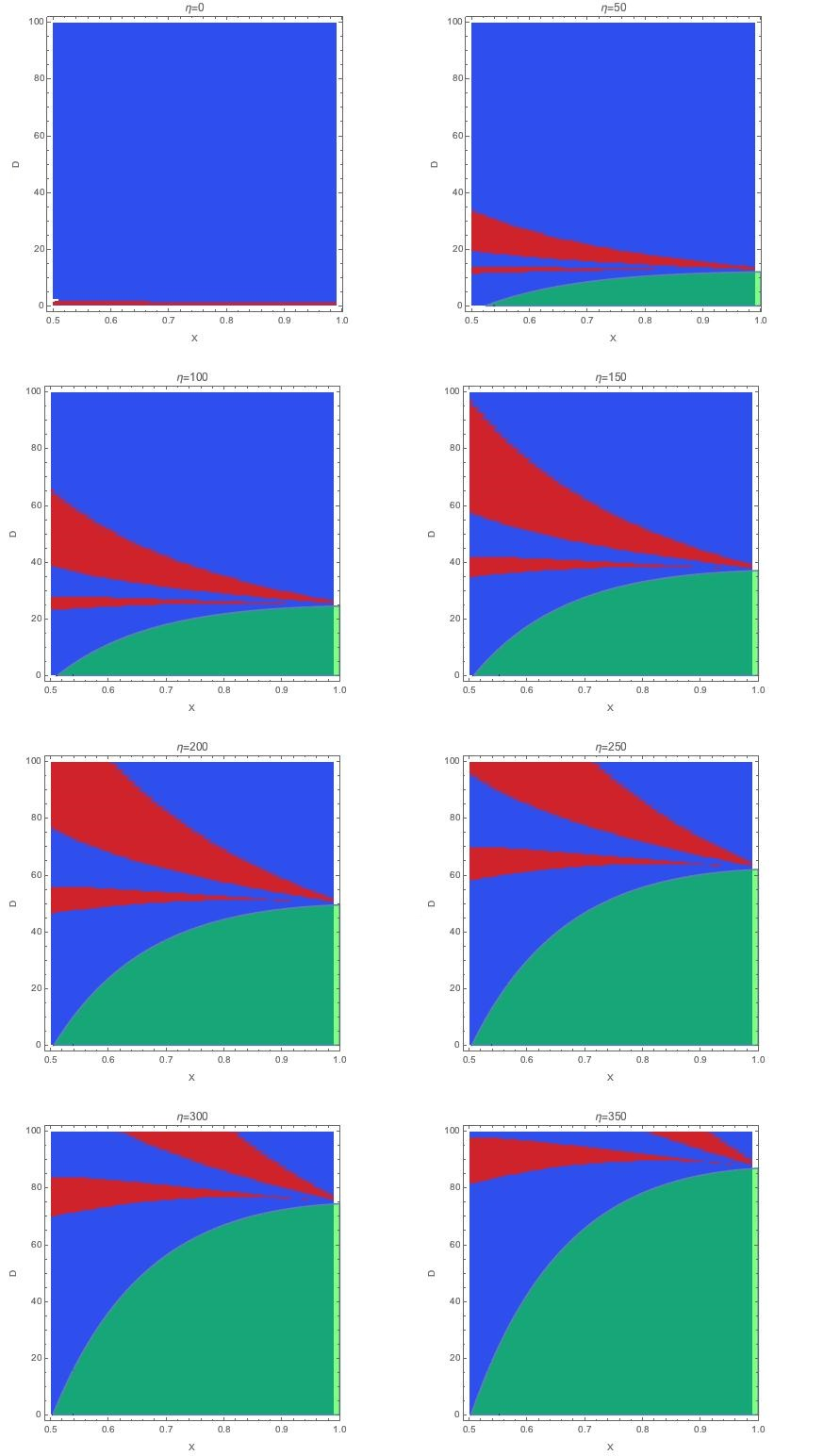}
\end{center}
\caption{\label{eigenfig} Stability analysis for our ansatz. The red region corresponds to parameters that make the real parts of all eigenvalues of $M^{i}_{j}$ negative in the matter era. The green region corresponds to parameters that yield physical values of $c$, $v$, and $\gamma$. As is, the scaling regimes we are predicting are clearly not attractors since the two regions do not overlap for $X<1$.}
\end{figure*}

Indeed, a direct application of the methodology described in section \ref{stable} reveals that it is not easy (if possible at all) to find stable non-trivial scaling for our heuristic choice of $f$, $f_{0}$, and $s$ as well as our ansatz for $d_{m}\left(\ell\right)$. This difficulty is illustrated in Fig. \ref{eigenfig}.

It should be noted, however, that checking stability requires knowing our energy-loss and multifractal dimension functions with more accuracy than if we just wanted to look for scaling solutions. The reason is that, since $M^{i}_{j}$ depends on derivatives of these functions, second-order corrections can have a first-order impact. As such, what this problem is telling us is not that our ansatze are bad first-order approximations, but rather that we need to go to higher orders if we want to draw conclusions from this sort of stability analysis.

\section{\label{fullev}Comparison with simulations}

\begin{figure*}[!h]
\includegraphics[width=3in]{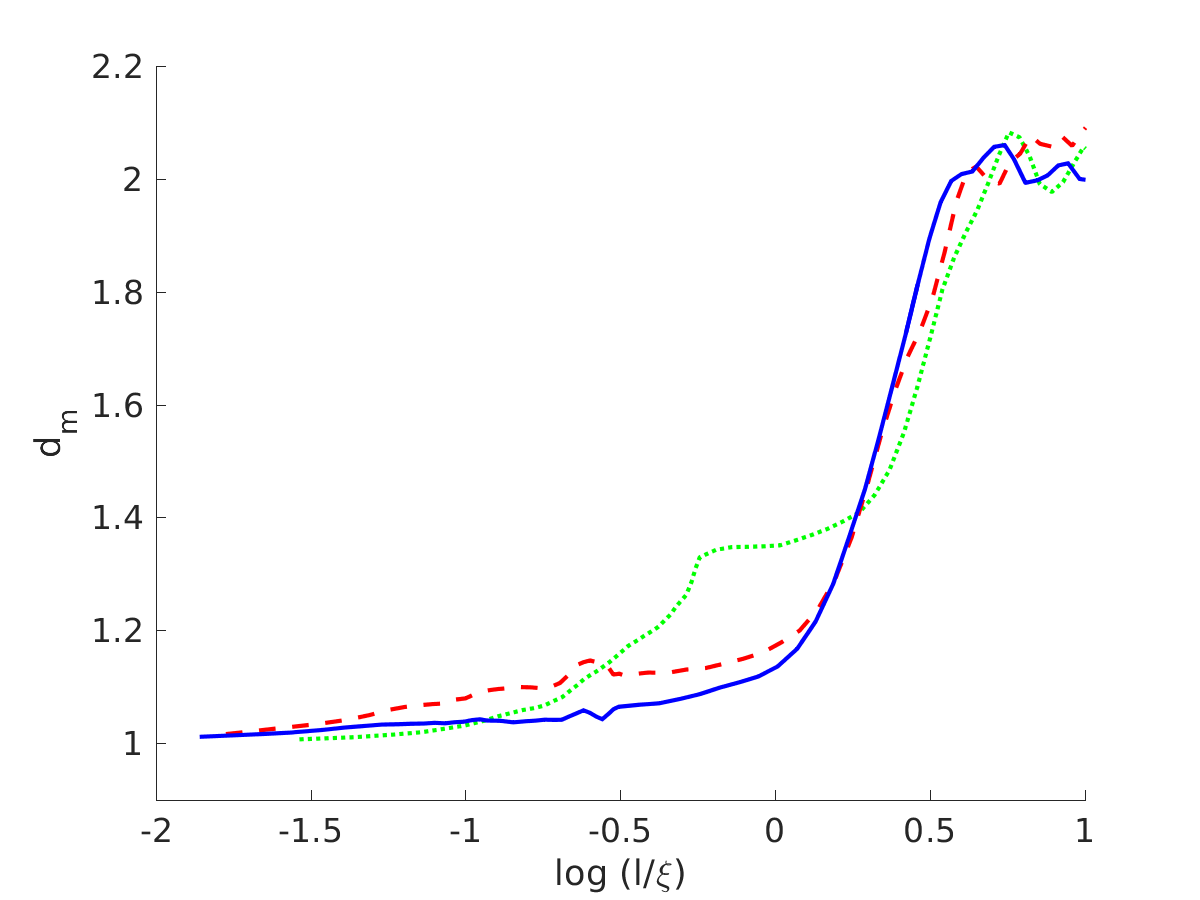}
\includegraphics[width=3in]{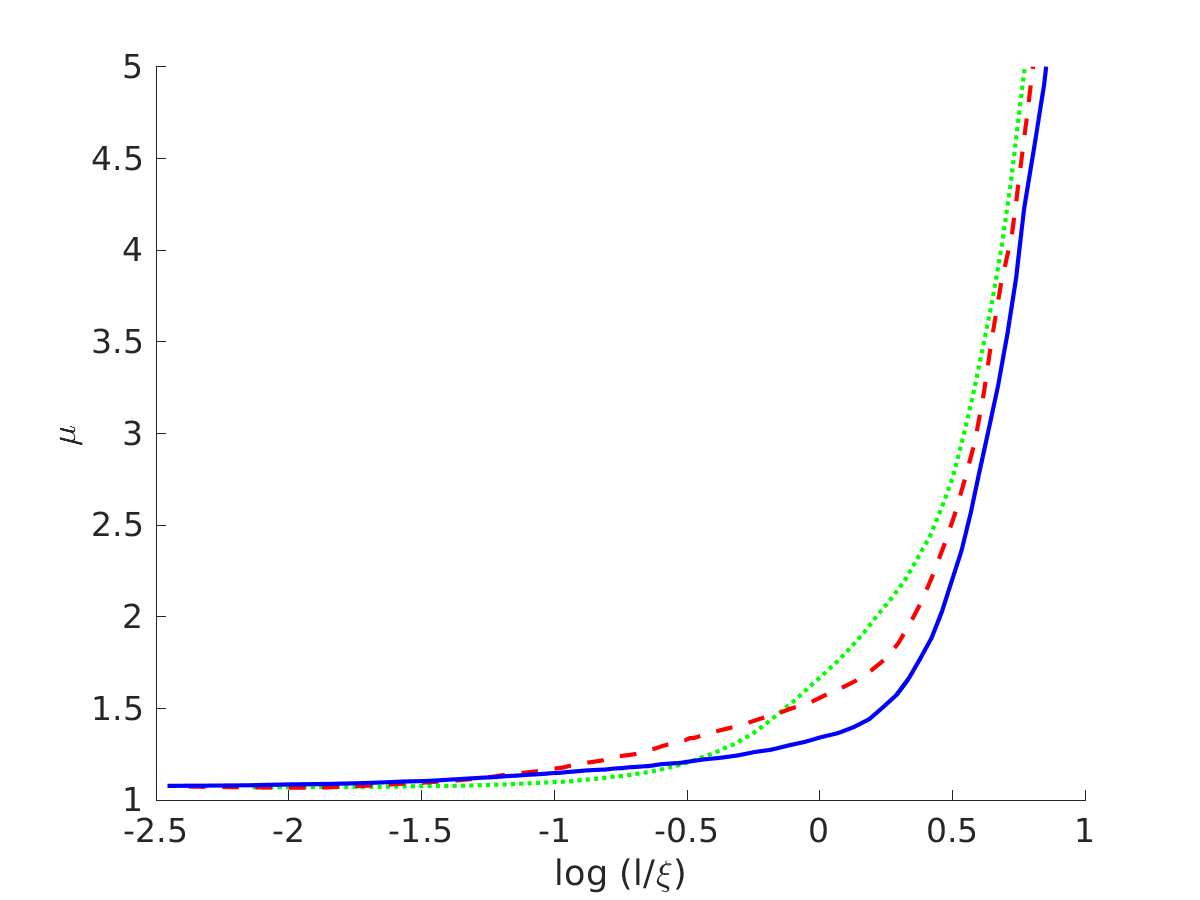}
\includegraphics[width=3in]{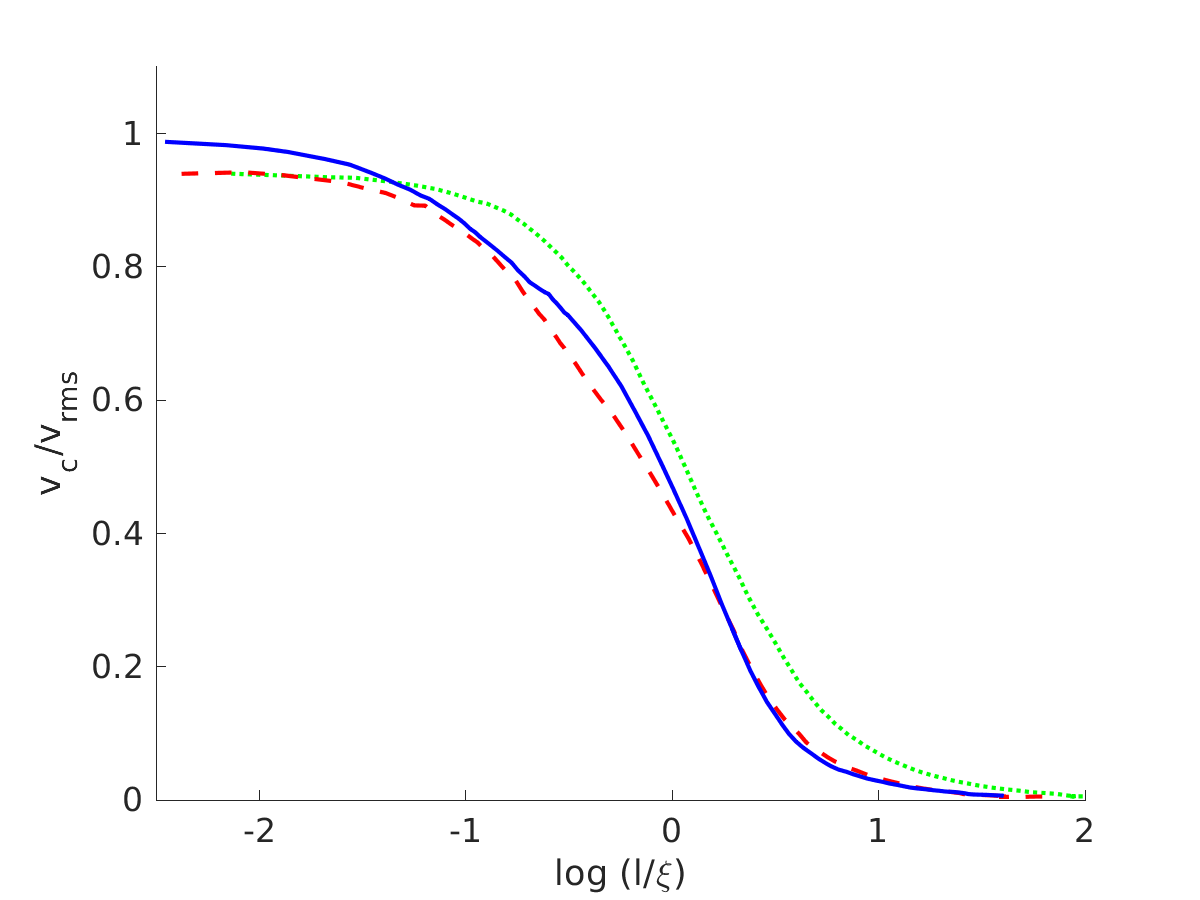}
\caption{\label{wigglyDMV}The behavior of the multifractal dimension, renormalized mass per unit length, and ratio of coherent and RMS velocities as a function of scale, for the final timestep of simulations in flat space (green dotted), radiation era (red dashed) and matter era (blue dotted).}
\end{figure*}

\begin{figure*}[!h]
\begin{center}
\includegraphics[width=3in]{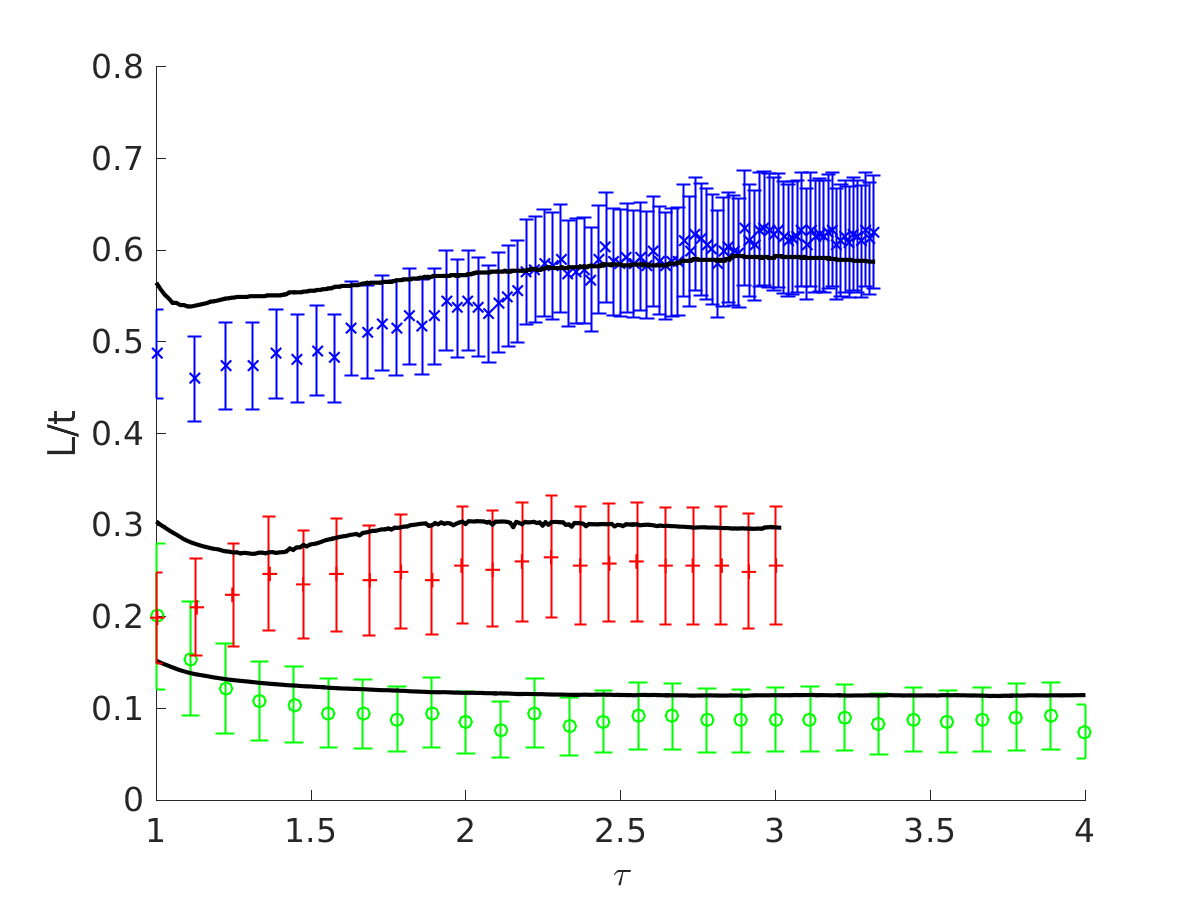}
\end{center}
\caption{\label{wigglyLXI}The behavior of the dimensionless lengthscale $L/t$, calculated from $L=\xi/\sqrt{\mu}$ using the values of $\xi$ and $\mu$ measured directly from the simulation box, in flat space (green data points), radiation era (red) and matter era (blue). Statistical error bars have been estimated from averaging values between neighboring timesteps (hence they are not independent). In all cases the black solid lines depict $L/t$ inferred from the measured total string energy in the simulation box.}
\end{figure*}

\begin{figure*}[!h]
\includegraphics[width=3in]{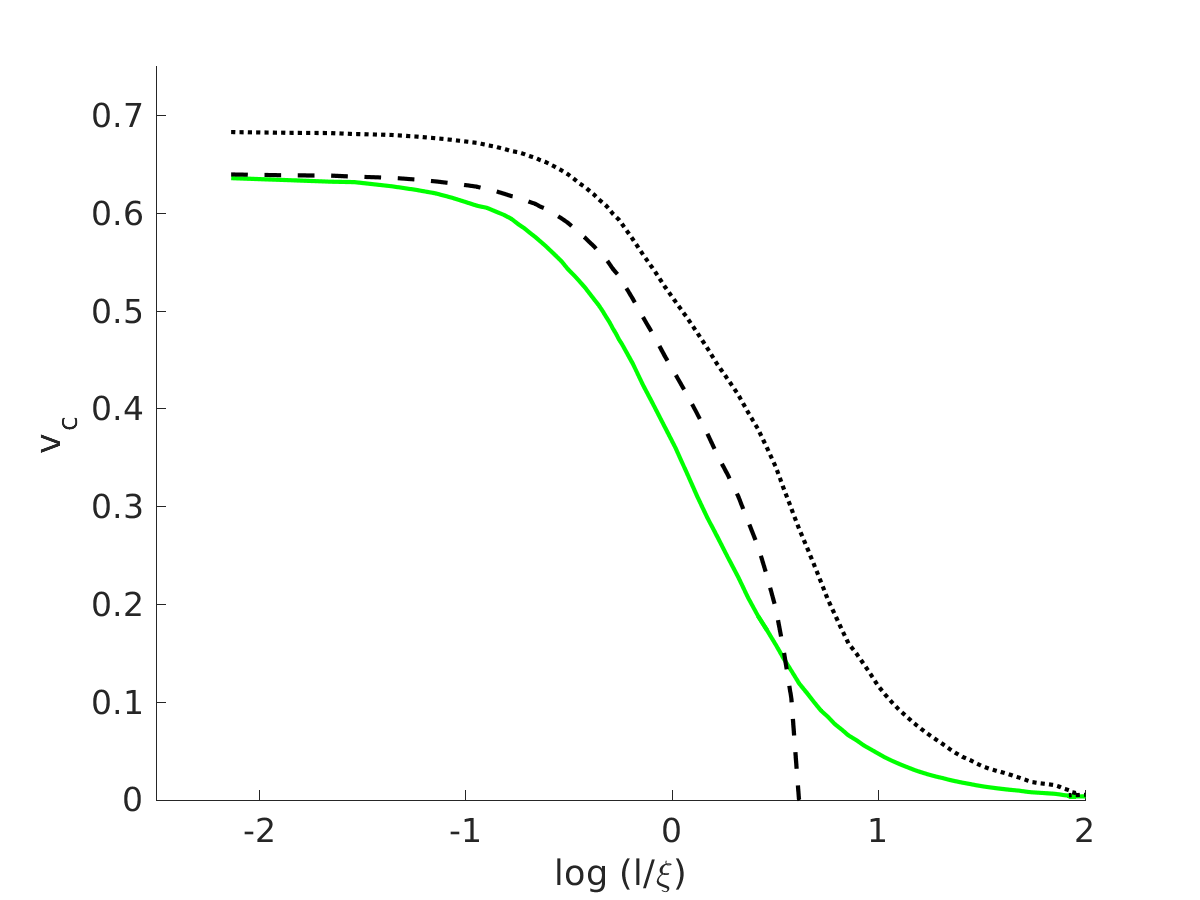}
\includegraphics[width=3in]{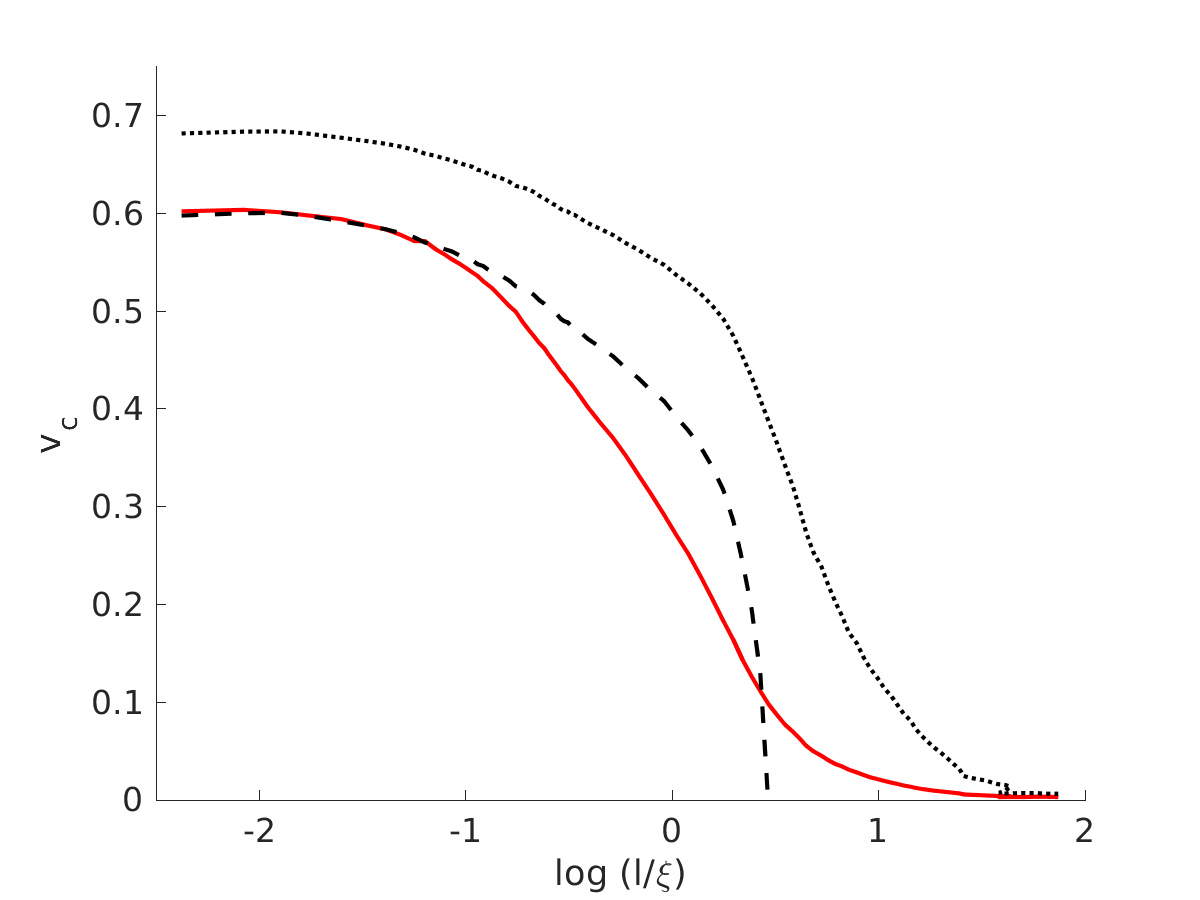}
\includegraphics[width=3in]{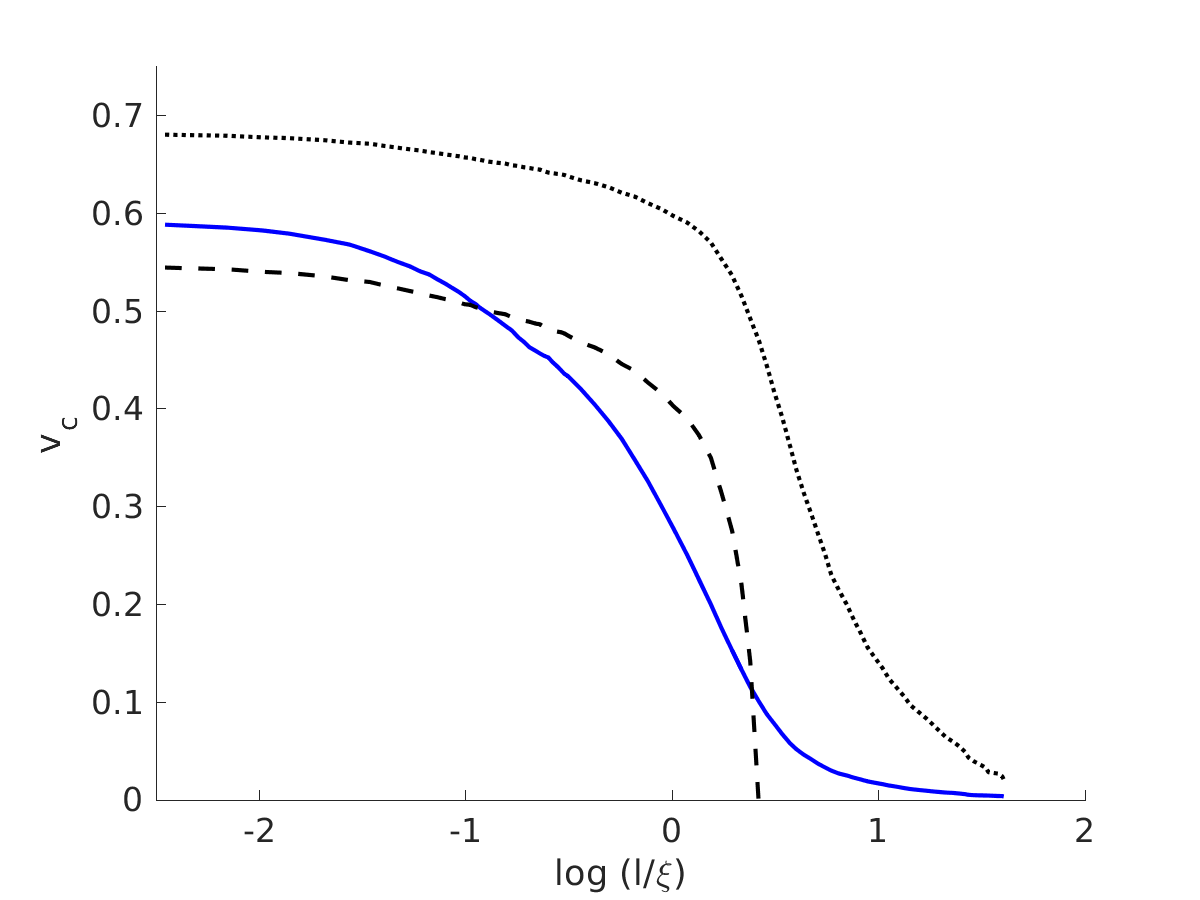}
\caption{\label{wigglyVCOH}The behavior of the average coherent string velocity as a function of scale, for the final timestep of simulations in flat space (solid green), radiation era (solid red) and matter era (solid blue). In all cases the black dashed lines depict the coherent velocity estimated using Eq.~\protect\ref{vrms1} while the black dotted lines depict the one estimated using Eq.~\protect\ref{vvrms}.}
\end{figure*}

Some data from the Goto-Nambu simulations first presented in \cite{FRAC} is shown in Figs. \ref{wigglyDMV}, \ref{wigglyLXI} and \ref{wigglyVCOH}. These are ultra-high resolution simulations, performed in the matter and radiation epochs as well as in flat (Minkowski) spacetime. The initial networks have resolutions of 75 points per correlation length (PPCL), and the simulations subsequently enforce a constant resolution in physical coordinates. Although computationally costly, this is mandatory to obtain accurate diagnostics of the small-scale properties of the network.

Figure \ref{wigglyDMV} shows the scale dependence of key properties for the final timestep of each simulation---respectively we have the multifractal dimension, the renormalized mass per unit length, and the coherent velocity. Note the similarity between the profiles for the different expansion rates (once lengths are re-scaled by the corresponding correlation length $\xi$). As emphasized in \cite{FRAC}, the main difference is the persistence of a significant amount of small-scale structure on scales slightly below the correlation length for the case of Minkowski space. In the expanding universe these structures gradually flow to smaller scales, but this does not happen in the absence of expansion: this interpretation is supported by the fact that on large scales (above the correlation length) the renormalized mass per unit length $\mu$ is larger in Minkowski space than in the expanding case, but the opposite happens for scales below about 1/3 of the correlation length.

Figure \ref{wigglyLXI} compare the values obtained from the simulations for the dimensionless lengthscale $L/t$ in two different ways: calculated from $L=\xi/\sqrt{\mu}$ using the values of $\xi$ and $\mu$ measured directly from the simulation (colored points with error bars, for each of the three epochs), and inferred from the measured total string energy in the simulation box (black line for each case). In the former case, the statistical error bars have been estimated from averaging values between neighboring timesteps (hence they are not independent). We find good overall agreement, although we see that the total string energy diagnostic gives values that are systematically high (though by a small amount) throughout the Minkowski and radiation era simulations as well as early in the matter era one. Is is encouraging that the agreement between the two is much better in the second half of the matter era simulation, where the network is expected to be scaling, as discussed in \cite{FRAC}.

Finally, Fig. \ref{wigglyVCOH} compares the behavior of the average coherent string velocity as a function of scale for the final timestep each simulations in flat space (solid color lines) to the coherent velocity estimated using Eq.~\eqref{vrms1} (solid dashed lines) and using Eq.~\protect\ref{vvrms} (solid dotted lines). One sees that the former provides a good fit on small scales but breaks down (as expected) on scales around 3 times that of the correlation length (thus, around the scale of the horizon). On the other hand the latter reproduces the overall shape of the curve reasonably well but systematically overestimates its values---by a value which is larger for faster expansion rates.

The asymptotic values of the key network parameters in these simulations are listed in Table \ref{table1}. These can be used for some preliminary calibration of the energy loss terms (which we will do shortly), although a full exploration of the parameter space (as was recently done for domain walls \cite{Rybak}) requires additional data that must come from future simulations.

\begin{table}
\centering
\begin{tabular}{|c|c|c|c|}
\hline
Parameter & Flat space & Radiation era & Matter era \\
\hline
$L/t$ & 0.10 & 0.27 & 0.62 \\
$v_{\rm rms}$ & 0.65 & 0.64 & 0.59 \\
$\xi/t$ & 0.13 & 0.31 & 0.70 \\
\hline
$\mu(\xi)$ & 1.61 & 1.42 & 1.26 \\
$v_{\ell}(\xi)$ & 0.35 & 0.35 & 0.35 \\
\hline
$\mu(\ell)$ from Eq.~\eqref{mu} & 1.69 & 1.32 & 1.27 \\
$v(\ell)$ from Eq.~\eqref{vrms1} & 0.38 & 0.50 & 0.44 \\
$v(\ell)$ from Eq.~\eqref{vvrms} & 0.51 & 0.60 & 0.62 \\
\hline
\end{tabular}
\caption{\label{table1}Asymptotic values of key network parameters in the simulations of \protect\cite{FRAC}. The first five lines are measured directly from simulations. Although no explicit error bars are provided, they are nominally expected to be around the ten percent level. The last three lines are inferred from the wiggly model, as discussed in the paper.}
\end{table}

The last three values in Table \ref{table1} are calculated by noting that $\ell$ must be the scale that makes $\xi\left(\ell\right)$ the correlation length. This way $\mu\left(\ell\right)$ is simply given by Eq.~\eqref{mu} and can be combined with $v_{rms}$ to yield $v(\ell)$ according to Eq.~\eqref{vrms1}. The equivalent result according to Eq.~\eqref{vvrms} is included for purely illustrational purposes (since, as has been discussed, we do not expect that to apply to these types of simulations).

It is interesting to notice that $\mu\left(\ell\right)$ calculated in this way is compatible with $\mu\left(\xi\right)$ taken directly from the simulations. This could be seen as evidence in favour of the natural identification $\ell=\xi$. Note also that, since $\mu\left(\ell\right)$ must be a non-decreasing function of $\ell$, the central values in Table \ref{table1} actually seem to favour $\ell>\xi$ in flat space and in the matter era. Nevertheless, this counter-intuitive apparent preference should not be too worrying as it is not statistically significant (after all, if $\ell$ truly is just $\xi$, then one would expect this sort of spread where some estimates of $\ell$ are above and some below the correlation length).

There is, however, at least one theoretical consequence of $\ell$ and $\xi$ being at least of the same order.
That is that, strictly speaking, we are not working with normal multifractal dimensions, as Eq.~\eqref{dm} has
only been shown to hold in the $\ell\ll\xi$ limit \cite{PAP1}. Nevertheless, this has no practical impact on our conclusions
as the simulations we have used to calibrate $d_{m}$ actually probe the left-hand side of Eq.~\eqref{dm} rather than the right one.

\begin{figure*}[!h]
\includegraphics[scale=0.4]{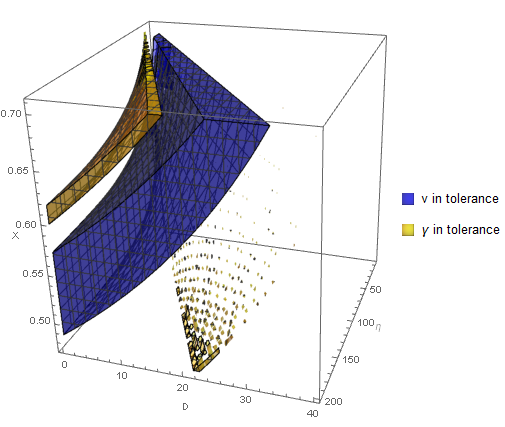} \includegraphics[scale=0.3]{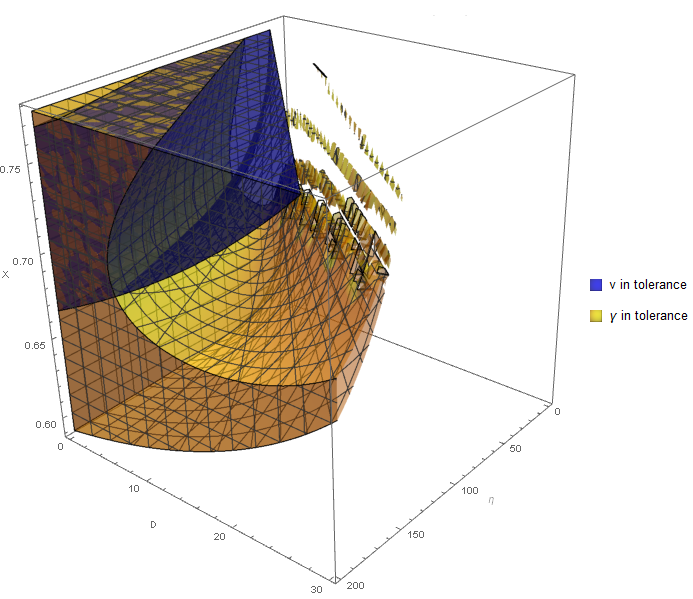}
\caption{\label{3dscaling}Model parameters that allow scaling in agreement with the results in Table \ref{table1} in the radiation era (top) and in the matter era (bottom). The blue region corresponds to scaling velocities allowed by the numerical uncertainty of our simulations and the yellow region is the equivalent for the correlation length. Interestingly, both constraints do not seem to be satisfiable in the radiation era, which seems to confirm our suspicion that strings are not approaching scaling in this era. In the matter era, it appears the overlap between the two regions is not bounded with respect to $\eta$.}
\end{figure*}

The graphs in Fig. \ref{3dscaling} show us which combinations of $\eta$, $D$, and $X_{s}$ (where $X_{s}$ can easily be related to $c$ when the other two are known) admit scaling regimes allowed by the results in Table \ref{table1}. The blue region in this figure corresponds to scaling values of $v\left(\ell\right)$ which are consistent with the values in Table \ref{table1}, and the yellow region is the analogous region concerning the correlation length. As $X$ is plotted in the range allowed by the uncertainty on $\mu\left(\xi\right)$ in the table, the allowed combinations of parameters are those in which the two regions overlap. (There would not be a qualitative difference if we did not use the $\ell=\xi$ identification and instead used the uncertainty on $\mu\left(\ell\right)$.)  These theoretical scaling values were obtained by a simple brute force implementation of the process described in subsection \ref{findscale}. Note also that the scaling regimes depicted here all come from choosing the same root of Eq.~\eqref{eq:varphi_X} as the other root yields unphysical values of $c$ in the matter era and too high velocities in the radiation era.

Interestingly, the two coloured regions in Fig. \ref{3dscaling} do not overlap in the radiation era, which supports our suspicions that a scaling regime is not being approached in that case. For the matter era, it is curious that allowed combinations of parameters seem to keep existing for arbitrarily large $\eta$ (corresponding to most energy lost to loops being in the form of small-scale loops). 

\section{\label{concl}Conclusions}

With the recent availability of high-quality CMB datasets and the forthcoming availability of comparable gravitational wave datasets, having realistic and accurate models of the evolution of networks of cosmic strings becomes a pressing problem. In this work we have taken further steps towards this goal. Specifically, we have built upon the mathematical formalism described in \cite{PAP1} for a wiggly extension of the VOS model for Goto-Nambu cosmic strings, which can describe the evolution of small-scale structure on string networks, and explored some of the consequences of this model.

Our analysis highlights the fact that the physical nature of the solutions on the model crucially depends on the dominant energy loss mechanisms for the network. Since at present these are still poorly understood, we have introduced a simple ansatz which tackles this problem in the context of an extended velocity-dependent one-scale model. We thus described a general procedure to determine all the scaling solutions admitted by a specific string model and studied their stability, enabling detailed comparisons with future numerical simulations.

Unfortunately, currently available Goto-Nambu and field theory simulations do not yet provide enough information on the small-scale properties of the network to enable a detailed comparison. (Naturally one expects that Got-Nambu simulations will be more useful in this regard, but field theory ones can also play a useful complementary role in the overall calabration of the model's large-scale properties.) The most useful currently available data is that from the Goto-Nambu simulations described in \cite{FRAC}. A comparison of our results with this data supports earlier (more qualitative) evidence that overall scaling of the network is easier to achieve in the matter era than in the radiation era. Still, the fact that a scaling solution can be reached does not {\it per se} ensure that such a solution is stable, and indeed our results show that imposing the requirement that a scaling regime be stable seems to notably constrain the allowed range of energy loss parameters.

In any case, a fully developed model for wiggly cosmic strings is now available. While it has several more free parameters than the original VOS model \cite{MS1,MS2,MS3,MS4}, we emphasize that recent advances in high-performance computing make a detailed calibration of the model's parameters a realistic possibility. Indeed this has been recently done for the analogous model for domain walls \cite{Rybak}, by comparing it to field theory simulations in universes with a range of fixed expansion rates as well as in the radiation-matter transition. In the case of cosmic strings, the possibility of comparing field theory and Goto-Nambu simulations is particularly exciting, both because it will make the calibration process more robust and because it should enable a clearer physical understanding of the relevance of the various energy loss mechanisms.

\section{Acknowledgments}
We are grateful to Ivan Rybak for helpful discussions on the subject of this work.

This work was done in the context of project PTDC/FIS/111725/2009 (FCT, Portugal), with additional support from grant UID/FIS/04434/2013. JV is supported by an STFC studentship. CJM is supported by an FCT Research Professorship, contract reference IF/00064/2012, funded by FCT/MCTES (Portugal) and POPH/FSE (EC).

This work was undertaken on the COSMOS Shared Memory system at DAMTP, University of Cambridge operated on behalf of the STFC DiRAC HPC Facility. This equipment is funded by BIS National E-infrastructure capital grant ST/J005673/1 and STFC grants ST/H008586/1, ST/K00333X/1.

\chapter{Can power spectrum observations rule out slow-roll inflation?}
\label{chap:SFSRI}

\begin{center}

J.P.P. Vieira,$^{1}$ Christian T. Byrnes,$^{1}$ and Antony Lewis$^{1}$\\[0.5cm]
$^{1}$Department of Physics \& Astronomy, University of Sussex, Brighton BN1 9QH, UK

\end{center}

\ \\

The spectral index of scalar perturbations is an important observable that allows us to learn about inflationary physics. In particular, a detection of a significant deviation from a constant spectral index could enable us to rule out
the simplest class of inflation models.
We investigate whether future observations could rule out canonical single-field slow-roll inflation given the parameters allowed
by current observational constraints. We find that future measurements of a constant running (or running of the running)
of the spectral index over currently available scales are unlikely to achieve this.
However, there remains a large region of parameter space (especially when considering the running of the running)
for falsifying the assumed class of slow-roll models if future observations accurately constrain a much
wider range of scales.

\newpage

\section{Introduction}

One of the main achievements of the recent era of precision cosmology
has been the increasing quality of measurements of the cosmic microwave
background (CMB) across the sky, for example by the
\emph{Planck} mission \cite{Planck_inf2015}. These have been invaluable
in constraining physics in the very early Universe. In particular,
these measurements can be used to measure the scale-dependence of the primordial power
spectrum, and have been instrumental in establishing
cosmic inflation as the most popular paradigm for the universe before the hot big bang.

Despite this success, so far only two perturbation parameters of relevance to inflationary models have been measured to be non-zero:
the amplitude of the scalar power spectrum and its spectral index, $n_{s}$. One consequence of this lack of measured
observables is a difficulty in differentiating between different specific models of inflation,
though the non-observation of primordial tensor modes already provides a powerful constraint on broad classes of inflationary models \cite{Martin:2013tda}.
Finding new measurable observables that could falsify some of the remaining allowed models
is one of the main goals of modern cosmology.

Although recent attempts at finding such observables have focused mostly on non-Gaussian signals
in higher-order correlation functions \cite{Vennin:2015egh,Renaux-Petel:2015bja},
there are still a few relevant quantities at the level of the power spectrum whose precision should be noticeably
improved by future probes \cite{Adshead:2010mc,Kohri:2013mxa,mudistorrunning,Munoz:2016owz,Sekiguchi:2017cdy}.
The running ($\alpha_{s}$) and the running of the
running ($\beta_{s}$) of the spectral index of scalar perturbations are examples of parameters that can be measured
more accurately in the future and are predicted to have very small magnitude
(compared to $n_{s}-1$) in the simplest classes of canonical single-field slow-roll inflation.
This is especially interesting because, even though
current constraints on these quantities are compatible with zero,
their best-fit values have an amplitude comparable to $n_{s}-1$ \cite{BKPanalysis,Planck_inf2015,running_running}. A future detection of $\alpha_s$ or $\beta_s$ could in principle
provide strong evidence against these simplest classes of inflationary models.

While a detection of $\alpha_{s}$ or $\beta_{s}$ at the same order as $n_s-1$ would rule out the simplest slow-roll models, the implications for the wider class of canonical single-field slow-roll inflation models
are less obvious and require a more general treatment.
In this paper, we study the more general implications using the well-studied formalism
for computing power spectra developed in Refs.~\cite{STEWART_RUN,DOD_STEWART,Choe_Gong_Stewart_second,Stewart_inverse1,Stewart_inverse2}.
Although we fall short of a completely generic conclusion,
our results are sufficient to show that it is much harder to rule out slow roll than the simplest arguments suggest.

Section \ref{sec:Generalised-Slow-Roll} of this paper is devoted
to motivating our treatment and introducing the formalism it is based on; section \ref{sec:Exploring-the-limits-of-slow-roll}
explains how to assess whether specific values of $\alpha_{s}$ and $\beta_{s}$ are compatible with
slow-roll inflation; and section \ref{sec:Results} presents the
results (with the main technical details of the calculations being left to the
appendices), including a comparison with current observational
bounds (effectively extending the analysis made with WMAP data in \cite{Easther:2006tv}). Finally, in section
\ref{sec:Conclusions}, we summarize our conclusions, including
a discussion of future prospects.

Throughout this work we assume a $\Lambda$CDM cosmology evolving according to general relativity seeded by fluctuations from single-field inflation, and use natural units with $c=\hbar=M_{P}^{2}=\left(8\pi G\right)^{-1}=1$.

\section{General slow-roll approximation\label{sec:Generalised-Slow-Roll}}

In canonical single-field inflation, the energy density of the Universe
is dominated by that of a scalar field, $\phi$ (the inflaton), and
thus the Hubble parameter of a flat FLRW metric is given by the first
Friedmann equation as
\begin{equation}
3H^{2}=\frac{1}{2}\dot{\phi}^{2}+V\left(\phi\right),\label{eq:Friedmann1}
\end{equation}
where $V$ is the inflaton potential and $H$ is the Hubble parameter.
The inflaton obeys the equation of motion
\begin{equation}
\ddot{\phi}+3H\dot{\phi}+V^{\prime}\left(\phi\right)=0,\label{eq:inflaton_eom}
\end{equation}
where the prime denotes differentiation with respect to argument (here with respect to $\phi$)
and the dot denotes differentiation with respect to time.

A simplifying assumption often used to study inflation models
is the slow-roll approximation, which states that the inflaton rolls
down its potential slowly enough that:
\begin{enumerate}
\item its kinetic energy is much less than its potential energy, i.e.,
\begin{equation}
\epsilon\equiv-\frac{\dot{H}}{H^{2}}=\frac{1}{2}\left(\frac{\dot{\phi}}{H}\right)^{2}\ll1;\label{eq:eps_small}
\end{equation}

\item $\ddot{\phi}$ can be neglected in Eq.~\eqref{eq:inflaton_eom}, i.e.,
\begin{equation}
\left|\delta_{1}\right|\equiv\left|\frac{\ddot{\phi}}{H\dot{\phi}}\right|\ll1.\label{eq:dealta1_small}
\end{equation}

\end{enumerate}
If this simplification is valid (which is the case for most models compatible with observations),
it is straightforward to compute the evolution of background quantities from the slow-roll equations
\begin{equation}
3H^{2}\simeq V,\label{eq:Friedmann1-1}
\end{equation}
\begin{equation}
3H\dot{\phi}+V^{\prime}\simeq0\label{eq:inflaton_eom-1}
\end{equation}
(which follow trivially from applying the slow-roll approximation to
Eqs. \eqref{eq:Friedmann1} and \eqref{eq:inflaton_eom}, respectively).

The quantities $\epsilon$ and $\delta_{1}$ defined above are known as the slow-roll parameters
(note that there are several popular alternative definitions and notations for $\delta_{1}$).
It is also possible to define ``higher-order'' slow-roll parameters, for example as
\begin{equation}
\delta_{n}\equiv\frac{1}{H^{n}\dot{\phi}}\frac{d^{n}\dot{\phi}}{dt^{n}}.\label{eq:delta_ndef}
\end{equation}
Although these parameters are not strictly important for establishing whether the
slow-roll approximation is valid, in practice it is often necessary to make assumptions regarding their relative smallness
in order to be able to compute the corresponding spectrum of scalar perturbations consistently to a given order.

\subsection{The scalar power spectrum in slow-roll inflation}

As previously noted by Stewart and Gong \cite{STEWART_RUN,STEWART_GONG},
the slow-roll approximation is not always sufficient
to accurately calculate the power spectrum of scalar perturbations.

The equation of motion for the Fourier modes of the scalar perturbations
is \cite{SL1993}
\begin{equation}
\frac{d^{2}\varphi_{k}}{d\xi^{2}}+\left(k^{2}-\frac{1}{z}\frac{d^{2}z}{d\xi^{2}}\right)\varphi_{k}=0,\label{eq:fourier_eom}
\end{equation}
where $z\equiv\frac{a\dot{\phi}}{H}$, the gauge-invariant curvature perturbation is $-\varphi_k / z$,
$\xi\equiv -\eta$ is minus the conformal time (varying from $\infty$
in the infinite past to $0$ in the infinite future),
and we assume asymptotic boundary conditions
\begin{equation}
\varphi_{k}\longrightarrow\begin{cases}
\frac{e^{ik\xi}}{\sqrt{2k}}, & k\xi\rightarrow\infty\\
A_{k}z, & k\xi\rightarrow0
\end{cases},\label{eq:fourier_boundary}
\end{equation}
 where $A_{k}$ is a constant for each wave vector $k$.

To keep track of the approximations
that will be needed, it is useful to use the rescaled variables
\begin{equation}
y\equiv\sqrt{2k}\varphi_{k},\label{eq:ydef}
\end{equation}
\begin{equation}
x\equiv k\xi.\label{eq:xdef}
\end{equation}
Using these we can  rewrite the equation of motion for each Fourier mode as
\begin{equation}
\frac{d^{2}y}{dx^{2}}+\left(1-\frac{2}{x^{2}}\right)y=\frac{g\left(\ln x\right)}{x^{2}}y,\label{eq:y_eom}
\end{equation}
where the important function $g$ is defined in terms of
\begin{equation}
f\left(\ln\xi\right)\equiv\frac{2\pi a\xi\dot{\phi}}{H}\label{eq:fdef}
\end{equation}
as
\begin{equation}
g\left(\ln x\right)\equiv\left[\frac{f^{\prime\prime}-3f^{\prime}}{f}\right]_{\xi=\frac{x}{k}}.\label{eq:gdef}
\end{equation}

The power spectrum can be straightforwardly (although not necessarily easily) calculated by solving
Eq.~\eqref{eq:y_eom} and then finding
\begin{equation}
\mathcal{P}\left(k\right)=\lim_{x\rightarrow0}\left|\frac{xy}{f}\right|^{2}.\label{eq:Ps_y}
\end{equation}
The homogeneous solution (for $g=0$),
\begin{equation}
y_{0}\left(x\right)=\left(1+\frac{i}{x}\right)e^{ix},\label{eq:y0}
\end{equation}
together with the relation (which is justified later in appendix \ref{sec:gappendix})
\begin{equation}
\xi=\frac{1}{aH}\left(1+\mathcal{O}\left(g\right)\right)\label{eq:eta_aH}
\end{equation}
lead, at zeroth order in $g$, to the simple scale-invariant\footnote{It can be seen
(for example through Eq.~\eqref{eq:Friedmann1}) that this result is divergent, as expected in a de Sitter background. This is not a problem
as all that matters is that when this becomes the leading contribution to a more
realistic power spectrum it is approximately scale-invariant (which is guaranteed by the slow-roll approximation).}
power spectrum
\begin{equation}
\mathcal{P}_{0}\left(k\right)=\lim_{x\rightarrow0}\left|\frac{i}{f}\right|^{2}=\frac{H^{4}}{\left(2\pi\dot{\phi}\right)^{2}}.\label{eq:standard_Ps}
\end{equation}

The standard slow-roll result can then be obtained by arguing that
in a more general slow-roll scenario (with small $g\neq0$) the leading
contribution to the power spectrum (with corrections being suppressed
by terms of order $g$) will still be given by Eq.~\eqref{eq:standard_Ps}
if the now non-constant terms are evaluated at some point around horizon crossing.

\subsection{The spectral index in general slow-roll inflation\label{specindex_GSRS}}

The slow-roll approximation has been sufficient
to derive the standard lowest-order result of Eq.~\eqref{eq:standard_Ps}.
However, to derive the standard first-order prediction for the spectral index \cite{Liddle:1992wi},
\begin{equation}
n_{s}-1\equiv\frac{d\ln\mathcal{P}}{d\ln k}=-4\epsilon-2\delta_{1}\label{eq:ns_1st_standard}
\end{equation}
(where the slow-roll parameters are to be evaluated around the time of horizon crossing),
 the first-order corrections to Eq.~\eqref{eq:standard_Ps}
must only give at most a second-order contribution to $n_{s}-1$,
which is not true in general. Ignoring those corrections, as is usually done, requires a hierarchy for higher-order slow-roll parameters such that~\cite{STEWART_RUN,DOD_STEWART}
\begin{equation}
\left|\delta_{n+1}\right|\ll\left|\delta_{n}\right|,\label{eq:usualhierarchy}
\end{equation}
which does not necessarily follow
from the ``vanilla'' slow-roll assumptions.

Assuming this hierarchy of slow-roll parameters, the leading-order prediction for the running of the spectral index becomes
\begin{equation}
\alpha_{s}\equiv\frac{dn_{s}}{d\ln k}=-2\delta_{2}-8\epsilon^{2}+2\delta_{1}^{2}-10\epsilon\delta_{1},\label{eq:alpha_def}
\end{equation}
so that (barring fine-tuning effects) $\alpha_{s} \sim {\cal{O}} \left(\left| n_{s}-1\right|^{2}\right)\ll\left| n_{s}-1\right|$,
which motivates the naive expectation that $\alpha_{s}$ be negligible in slow-roll inflation. Mutatis mutandis, it can be seen that the equivalent expectation for the running of the running,
\begin{equation}
\beta_{s}\equiv\frac{d^{2}n_{s}}{d\ln k^{2}},\label{eq:beta_def}
\end{equation}
is that $\left|\beta_{s}\right|\sim{\cal{O}}\left(\left| n_{s}-1\right|^{3}\right)\ll\left| \alpha_{s}\right|\ll\left|n_{s}-1\right|$.

\begin{figure}[h]
\begin{center}
\includegraphics[scale=0.9]{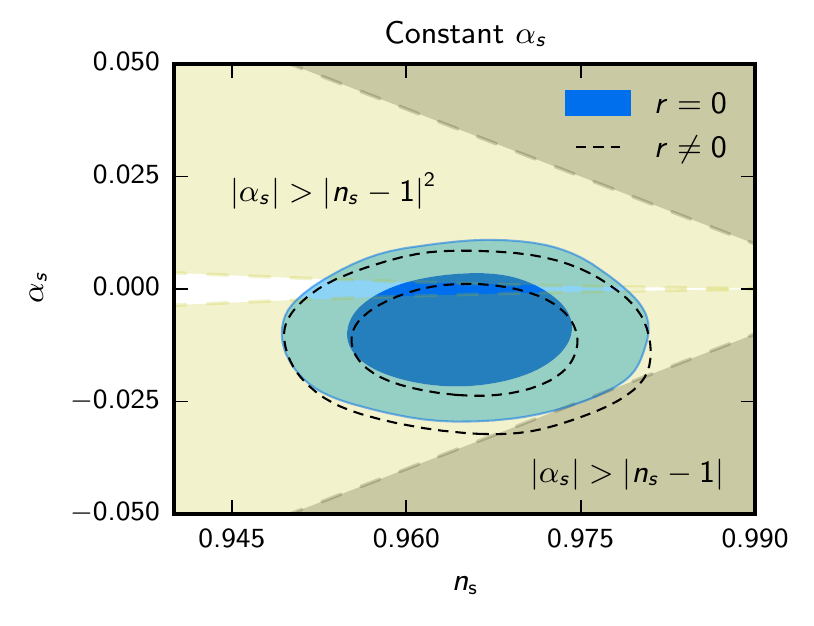}\includegraphics[scale=0.9]{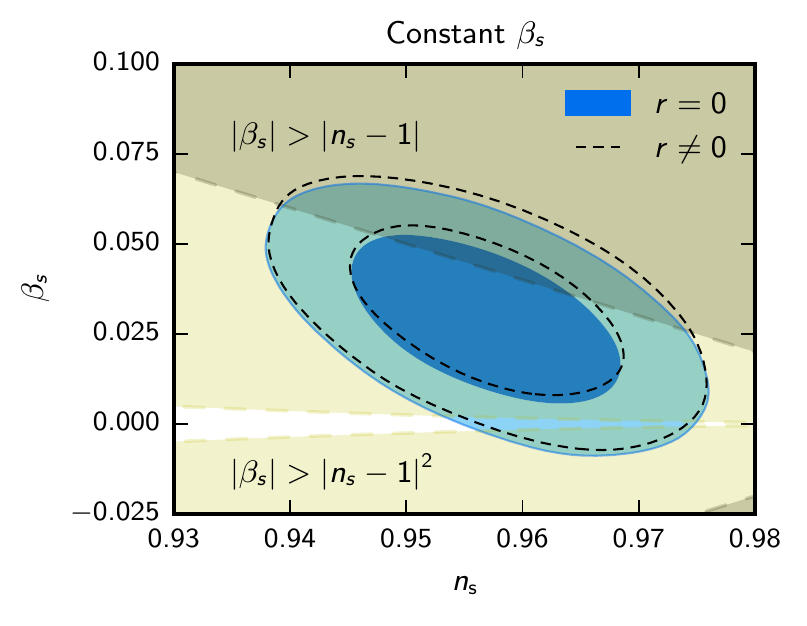}
\caption{\label{fig:motivation} Constraints  from \emph{Planck} 2015 TT+lowTEB~\cite{Planck_inf2015} and BICEP-Keck~\cite{Array:2015xqh} on a constant $\alpha_{s}$ (left) and
a constant $\beta_{s}$ (marginalized over $\alpha_{s}$ at the pivot scale; right), both against $n_{s}$ at the pivot scale.
Dashed black contours assume a null tensor-to-scalar ratio, $r$, whereas blue contours marginalize over it.
The light shaded region corresponds to the part of the parameter space where the quantity in the vertical axis becomes greater than
$\left|n_{s}-1\right|^{2}$ and the dark shaded region is where it becomes greater
than $\left|n_{s}-1\right|$. The naive expectation is that the true value of $\alpha_{s}$ (left) should be close to the boundary of the
unshaded region and far away from the dark shaded region, whereas that of $\beta_{s}$ (right) should be well inside the
unshaded region. Current constraints allow a much greater area of the parameter space.}
\end{center}
\end{figure}

As is shown in figure \ref{fig:motivation}, although current constraints are consistent with small $\alpha_{s}$ and $\beta_{s}$ as predicted by the naive hierarchy, much larger values are still currently allowed. Indeed, the
posteriors currently peak substantially away from zero, especially for $\beta_{s}$ (largely due to the low-$\ell$ feature in the CMB temperature power spectrum \cite{Ade:2013zuv}).
Improved future constraints\footnote{Note that next-generation
missions may improve these bounds by about an order of magnitude \cite{Munoz:2016owz,Kohri:2013mxa}.} on $\alpha_{s}$ and $\beta_{s}$ that peak away from zero could rule out
the simplest class of inflationary models (characterized by the slow-roll approximation
and the hierarchy in Eq.~\eqref{eq:usualhierarchy}), but a more general statement about
the wider class of canonical single-field slow-roll inflation models requires a more general treatment.

A few ways to approach modelling a more general slow-roll scenario are available in the literature \cite{STEWART_RUN,Choe_Gong_Stewart_second,DVORKIN_GSR,Stewart_inverse2,
Stewart_inverse1,GSR_inflEFT,GSRfeatures,GSRtensor,GSRinin,Adshead:2013zfa}.
In this work, we use the results from Ref.~\cite{Stewart_inverse2}
(which in turn use the results from Ref.~\cite{Choe_Gong_Stewart_second}), which we briefly review.

To solve for the power spectrum (Eq.~\eqref{eq:Ps_y}) we need to solve for the modes $y$.
The second-order linear differential equation in Eq.~\eqref{eq:y_eom} can be solved for $y$ using Green's functions, with the solution satisfying the boundary conditions of Eq.~\eqref{eq:fourier_boundary} given implicitly by
\begin{equation}
y\left(x\right)=y_{0}\left(x\right)+\frac{i}{2}\intop_{x}^{\infty}\frac{du}{u^{2}}g\left(\ln u\right)\left[y_{0}^{*}\left(u\right)y_{0}\left(x\right)-y_{0}^{*}\left(x\right)y_{0}\left(u\right)\right]y\left(u\right).\label{eq:integral_y}
\end{equation}
This can be solved iteratively for $y$ to successively higher order in $g$ (assuming $\left|g\right|<1$) by substituting
the previous order result into the right-hand
side of Eq.~\eqref{eq:integral_y} (starting with $y\left(u\right)=y_{0}\left(u\right))$.
The result for the power spectrum at the desired order can then be obtained by substituting into Eq.~\eqref{eq:Ps_y}
and simplifying as much as possible. The result for the scalar power spectrum correct to quadratic order in $g$ is then \cite{Choe_Gong_Stewart_second}
\begin{multline}
\ln\mathcal{P}\left(\ln k\right)=\intop_{0}^{\infty}\frac{d\xi}{\xi}\left[-k\xi W^{\prime}\left(k\xi\right)\right]\left[\ln\frac{1}{f\left(\ln\xi\right)^{2}}+\frac{2}{3}\frac{f^{\prime}\left(\ln\xi\right)}{f\left(\ln\xi\right)}\right]+\frac{\pi^{2}}{2}\left[\intop_{0}^{\infty}\frac{d\xi}{\xi}m\left(k\xi\right)\frac{f^{\prime}\left(\ln\xi\right)}{f\left(\ln\xi\right)}\right]^{2}
\\
-2\pi\intop_{0}^{\infty}\frac{d\xi}{\xi}m\left(k\xi\right)\frac{f^{\prime}\left(\ln\xi\right)}{f\left(\ln\xi\right)}\intop_{\xi}^{\infty}\frac{d\zeta}{\zeta}\frac{1}{k\zeta}\frac{f^{\prime}\left(\ln\zeta\right)}{f\left(\ln\zeta\right)}+{\cal{O}}\left(g^{3}\right),\label{eq:P_f_2nd}
\end{multline}
where $W$ and $m$ are window functions defined by
\begin{equation}
W\left(x\right)=\frac{3\sin\left(2x\right)}{x^{3}}-\frac{3\cos\left(2x\right)}{x^{2}}-\frac{3\sin\left(2x\right)}{2x}-1\label{eq:Wdef}
\end{equation}
 and
\begin{equation}
m\left(x\right)=\frac{2}{\pi}\left[\frac{1}{x}-\frac{\cos\left(2x\right)}{x}-\sin\left(2x\right)\right].\label{eq:mdef}
\end{equation}

In this paper we are interested in relating properties of the observable power spectrum to those of the inflationary model,
so we need the inverse version of this result, which can be shown to be \cite{Stewart_inverse2}
\newform{
\begin{multline}
\ln\frac{1}{f\left(\ln\xi\right)^{2}}= \intop_{0}^{\infty}\frac{dk}{k}m\left(k\xi\right)\ln\mathcal{P}\left(\ln k\right)
-\frac{\pi^2}{8}\intop_{0}^{\infty}\frac{dk}{k}m\left(k\xi\right)
\left[\intop_{0}^{\infty}\frac{dl}{l}
 \frac{\mathcal{P}'(\ln l)}{\mathcal{P}(\ln l)}
 \intop_{0}^{\infty}\frac{d\zeta}{\zeta}m\left(k\zeta\right)m\left(l\zeta\right)
 \right]^{2}
\\+\frac{\pi}{2}\intop_{0}^{\infty}\frac{dl}{l} \frac{\mathcal{P}'(\ln l)}{\mathcal{P}(\ln l)} \intop_{0}^{\infty}\frac{dq}{q}
\frac{\mathcal{P}'(\ln q)}{\mathcal{P}(\ln q)}
\intop_{0}^{\infty}\frac{d\zeta}{\zeta}m\left(l\zeta\right)\intop_{0}^{\infty}\frac{dk}{k^{2}}m\left(k\xi\right)m\left(k\zeta\right)\intop_{\zeta}^{\infty}\frac{d\chi}{\chi^{2}}m\left(q\chi\right).
\label{eq:inv2_formula}
\end{multline}
}{
\begin{multline}
\ln\frac{1}{f\left(\ln\xi\right)^{2}}=\intop_{0}^{\infty}\frac{dk}{k}m\left(k\xi\right)\ln\mathcal{P}\left(\ln k\right)
\\
-\frac{1}{2\pi^{2}}\intop_{0}^{\infty}\frac{dk}{k}m\left(k\xi\right)\left[\intop_{0}^{\infty}\frac{dl}{l}\ln\left|\frac{k+l}{k-l}\right|\frac{\mathcal{P}^{\prime}\left(\ln l\right)}{\mathcal{P}\left(\ln l\right)}\right]^{2}
\\
+\intop_{0}^{\infty}\frac{dl}{l}\intop_{0}^{\infty}\frac{dq}{q}M\left(l\xi,q\xi\right)\frac{\mathcal{P}^{\prime}\left(\ln l\right)}{\mathcal{P}\left(\ln l\right)}\frac{\mathcal{P}^{\prime}\left(\ln q\right)}{\mathcal{P}\left(\ln q\right)},\label{eq:inv2_formula}
\end{multline}
where we have introduced the new window function $M$, which can be
written as
\begin{equation}
M\left(x,y\right)=\frac{2}{\pi^{2}xy}\left[h\left(x\right)+h\left(y\right)-\frac{1}{2}h\left(x-y\right)-\frac{1}{2}h\left(x+y\right)\right],\label{eq:Mdef}
\end{equation}
where we have defined
\begin{equation}
h\left(x\right)=x\,\mathrm{Si}\left(2x\right)+\frac{\cos\left(2x\right)}{2}-\frac{1}{2},\label{eq:h_g_M}
\end{equation}
where $\mathrm{Si}$ stands for the sine integral function,
\begin{equation}
\mathrm{Si}\left(x\right)\equiv\intop_{0}^{x}\frac{\sin t}{t}dt.\label{eq:sinint}
\end{equation}
}

\section{Exploring the limits of slow-roll\label{sec:Exploring-the-limits-of-slow-roll}}

\subsection{How slow is slow-roll?}

To assess how much running there can be in slow-roll inflation, we would like
some objective criteria to decide whether any given
inflationary model is slow-roll or not.
The ``$\ll$'' signs in Eqs. \eqref{eq:eps_small}-\eqref{eq:dealta1_small}
defining the slow-roll approximations do not allow a clear distinction unless the numbers being compared are
orders of magnitude apart. To make matters worse, Eq.~\eqref{eq:dealta1_small}
has been defined in the literature in terms of a number of slightly
different slow-roll parameters (usually referred to as $\eta$), all
of which would lead to different classifications of borderline
cases even if we were to decide on an objective meaning for ``$\ll$''
in these equations.

When faced with this sort of problem it is important not to get lost
in an overly semantic discussion. One pragmatic reason to care about whether a model falls under the
category of slow-roll is simply to know whether the power spectrum can be straightforwardly computed
using results like Eq.~\eqref{eq:standard_Ps} and Eq.~\eqref{eq:P_f_2nd}.
Therefore, from the perspective of this work, the best way to define
slow-roll is in terms of a quantity that can quantify how precise
this formula actually is. From the derivation, the
most natural quantity appears to be the parameter $g$. Unfortunately, this will result in a slightly stronger definition than using
just the slow-roll approximation, as it discards scenarios in which $\delta_{2}$ is large
but $\epsilon$ and $\delta_{1}$ remain small (see appendix \ref{sec:gappendix}).
Nevertheless, it is a weak enough definition that we will be able to
qualitatively improve on the simplistic constraints in subsection \ref{specindex_GSRS}\footnote{
To calculate the power spectra for specific slow-roll potentials,
one could always resort to the more general formalism of Generalized Slow-Roll \cite{DVORKIN_GSR},
which relies on a weaker assumption than the slow-roll approximation (allowing for even $\delta_{1}$ to
become large for short periods of time). However, our analysis would be much more complicated
in that context, both due to difficulties in defining slow-roll (which is the regime we are interested in here)
and due to the added difficulties in solving the inverse problem of finding the model that corresponds
to a given power spectrum.}.

Instead of committing to any arbitrary definition of what a ``very small'' number is, we show,
for each combination of observable parameter values, how large $g$ can become during the period of time
in which observable scales crossed the horizon. The reader can not only decide which values are
``not small'' on his/her own, but also have a good understanding of the meaning of any specific choice:
the larger the allowed values, the less accurate our formulas.

\subsection{Outline of the method}

We start by parameterizing the observed scalar power spectrum as
\begin{equation}
\ln\mathcal{P}\left(\ln k\right)=\sum_{n=0}^{N}\frac{\beta_{n}}{n!}\left(\ln\frac{k}{k_{0}}\right)^{n},\label{eq:P_ansatz}
\end{equation}
where $k_{0}$ is a pivot scale and the $\beta_{n}$ coefficients
are to be constrained by observations. Of course,
\begin{equation}
\begin{split}
& \beta_{0}\equiv\ln\mathcal{P}_{0} \\
& \beta_{1}\equiv n_{s}-1 \\
& \beta_{2}\equiv\alpha_{s} \\
& \beta_{3}\equiv\beta_{s}
\label{eq:betans}
\end{split}
\end{equation}
where $\mathcal{P}_{0}$ is the magnitude of the power spectrum at
the pivot scale. For the purposes of this work, we will be interested
in the cases with $N=2$ and $N=3$, for which $\beta_{N}$ have already
been constrained by the \emph{Planck} collaboration \cite{Planck_inf2015}\footnote{
Other works \cite{running_running} have claimed slightly more dramatic constraints for the
$N=3$ case.}. A natural extension of our calculations is sufficient to deal with cases with higher $N$ should observational constraints
on higher-order runnings become available (it has been claimed such constraints could come from minihalo effects on 21cm fluctuations \cite{Sekiguchi:2017cdy}). Likewise, radically different parameterizations of the power spectrum
can be incorporated by making the appropriate changes to Eq.~\eqref{eq:P_ansatz}.

For each point in the $\left(\beta_{0},\beta_{1},...,\beta_{N}\right)$
parameter space, we want to know to what extent a canonical single-field inflation model
must violate slow-roll during the interval of time during which observable scales left the horizon
(i.e., how large its respective $g$ function must become during that time).

We proceed by defining a $g\left(\ln x\right)$ for every $k$ by inserting the
power spectrum from Eq.~\eqref{eq:P_ansatz}\footnote{We can ignore the term
with $\beta_{0}$ since, from Eq.~\eqref{eq:inv2_formula},
it only contributes to a proportionality constant in $f$, and thus
has no effect on $g$.%
} into Eq.~\eqref{eq:inv2_formula}, and then the resulting
$f\left(\ln\xi\right)$ into Eq.~\eqref{eq:gdef}.
The main obstacle in the way of this calculation is the computation
of the integrals in Eq.~\eqref{eq:inv2_formula}
when the power spectrum is a polynomial in $\ln \frac{k}{k_{0}}$, as we assume (in Eq.~\eqref{eq:P_ansatz}).
To solve these integrals for power spectra with non-null $\beta_{s}$, we extended
known results for the standard hierarchy in the slow-roll approximation \cite{Stewart_inverse2} (see
appendix \ref{sec:Solving-the-integrals}).
The results are polynomials in $\ln\left(k_{0}\xi\right)$, so $g$ can then be found straightforwardly
from Eq.~\eqref{eq:gdef} by differentiation.

Once these $g$ functions have been found for $k$ corresponding to observable scales, we check the absolute value of $g$
at $x=1+\epsilon=1+\frac{r}{16}$, corresponding to the time of horizon-crossing to leading order in $\epsilon$
(see, e.g., Eqs.~\eqref{eq:eta_epsbar} and \eqref{eq:eps_bar_slow_final} in appendix \ref{sec:gappendix})\footnote{
The reason we are justified in resorting to a first-order result after using second-order results up until this point
is that $r$ is already observationally constrained to be so small that even the leading order term has
no significant effect on our constraints.}.

\section{Results\label{sec:Results}}

The method described in section \ref{sec:Exploring-the-limits-of-slow-roll}
was implemented in a Python code using the results from appendix \ref{sec:Solving-the-integrals}.
This allowed us to draw contour plots indicating how large $g$ can
get during the relevant epochs for different pivot values of $n_{s}$,
$\alpha_{s}$, and $\beta_{s}$, assuming that Eq.~\eqref{eq:P_ansatz}
holds for a specific range of observable scales. In this work we present results
for three different ranges of observable scales, from $k_{\rm min}=10^{-3}\mathrm{Mpc^{-1}}$ (set by the largest scales that can be
reasonably well measured) up to: $k_{\rm max}=0.3\mathrm{Mpc^{-1}}$ (spanning about 6 efoldings), roughly
corresponding to the smallest scale well constrained by Planck;
$k_{\rm max}=100\mathrm{Mpc^{-1}}$ (spanning about 12 efoldings), roughly corresponding
to a future constraint from 21cm observations \cite{Loeb:2003ya,Sekiguchi:2017cdy}; and $k_{\rm max}=10^4\mathrm{Mpc^{-1}}$
(spanning about 16 efoldings), roughly corresponding to the smallest scale constrained by spectral distortions \cite{Chluba:2012we,Khatri:2013dha}\footnote{We note that the supernova lensing dispersion can also probe the averaged value of the power spectrum on small scales, down to $k_{\rm max}\gtrsim 100 \mathrm{Mpc^{-1}}$, but there is a degeneracy with the effect of baryons on the small-scale low-redshift power spectrum \cite{Ben-Dayan:2015zha}.}.
The pivot scale is taken to be $k_{0}=0.05\mathrm{Mpc^{-1}}$, the \emph{Planck} pivot scale\footnote{Note
that this value is only important for including observational constraints in our plots. Naturally, if one merely wanted to know
how large a constant $\alpha_{s}$ or $\beta_{s}$ are allowed to be over a certain range of scales,
the pivot scale would be irrelevant (for example because it plays no role in Eq.~\eqref{eq:inv2_formula}).}.

In order to make statements about the status of this class of canonical single-field slow-roll inflation,
we use CosmoMC \cite{Lewis:2002ah,Lewis:2013hha} to superimpose
current constraints from \emph{Planck} 2015 data (temperature plus low-$\ell$ polarization, TT + lowTEB~\cite{Planck_inf2015})
and the latest BICEP-KECK-\emph{Planck} joint analysis~\cite{Array:2015xqh}, showing the
 $1\sigma$ and $2\sigma$ allowed regions.
Additionally, we plot contours for the inferred maximum values of $g$ from current observations
against $\alpha_{s}$ and $\beta_{s}$.

\subsection{Constant running ($N=2$)}

\begin{figure}[h]
\begin{center}
\includegraphics[scale=0.9]{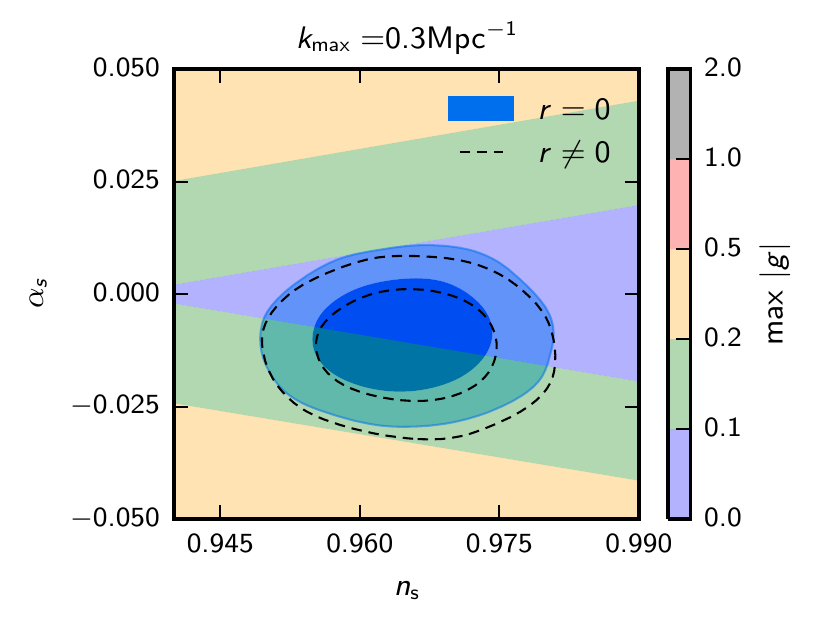}
\includegraphics[scale=0.9]{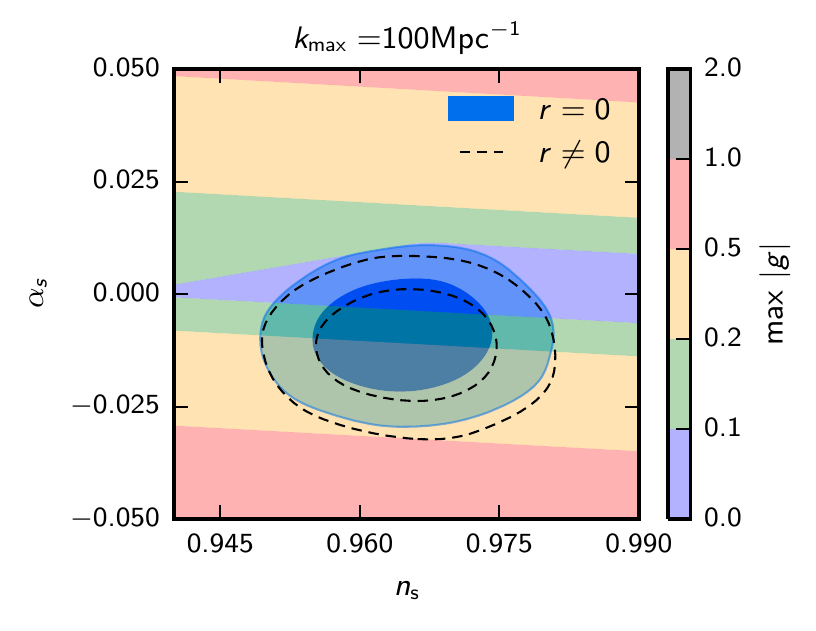}\includegraphics[scale=0.9]{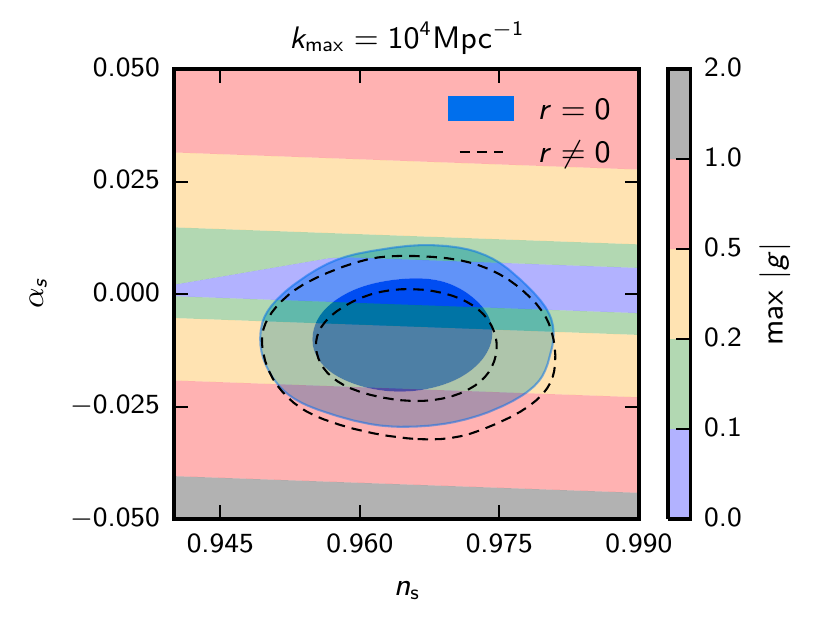}
\caption{\label{fig:alpha_fig}
Slow-roll and observational constraints on parameterizations of the power spectrum with a constant $\alpha_{s}$.
The observational contours are \emph{Planck} 2015 TT+lowTEB and joint BICEP-Keck-\emph{Planck} constraints for a constant $\alpha_{s}$ and $n_{s}$ at the pivot scale. Blue contours assume a null tensor-to-scalar ratio, $r$, whereas dashed black contours marginalise over allowed values of $r\geq 0$.
The coloured areas indicate the maximum magnitude of $g$ during the interval of time during which constrained scales
left the horizon.
Note that for $g>1$ our method breaks down as Eq.~\eqref{eq:P_f_2nd} ceases to be valid.}
\end{center}
\end{figure}

If we limit ourselves to the case with constant $\alpha_{s}$ (corresponding
to $N=2$ in Eq.~\eqref{eq:P_ansatz}) we have only two relevant observables:
$\alpha_{s}$ and $n_{s}$ at the pivot scale. The corresponding plots for the magnitude
of $g$ can be found in figure \ref{fig:alpha_fig}.

For the currently constrained range of scales even the 2$\sigma$ observational contours never go
beyond the $\left|g\right|<0.2$ line (which is still comfortably much less than unity). Even our
futuristic scenario with $k_{\rm max}=100\rm{Mpc^{-1}}$  has the 2$\sigma$ contour being well inside
the $\left|g\right|<0.5$ region (which corresponds to a borderline case for which the designation of ``slow-roll''
is rather dubious, but which still does not allow us to make a very strong statement\footnote{Note that
for such high values of $\left|g\right|$ we also need to worry about corrections to
Eq.~\eqref{eq:P_f_2nd} possibly becoming comparable to the observational uncertainty for the power spectrum
at the (futuristic) scale at which this maximum value is reached.}).
Only a futuristic scenario with $k_{\rm  max}=10^4\rm{Mpc^{-1}}$ would permit a measurement of constant $\alpha_s$ to
provide a strong test of slow roll. However, the usual constraints from $\mu$- and $y$-type
spectral distortions would depend on integrals of the power spectrum over the range
$1{\rm Mpc}^{-1}\lesssim k\lesssim 10^4{\rm Mpc}^{-1}$, and cannot on their own establish
the constancy of $\alpha_s$ (even if they could accurately measure $\alpha_s$ provided it is assumed to be constant \cite{Chluba:2012we}). Nevertheless, smaller residual distortions of a different type might provide some information on
the shape of the power spectrum \cite{Chluba:2013pya}.

\begin{figure}[h]
\begin{center}
\includegraphics[scale=0.9]{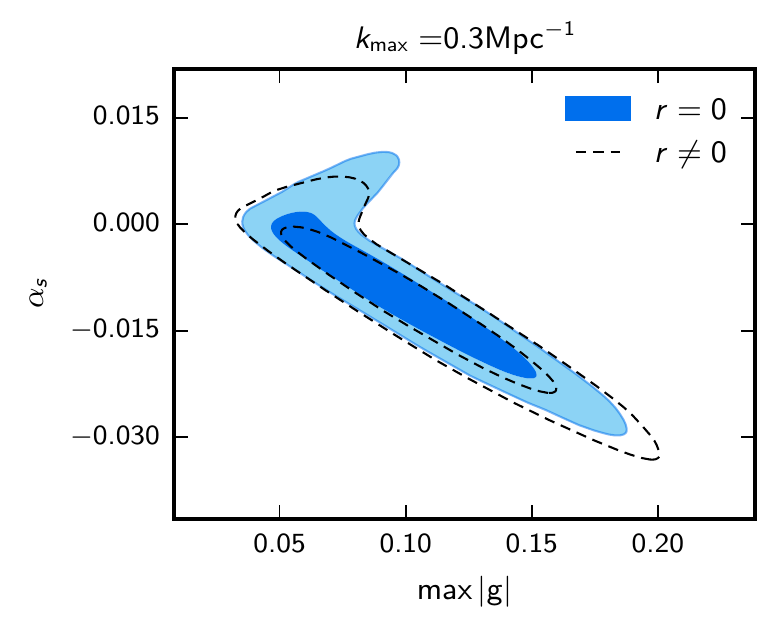}
\includegraphics[scale=0.9]{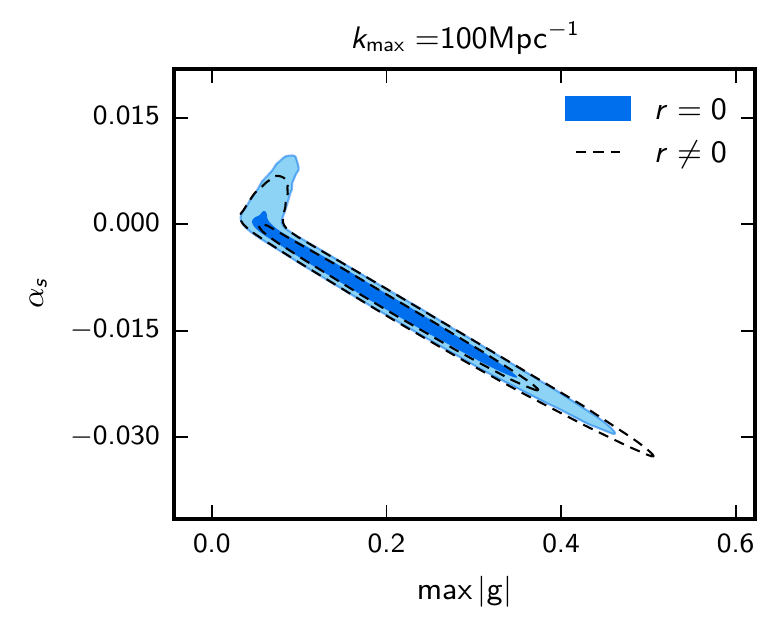}\includegraphics[scale=0.9]{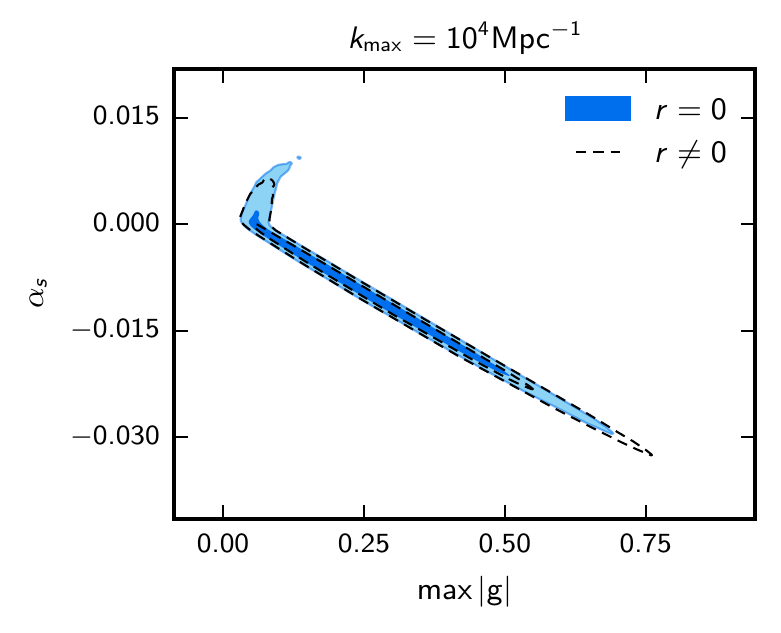}
\caption{\label{fig:galpha_fig}Bounds, over the constrained ranges of scales, on the maximum magnitude of $g$ inferred from
\emph{Planck} 2015 TT+lowTEB and joint BICEP-Keck-\emph{Planck} constraints on a constant $\alpha_{s}$ marginalized over $n_{s}$ at the pivot scale.
Filled contours assume $r=0$ whereas dashed lines marginalize over allowed values of $r\geq 0$.}
\end{center}
\end{figure}

These conclusions are confirmed (and more easily seen) in the plots in figure \ref{fig:galpha_fig}, which show the bounds
on the maximum magnitude of $g$ inferred from the bounds on the running and the spectral index. Note that their asymmetric
boomerang shape is due to the modulus sign in $\left|g\right|$ as well as the significant deviation of the \emph{Planck} best-fit value
for $\alpha_{s}$ from zero.

\subsection{Constant running of the running ($N=3$)}

\begin{figure}[h]
\begin{center}
\includegraphics[scale=0.9]{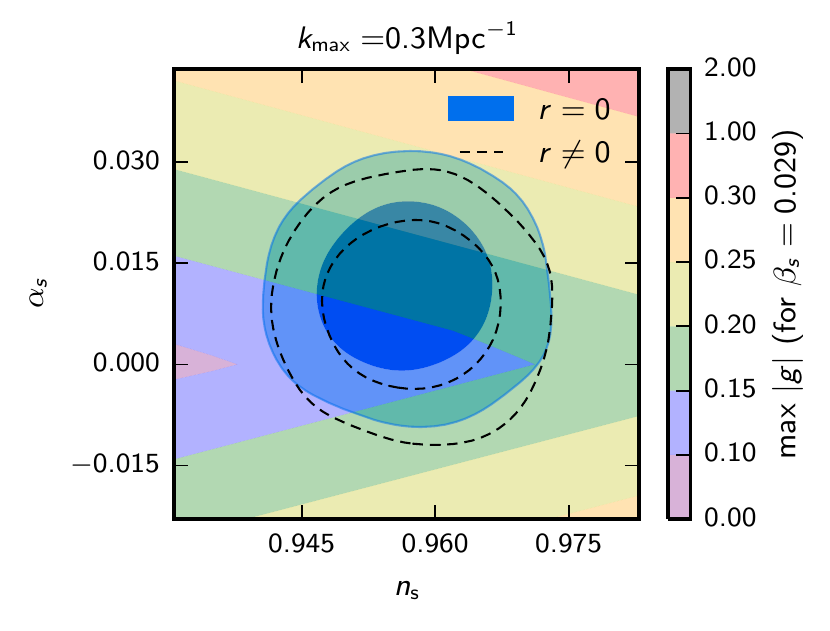}\includegraphics[scale=0.9]{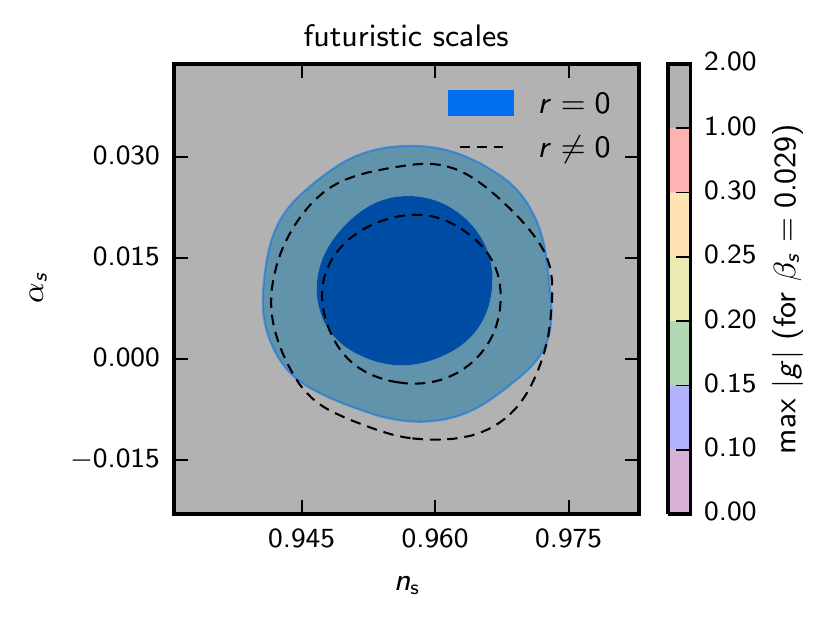}
\caption{\label{fig:beta_fig_pess}Slow-roll and observational constraints on parameterizations of the power spectrum with a constant $\beta_{s}=0.029$.
The observational contours are \emph{Planck} 2015 TT+lowTEB and joint
BICEP-Keck-\emph{Planck} constraints for $\alpha_{s}$
and $n_{s}$ at the pivot scale. Blue contours assume a null tensor-to-scalar ratio,
$r$,  whereas dashed black contours marginalise over allowed values of $r\geq 0$.
The coloured areas indicate the maximum magnitude of $g$ during the interval of time during which the constrained scales
left the horizon (``futuristic scales'' denoting both $k_{\rm max}=100\rm{Mpc^{-1}}$ and $k_{\rm max}=10^4\rm{Mpc^{-1}}$).
Note that for $g>1$
(as is the case everywhere on the plot on the right-hand side) our method breaks down as
Eq.~\eqref{eq:P_f_2nd} ceases to be valid.}
\end{center}
\end{figure}

If we allow the running to vary with a constant $\beta_{s}$ (corresponding to $N=3$ in Eq.~\eqref{eq:P_ansatz}) we have three relevant observables: the constant $\beta_{s}$, as well as the values of $\alpha_{s}$ and
$n_{s}$ at the pivot scale. In order to illustrate typical constraints, we present the plots
corresponding to $\beta_{s}=0.029$ (the \emph{Planck} best fit) in figure \ref{fig:beta_fig_pess} (higher values of $\beta_{s}$
would result in a more dramatic version of these plots, whereas lower values would yield
plots more similar to those in figure \ref{fig:alpha_fig}).

\begin{figure}[h]
\begin{center}
\includegraphics[scale=0.9]{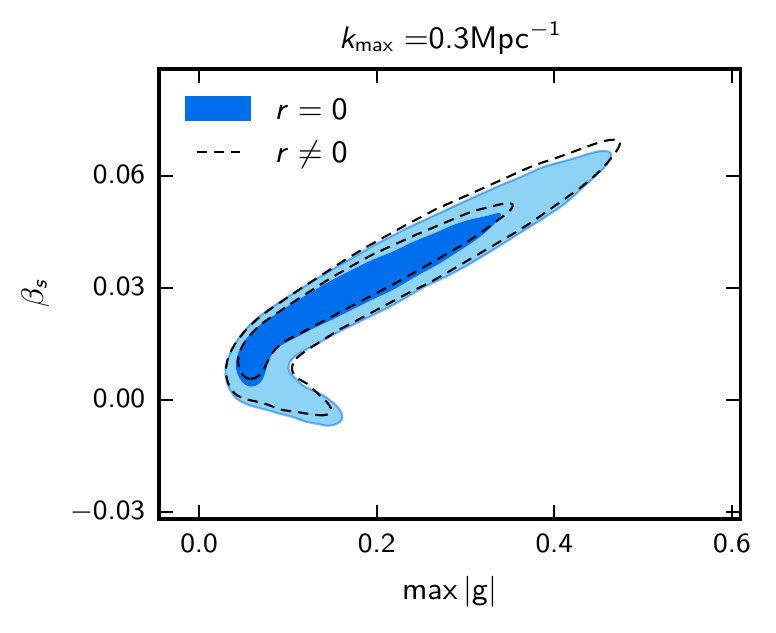}
\includegraphics[scale=0.9]{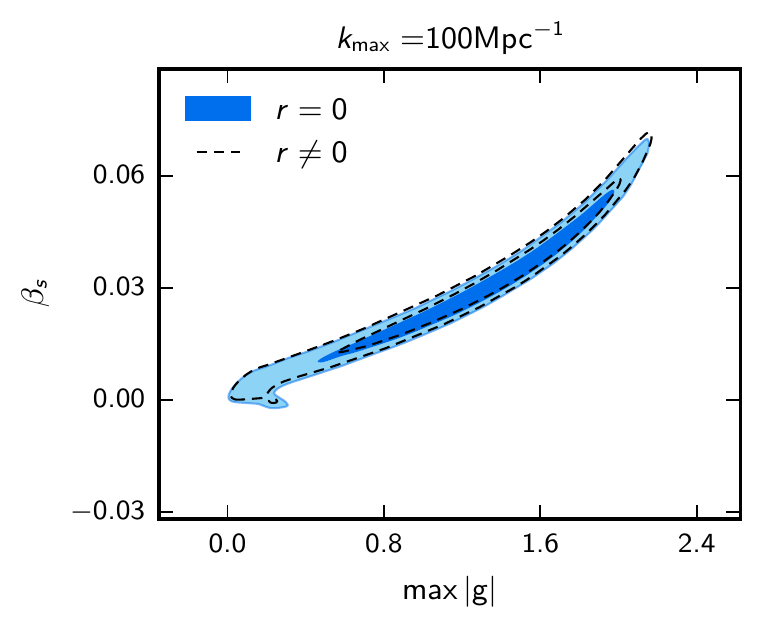}\includegraphics[scale=0.9]{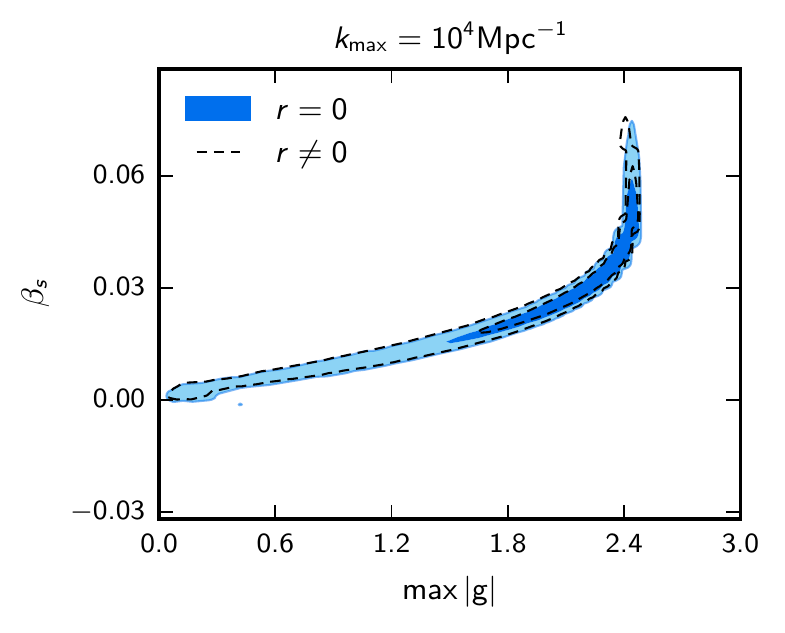}
\caption{\label{fig:beta_gfig}Bounds, over the constrained ranges of scales, on the maximum magnitude of $g$ inferred from
\emph{Planck} 2015 TT+lowTEB and BICEP-Keck constraints on a constant $\beta_{s}$ marginalized over $\alpha_{s}$ and
$n_{s}$ at the pivot scale. Filled contours assume $r=0$ whereas dashed lines marginalize over allowed values of $r\geq 0$.
Note that for $g>1$ our method breaks down as Eq.~\eqref{eq:P_f_2nd} ceases to be valid.}
\end{center}
\end{figure}

To comment more generally on whether this class of slow-roll models can be ruled out by measuring $\beta_{s}$ over the range of its currently allowed possible values, it is easier to focus on the
 constraints on the maximum magnitude of $g$ shown in figure \ref{fig:beta_gfig}
(since they conveniently reduce the relevant three-dimensional information to simple two-dimensional contours).
The current preference for $\beta_s\ne0$ is driven by large scales, but small-scale data is consistent with constant spectral index, so as more small-scale data is added it is plausible that constraints on $\beta_s$ will converge to be closer to zero in the future.
However, if they do not, it is quite possible that a future detection of non-zero running of the running could significantly
disfavour this class of single-field slow-roll inflation, but only if information on a slightly wider range of scales is obtained
(about an extra efolding should suffice for large values of $\beta_s$ to clearly lead to high values of $\left|g\right|$, given how
some are already at the borderline $\left|g\right|\sim 0.5$.).
In particular, a future detection near \emph{Planck}'s current best fit ($\beta_{s}=0.029$)
could clearly rule out this class of slow roll.

That the larger values of constant $\beta_s$ would rule out simple slow-roll inflation models should not be a surprise.
An intuitive argument for this uses the fact that, under fairly general assumptions, to leading order in slow roll,
$n_{s}-1$ can be written in a simpler form as a sum of small $\delta_n$ parameters (of which
Eq.~\eqref{eq:ns_1st_standard} is a truncation) \cite{STEWART_RUN}.
If $\beta_s = \mathcal{O}(0.05)$ and constant, $n_s$ would change by $\mathcal{O}(1)$ over the observable range
of scales, implying that this form of $n_s-1$ cannot be valid everywhere.

\subsection{Consequences for the power spectrum}

It is interesting to consider what current data say about the allowed range for
the small-scale power spectrum that could be observed by future data.
Assuming that the parameterization we have used (with constant $\beta_s$) can be extended, current Planck constraints with non-zero $\beta_s$ allow the power spectrum to grow to order unity at the smallest scales we consider (which would already be ruled out by other probes \cite{Chluba:2012we,Carr:2009jm}). Therefore, it is instructive to see how the requirement of slow roll
(as defined by a maximum
allowed magnitude of $g$ over constrained scales) would affect this extrapolation, and how that compares with the effect of the
naive expectations resulting from the imposition of the usual hierarchy on slow-roll parameters.

\begin{figure}[h]
\includegraphics[scale=1.0]{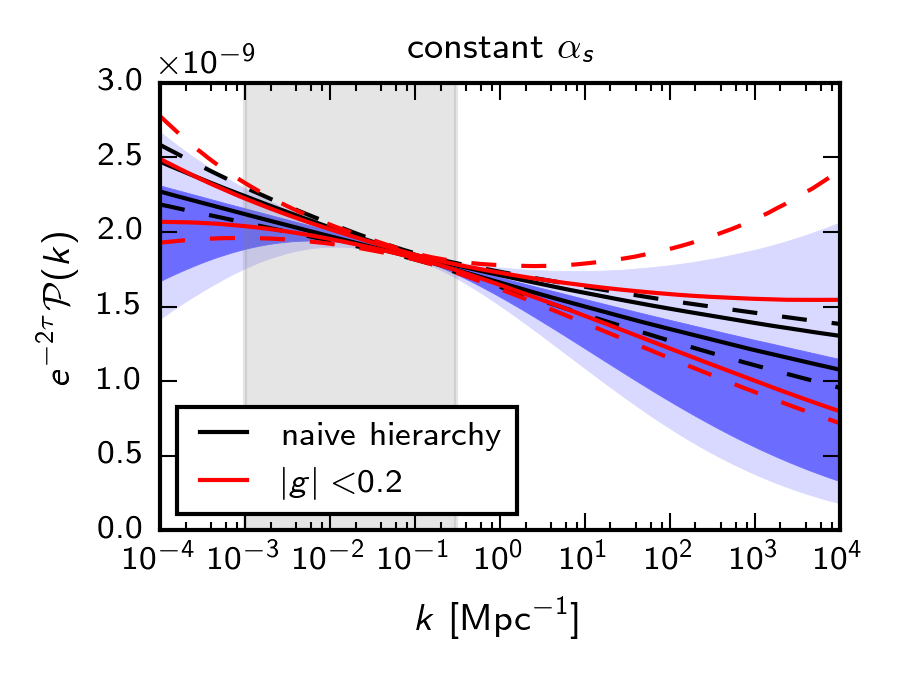}
\includegraphics[scale=1.0]{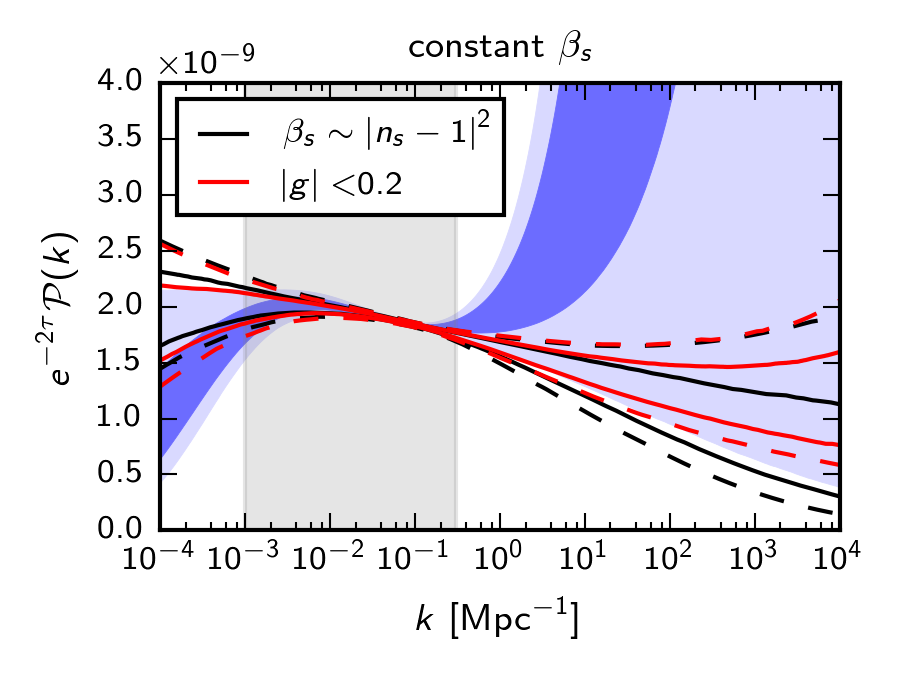}
\caption{\label{fig:Pkplots}Consequences of the imposition of slow roll (defined by the smallness of $g$) for the power spectrum
scaled by $e^{-2\tau}$, where $\tau$ is the optical depth (whose value affects the amplitude of the spectrum, but not its shape).
The blue contours represent the 68\% (dark blue) and 95\% (light blue) limits on the allowed values of the power spectrum (rescaled
by a factor of $e^{-2\tau}$) extrapolated from \emph{Planck} 2015 TT+lowTEB constraints (over gray shaded scales) assuming a
constant $\alpha_s$ (left) and a constant $\beta_{s}$ (right), for different values of $k$.
The solid and dashed red contours represent the 68\% and 95\% limits on the fraction of these spectra
for which $\left|g\right|<0.2$ for the range of scales corresponding to $10^{-3}\rm{Mpc^{-1}}<k<10^4\rm{Mpc^{-1}}$.
The solid and dashed black contours represent the 68\% and 95\% limits on the fraction of these spectra corresponding
to the unshaded regions in figure \ref{fig:motivation}
(note that for the plot on the right the limits of this region already violate the naive expectation for the magnitude of $\beta_s$).
}
\end{figure}

Our inferred constraints on the power spectrum are shown in
figure \ref{fig:Pkplots}: the assumption of slow roll leads to significantly tilted and  narrower bounds on the
small-scale power spectrum (compared to assuming only Planck constraints), especially for the case of constant $\beta_s$.

\section{Conclusions\label{sec:Conclusions}}

We devised a straightforward method to assess whether
specific observed values of the running (of the running) of the spectral index
are consistent with canonical single-field slow-roll inflation.
 We showed that slow roll is much harder to discard than simple expectations based on a hierarchy of slow-roll parameters suggest,
and in particular that for constant running any of the currently allowed values would not necessarily imply a violation of slow roll
over observable scales. However, a detection of constant $\beta_{s}$ significantly away from zero could be much more powerful\footnote{This is partly because current constraints
allow for larger constant $\beta_{s}$ than constant $\alpha_{s}$. However, mostly, it is because allowing significant higher-order
runnings implies allowing significant higher-order slow-roll parameters, which naturally makes $g$ vary faster.
In other words, $g\left(\ln\xi\right)$ computed from $f\left(\ln\xi\right)$ in Eq.~\eqref{eq:inv2_formula}
is a polynomial in $\ln\left(\xi\right)$ whose order is higher if the power spectrum has higher-order runnings.
This can also be seen from the different rates of deviation between the blue and the red limits on plots in figure \ref{fig:Pkplots}.}:
a firmer detection over currently-available scales could be enough to restrict slow-roll inflation to a region of borderline validity, and future data over a wider range of scales could invalidate slow roll for the simple parameterization of the power spectrum assumed.


There are, however, a couple of limitations of our approach:

 - Firstly, we redefined slow-roll as meaning $\left|g\right|\ll1$, which, despite not assuming any hierarchy of slow-roll parameters,
is a somewhat stronger condition than the general definition of $\epsilon,\left|\delta_{1}\right|\ll1$
(see the conclusions of appendix \ref{sec:gappendix}). Nevertheless,
this still corresponds to a very simple and wide class of models,
including all the ones which make the wider class so popular (slow-roll formulae for the power spectrum should break down for the models left out).

 - Secondly, we follow a constructive approach: for each specific combination of
observable parameters $\{n_{s},\alpha_{s},\beta_{s},...\}$ we find a function $g\left(\ln\xi\right)$ which generates them
and check whether it breaks our weaker definition of slow-roll during the time during which observably measurable scales
crossed the horizon. Inevitably, we can only find (or fail to find) examples
of models which generate power spectra of the specific kind assumed (in the case of this work, with constant
$\alpha_{s}$  or $\beta_{s}$). In the case of constant $\alpha_s$, an existence proof is sufficient to demonstrate that models
do not violate slow roll. However, in cases where slow-roll is violated, our assumption is restrictive and different parameterizations
(e.g., involving oscillatory features, or large-scale features) might lead to different conclusions.
It would be  straightforward to generalize our method to constrain both higher-order runnings and completely
different parameterizations by making appropriate changes to Eq.~\eqref{eq:P_ansatz}.

Due to its smallness, the tensor-to-scalar ratio does not noticeably affect our results.

\section{Acknowledgements}

We thank Ewan Stewart and Daniel Passos for helpful comments.

JV is supported by an STFC studentship, CB is supported by a Royal
Society University Research Fellowship, and AL acknowledges support
from the European Research Council under the European
Union's  Seventh  Framework  Programme  (FP/2007-2013)  /
ERC Grant Agreement No.  [616170]
and from the Science and Technology Facilities Council {[}grant number ST/L000652/1{]}.

\chapter{Conclusions and Discussion}
\label{chap:conc}

This is an exciting time for cosmology in general and for the study of the early Universe in particular. The observational advances of the last few decades have firmly established $\Lambda$CDM as the standard  model of cosmology and led to inflation arising as the dominant paradigm in the primordial Universe. In the next decades, new observations will enable us to test these ideas like never before, allowing for a better understanding of not just cosmic history but also of particle physics at very high energies. However, in order to make full use of these future measurements, our theoretical understanding of important physical mechanisms in these theories must also be improved. In particular, it is paramount to know which observational tests can distinguish between which different early Universe scenarios, and it is necessary to learn more about possible exotic signatures of some of these scenarios (like topological defects).

This thesis presents three distinct projects which aim to contribute to this improved understanding in three distinct ways. In this chapter, we briefly review the main findings and prospects for future work of each.

\section{Negative absolute temperatures in cosmology}

In chapter \ref{chap:NAT}, we presented the first (thermodynamically) consistent discussion of the possible role of negative absolute temperatures (NAT) in cosmology. Aiming to be as general (i.e., model-independent) as possible, we postulated that NAT are possible on cosmological scales due to the energy of single-particle states being bounded from above\footnote{Technically, our work requires only that this be true for one specific species of particles, although it seems unlikely that such a feature should not be universal in realistic scenarios.} but made no a priori assumptions about the origin and scale of this cut-off. For the sake of simplicity, we focused on the special case in which the Universe is dominated by a single component which can be in thermal equilibrium at NAT and investigated how it should evolve according to Friedmann's equations. We also assumed that the particles this component is made of, temperons, are fermions whose number is not conserved --- both because we see in appendix \ref{sec:The-problem-of-mu} that otherwise only very special (and not very relevant) models, if any, may permit thermalisation at NAT, and because this allows for their statistics to be described in terms of holes rather than particles (which leads to a convenient correspondence between quantities at positive and negative temperatures).

In the end, we found NAT led to two physically relevant solutions: an inflationary epoch and a bouncing Universe, depending on whether the Universe is initially expanding or contracting (the latter case not even requiring that NAT ever actually occur, but only that the cut-off allows infinite temperature). Whether either of them are viable as realistic models of the very early Universe, however, is still unclear. In both situations, observational predictions seem to require an out of equilibrium treatment which is likely not possible in this type of model-independent approach. Moreover, it seems that the most appealing of the two solutions (the inflationary one) may additionally require an extra curvaton-like component in order to fit current CMB observations --- which is a major blow to the relative simplicity of the idea, unless this component can be seen to naturally arise in specific NAT scenarios.

Necessarily, future work on NATive cosmology will have to address the shortcomings of our results, particularly concerning the end of inflation and perturbation generation. Given that it is unlikely that this may be done in a model-independent way, it will also be necessary to better understand under which conditions specific models with energy cut-offs can lead to NAT (instead of just not allowing thermalisation at too high energies). The following insights on this thermalisation issue are informed by ongoing work in collaboration with Djuna Croon (Dartmouth College) and Sonali Mohapatra (University of Sussex).

One basic question that must be properly answered in the context of any specific model with the sort of cut-off we are interested in is: where does the negative pressure come from? In the usual picture of particles ``in a box'' the pressure is just the total force per unit area exerted by the particles on the walls of the box, and therefore it can never be negative. This tells us that if we were to just take the states of a normal theory of particle physics and postulate that states above a certain cut-off aren't accessible anymore we wouldn't be able to achieve NAT. Of course, as we mention in subsection \ref{sub:The-end-of-NAT}, this conceptual picture is expected to break down at high energies, at which we have suggested that an alternative sort of picture, based on a deformed version of the momentum operator, might offer a principled solution to this problem. However, we should not expect such a method (or any other method) to allow thermalisation for all temperatures $T<0$. This is because, writing the ``mechanical'' pressure as
\begin{equation}
P_\mathrm{mech}=\intop^\Lambda_m\frac{g\left(\epsilon\right) p\left(\epsilon\right)}{e^{\beta\epsilon}+1}d\epsilon,\label{Pmechgen}
\end{equation}
where $p\left(\epsilon\right)\rightarrow \frac{\epsilon^2-m^2}{3\epsilon}$ at low energies (to recover Eq.~\eqref{eq:P_standard}) but whose full form should depend on how the cut-off is implemented, we find that the requirement of thermalisation at $\beta=-\infty$ implies
\begin{equation}
\intop^\Lambda_m g\left(\epsilon\right) p\left(\epsilon\right) d\epsilon =-\rho_\mathrm{max},\label{thermalinfty}
\end{equation}
whereas thermalisation at $\beta=0^\pm$ implies
\begin{equation}
\intop^\Lambda_m g\left(\epsilon\right) p\left(\epsilon\right) d\epsilon =\pm\infty.\label{thermalzero}
\end{equation}

While this still allows for cosmologically relevant phenomena (which all take place either in the vicinity of $\beta=-\infty$ or of $\beta=0$), it is an indication that the conditions for thermalisation at NAT seem to be much more constraining than originally realised; and thus full calculations in specific working models are needed. Nevertheless, there is still good reason to think that thermalisation should be possible at least close to $\beta=-\infty$ because the thermodynamical solution there makes physical sense: $P=-\rho_\mathrm{max}$ corresponds to the term that cancels the temperon contribution to the cosmological constant\footnote{In other words, the quantum zero-point contribution of the ``vacuum'' to the pressure is cancelled out when there is no ``vacuum'' left because all states are completely filled. In a sense, it seems that it is the ``hole vacuum'' that is truly empty (as opposed to the ``fermion vacuum''), as its total energy and pressure both vanish.} \cite{MSC_thermalQFT,thermalQFT_lectnotes}.

\section{Modelling small-scale structure in cosmic strings}

In chapter \ref{chap:wiggly}, we presented a new model for the evolution of cosmic string networks which takes into account both the dynamics of small wiggles and their effect on the evolution of large-scale string properties. While the basic formalism for taking into account these small-scale effects was provided by previous work \cite{PAP1}, there are important parameters (notably energy loss parameters) whose functional form is not adequately constrained by current simulations --- and for those we had to settle for more or less well-motivated ansatze.

The result was a partial success. It is encouraging that parameters derived from matter era simulations are consistent with the type of scaling allowed by this model. Likewise, it is potentially positive that the same agreement is not found in the radiation era, when it is not clear that scaling has been reached. Nevertheless, a full explanation of these scaling phenomena in the context of this model warrants a detailed study of the stability of scaling in the different epochs --- which requires more knowledge of some poorly understood parameters, including fractal dimensions as well as energy loss parameters.

A main focus of future work will therefore have to be on understanding energy loss mechanisms and their relation to the multifractal structure of cosmic string networks. This is necessary to make this model sufficiently complete to enable a detailed study of how small-scale structure should affect observational bounds on models which lead to cosmic string formation\footnote{An example of the kind of study that can be done in the case of CMB observables is given by the work of Rybak \emph{et al} \cite{Rybak:2017yfu}, where a version of this model with $\ell=const$ was shown to lead to laxer constraints than those for Goto-Nambu strings.}. This is particularly relevant now in the advent of gravitational wave cosmology, when the detection of a stochastic gravitational wave background due to cosmic strings is becoming a real possibility (e.g., with an experiment like LISA) --- and could in the future potentially provide precious information on the scale of inflation \cite{BLANCOPILLADO2018392}.

Given the advances in high-performance computing since 2006, when the simulations used in this work came out \cite{FRAC}, it is expected that upcoming high-resolution simulations will enable a more rigorous calibration of the relevant free parameters. Such a calibration would be analogous to recent work on domain walls \cite{Rybak}, with the added complication of the need for a study of multifractality.

\section{Testing slow-roll inflation}

In chapter \ref{chap:SFSRI}, it was shown that the implications to the validity of the slow-roll approximation of an eventual detection of significant running (or running of the running) of the spectral index of scalar perturbations are less straightforward than the simplest arguments suggest. While this is not a new realisation \cite{Easther:2006tv}, this work studies these implications more generally and in light of more accurate data than ever before (at least as far as the authors are aware).

At the core of this work is a method for finding the family of models that corresponds to a given power spectrum given a relatively general definition of slow roll (somewhat stronger than just the standard definition of slow-roll, but weaker than imposing the usual hierarchy to the slow-roll parameters). Using this method, it is found that slow roll can in fact account for significantly larger regions of the available parameter space than the usually assumed slow roll hierarchy. In particular, it is shown that it is unlikely that any future detections of running or running of the running over currently constrained scales will be able to falsify the assumption of slow roll in canonical single-field scenarios.

Naturally, prospects for testing slow roll with the running and the running of the running are better for future observations constraining much wider ranges of scales (and thus more e-folds of cosmic history). While the size of the observable Universe sets a hard limit on the largest scales that can be observed, future probes are expected to considerably improve constraints on the smallest scales. In this work we focus mainly on two types of future probes: 21cm cosmology and spectral distortions. The former is expected to roughly double the number of e-folds over which the power spectrum is well constrained (relative to what is currently done with CMB observations) and the latter is expected to almost triple it (although usually via integrated constraints which are not generally able to distinguish the scale dependence of the power spectrum). Given that large values of the relevant runnings of the spectral index are much harder to maintain for longer periods of time, this increase in the range of constrained scales is critical for future tests of slow roll using this sort of method. In fact, it turns out that these future probes would not even need to reduce current error bars on the runnings\footnote{Which they are expected to do.} (but merely to maintain the current central values) in order for the simple slow-roll models considered to be falsified.

Perhaps the most concerning limitation of this approach is the need for a specific parameterisation of the power spectrum to be assumed. Even though it should be straightforward to modify this analysis for radically different types of scale dependences (e.g., with oscillatory features), it gives no rigorous way to decisively rule out a region of parameter space in this class of slow roll --- as it relies on an explicit construction to provide examples of models, but says nothing about whether other types of examples may exist when this construction fails. In other words, this method is good for showing that slow roll can account for power spectra which are not obviously consistent with slow roll due to their unusual scale dependence, but it can never show that a given power spectrum (given observational uncertainties) is inconsistent with slow roll.

Future work should try to surpass this limitation, especially if evidence for large running of the running does not decrease as smaller scales are better constrained. Ideally, future versions of this method should be able to point to the family of models consistent with observations which leads to the smallest variation in $\left|g\right|$ over constrained scales, instead of assuming any specific parameterisation of the power spectrum. Possibly, this may require an extra assumption on the hierarchy of slow-roll parameters\footnote{One natural such assumption would be that only a small number of slow-roll parameters are non-null to the desired accuracy in slow roll, similarly to what is done in Easther and Peiris \cite{Easther:2006tv} (but possibly allowing more relevant parameters).}, leading to stronger statements about a stronger definition of slow roll.

\clearpage
\phantomsection
\addcontentsline{toc}{chapter}{Bibliography}
\bibliographystyle{plain}
\bibliography{PhD_BIB,paper2}

\begin{thebibliography}{100}

\bibitem{Achucarro:2008fn}
A.~Achucarro and C.~J. A.~P. Martins.
\newblock {Cosmic Strings}.
\newblock {\em Meyers, Robert (Ed.) Encyclopedia of Complexity and Systems
  Science, Springer New York}, 2009.
\newblock arXiv:0811.1277.

\bibitem{GSRinin}
Ana Achucarro, Vicente Atal, Bin Hu, Pablo Ortiz, and Jesus Torrado.
\newblock {Inflation with moderately sharp features in the speed of sound:
  Generalized slow roll and in-in formalism for power spectrum and bispectrum}.
\newblock {\em Phys. Rev.}, D90(2):023511, 2014.
\newblock arXiv:1404.7522.

\bibitem{Adam:2015rua}
R.~Adam et~al.
\newblock {Planck 2015 results. I. Overview of products and scientific
  results}.
\newblock {\em Astron. Astrophys.}, 594:A1, 2016.
\newblock arXiv:1502.01582.

\bibitem{Ade:2013zuv}
P.~A.~R. Ade et~al.
\newblock {Planck 2013 results. XVI. Cosmological parameters}.
\newblock {\em Astron. Astrophys.}, 571:A16, 2014.
\newblock arXiv:1303.5076.

\bibitem{Planck}
P.~A.~R. Ade et~al.
\newblock {Planck 2013 results. XXV. Searches for cosmic strings and other
  topological defects}.
\newblock {\em Astron. Astrophys.}, 571:A25, 2014.
\newblock arXiv:1303.5085.

\bibitem{Array:2015xqh}
P.~A.~R. Ade et~al.
\newblock {Improved Constraints on Cosmology and Foregrounds from BICEP2 and
  Keck Array Cosmic Microwave Background Data with Inclusion of 95 GHz Band}.
\newblock {\em Phys. Rev. Lett.}, 116:031302, 2016.
\newblock arXiv:1510.09217.

\bibitem{Ade:2015xua}
P.~A.~R. Ade et~al.
\newblock {Planck 2015 results. XIII. Cosmological parameters}.
\newblock {\em Astron. Astrophys.}, 594:A13, 2016.
\newblock arXiv:1502.01589.

\bibitem{Planck_inf2015}
P.~A.~R. Ade et~al.
\newblock {Planck 2015 results. XX. Constraints on inflation}.
\newblock {\em Astron. Astrophys.}, 594:A20, 2016.
\newblock arXiv:1502.02114.

\bibitem{BKPanalysis}
P.A.R. Ade et~al.
\newblock {Joint Analysis of BICEP2/$Keck Array$ and $Planck$ Data}.
\newblock {\em Phys.Rev.Lett.}, 114:101301, 2015.
\newblock arXiv:1502.00612.

\bibitem{Adshead:2010mc}
Peter Adshead, Richard Easther, Jonathan Pritchard, and Abraham Loeb.
\newblock {Inflation and the Scale Dependent Spectral Index: Prospects and
  Strategies}.
\newblock {\em JCAP}, 1102:021, 2011.
\newblock arXiv:1007.3748.

\bibitem{Adshead:2013zfa}
Peter Adshead, Wayne Hu, and Vinícius Miranda.
\newblock {Bispectrum in Single-Field Inflation Beyond Slow-Roll}.
\newblock {\em Phys. Rev.}, D88(2):023507, 2013.
\newblock arXiv:1303.7004.

\bibitem{1990PhRvL..65.1705A}
B.~{Allen} and R.~R. {Caldwell}.
\newblock {Generation of structure on a cosmic-string network}.
\newblock {\em Physical Review Letters}, 65:1705--1708, October 1990.

\bibitem{AS}
B.~Allen and E.~P.~S. Shellard.
\newblock Cosmic string evolution: A numerical simulation.
\newblock {\em Phys. Rev. Lett.}, 64:119--122, 1990.

\bibitem{1948PhRv...74.1577A}
R.~A. {Alpher}.
\newblock {A Neutron-Capture Theory of the Formation and Relative Abundance of
  the Elements}.
\newblock {\em Physical Review}, 74:1577--1589, December 1948.

\bibitem{abc}
R.~A. {Alpher}, H.~{Bethe}, and G.~{Gamow}.
\newblock {The Origin of Chemical Elements}.
\newblock {\em Physical Review}, 73:803--804, April 1948.

\bibitem{1948Natur.162..774A}
R.~A. {Alpher} and R.~{Herman}.
\newblock {Evolution of the Universe}.
\newblock {\em Nature}, 162:774--775, November 1948.

\bibitem{IBSUSJ}
Simon Appolloni.
\newblock \''repugnant\'', \''not repugnant at all\'': How the respective
  epistemic attitudes of georges lemaitre and sir arthur eddington influenced
  how each approached the idea of a beginning of the universe.
\newblock {\em IBSU Scientific Journal}, 5(1):19--44, 2011.

\bibitem{ACK}
Daren Austin, Edmund~J. Copeland, and T.~W.~B. Kibble.
\newblock Evolution of cosmic string configurations.
\newblock {\em Phys. Rev.}, D48:5594--5627, 1993.
\newblock arXiv:hep-ph/9307325.

\bibitem{Bamba:2015uma}
Kazuharu Bamba and Sergei~D. Odintsov.
\newblock {Inflationary cosmology in modified gravity theories}.
\newblock {\em Symmetry}, 7(1):220--240, 2015.
\newblock arXiv:1503.00442.

\bibitem{Ben-Dayan:2015zha}
Ido Ben-Dayan and Ryuichi Takahashi.
\newblock {Constraints on small-scale cosmological fluctuations from SNe
  lensing dispersion}.
\newblock {\em Mon. Not. Roy. Astron. Soc.}, 455(1):552--562, 2016.
\newblock arXiv:1504.07273.

\bibitem{Bengaly:2016amk}
C.~A.~P. Bengaly, A.~Bernui, J.~S. Alcaniz, H.~S. Xavier, and C.~P. Novaes.
\newblock {Is there evidence for anomalous dipole anisotropy in the large-scale
  structure?}
\newblock {\em Mon. Not. Roy. Astron. Soc.}, 464(1):768--774, 2017.
\newblock arXiv:1606.06751.

\bibitem{2013ApJS..208...20B}
C.~L. {Bennett}, D.~{Larson}, J.~L. {Weiland}, N.~{Jarosik}, G.~{Hinshaw},
  N.~{Odegard}, K.~M. {Smith}, R.~S. {Hill}, B.~{Gold}, M.~{Halpern},
  E.~{Komatsu}, M.~R. {Nolta}, L.~{Page}, D.~N. {Spergel}, E.~{Wollack},
  J.~{Dunkley}, A.~{Kogut}, M.~{Limon}, S.~S. {Meyer}, G.~S. {Tucker}, and
  E.~L. {Wright}.
\newblock {Nine-year Wilkinson Microwave Anisotropy Probe (WMAP) Observations:
  Final Maps and Results}.
\newblock {\em Astrophys. J. Suppl.}, 208:20, October 2013.
\newblock arXiv:1212.5225.

\bibitem{BB}
David~P. Bennett and Francois~R. Bouchet.
\newblock High resolution simulations of cosmic string evolution. 1. network
  evolution.
\newblock {\em Phys. Rev.}, D41:2408, 1990.

\bibitem{Blanco}
Jose~J. Blanco-Pillado, Ken~D. Olum, and Benjamin Shlaer.
\newblock {Large parallel cosmic string simulations: New results on loop
  production}.
\newblock {\em Phys. Rev.}, D83:083514, 2011.
\newblock arXiv:1101.5173.

\bibitem{blundell}
S.~Blundell and K.M. Blundell.
\newblock {\em Concepts in Thermal Physics}.
\newblock Oxford University Press, 2006.

\bibitem{discrete_heisenberg}
Martin Bojowald and Achim Kempf.
\newblock {Generalized uncertainty principles and localization of a particle in
  discrete space}.
\newblock {\em Phys. Rev.}, D86:085017, 2012.
\newblock arXiv:1112.0994.

\bibitem{AlHaythamWorkshop}
H.-E. {Bouali}, M.~{Zghal}, and Z.~{Ben Lakhdar}.
\newblock {Popularisation of optical phenomena: establishing the first Ibn
  Al-Haytham workshop on photography}.
\newblock In {\em Society of Photo-Optical Instrumentation Engineers (SPIE)
  Conference Series}, volume 9664, page 966422, October 2005.

\bibitem{BraunNAT}
S.~Braun, J.~P. Ronzheimer, M.~Schreiber, S.~S. Hodgman, T.~Rom, I.~Bloch, and
  U.~Schneider.
\newblock Negative absolute temperature for motional degrees of freedom.
\newblock {\em Science}, 339(6115):52--55, 2013.
\newblock arXiv:1211.0545.

\bibitem{NEGVISC}
Iver Brevik and Øyvind Grøn.
\newblock {Universe Models with Negative Bulk Viscosity}.
\newblock {\em Astrophys. Space Sci.}, 347:399--404, 2013.
\newblock arXiv:1306.5634.

\bibitem{RevModPhys.39.883}
Stephen~G. Brush.
\newblock History of the lenz-ising model.
\newblock {\em Rev. Mod. Phys.}, 39:883--893, Oct 1967.

\bibitem{Buchbinder:2007ad}
Evgeny~I. Buchbinder, Justin Khoury, and Burt~A. Ovrut.
\newblock {New Ekpyrotic cosmology}.
\newblock {\em Phys. Rev.}, D76:123503, 2007.
\newblock arXiv:hep-th/0702154.

\bibitem{dispute_boltzgibbs}
P.~{Buonsante}, R.~{Franzosi}, and A.~{Smerzi}.
\newblock {On the dispute between Boltzmann and Gibbs entropy}.
\newblock {\em Annals of Physics}, 375:414--434, December 2016.
\newblock arXiv:1601.01509.

\bibitem{1957RvMP...29..547B}
E.~M. {Burbidge}, G.~R. {Burbidge}, W.~A. {Fowler}, and F.~{Hoyle}.
\newblock {Synthesis of the Elements in Stars}.
\newblock {\em Reviews of Modern Physics}, 29:547--650, 1957.

\bibitem{reviewNG_multif}
Christian~T. Byrnes and Ki-Young Choi.
\newblock {Review of local non-Gaussianity from multi-field inflation}.
\newblock {\em Adv. Astron.}, 2010:724525, 2010.
\newblock arXiv:1002.3110.

\bibitem{running_running}
Giovanni Cabass, Eleonora Di~Valentino, Alessandro Melchiorri, Enrico Pajer,
  and Joseph Silk.
\newblock {Constraints on the running of the running of the scalar tilt from
  CMB anisotropies and spectral distortions}.
\newblock {\em Phys. Rev.}, D94(2):023523, 2016.
\newblock arXiv:1605.00209.

\bibitem{mudistorrunning}
Giovanni Cabass, Alessandro Melchiorri, and Enrico Pajer.
\newblock {$\mu$ distortions or running: A guaranteed discovery from CMB
  spectrometry}.
\newblock {\em Phys. Rev.}, D93(8):083515, 2016.
\newblock arXiv:1602.05578.

\bibitem{BigRip}
Robert~R. Caldwell, Marc Kamionkowski, and Nevin~N. Weinberg.
\newblock {Phantom energy and cosmic doomsday}.
\newblock {\em Phys. Rev. Lett.}, 91:071301, 2003.
\newblock arXiv:astro-ph/0302506.

\bibitem{Carr:2009jm}
B.~J. Carr, Kazunori Kohri, Yuuiti Sendouda, and Jun'ichi Yokoyama.
\newblock {New cosmological constraints on primordial black holes}.
\newblock {\em Phys. Rev.}, D81:104019, 2010.
\newblock arXiv:0912.5297.

\bibitem{CARTERA}
Brandon Carter.
\newblock Transonic elastic model for wiggly goto-nambu string.
\newblock {\em Phys. Rev. Lett.}, 74:3098--3101, 1995.
\newblock arXiv:hep-th/9411231.

\bibitem{thermo_perturb}
Bin Chen, Yi~Wang, and Wei Xue.
\newblock {Inflationary nonGaussianity from thermal fluctuations}.
\newblock {\em JCAP}, 0805:014, 2008.
\newblock arXiv:0712.2345.

\bibitem{Chluba:2011hw}
J.~Chluba and R.~A. Sunyaev.
\newblock {The evolution of CMB spectral distortions in the early Universe}.
\newblock {\em Mon. Not. Roy. Astron. Soc.}, 419:1294--1314, 2012.
\newblock arXiv:1109.6552.

\bibitem{Chluba:2013vsa}
Jens Chluba.
\newblock {Green's function of the cosmological thermalization problem}.
\newblock {\em Mon. Not. Roy. Astron. Soc.}, 434:352, 2013.
\newblock arXiv:1304.6120.

\bibitem{Chluba:2014sma}
Jens Chluba.
\newblock {Science with CMB spectral distortions}.
\newblock In {\em {Proceedings, 49th Rencontres de Moriond on Cosmology: La
  Thuile, Italy, March 15-22, 2014}}, pages 327--334, 2014.
\newblock arXiv:1405.6938.

\bibitem{Chluba:2012we}
Jens Chluba, Adrienne~L. Erickcek, and Ido Ben-Dayan.
\newblock {Probing the inflaton: Small-scale power spectrum constraints from
  measurements of the CMB energy spectrum}.
\newblock {\em Astrophys. J.}, 758:76, 2012.
\newblock arXiv:1203.2681.

\bibitem{Chluba:2013pya}
Jens Chluba and Donghui Jeong.
\newblock {Teasing bits of information out of the CMB energy spectrum}.
\newblock {\em Mon. Not. Roy. Astron. Soc.}, 438(3):2065--2082, 2014.
\newblock arXiv:1306.5751.

\bibitem{Choe_Gong_Stewart_second}
Jeongyeol Choe, Jinn-Ouk Gong, and Ewan~D. Stewart.
\newblock {Second order general slow-roll power spectrum}.
\newblock {\em JCAP}, 0407:012, 2004.
\newblock arXiv:hep-ph/0405155.

\bibitem{Clarkson:1999zq}
Christopher~A. Clarkson.
\newblock {\em {On the observational characteristics of inhomogeneous
  cosmologies: Undermining the cosmological principle or have cosmologists put
  all their EGS in one basket?}}
\newblock PhD thesis, Glasgow U., 1999.
\newblock arXiv:astro-ph/0008089.

\bibitem{Copeland:2009ga}
Edmund~J. Copeland and T.~W.~B. Kibble.
\newblock {Cosmic Strings and Superstrings}.
\newblock {\em Proc. Roy. Soc. Lond.}, A466:623--657, 2010.
\newblock arXiv:0911.1345.

\bibitem{DamourVilenkin}
Thibault Damour and Alexander Vilenkin.
\newblock {Gravitational radiation from cosmic (super)strings: Bursts,
  stochastic background, and observational windows}.
\newblock {\em Phys. Rev.}, D71:063510, 2005.
\newblock arXiv:hep-th/0410222.

\bibitem{DelPopolo:2016emo}
Antonino Del~Popolo and Morgan Le~Delliou.
\newblock {Small scale problems of the $\Lambda$CDM model: a short review}.
\newblock {\em Galaxies}, 5(1):17, 2017.
\newblock arXiv:1606.07790.

\bibitem{DIRAC_CONST}
P.~A.~M. Dirac.
\newblock The cosmological constants.
\newblock {\em Nature}, 139:323, 1937.

\bibitem{DOD_STEWART}
Scott Dodelson and Ewan Stewart.
\newblock {Scale dependent spectral index in slow roll inflation}.
\newblock {\em Phys. Rev.}, D65:101301, 2002.
\newblock arXiv:astro-ph/0109354.

\bibitem{1964SPhD....9..111D}
A.~G. {Doroshkevich} and I.~D. {Novikov}.
\newblock {Mean Density of Radiation in the Metagalaxy and Certain Problems in
  Relativistic Cosmology}.
\newblock {\em Soviet Physics Doklady}, 9:111, August 1964.

\bibitem{Repy_comment_inconsist_NAT}
J.~{Dunkel} and S.~{Hilbert}.
\newblock {Reply to Schneider et al. [arXiv:1407.4127v1]}.
\newblock August 2014.
\newblock arXiv:1408.5392.

\bibitem{inconsist_NAT}
Jorn Dunkel and Stefan Hilbert.
\newblock Consistent thermostatistics forbids negative absolute temperatures.
\newblock {\em Nature Physics}, 10:67--72, 2014.

\bibitem{inhom_reheating_dvali}
Gia Dvali, Andrei Gruzinov, and Matias Zaldarriaga.
\newblock {A new mechanism for generating density perturbations from
  inflation}.
\newblock {\em Phys. Rev.}, D69:023505, 2004.
\newblock arXiv:astro-ph/0303591.

\bibitem{DVORKIN_GSR}
Cora Dvorkin and Wayne Hu.
\newblock {Generalized Slow Roll for Large Power Spectrum Features}.
\newblock {\em Phys. Rev.}, D81:023518, 2010.
\newblock arXiv:0910.2237.

\bibitem{Easther:2006tv}
Richard Easther and Hiranya Peiris.
\newblock {Implications of a Running Spectral Index for Slow Roll Inflation}.
\newblock {\em JCAP}, 0609:010, 2006.
\newblock arXiv:astro-ph/0604214.

\bibitem{1930MNRAS..90..668E}
A.~S. {Eddington}.
\newblock {On the instability of Einstein's spherical world}.
\newblock {\em Monthly Notices of the Royal Astronomical Society}, 90:668--678,
  May 1930.

\bibitem{1931Natur.127..447E}
A.~S. {Eddington}.
\newblock {The End of the World: from the Standpoint of Mathematical Physics.}
\newblock {\em Nature}, 127:447--453, March 1931.

\bibitem{1917SPAW.......142E}
A.~{Einstein}.
\newblock {Kosmologische Betrachtungen zur allgemeinen
  Relativit{\"a}tstheorie}.
\newblock {\em Sitzungsberichte der K{\"o}niglich Preu{\ss}ischen Akademie der
  Wissenschaften (Berlin), Seite 142-152.}, 1917.
\newblock The Digital Einstein papers:
  \url{http://einsteinpapers.press.princeton.edu/vol6-trans/433}.

\bibitem{Einstein1923}
Albert Einstein.
\newblock Note to the paper by a. friedmann 'on the curvature of space'.
\newblock {\em Zeitschrift für Physik}, 16:228, 1923.

\bibitem{NAT_turbul_vortex}
G.~L. Eyink and H.~Spohn.
\newblock Negative-temperature states and large-scale, long-lived vortices in
  two-dimensional turbulence.
\newblock {\em Journal of Statistical Physics}, 70(3):833--886, 1993.

\bibitem{Boltz_vs_Gibbs}
L.~{Ferrari}.
\newblock {Boltzmann vs Gibbs: a finite-size match}.
\newblock January 2015.
\newblock arXiv:1501.04566.

\bibitem{Gibbs_Boltzmann_Teq}
D.~{Frenkel} and P.~B. {Warren}.
\newblock {Gibbs, Boltzmann, and negative temperatures}.
\newblock {\em American Journal of Physics}, 83:163--170, February 2015.
\newblock arXiv:1403.4299.

\bibitem{sep-fine-tuning}
Simon Friederich.
\newblock Fine-tuning.
\newblock In Edward~N. Zalta, editor, {\em The Stanford Encyclopedia of
  Philosophy}. Metaphysics Research Lab, Stanford University, spring 2018
  edition, 2018.

\bibitem{1922ZPhy...10..377F}
A.~{Friedmann}.
\newblock {{\"U}ber die Kr{\"u}mmung des Raumes}.
\newblock {\em Zeitschrift fur Physik}, 10:377--386, 1922.
\newblock English version available at
  \url{http://www.ymambrini.com/My_World/History_files/Friedman_1922.pdf}.

\bibitem{inform_gibbshertz}
A.~{Gagliardi} and A.~{Pecchia}.
\newblock {How to reconcile Information theory and Gibbs-Herz entropy for
  inverted populated systems}.
\newblock March 2015.
\newblock arXiv:1503.02824.

\bibitem{1946PhRv...70..572G}
G.~{Gamow}.
\newblock {Expanding Universe and the Origin of Elements}.
\newblock {\em Physical Review}, 70:572--573, October 1946.

\bibitem{Garay_min_length}
Luis~J. Garay.
\newblock Quantum gravity and minimum length.
\newblock {\em Int. J. Mod. Phys.}, A10:145--166, 1995.
\newblock arXiv:gr-qc/9403008.

\bibitem{GEYER}
H.~B.. Geyer.
\newblock {\em Field Theory, Topology and Condensed matter Physics}.
\newblock Springer-Verlag, New York, U.S.A., 1995.

\bibitem{Golovnev:2008cf}
Alexey Golovnev, Viatcheslav Mukhanov, and Vitaly Vanchurin.
\newblock {Vector Inflation}.
\newblock {\em JCAP}, 0806:009, 2008.
\newblock arXiv:0802.2068.

\bibitem{STEWART_GONG}
Jinn-Ouk Gong and Ewan~D. Stewart.
\newblock {The Density perturbation power spectrum to second order corrections
  in the slow roll expansion}.
\newblock {\em Phys. Lett.}, B510:1--9, 2001.
\newblock arXiv:astro-ph/0101225.

\bibitem{Phantom_thermo}
Pedro~F. Gonzalez-Diaz and Carmen~L. Siguenza.
\newblock {Phantom thermodynamics}.
\newblock {\em Nucl. Phys.}, B697:363--386, 2004.
\newblock arXiv:astro-ph/0407421.

\bibitem{Guth:2007ng}
Alan~H. Guth.
\newblock {Eternal inflation and its implications}.
\newblock {\em J. Phys.}, A40:6811--6826, 2007.
\newblock arXiv:hep-th/0702178.

\bibitem{thermo_isolated}
S.~{Hilbert}, P.~{H{\"a}nggi}, and J.~{Dunkel}.
\newblock {Thermodynamic laws in isolated systems}.
\newblock {\em Phys. Rev.}, E90(6):062116, December 2014.
\newblock arXiv:1408.5382.

\bibitem{Stuckey}
Mark Hindmarsh, Stephanie Stuckey, and Neil Bevis.
\newblock {Abelian Higgs Cosmic Strings: Small Scale Structure and Loops}.
\newblock {\em Phys. Rev.}, D79:123504, 2009.
\newblock arXiv:0812.1929.

\bibitem{Hiramatsu}
Takashi Hiramatsu, Yuuiti Sendouda, Keitaro Takahashi, Daisuke Yamauchi, and
  Chul-Moon Yoo.
\newblock {Type-I cosmic string network}.
\newblock {\em Phys. Rev.}, D88(8):085021, 2013.
\newblock arXiv:1307.0308.

\bibitem{Hitchin413}
Nigel~J. Hitchin.
\newblock Arthur geoffrey walker. 17 july 1909 {\textemdash} 31 march 2001.
\newblock {\em Biographical Memoirs of Fellows of the Royal Society},
  52:413--421, 2006.

\bibitem{min_length_review}
Sabine Hossenfelder.
\newblock {Minimal Length Scale Scenarios for Quantum Gravity}.
\newblock {\em Living Rev. Rel.}, 16:2, 2013.
\newblock arXiv:1203.6191.

\bibitem{1946MNRAS.106..343H}
F.~{Hoyle}.
\newblock {The synthesis of the elements from hydrogen}.
\newblock {\em Monthly Notices of the Royal Astronomical Society}, 106:343,
  1946.

\bibitem{1948MNRAS.108..372H}
F.~{Hoyle}.
\newblock {A New Model for the Expanding Universe}.
\newblock {\em Monthly Notices of the Royal Astronomical Society}, 108:372,
  1948.

\bibitem{1954ApJS....1..121H}
F.~{Hoyle}.
\newblock {On Nuclear Reactions Occuring in Very Hot STARS.I. the Synthesis of
  Elements from Carbon to Nickel.}
\newblock {\em Astrophys. J. Suppl.}, 1:121, September 1954.

\bibitem{GSRtensor}
Wayne Hu.
\newblock {Generalized slow roll for tensor fluctuations}.
\newblock {\em Phys. Rev.}, D89(12):123503, 2014.
\newblock arXiv:1405.2020.

\bibitem{1929PNAS...15..168H}
E.~{Hubble}.
\newblock {A Relation between Distance and Radial Velocity among Extra-Galactic
  Nebulae}.
\newblock {\em Proceedings of the National Academy of Science}, 15:168--173,
  March 1929.

\bibitem{1931ApJ....74...43H}
E.~{Hubble} and M.~L. {Humason}.
\newblock {The Velocity-Distance Relation among Extra-Galactic Nebulae}.
\newblock {\em Astrophys. J.}, 74:43, July 1931.

\bibitem{infchall}
Anna Ijjas, Paul~J. Steinhardt, and Abraham Loeb.
\newblock Cosmic inflation theory faces challenges.
\newblock Scientific American, February 2017.

\bibitem{Stewart_inverse2}
Minu Joy and Ewan~D. Stewart.
\newblock {From the spectrum to inflation: a second order inverse formula for
  the general slow-roll spectrum}.
\newblock {\em JCAP}, 0602:005, 2006.
\newblock arXiv:astro-ph/0511476.

\bibitem{Stewart_inverse1}
Minu Joy, Ewan~D. Stewart, Jinn-Ouk Gong, and Hyun-Chul Lee.
\newblock {From the spectrum to inflation: An Inverse formula for the general
  slow-roll spectrum}.
\newblock {\em JCAP}, 0504:012, 2005.
\newblock arXiv:astro-ph/0501659.

\bibitem{Khatri:2013dha}
Rishi Khatri and Rashid~A. Sunyaev.
\newblock {Forecasts for CMB $\mu$ and $i$-type spectral distortion constraints
  on the primordial power spectrum on scales $8 \lesssim k \lesssim 10^{4}
  Mpc^{-1}$ with the future Pixie-like experiments}.
\newblock {\em JCAP}, 1306:026, 2013.
\newblock arXiv:1303.7212.

\bibitem{2001PhRvD..64l3522K}
J.~{Khoury}, B.~A. {Ovrut}, P.~J. {Steinhardt}, and N.~{Turok}.
\newblock {Ekpyrotic universe: Colliding branes and the origin of the hot big
  bang}.
\newblock {\em Phys. Rev. D}, 64(12):123522, December 2001.
\newblock arXiv:hep-th/0103239.

\bibitem{1976JPhA....9.1387K}
T.~W.~B. {Kibble}.
\newblock {Topology of cosmic domains and strings}.
\newblock {\em Journal of Physics A Mathematical General}, 9:1387--1398, August
  1976.

\bibitem{KIB}
T.~W.~B. Kibble.
\newblock Evolution of a system of cosmic strings.
\newblock {\em Nucl. Phys.}, B252:227, 1985.

\bibitem{Kinney:2005vj}
William~H. Kinney.
\newblock {Horizon crossing and inflation with large eta}.
\newblock {\em Phys. Rev.}, D72:023515, 2005.
\newblock arXiv:gr-qc/0503017.

\bibitem{Kohri:2013mxa}
Kazunori Kohri, Yoshihiko Oyama, Toyokazu Sekiguchi, and Tomo Takahashi.
\newblock {Precise Measurements of Primordial Power Spectrum with 21 cm
  Fluctuations}.
\newblock {\em JCAP}, 1310:065, 2013.
\newblock arXiv:1303.1688.

\bibitem{pittphilsci9062}
Helge Kragh.
\newblock ?the most philosophically of all the sciences?: Karl popper and
  physical cosmology, March 2012.
\newblock PhilSci archive, item ID: 9062.

\bibitem{PDG_review}
Ofer Lahav and Andrew~R Liddle.
\newblock {The Cosmological Parameters 2014}.
\newblock {\em The Review of Particle Physics (Particle Data Group)}, 2014.
\newblock arXiv:1401.1389.

\bibitem{run_bounce}
Jean-Luc Lehners and Edward Wilson-Ewing.
\newblock {Running of the scalar spectral index in bouncing cosmologies}.
\newblock {\em JCAP}, 1510(10):038, 2015.
\newblock arXiv:1507.08112.

\bibitem{1927ASSB...47...49L}
G.~{Lema{\^i}tre}.
\newblock {Un Univers homog{\`e}ne de masse constante et de rayon croissant
  rendant compte de la vitesse radiale des n{\'e}buleuses extra-galactiques}.
\newblock {\em Annales de la Soci{\'e}t{\'e} Scientifique de Bruxelles},
  47:49--59, 1927.
\newblock \url{http://adsabs.harvard.edu/full/1927ASSB...47...49L}.

\bibitem{1931Natur.127..706L}
G.~{Lema{\^i}tre}.
\newblock {The Beginning of the World from the Point of View of Quantum
  Theory.}
\newblock {\em Nature}, 127:706, May 1931.

\bibitem{EU_lectnotes}
Antony Lewis.
\newblock Early universe: Inflation and the generation of fluctuations.
\newblock Lecture Notes.
\newblock \url{https://cosmologist.info/teaching/EU/}.

\bibitem{Lewis:2013hha}
Antony Lewis.
\newblock {Efficient sampling of fast and slow cosmological parameters}.
\newblock {\em Phys. Rev.}, D87(10):103529, 2013.
\newblock arXiv:1304.4473.

\bibitem{Lewis:2002ah}
Antony Lewis and Sarah Bridle.
\newblock {Cosmological parameters from CMB and other data: A Monte Carlo
  approach}.
\newblock {\em Phys. Rev.}, D66:103511, 2002.
\newblock arXiv:astro-ph/0205436.

\bibitem{Liddle:1992wi}
Andrew~R. Liddle and David~H. Lyth.
\newblock {COBE, gravitational waves, inflation and extended inflation}.
\newblock {\em Phys. Lett.}, B291:391--398, 1992.
\newblock arXiv:astro-ph/9208007.

\bibitem{thermo_phantom_mu1}
J.~A.~S. Lima and S.~H. Pereira.
\newblock {Chemical Potential and the Nature of the Dark Energy: The case of
  phantom}.
\newblock {\em Phys. Rev.}, D78:083504, 2008.
\newblock arXiv:0801.0323.

\bibitem{Loeb:2003ya}
Abraham Loeb and Matias Zaldarriaga.
\newblock {Measuring the small - scale power spectrum of cosmic density
  fluctuations through 21 cm tomography prior to the epoch of structure
  formation}.
\newblock {\em Phys. Rev. Lett.}, 92:211301, 2004.
\newblock arXiv:astro-ph/0312134.

\bibitem{Lyth:2009zz}
David~H. Lyth and Andrew~R. Liddle.
\newblock {\em {The primordial density perturbation: Cosmology, inflation and
  the origin of structure}}.
\newblock Cambridge University Press, 2009.

\bibitem{Undagoitia:2015gya}
Teresa Marrodán~Undagoitia and Ludwig Rauch.
\newblock {Dark matter direct-detection experiments}.
\newblock {\em J. Phys.}, G43(1):013001, 2016.
\newblock arXiv:1509.08767.

\bibitem{Martin2014}
Jerome Martin, Christophe Ringeval, and Vincent Vennin.
\newblock {Encyclopaedia Inflationaris}.
\newblock {\em Phys. Dark Univ.}, 5-6:75--235, 2014.
\newblock arXiv:1303.3787.

\bibitem{Martin:2013tda}
Jérôme Martin, Christophe Ringeval, Roberto Trotta, and Vincent Vennin.
\newblock {The Best Inflationary Models After Planck}.
\newblock {\em JCAP}, 1403:039, 2014.
\newblock arXiv:1312.3529.

\bibitem{PHD}
C.~J. A.~P. Martins.
\newblock {\em Quantitative String Evolution}.
\newblock Ph.D. Thesis, Cambridge University, 1997.

\bibitem{MS4}
C.~J. A.~P. Martins, J.~N. Moore, and E.~P.~S. Shellard.
\newblock A unified model for vortex-string network evolution.
\newblock {\em Phys. Rev. Lett.}, 92:251601, 2004.
\newblock arXiv:hep-ph/0310255.

\bibitem{Rybak}
C.~J. A.~P. Martins, I.~{\relax Yu}. Rybak, A.~Avgoustidis, and E.~P.~S.
  Shellard.
\newblock {Extending the velocity-dependent one-scale model for domain walls}.
\newblock {\em Phys. Rev.}, D93(4):043534, 2016.
\newblock arXiv:1602.01322.

\bibitem{MS2}
C.~J. A.~P. Martins and E.~P.~S. Shellard.
\newblock Quantitative string evolution.
\newblock {\em Phys. Rev.}, D54:2535--2556, 1996.
\newblock arXiv:hep-ph/9602271.

\bibitem{MS1}
C.~J. A.~P. Martins and E.~P.~S. Shellard.
\newblock String evolution with friction.
\newblock {\em Phys. Rev.}, D53:575--579, 1996.
\newblock arXiv:hep-ph/9507335.

\bibitem{SUPERC1}
C.~J. A.~P. Martins and E.~P.~S. Shellard.
\newblock Evolution of superconducting string currents.
\newblock {\em Phys. Lett.}, B432:58--64, 1998.
\newblock arXiv:hep-ph/9706533.

\bibitem{SUPERC2}
C.~J. A.~P. Martins and E.~P.~S. Shellard.
\newblock Vorton formation.
\newblock {\em Phys. Rev.}, D57:7155--7176, 1998.
\newblock arXiv:hep-ph/9804378.

\bibitem{MS3}
C.~J. A.~P. Martins and E.~P.~S. Shellard.
\newblock Extending the velocity-dependent one-scale string evolution model.
\newblock {\em Phys. Rev.}, D65:043514, 2002.
\newblock arXiv:hep-ph/0003298.

\bibitem{FRAC}
C.~J. A.~P. Martins and E.~P.~S. Shellard.
\newblock Fractal properties and small-scale structure of cosmic string
  networks.
\newblock {\em Phys. Rev.}, D73:043515, 2006.
\newblock arXiv:astro-ph/0511792.

\bibitem{PAP1}
C.~J. A.~P. Martins, E.~P.~S. Shellard, and J.~P.~P. Vieira.
\newblock {Models for Small-Scale Structure on Cosmic Strings: Mathematical
  Formalism}.
\newblock {\em Phys. Rev.}, D90(4):043518, 2014.
\newblock arXiv:1405.7722.

\bibitem{1940PASP...52..187M}
A.~{McKellar}.
\newblock {Evidence for the Molecular Origin of Some Hitherto Unidentified
  Interstellar Lines}.
\newblock {\em Publications of the Astronomical Society of the Pacific},
  52:187, June 1940.

\bibitem{COND1}
N.~D. Mermin.
\newblock {\em Rev. Mod. Phys.}, 51:591, 1979.

\bibitem{1935rgws.book.....M}
E.~A. {Milne}.
\newblock {\em {Relativity, gravitation and world-structure}}.
\newblock The Clarendon press, 1935.

\bibitem{1937RSPSA.158..324M}
E.~A. {Milne}.
\newblock {Kinematics, Dynamics, and the Scale of Time}.
\newblock {\em Proceedings of the Royal Society of London Series A},
  158:324--348, January 1937.

\bibitem{sep-ibn-bajja}
Josép~Puig Montada.
\newblock Ibn bâjja [avempace].
\newblock In Edward~N. Zalta, editor, {\em The Stanford Encyclopedia of
  Philosophy}. Metaphysics Research Lab, Stanford University, spring 2018
  edition, 2018.

\bibitem{ABELIAN}
J.~N. Moore, E.~P.~S. Shellard, and C.~J. A.~P. Martins.
\newblock On the evolution of abelian-higgs string networks.
\newblock {\em Phys. Rev.}, D65:023503, 2002.
\newblock arXiv:hep-ph/0107171.

\bibitem{GSRfeatures}
Hayato Motohashi and Wayne Hu.
\newblock {Running from Features: Optimized Evaluation of Inflationary Power
  Spectra}.
\newblock {\em Phys. Rev.}, D92(4):043501, 2015.
\newblock arXiv:1503.04810.

\bibitem{GSR_inflEFT}
Hayato Motohashi and Wayne Hu.
\newblock {Generalized Slow Roll in the Unified Effective Field Theory of
  Inflation}.
\newblock {\em Phys. Rev.}, D96(2):023502, 2017.
\newblock arXiv:1704.01128.

\bibitem{Munoz:2016owz}
Julian~B. Mu\~{n}oz, Ely~D. Kovetz, Alvise Raccanelli, Marc Kamionkowski, and
  Joseph Silk.
\newblock {Towards a measurement of the spectral runnings}.
\newblock {\em JCAP}, 1705:032, 2017.
\newblock arXiv:1611.05883.

\bibitem{mukperturb}
V.F. Mukhanov, H.A. Feldman, and R.H. Brandenberger.
\newblock Theory of cosmological perturbations.
\newblock {\em Physics Reports}, 215(5):203 -- 333, 1992.

\bibitem{thermo_phantom2}
Yun~Soo Myung.
\newblock {Does phantom thermodynamics exist?}
\newblock {\em Phys. Lett.}, B671:216--218, 2009.
\newblock arXiv:0810.4385.

\bibitem{1967Natur.216...43N}
J.~V. {Narlikar} and N.~C. {Wickramasinghe}.
\newblock {Microwave Background in a Steady-state Universe}.
\newblock {\em Nature}, 216:43--44, October 1967.

\bibitem{Nojiri:2017ncd}
S.~Nojiri, S.~D. Odintsov, and V.~K. Oikonomou.
\newblock {Modified Gravity Theories on a Nutshell: Inflation, Bounce and
  Late-time Evolution}.
\newblock {\em Phys. Rept.}, 692:1--104, 2017.
\newblock arXiv:1705.11098.

\bibitem{GamowEncyclopaediaBritannica2017}
The~Editors of~Encyclopædia~Britannica.
\newblock George gamow, October 2017.
\newblock \url{https://www.britannica.com/biography/George-Gamow}.

\bibitem{VVO}
Ken~D. Olum and Vitaly Vanchurin.
\newblock Cosmic string loops in the expanding universe.
\newblock {\em Phys. Rev.}, D75:063521, 2007.
\newblock arXiv:astro-ph/0610419.

\bibitem{ORaifeartaigh2017}
Cormac O'Raifeartaigh.
\newblock Albert einstein and the origins of modern cosmology.
\newblock {\em Physics Today}, February 2017.
\newblock \url{http://physicstoday.scitation.org/do/10.1063/PT.5.9085/full/}.

\bibitem{ORaifeartaigh:2017uct}
Cormac O'Raifeartaigh, Michael O'Keeffe, Werner Nahm, and Simon Mitton.
\newblock {Einstein's 1917 static model of the universe: a centennial review}.
\newblock {\em Eur. Phys. J.}, H42(3):431--474, 2017.
\newblock arXiv:1701.07261.

\bibitem{thermo_phantom_mu3}
J.~A. de~Freitas Pacheco.
\newblock {Comments on Accretion of Phantom Fields by Black Holes and the
  Generalized Second Law}.
\newblock 2008.
\newblock arXiv:0808.1863.

\bibitem{phantomGUP}
Andronikos Paliathanasis, Supriya Pan, and Souvik Pramanik.
\newblock {Scalar field cosmology modified by the Generalized Uncertainty
  Principle}.
\newblock {\em Class. Quant. Grav.}, 32(24):245006, 2015.
\newblock arXiv:1508.06543.

\bibitem{1991Natur.352..769P}
P.~J.~E. {Peebles}, D.~N. {Schramm}, R.~G. {Kron}, and E.~L. {Turner}.
\newblock {The case for the relativistic hot big bang cosmology}.
\newblock {\em Nature}, 352:769--776, August 1991.

\bibitem{Penzias_NL}
Arno~A. Penzias.
\newblock The origin of elements.
\newblock Nobel Lecture, December 1978.
\newblock
  \url{https://www.nobelprize.org/nobel_prizes/physics/laureates/1978/penzias-lecture.pdf}.

\bibitem{thermo_phantom_mu2}
S.~H. Pereira and J.~A.~S. Lima.
\newblock {On Phantom Thermodynamics}.
\newblock {\em Phys. Lett.}, B669:266--270, 2008.
\newblock arXiv:0806.0682.

\bibitem{2012arXiv1201.3942P}
A.~H.~G. {Peter}.
\newblock {Dark Matter: A Brief Review}.
\newblock January 2012.
\newblock arXiv:1201.3942.

\bibitem{POLR}
Joseph Polchinski and Jorge~V. Rocha.
\newblock Cosmic string structure at the gravitational radiation scale.
\newblock {\em Phys. Rev.}, D75:123503, 2007.
\newblock arXiv:gr-qc/0702055.

\bibitem{0038-5670-6-4-E01}
P~Ya Polubarinova-Kochina.
\newblock Aleksandr aleksandrovich fridman (on the seventy-fifth anniversary of
  his birth).
\newblock {\em Soviet Physics Uspekhi}, 6(4):467, 1964.

\bibitem{NuclearNAT}
E.~M. Purcell and R.~V. Pound.
\newblock A nuclear spin system at negative temperature.
\newblock {\em Phys. Rev.}, 81:279--280, Jan 1951.

\bibitem{RamseyNAT}
Norman~F. Ramsey.
\newblock Thermodynamics and statistical mechanics at negative absolute
  temperatures.
\newblock {\em Phys. Rev.}, 103:20--28, Jul 1956.

\bibitem{Renaux-Petel:2015bja}
Sébastien Renaux-Petel.
\newblock {Primordial non-Gaussianities after Planck 2015: an introductory
  review}.
\newblock {\em Comptes Rendus Physique}, 16:969--985, 2015.
\newblock arXiv:1508.06740.

\bibitem{RSB}
Christophe Ringeval, Mairi Sakellariadou, and Francois Bouchet.
\newblock Cosmological evolution of cosmic string loops.
\newblock {\em JCAP}, 0702:023, 2007.
\newblock arXiv:astro-ph/0511646.

\bibitem{Romano:2015vxz}
Antonio~Enea Romano, Sander Mooij, and Misao Sasaki.
\newblock {Adiabaticity and gravity theory independent conservation laws for
  cosmological perturbations}.
\newblock {\em Phys. Lett.}, B755:464--468, 2016.
\newblock arXiv:1512.05757.

\bibitem{NO_NEGT}
V.~{Romero-Roch{\'{\i}}n}.
\newblock {Nonexistence of equilibrium states at absolute negative
  temperatures}.
\newblock {\em Phys. Rev.}, E88(2):022144, August 2013.
\newblock arXiv:1301.0852.

\bibitem{Rybak:2017yfu}
I.~{\relax Yu}. Rybak, A.~Avgoustidis, and C.~J. A.~P. Martins.
\newblock {Semianalytic calculation of cosmic microwave background anisotropies
  from wiggly and superconducting cosmic strings}.
\newblock {\em Phys. Rev.}, D96(10):103535, 2017.
\newblock arXiv:1709.01839.

\bibitem{Saadeh:2016sak}
Daniela Saadeh, Stephen~M. Feeney, Andrew Pontzen, Hiranya~V. Peiris, and
  Jason~D. McEwen.
\newblock {How isotropic is the Universe?}
\newblock {\em Phys. Rev. Lett.}, 117(13):131302, 2016.
\newblock arXiv:1605.07178.

\bibitem{COND2}
M.~Salomaa and G.~Volovik.
\newblock {\em Rev. Mod. Phys.}, 59:533, 1987.

\bibitem{Sarangi:2002yt}
Saswat Sarangi and S.~H.~Henry Tye.
\newblock {Cosmic string production towards the end of brane inflation}.
\newblock {\em Phys. Lett.}, B536:185--192, 2002.
\newblock arXiv:hep-th/0204074.

\bibitem{scalarNATmail}
Emmanuel~N. Saridakis, Pedro~F. Gonzalez-Diaz, and Carmen~L. Siguenza.
\newblock {Unified dark energy thermodynamics: varying w and the -1-crossing}.
\newblock {\em Class. Quant. Grav.}, 26:165003, 2009.
\newblock arXiv:0901.1213.

\bibitem{thermalQFT_lectnotes}
Andreas Schmitt.
\newblock Thermal field theory.
\newblock Lecture Notes, 2013.
\newblock \url{http://hep.itp.tuwien.ac.at/~aschmitt/thermal13.pdf}.

\bibitem{comment_consist_NAT}
U.~{Schneider}, S.~{Mandt}, A.~{Rapp}, S.~{Braun}, H.~{Weimer}, I.~{Bloch}, and
  A.~{Rosch}.
\newblock {Comment on ''Consistent thermostatistics forbids negative absolute
  temperatures''}.
\newblock July 2014.
\newblock arXiv:1407.4127.

\bibitem{Sekiguchi:2017cdy}
Toyokazu Sekiguchi, Tomo Takahashi, Hiroyuki Tashiro, and Shuichiro Yokoyama.
\newblock {21 cm Angular Power Spectrum from Minihalos as a Probe of Primordial
  Spectral Runnings}.
\newblock {\em JCAP}, 1802(02):053, 2018.
\newblock arXiv:1705.00405.

\bibitem{thermo_phantom_mu4}
H.~H.~B. Silva, R.~Silva, R.~S. Gonçalves, Zong-Hong Zhu, and J.~S. Alcaniz.
\newblock {General treatment for dark energy thermodynamics}.
\newblock {\em Phys. Rev.}, D88:127302, 2013.
\newblock arXiv:1312.3216.

\bibitem{1917PAPhS..56..403S}
V.~M. {Slipher}.
\newblock {Nebulae}.
\newblock {\em Proceedings of the American Philosophical Society}, 56:403--409,
  1917.

\bibitem{STEWART_RUN}
Ewan~D. Stewart.
\newblock {The Spectrum of density perturbations produced during inflation to
  leading order in a general slow roll approximation}.
\newblock {\em Phys. Rev.}, D65:103508, 2002.
\newblock arXiv:astro-ph/0110322.

\bibitem{SL1993}
Ewan~D. Stewart and David~H. Lyth.
\newblock {A More accurate analytic calculation of the spectrum of cosmological
  perturbations produced during inflation}.
\newblock {\em Phys. Lett.}, B302:171--175, 1993.
\newblock arXiv:gr-qc/9302019.

\bibitem{TAKAYASU}
H.~Takayasu.
\newblock {\em Fractals in the Physical Sciences}.
\newblock Manchester University Press, Manchester, U. K., 1990.

\bibitem{1934PNAS...20..169T}
R.~C. {Tolman}.
\newblock {Effect of Inhomogeneity on Cosmological Models}.
\newblock {\em Proceedings of the National Academy of Science}, 20:169--176,
  March 1934.

\bibitem{1934rtc..book.....T}
R.~C. {Tolman}.
\newblock {\em {Relativity, Thermodynamics, and Cosmology}}.
\newblock Clarendon Press, 1934.

\bibitem{Vennin:2015egh}
Vincent Vennin, Kazuya Koyama, and David Wands.
\newblock {Inflation with an extra light scalar field after Planck}.
\newblock {\em JCAP}, 1603(03):024, 2016.
\newblock arXiv:1512.03403.

\bibitem{VSH}
A.~Vilenkin and E.~P.~S. Shellard.
\newblock {\em Cosmic Strings and other Topological Defects}.
\newblock Cambridge University Press, Cambridge, U.K., 1994.

\bibitem{WIG1}
Alexander Vilenkin.
\newblock Effect of small scale structure on the dynamics of cosmic strings.
\newblock {\em Phys. Rev.}, D41:3038, 1990.

\bibitem{local_NG_inflation}
David Wands.
\newblock {Local non-Gaussianity from inflation}.
\newblock {\em Class. Quant. Grav.}, 27:124002, 2010.
\newblock arXiv:1004.0818.

\bibitem{WANDS_SEPARATE}
David Wands, Karim~A. Malik, David~H. Lyth, and Andrew~R. Liddle.
\newblock {A New approach to the evolution of cosmological perturbations on
  large scales}.
\newblock {\em Phys. Rev.}, D62:043527, 2000.
\newblock arXiv:astro-ph/0003278.

\bibitem{White:1999nh}
Martin~J. White.
\newblock {Anisotropies in the CMB}.
\newblock In {\em {American Physical Society (APS) Meeting of the Division of
  Particles and Fields (DPF 99) Los Angeles, California, January 5-9, 1999}},
  1999.
\newblock arXiv:astro-ph/9903232.

\bibitem{MSC_thermalQFT}
Yuhao Yang.
\newblock An introduction to thermal field theory.
\newblock Master's thesis, Imperial College London, 2011.
\newblock
  \url{https://www.imperial.ac.uk/media/imperial-college/research-centres-and-groups/theoretical-physics/msc/dissertations/2011/Yuhao-Yang-Dissertation.pdf}.

\bibitem{1985Natur.317..505Z}
W.~H. {Zurek}.
\newblock {Cosmological experiments in superfluid helium?}
\newblock {\em Nature}, 317:505--508, October 1985.

\bibitem{1929PNAS...15..773Z}
F.~{Zwicky}.
\newblock {On the Red Shift of Spectral Lines through Interstellar Space}.
\newblock {\em Proceedings of the National Academy of Science}, 15:773--779,
  October 1929.

\end{thebibliography}

\newpage
\appendix
\addcontentsline{toc}{chapter}{Appendices}

\chapter{Cosmology with Negative Absolute Temperatures}

\section{Problems maintaining negative temperatures with number conservation and bosons}
\label{sec:The-problem-of-mu}

Let us start by assuming number conservation so that we can explore the kind of problems it causes.
In an FLRW Universe with
scale factor $a$ and Hubble parameter $H$, there will then be two
equations governing the dynamics of these functions,
the continuity equation
\begin{equation}
\dot{\rho}=-3H\left(\rho+P\right)\label{eq:continuity_FLRW}
\end{equation}
and the number conservation equation
\begin{equation}
\frac{\dot{n}}{n}=-3H.\label{eq:n_conservation}
\end{equation}
Equivalently, one can make use of the symmetries in Eqs.~\eqref{eq:rho_holes}, \eqref{eq:n_holes}, and \eqref{eq:P_holes}
to rewrite these for holes as
\begin{equation}
\dot{\rho_{h}}=-3H\left(\rho_{h}+P_{h}\right)\label{eq:continuity_FLRW_holes}
\end{equation}
\begin{equation}
\dot{n_{h}}=3H\left(n_{\rm{max}}-n_{h}\right) \label{eq:n_conservation_holes}
\end{equation}
where the subscript $h$ denotes a quantity relative to holes (with $T_{h}=-T$).
Note that the formal equivalence between Eqs.~\eqref{eq:continuity_FLRW} and \eqref{eq:continuity_FLRW_holes}
is due to the fact that they both express the constraint that the entropy be conserved
and entropy cannot distinguish between particles and holes (since it is purely combinatorial).

Suppose now that the Universe is filled with a temperon gas at NAT with $m\gg T_{h}$. If
the very low-energy holes behave like normal (pressureless) matter then $\rho_{h}=m n_{h}$
and the previous equations are reduced to
\begin{equation}
\dot{\rho_{h}}=-3H\rho_{h}\label{eq:continuity_FLRW_holes_bad}
\end{equation}
\begin{equation}
\dot{\rho_{h}}=3Hm n_{\rm{max}}-3H\rho_{h} \label{eq:n_conservation_holes_bad}
\end{equation}
which is an inconsistent system as long as $H\neq 0$. This would mean
that in this situation equilibrium could not be maintained during the expansion.

Of course, this problem assumes a specific low-energy form of $g\left(\epsilon\right)$ and
thus it can by no means be considered a refutation of the $\mu \neq 0$ case.
Nevertheless, it is a difficulty that has to be taken into account and which raises questions about
how model-independent (i.e., how independent of $g\left(\epsilon\right)$) such an analysis can be.
In addition, if we just assume $g\left(\epsilon\right)$ is whatever is necessary
to make this system of equations consistent, we have to live with the fact
that there are possible situations in which the energy density will be increasing
while the number density decreases (and vice-versa, if the Universe is contracting),
since $\dot{\rho}/\dot{n}$ has the same sign as $\rho +P$ and $\rho$ is bounded
whereas $P$ is not.

Moreover, it can be easily seen that, even accepting these odd behaviours,
such a solution can never be consistent in all situations.
For example, consider the case where temperons are massless fermions at $T=0^{-}$.
In this situation, the energy density must be constant and equal to its maximum possible value,
whereas the number density must vary according to $H$. This is clearly absurd
as there is no way the system can be at $\rho = \rho_{\rm{max}}$ unless all states are filled.

Note that once we restrict ourselves to the study of cases without number conservation
it becomes clear that we cannot use bosons: without number conservation,
the energy of the system is no longer bounded from above, which makes
NAT impossible.

\section{Thermal Perturbation Generation\label{sec:Thermal-Perturbation-Generation}}

A system in equilibrium will in general have thermal fluctuations.
Here we consider the case where the Universe is dominated by temperons in thermal equilibrium,
and calculate the density and curvature perturbations produced. We focus on the case where holes behave
like radiation as $\beta\rightarrow-\infty$%
\footnote{Ignoring any contributions from whatever process might be responsible
for ending inflation.%
}.

The main difference between perturbations here and in the standard
inflationary scenario is that density perturbations here are produced
due to classical thermal fluctuations rather than by quantum effects.
The basic methodology used in this subsection is therefore essentially
the same as the one used to work out thermal fluctuations in models
such as chain inflation or warm inflation in the very weakly dissipative
regime \cite{thermo_perturb}.

\subsection{Moments in position space}

For a canonical thermal system with volume $V$, the $n$-th moment of the energy density
distribution is given by
\begin{equation}
\left\langle \rho^{n}\right\rangle =\frac{\left\langle E^{n}\right\rangle }{V^{n}}=\frac{1}{Z}\left(-\frac{1}{V}\right)^{n}\frac{\partial^{n}Z}{\partial\beta^{n}}\label{eq:moment_density}
\end{equation}
where $Z$ is the partition function as given by Eq.~\eqref{eq:Zg_fermions}
with $\mu=0$. Making the substitution $\partial_{\alpha}\equiv-V^{-1}\partial_{\beta}$,
we can find the simple recursive relation
\begin{equation}
\left\langle \rho^{n+1}\right\rangle =\left[\left\langle \rho\right\rangle +\partial_{\alpha}\right]\left\langle \rho^{n}\right\rangle \label{eq:recursive_moment_density}
\end{equation}
which we can then use to find an analogous formula for the moments
of $\delta\rho=\rho-\left\langle \rho\right\rangle $. Taking a derivative
with respect to $\alpha$ of
\begin{equation}
\left\langle \left(\delta\rho\right)^{n}\right\rangle =\sum_{k=0}^{n}\frac{n!}{\left(n-k\right)!k!}\left(-\left\langle \rho\right\rangle \right)^{n-k}\left\langle \rho^{k}\right\rangle \label{eq:moment_deltarho}
\end{equation}
and then making use of Eq.~\eqref{eq:recursive_moment_density} yields

\begin{equation}
\left\langle \left(\delta\rho\right)^{n+1}\right\rangle =\partial_{\alpha}\left\langle \left(\delta\rho\right)^{n}\right\rangle +n\partial_{\alpha}\left\langle \rho\right\rangle \left\langle \left(\delta\rho\right)^{n-1}\right\rangle .\label{eq:recursive_moment_deltarho}
\end{equation}

Eq.~\eqref{eq:recursive_moment_deltarho} has the interesting feature
of separating contributions from even and odd momenta in the right-hand
side. Because of it, and since $\left\langle \delta\rho\right\rangle =0$,
if we assume that $\left\langle \left(\delta\rho\right)^{3}\right\rangle =\partial_{\alpha}^{2}\left\langle \rho\right\rangle =0$
then (as can be easily checked by induction) every odd moment has
to be zero and $\left\langle \left(\delta\rho\right)^{2n}\right\rangle =\left(2n-1\right)!!\left\langle \left(\delta\rho\right)^{2}\right\rangle ^{n}$,
corresponding to exactly Gaussian perturbations whose statistics depend
only on the size of the thermal system $V$ (and not on $\beta$).
In other words, in this scenario Gaussianity is equivalent to the
third moment of $\delta\rho$ being null and to the second (and indeed
every even) moment being independent of $\beta$. Since
\begin{equation}
\left\langle \left(\delta\rho\right)^{3}\right\rangle =\partial_{\alpha}^{2}\left\langle \rho\right\rangle =\frac{1}{V^{2}}\intop g\left(\epsilon\right)\epsilon^{3}\left[\frac{e^{2\beta\epsilon}-e^{\beta\epsilon}}{\left(e^{\beta\epsilon}+1\right)^{3}}\right]d\epsilon,\label{eq:3rd_moment}
\end{equation}
then density perturbations must always be exactly Gaussian for $\beta=0$
(at the bounce) and for the attractors $\beta=\pm\infty$. This exact
Gaussianity at the attractors, however, is misleading since it corresponds
to a limit in which density perturbations must vanish --- recall that
from Eq.~\eqref{eq:rho_holes} we have $\delta\rho=-\delta\rho_{\rm holes}$
(where the temperature of holes is symmetric to that of particles)
and $\delta\rho$ has to be zero for $T=0^{+}$ since at that temperature
the density itself is zero. We should thus look at the relevant perturbation,
the curvature perturbation, which can be written in terms of the density
perturbation in the zero curvature frame (in which the previous
calculations make sense as the shape of the ``box'' is not perturbed)
as
\begin{equation}
\zeta=\frac{1}{3}\frac{\delta\rho}{\left\langle \rho\right\rangle +\left\langle P\right\rangle }.\label{eq:zeta_def}
\end{equation}
In the case of a bounce, this also shows that even the Gaussian perturbations
are less interesting than one might think, since despite
the numerator being non-zero the denominator diverges, making the
relevant curvature perturbation negligible.

The variance of the curvature perturbation can now be found to be
given by
\begin{equation}
\left\langle \zeta^{2}\right\rangle =\frac{1}{9}\frac{\left\langle \delta\rho^{2}\right\rangle }{\left(\rho+P\right)^{2}}=-\frac{1}{9V}\frac{\partial_{\beta}\rho}{\left(\rho+P\right)^{2}},\label{eq:variance_form}
\end{equation}
where for simplicity we are using $\rho$ and $P$ interchangeably
with their averages.

The most interesting limit for Eq.~\eqref{eq:variance_form} is when
$\beta\rightarrow-\infty$ as the spacetime then tends towards unperturbed
de Sitter, yet the curvature perturbation does not necessarily tend
to zero as the denominator in the right-hand side also goes to zero
in this limit%
\footnote{These calculations may not even be physically
meaningful too close to that limit, since then most Hubble volumes
will have no holes and will be indistinguishable from de Sitter space,
for which $\zeta$ is not well defined since there isn't a unique
constant-density frame.
There could also be additional effects, for example if the equilibration time is not
negligibly smaller than the Hubble time the density perturbation could be dominated by
fluctuations in the equilibration process.
}. For example, if holes behave like radiation, the denominator vanishes
at a faster rate than the numerator, causing $\zeta$ to diverge as
(see table \ref{tab:Summary-of-relevant-rad} for a breakdown of the
relevant terms in Eq.~\eqref{eq:variance_form} in this limit)
\begin{equation}
\left\langle \zeta^{2}\right\rangle =\left(\frac{15}{7\pi^{2}g}\right)^{1/4}\frac{\left(\rho_{\mathrm{max}}-\rho\right)^{-3/4}}{2V}.\label{eq:variance_zeta}
\end{equation}

\begin{table}[h]
\begin{centering}
\begin{tabular}{|c|c|c|}
\hline
 & $\rho_{h}=\rho_{\mathrm{max}}-\rho$ & $a_{\star}=aV_{0}^{-1/3}$\tabularnewline
\hline
\hline
$\rho+P$ & $-\frac{4}{3}\rho_{h}$ & $-\frac{4}{3}a_{\star}^{-4}$\tabularnewline
\hline
$\beta$ & -$\left(\frac{7\pi^{2}}{240}\frac{g}{\rho_{h}}\right)^{1/4}$ & $-\left(\frac{7\pi^{2}}{240}g\right)^{1/4}a_{\star}$\tabularnewline
\hline
$\left\langle \left(\delta\rho\right)^{2}\right\rangle $ & $\frac{8}{V}\left(\frac{15}{7\pi^{2}g}\right)^{1/4}\rho_{h}^{5/4}$ & $\frac{8}{V}\left(\frac{15}{7\pi^{2}g}\right)^{1/4}a_{\star}^{-5}$\tabularnewline
\hline
$\left\langle \zeta^{2}\right\rangle $ & $\frac{1}{2V}\left(\frac{15}{7\pi^{2}g}\right)^{1/4}\rho_{h}^{-3/4}$ & $\frac{1}{2V}\left(\frac{15}{7\pi^{2}g}\right)^{1/4}a_{\star}^{3}$\tabularnewline
\hline
\end{tabular}
\par\end{centering}

\caption{\label{tab:Summary-of-relevant-rad}Summary of relevant quantities
as functions of hole energy density, $\rho_{h}$, and rescaled scale
factor, $a_{\star}=\rho_{h}^{-1/4}$, when holes behave like radiation.
The row for $\beta$ comes from applying the standard result for the
fermion energy density to $\rho_{h}$. $V_{0}$ is the integration
constant used later in Eq.~\eqref{eq:final_P_spectrum}.}
\end{table}

\subsection{The thermal power spectrum}

In order to turn the results from the previous section into predictions
for the power spectrum, we must introduce a couple of mathematical
objects and endure some integral manipulations.

Let us start by considering the average density fluctuation in the
vicinity of a point,
\begin{multline}
\delta\rho_{\mathbf{x_{0}}}\left(r\right)\equiv\frac{1}{V_{r}}\intop_{S_{r,\mathrm{x_{0}}}}d^{3}{\bf x}\delta\rho\left(\mathbf{x}\right)=\frac{1}{V_{r}}\intop_{S_{r}}d^{3}{\bf x}\delta\rho\left(\mathbf{x_{0}+x}\right)
\\=\frac{1}{V_{r}}\intop\frac{d^{3}{\bf k}}{\left(2\pi\right)^{3/2}}\delta\rho_{{\bf k}}\intop_{S_{r}}d^{3}{\bf x}e^{i{\bf k\cdot\left(x_{0}+x\right)}},\label{eq:local_drhor}
\end{multline}
where $S_{r,{\bf x_{0}}}$ is the sphere of comoving radius $r$ centred
around $\mathbf{x_{0}}$, $S_{r}=S_{r,{\bf 0}}$, and $V_{r}=\frac{4}{3}\pi r^{3}$.

We can define the average power of this quantity as
\begin{equation}
\overline{\delta\rho}^{2}\left(r\right)\equiv\lim_{R\rightarrow\infty}\frac{1}{V_{R}}\intop_{S_{R}}d^{3}{\bf x_{0}}\left|\delta\rho_{{\bf x_{0}}}\left(r\right)\right|^{2}\label{eq:deltarhobar}
\end{equation}
which can be rewritten as
\begin{equation}
\overline{\delta\rho}^{2}\left(r\right)=\lim_{R\rightarrow\infty}\intop\frac{d^{3}{\bf k}}{\left(2\pi\right)^{3/2}}\intop\frac{d^{3}{\bf k}^{\prime}}{\left(2\pi\right)^{3/2}}
\times\delta\rho_{{\bf k}}\delta\rho_{{\bf k}^{\prime}}W_{r}\left(k\right)W_{r}\left(k^{\prime}\right)W_{R}\left(\left\Vert {\bf k+k^{\prime}}\right\Vert \right),\label{eq:rhobar_beforeav}
\end{equation}
where we have used the window function defined as
\begin{equation}
W_{r}\left(k\right)\equiv\frac{1}{V_{r}}\intop_{S_{r}}d^{3}{\bf x}e^{i{\bf x\cdot k}}=3\frac{\sin\left(k r\right)-k r \cos\left(k r\right)}{\left(k r\right)^{3}}.\label{eq:window}
\end{equation}

Taking the expected value on both sides and using the usual definition $\left\langle \delta\rho_{{\bf k}}\delta\rho_{{\bf k^{\prime}}}\right\rangle \equiv \delta\left({\bf k+k^{\prime}}\right) P_{\delta\rho}\left(k\right)$
then yields
\begin{equation}
\left\langle \overline{\delta\rho}^{2}\left(r\right)\right\rangle =\intop\frac{d^{3}{\bf k}}{\left(2\pi\right)^{3}}\left|W_{r}\left(k\right)\right|^{2}P_{\delta\rho}\left(k\right).\label{eq:bar_spectrum}
\end{equation}
Alternatively, using the usual definition of the power spectrum
${\cal P}\left(k\right)=k^{3}P\left(k\right)/\left(2\pi^{2}\right)$, this is
\begin{equation}
\left\langle \overline{\delta\rho}^{2}\left(r\right)\right\rangle =\intop_{0}^{\infty}\frac{dk}{k}\left|W_{r}\left(k\right)\right|^{2}\mathcal{P}_{\delta\rho}\left(k\right).\label{eq:bar_calspectrum}
\end{equation}

Provided that $P_{\delta\rho}\left(k\right)$ doesn't diverge faster
than $k^{-3}$ as $k\rightarrow0$, the integral in Eq.~\eqref{eq:bar_spectrum}
is dominated by $k\sim r^{-1}$ and \cite{mukperturb}
\begin{equation}
P_{\delta\rho}\left(k\right)\sim \frac{\left(2\pi\right)^{3}}{k^{3}}\left\langle \overline{\delta\rho}^{2}\left(k^{-1}\right)\right\rangle .\label{eq:calspectrum_variance}
\end{equation}

Assuming there is a \emph{thermal horizon}%
\footnote{
In the literature, some measure of the typical wavelength of a particle (usually a photon) in the thermal
system has been used as the thermal horizon, although \cite{thermo_perturb}
note it can in principle be any scale between that and the \emph{acoustic
horizon}, $c_{s}H^{-1}$.%
}, $L_{\rm th}$, corresponding to the physical distance beyond which there can
be no thermal correlation, the actual observed density power spectrum
can be calculated from Eq.~\eqref{eq:calspectrum_variance} evaluated
around thermal horizon exit, when $k\sim a/L_{\rm th}$. Using $\left\langle \overline{\delta\rho}^{2}\left(L/a\right)\right\rangle =\left\langle \left(\delta\rho\right)^{2}\right\rangle_{V_{\rm th}} $
we conclude that
\begin{equation}
\left\langle \overline{\delta\rho}^{2}\left(r\right)\right\rangle =\left\langle \left(\delta\rho\right)^{2}\right\rangle _{V=a^{3}V_{r}} ,\label{eq:ergodic_rms}
\end{equation}
 and hence
\begin{equation}
P_{\zeta}\left(k\sim a/L_{\rm th}\right)\sim\frac{\left(2\pi\right)^{3} L_{\rm th}^{3}}{9a^{3}\left(\rho+P\right)^{2}}\left\langle \delta\rho^{2}\right\rangle _{V=\frac{4\pi}{3}L_{\rm th}^{3}}\sim-\frac{\left(2\pi\right)^{3}}{12 a^{3}}\frac{\partial_{\beta}\rho}{\left(\rho+P\right)^{2}},\label{eq:P_spectrum_general_all}
\end{equation}
where everything is evaluated around thermal horizon crossing.

If we further assume holes behave like radiation, we can use this
together with equation \eqref{eq:variance_zeta} and immediately get
\begin{equation}
P_{\zeta}\left(k\right)\sim\frac{3}{8}\left(\frac{15}{7\pi^{6}g}\right)^{1/4}\frac{\left(2\pi\right)^{3}}{V_{0}}, \label{eq:final_P_spectrum}
\end{equation}
where $V_0 \equiv a^{3}\left(\rho_{\mathrm{max}}-\rho\right)^{3/4}$
is a constant thanks to Eq.~\eqref{eq:DN_radiationhole}.
Note that this corresponds to
a white noise spectrum with $n_{s}=4$.

If the NAT model were to describe a realistic cosmology, we would need the power spectrum to be (approximately) scale-invariant,
at least in the attractor $\beta\rightarrow-\infty$ limit. Unfortunately,
we can show that is not necessarily possible even if we allow drastic
departures from Eq.~\eqref{eq:DOS_Lambda}. From Eq.~\eqref{eq:calspectrum_variance}
it is clear that the spectrum will be scale invariant if and only
if $\left\langle \overline{\delta\rho}^{2}\left(k^{-1}\right)\right\rangle $
is independent of $k$. In other words, the power spectrum can be
written as
\begin{equation}
{\cal P}_{\zeta}\left(k\right)\sim-\frac{4\pi}{9V}\frac{\partial_{\beta}\rho}{\left(\rho+P\right)^{2}},\label{eq:power_spectrum_general_thermal}
\end{equation}
which is a constant if and only if
\begin{equation}
\partial_{k}\left(V^{-1}\frac{\partial_{\beta}\rho}{\left(\rho+P\right)^{2}}\right)=0.\label{eq:scale_inv_condition}
\end{equation}
\newline\newline Assuming, as before, that $V\propto L_{\rm th}^{3}=\text{const}$, this can be
rewritten as
\begin{equation}
\frac{\partial_{\beta}\rho^{-1}}{\left(1+w\right)^{2}}=\text{const}.\label{eq:scale_inv_w}
\end{equation}
Note that if $w_{h}$ is the equation of state of holes then Eqs.~\eqref{eq:rho_holes} and \eqref{eq:P_holes_mu0} yield $\rho+P=-\rho_{h}-P_{h}$
(where the subscript $h$ once again denotes holes) and thus Eq.~\eqref{eq:scale_inv_w}
is equivalent to
\begin{equation}
\frac{\partial_{\beta_{h}}\rho_{h}^{-1}}{\left(1+w_{h}\right)^{2}}=\text{const}.\label{eq:scale_inv_holes}
\end{equation}
Consequently, if $w_{h}=\text{const}$ then in this large $\beta>0$ limit
\begin{equation}
\rho=\frac{1}{A+B\beta}, \label{eq:weirdrho}
\end{equation}
 where $A$ and $B$ are positive constants of integration --- they
have to be positive because $B$ is related to the power spectrum
by ${\cal P}_{\zeta}\left(k\right)\sim\frac{4\pi}{9V}\frac{B}{\left(1+w_{h}\right)^{2}}$.
If we now equate the right-hand sides of Eq.~\eqref{eq:rho_general}
and Eq.~\eqref{eq:weirdrho} and take a derivative with respect to $T$
at $T=0^{+}$ we get
\begin{equation}
\lim_{\beta\rightarrow\infty}\beta^{2}\int\frac{g\left(\epsilon\right)\epsilon^{2}e^{\beta\epsilon}}{\left(e^{\beta\epsilon}+1\right)^{2}}d\epsilon=1\label{eq:absurd}
\end{equation}
which is absurd since the left-hand side should be zero as long as
there is a finite total number of one-particle states.

\chapter{Can power spectrum observations rule out slow-roll inflation?}

\section{$g$ and the slow-roll parameters\label{sec:gappendix}}

In this work, slow-roll is tested via the $g$ function defined in
Eq.~\eqref{eq:gdef} rather than directly via the slow-roll parameters. We relate the two here.

\subsection{Slow-roll parameters from conformal time}

The main difficulty in relating $g$ to the slow-roll parameters stems from the terms in $g$ which are related to the conformal time.
We thus start by manipulating the usual expression for (minus) the conformal time,
\begin{equation}
\xi\left(t\right)=\intop_{t}^{\infty}\frac{dt^{'}}{a\left(t^{'}\right)}=\intop_{a\left(t\right)}^{\infty}\frac{da}{Ha^{2}}=\frac{1}{a\left(t\right)H\left(t\right)}-\intop_{a\left(t\right)}^{\infty}\frac{\dot{H}}{H^{2}}\frac{da}{a\dot{a}}=\frac{1}{a\left(t\right)H\left(t\right)}-\intop_{a\left(t\right)}^{\infty}\frac{\dot{H}}{H^{2}}\frac{da}{a^{2}H}.\label{eq:eta_(re)def}
\end{equation}
Using Eq.~\eqref{eq:eps_small}, we write it as
\begin{equation}
\xi=\frac{1}{aH}\left[1+\bar{\epsilon}\right],\label{eq:eta_epsbar}
\end{equation}
where $\bar{\epsilon}$ is given by
\begin{equation}
\bar{\epsilon}\left(\xi\right)\equiv a\left(\xi\right)H\left(\xi\right)\intop_{0}^{\xi}\epsilon(\tilde{\xi})d\tilde{\xi}=\frac{1}{\left\langle \epsilon\right\rangle _{\xi}^{-1}-1},\label{eq:eps_bar}
\end{equation}
$\left\langle \epsilon\right\rangle $ being the conformal time average
of $\epsilon$ at a given instant, defined as
\begin{equation}
\left\langle \epsilon\right\rangle _{\xi}\equiv\frac{1}{\xi}\intop_{0}^{\xi}\epsilon(\tilde{\xi})d\tilde{\xi.}\label{eq:average_eps}
\end{equation}

From Eq.~\eqref{eq:eps_small}, it can be easily seen that
\begin{equation}
\frac{d\epsilon}{d\ln\xi}=-\left(1+\bar{\epsilon}\right)\left(2\epsilon^{2}+2\epsilon\delta_{1}\right),\label{eq:deps}
\end{equation}
so variations in $\epsilon$ are second-order in slow roll and thus $\left\langle \epsilon\right\rangle $ and $\bar{\epsilon}$
are not expected to differ from $\epsilon$ at first order.

In fact, if we further assume that $\epsilon$ is well-behaved (in the sense that it can
be expressed as a Taylor series in the domain of integration of Eq.
\eqref{eq:average_eps}%
\footnote{Since the integration
domain for $\left\langle \epsilon\right\rangle $ stretches all the
way to the infinite future, a drastic departure from slow-roll at
(or even after) the end of inflation may cause this assumption to
be violated - potentially leading to $\left|\bar{\epsilon}-\epsilon\right|$ being larger
than expected. However, as long as this violation is far
enough in the future, for our purposes we can always ignore it and
pretend that slow-roll goes on forever (or alternatively stop the
integration at a very distant point before slow-roll is violated)
since we know that the curvature perturbation is conserved on very
large superhorizon scales \cite{WANDS_SEPARATE}.\label{fn:far_future}%
}), we can write $\left\langle \epsilon\right\rangle$ as
\begin{equation}
\left\langle \epsilon\right\rangle_{\xi} =\epsilon+\sum_{n=1}^{\infty}\frac{d^{n}\epsilon}{d\xi^{n}}\frac{\left(-\xi\right)^{n}}{\left(n+1\right)!}=\epsilon+\sum_{n=1}^{\infty}\left(-aH\right)^{-n}\frac{d^{n}\epsilon}{d\xi^{n}}\frac{\left(1+\bar{\epsilon}\right)^{n}}{\left(n+1\right)!}.\label{eq:epsaverageTaylor}
\end{equation}
Combining this with Eq.~\eqref{eq:deps} and its equivalent for $\delta_{n}$,
\begin{equation}
\frac{d\delta_{n}}{d\ln\xi}=-\left(1+\bar{\epsilon}\right)\left(\delta_{n+1}+n\epsilon\delta_{n}-n\delta_{1}\delta_{n}\right),\label{eq:ddeltan}
\end{equation}
it can be seen that, to second order in the slow-roll parameters,
\begin{equation}
\left\langle \epsilon\right\rangle_{\xi} \approx\epsilon+2\left(\epsilon+\sum_{p=1}^{\infty}\delta_{p}\right)\epsilon\label{eq:av_eps_final}
\end{equation}
(the right-hand side being evaluated at minus conformal time $\xi$) and
\begin{equation}
\bar{\epsilon}\approx\epsilon\left(1+3\epsilon+2\sum_{p=1}^{\infty}\delta_{p}\right).\label{eq:eps_bar_slow_final}
\end{equation}

\subsection{$g$ from $f$}

Using these results, the $f$ function defined in Eq.~\eqref{eq:fdef}
can be written as
\begin{equation}
f\left(\ln x\right)=2\pi\frac{\dot{\phi}}{H^{2}}\left[1+\bar{\epsilon}\right].\label{eq:f_epsbar}
\end{equation}
Now, using
\begin{equation}
\frac{d\ln\dot{\phi}}{d\ln\xi}=-\left(1+\bar{\epsilon}\right)\delta_{1},\label{eq:dlnphidot}
\end{equation}
\begin{equation}
\frac{d\ln H}{d\ln\xi}=\left(1+\bar{\epsilon}\right)\epsilon,\label{eq:dlnH}
\end{equation}
and
\begin{equation}
\frac{d\bar{\epsilon}}{d\ln\xi}=\left(1+\bar{\epsilon}\right)\left(\epsilon-\bar{\epsilon}+\epsilon\bar{\epsilon}\right),\label{eq:depsbar}
\end{equation}
we can find
\begin{equation}
\frac{d\ln f}{d\ln\xi}=-\bar{\epsilon}-\epsilon-\delta_{1}-\bar{\epsilon}\epsilon-\bar{\epsilon}\delta_{1}.\label{eq:dlnf}
\end{equation}
In addition, using also Eqs.~\eqref{eq:deps} and \eqref{eq:ddeltan}, we can find
\begin{equation}
\frac{d^{2}\ln f}{d\ln\xi^{2}}=\left(1+\bar{\epsilon}\right)\left(\bar{\epsilon}-\epsilon+\delta_{2}+\bar{\epsilon}\delta_{1}+\bar{\epsilon}\delta_{2}+\epsilon^{2}+2\epsilon\delta_{1}-\delta_{1}^{2}+\bar{\epsilon}\epsilon^{2}-\bar{\epsilon}\delta_{1}^{2}\right).\label{eq:d2lnf}
\end{equation}
Finally, using the definition of $g$ (Eq.~\eqref{eq:gdef}), we have
\begin{multline}
g\left(\ln x\right)=\left[
4\bar{\epsilon}+2\epsilon+3\delta_1+\delta_2+2\bar{\epsilon}^2+4\bar{\epsilon}\epsilon+5\bar{\epsilon}\delta_1
+2\bar{\epsilon}\delta_2+2\epsilon^2+4\epsilon\delta_1+2\bar{\epsilon}^2\epsilon+2\bar{\epsilon}^2\delta_1\right.\\
\left.+\bar{\epsilon}^2\delta_2
+4\bar{\epsilon}\epsilon^2+6\bar{\epsilon}\epsilon\delta_1+\bar{\epsilon}\delta_1^2+2\bar{\epsilon}^2\epsilon^2
+2\bar{\epsilon}^2\epsilon\delta_1
+\bar{\epsilon}^2\delta_1^2\right]_{\xi=\frac{x}{k}}.\label{eq:gredef}
\end{multline}

\section{Evaluating the integrals\label{sec:Solving-the-integrals}}

We shall see how each of the integrals in Eq.~\eqref{eq:inv2_formula}
can be calculated in a straightforward (albeit tedious) manner when
assuming Eq.~\eqref{eq:P_ansatz}.

\subsection{$\mathcal{I}_1(\xi)\equiv \intop_{0}^{\infty}\frac{dk}{k}m\left(k\xi\right)\ln\mathcal{P}\left(\ln k\right)$\label{sub:1st_int}}

Assuming Eq.~\eqref{eq:P_ansatz}, this term can be rewritten as
\begin{equation}
\mathcal{I}_1(\xi) = \sum_{n=0}^{N}\frac{\beta_{n}}{n!}\intop_{0}^{\infty}\frac{dk}{k}m\left(k\xi\right)\left(\ln\frac{k}{k_{0}}\right)^{n}\equiv\sum_{n=0}^{N}\frac{\beta_{n}}{n!}I_{n}\left(k_{0}\xi\right),\label{eq:integral1_expand}
\end{equation}
where we have defined
\begin{equation}
I_{n}\left(y\right)\equiv\intop_{0}^{\infty}\frac{dx}{x}m\left(x\right)\left(\ln\frac{x}{y}\right)^{n}=\sum_{k=0}^{n}\left(\begin{array}{c}
n\\
k
\end{array}\right)I_{n-k}\left(1\right)\left(-\ln y\right)^{k}=\sum_{k=0}^{n}\left(\begin{array}{c}
n\\
k
\end{array}\right)I_{k}\left(1\right)\left(-\ln y\right)^{n-k}.\label{eq:I_n_def}
\end{equation}

One way of iteratively computing the constant terms $I_{k}\left(1\right)$
is by considering the more general family of integrals,
\begin{equation}
\tilde{I}_{k}\left(\alpha\right)\equiv\intop_{0}^{\infty}\frac{dx}{x}m\left(x\right)\left(\ln x\right)^{k}x^{\alpha},\label{eq:Itil_def}
\end{equation}
which are related to the terms we want to compute by
\begin{equation}
\tilde{I}_{k}\left(0\right)=I_{k}\left(1\right)\label{eq:Itil_I}.
\end{equation}
The $\tilde{I}_{k}$ obey the recursive formula
\begin{equation}
\frac{\partial\tilde{I}_{k}\left(\alpha\right)}{\partial\alpha}=\tilde{I}_{k+1}\left(\alpha\right),\label{eq:Itil_recursive}
\end{equation}
which gives us a simple way to generate all the integrals we are interested
in (since we are not interested in non-integer $n$). The recursion can start from $\tilde{I}_{0}$, which can be shown to be%
\footnote{For example by first calculating the indefinite version of the corresponding
integral by expressing the trigonometric functions in Eq.~\eqref{eq:mdef}
as combinations of complex exponentials and then using the definition
of the incomplete gamma function, before taking the relevant limits
of the result.%
} the continuous version of
\begin{equation}
\tilde{I}_{0}\left(\alpha\right)=-\frac{2^{1-\alpha}}{\pi}\left(1+\alpha\right)\Gamma\left(\alpha-1\right)\sin\left(\frac{\pi}{2}\alpha\right).\label{eq:Itil_0}
\end{equation}

Putting all of this together, we can finally write the relevant integrals
up to $N=3$ as
\begin{equation}
I_{0}\left(k_{0}\xi\right)=1\label{eq:I0}
\end{equation}
\begin{equation}
I_{1}\left(k_{0}\xi\right)=-\ln\left(k_{0}\xi\right)-\gamma+2-\ln2\label{eq:I1}
\end{equation}
\begin{equation}
I_{2}\left(k_{0}\xi\right)=\ln^{2}\left(k_{0}\xi\right)+\left(-4+2\gamma+2\ln2\right)\ln\left(k_{0}\xi\right)+\frac{\pi^{2}}{12}+\gamma\left(-4+\gamma+2\ln2\right)+\left(\ln2-2\right)^{2}\label{eq:I2}
\end{equation}
\begin{multline}
I_{3}\left(k_{0}\xi\right)=-\ln^{3}\left(k_{0}\xi\right)+3\left(2-\gamma-\ln2\right)\ln^{2}\left(k_{0}\xi\right)\\-\frac{1}{4}\left(12\left(\gamma-2\right)^{2}+\pi^{2}+12\ln2\left(-4+2\gamma+\ln2\right)\right)\ln\left(k_{0}\xi\right)\\
+\frac{1}{4}\left(48+2\pi^{2}-8\zeta\left(3\right)-\gamma\left(48+4\left(\gamma-6\right)\gamma+\pi^{2}\right)^{^{^{}}}\right.\\
\left.-4\ln^{3}2+24\ln^{2}2-12\gamma\ln^{2}2-\left(12\left(\gamma-2\right)^{2}+\pi^{2}\right)\ln2\right),\label{eq:I3}
\end{multline}
where $\gamma\simeq0.5772$ is the Euler-Mascheroni constant and $\zeta$
is the Riemann zeta function, the next integral (which it turns out
will be relevant further ahead) being given by
\begin{multline}
I_{4}\left(k_{0}\xi\right)=\ln^{4}\left(k_{0}\xi\right)+4\left(-2+\gamma+\ln2\right)\ln^{3}\left(k_{0}\xi\right)\\
+\frac{1}{2}\left(12\gamma^{2}+\pi^{2}+24\gamma\left(\ln2-2\right)+12\left(\ln2-2\right)^{2}\right)\ln^{2}\left(k_{0}\xi\right)\\
+\left(8\zeta\left(3\right)-48+4\ln^{3}2-24\ln^{2}2+48\ln2+4\gamma^{3} {}^{^{^{^{}}}}\right.\\
\left.
+\gamma\left(\pi^{2}+12\left(\ln2-2\right)^{2}\right)+12\gamma^{2}\left(\ln2-2\right)+\pi^{2}\left(\ln2-2\right)\right)\times\ln\left(k_{0}\xi\right)\\
-16\zeta\left(3\right)+\gamma\left(4\left(2\zeta\left(3\right)-12+\ln^{3}2-6\ln^{2}2+12\ln2\right)+\pi^{2}\left(\ln2-2\right)\right)\\
+8\zeta\left(3\right)\ln2+\frac{19\pi^{4}}{240}+2\pi^{2}+\gamma^{4}+48+\ln^{4}2-8\ln^{3}2\\
+\frac{1}{2}\pi^{2}\ln^{2}2+24\ln^{2}2-2\pi^{2}\ln2-48\ln2+4\gamma^{3}\left(\ln2-2\right)+\frac{1}{2}\gamma^{2}\left(\pi^{2}+12\left(\ln2-2\right)^{2}\right).\label{eq:I4}
\end{multline}

\subsection{\textmd{\normalsize{
\newform{
$\mathcal{I}_2(\xi)\equiv -\frac{\pi^2}{8}\intop_{0}^{\infty}\frac{dk}{k}m\left(k\xi\right)
\left[\intop_{0}^{\infty}\frac{dl}{l}
 \frac{\mathcal{P}'(\ln l)}{\mathcal{P}(\ln l)}
 \intop_{0}^{\infty}\frac{d\zeta}{\zeta}m\left(k\zeta\right)m\left(l\zeta\right)
 \right]^{2}$
}{
$\mathcal{I}_2(\xi)\equiv -\frac{1}{2\pi^{2}}\intop_{0}^{\infty}\frac{dk}{k}m\left(k\xi\right)\left[\intop_{0}^{\infty}\frac{dl}{l}\ln\left|\frac{k+l}{k-l}\right|\frac{\mathcal{P}^{\prime}\left(\ln l\right)}{\mathcal{P}\left(\ln l\right)}\right]^{2}$
}}}}

Assuming Eq.~\eqref{eq:P_ansatz}, this term can be written as
\newform{
\begin{equation}
\mathcal{I}_2(\xi) = -\frac{\pi^2}{8}\intop_{0}^{\infty}\frac{dk}{k}m\left(k\xi\right)
\left[
\sum_{n=0}^{N-1}\frac{\beta_{n+1}}{n!}
\intop_{0}^{\infty}\frac{dl}{l}
 \intop_{0}^{\infty}\frac{d\zeta}{\zeta}m\left(k\zeta\right)m\left(l\zeta\right)
 \left(\ln\frac{l}{k_{0}}\right)^{n}
 \right]^{2}.\label{eq:integral2_expand}
\end{equation}
}{
\begin{equation}
-\frac{1}{2\pi^{2}}\intop_{0}^{\infty}\frac{dk}{k}m\left(k\xi\right)\left[\sum_{n=0}^{N-1}\frac{\beta_{n+1}}{n!}\intop_{0}^{\infty}\frac{dl}{l}\ln\left|\frac{k+l}{k-l}\right|\left(\ln\frac{l}{k_{0}}\right)^{n}\right]^{2}.\label{eq:integral2_expand}
\end{equation}
}
\newform{
It is convenient to focus first on the integral being squared, which we can write as a sum of terms of the form
\begin{equation}
\intop_{0}^{\infty}\frac{dl}{l}\intop_{0}^{\infty}\frac{d\zeta}{\zeta}m\left(k\zeta\right)m\left(l\zeta\right)\left(\ln\frac{l}{k_{0}}\right)^{n}
=\intop_{0}^{\infty}\frac{d\zeta}{\zeta}m\left(k\zeta\right)I_{n}\left(k_{0}\zeta\right),\label{eq:logabs_I}
\end{equation}
}{
It is convenient to focus first on the integral being squared.
Using
the known relation \cite{Stewart_inverse2}
\begin{equation}
\intop_{0}^{\infty}\frac{d\zeta}{\zeta}m\left(k\zeta\right)m\left(l\zeta\right)=\frac{2}{\pi^{2}}\ln\left|\frac{k+l}{k-l}\right|\label{eq:logabs_integral}
\end{equation}
we can write
\begin{equation}
\intop_{0}^{\infty}\frac{dl}{l}\ln\left|\frac{k+l}{k-l}\right|\left(\ln\frac{l}{k_{0}}\right)^{n}
=
\frac{\pi^{2}}{2}\intop_{0}^{\infty}\frac{dl}{l}\intop_{0}^{\infty}\frac{d\zeta}{\zeta}m\left(k\zeta\right)m\left(l\zeta\right)\left(\ln\frac{l}{k_{0}}\right)^{n}=\frac{\pi^{2}}{2}\intop_{0}^{\infty}\frac{d\zeta}{\zeta}m\left(k\zeta\right)I_{n}\left(k_{0}\zeta\right),\label{eq:logabs_I}
\end{equation}
}
where we changed the order of integration and
used the definition of $I_{n}$ from Eq.~\eqref{eq:I_n_def}. Given
that these functions can be quite messy in appearance, but are always
polynomials in $\ln\left(k_{0}\xi\right)$, we write them as
\begin{equation}
I_{n}\left(k_{0}\xi\right)\equiv\sum_{i=0}^{n}c_{n}\left[i\right]\ln^{i}\left(k_{0}\xi\right),\label{eq:cn_def}
\end{equation}
where the $c_{n}\left[i\right]$ coefficients can be found as described in subsection \ref{sub:1st_int} (the relevant ones
being trivially obtained by comparison with Eqs.
\eqref{eq:I0}, \eqref{eq:I1}, \eqref{eq:I2}, \eqref{eq:I3}, and \eqref{eq:I4}).
This ``inner'' integral thus becomes
\newform{
\begin{equation}
\intop_{0}^{\infty}\frac{d\zeta}{\zeta}m\left(k\zeta\right)I_{n}\left(k_{0}\zeta\right)=
\sum_{i=0}^{n}c_{n}\left[i\right]\intop_{0}^{\infty}\frac{d\zeta}{\zeta}m\left(k\zeta\right)\ln^{i}\left(k_{0}\zeta\right)=\sum_{i=0}^{n}\sum_{j=0}^{i}c_{n}\left[i\right]c_{i}\left[j\right]\left(\ln\frac{k}{k_{0}}\right)^{j}\label{eq:logabscn2}.
\end{equation}
Substituting this into Eq.~\eqref{eq:integral2_expand} and changing the order of summation  we are left with
\begin{equation}
\sum_{n=0}^{N-1}\frac{\beta_{n+1}}{n!}
\intop_{0}^{\infty}\frac{dl}{l}
 \intop_{0}^{\infty}\frac{d\zeta}{\zeta}m\left(k\zeta\right)m\left(l\zeta\right)
 \left(\ln\frac{l}{k_{0}}\right)^{n}
=\sum_{j=0}^{N-1}\tilde{c}_{N}\left[j\right]\left(\ln\frac{k}{k_{0}}\right)^{j},\label{eq:inner_int}
\end{equation}
where we have defined
\begin{equation}
\tilde{c}_{N}\left[j\right]\equiv\sum_{n=j}^{N-1}\sum_{i=j}^{n}\frac{\beta_{n+1}}{n!}c_{n}\left[i\right]c_{i}\left[j\right].\label{eq:cNtil_def}
\end{equation}
Finally, we can tackle the full double integral, writing
\begin{equation}
\mathcal{I}_2(\xi) = -\frac{\pi^2}{8}\sum_{i=0}^{N-1}\sum_{j=0}^{N-1}\tilde{c}_{N}\left[i\right]\tilde{c}_{N}\left[j\right]\intop_{0}^{\infty}\frac{dk}{k}m\left(k\xi\right)\left(\ln\frac{k}{k_{0}}\right)^{i+j}\label{eq:outer_int_expand}
\end{equation}
which, using Eq.~\eqref{eq:cn_def} once more, simplifies to
\begin{equation}
\mathcal{I}_2(\xi) = -\frac{\pi^2}{8}\sum_{i=0}^{N-1}\sum_{j=0}^{N-1}\sum_{s=0}^{i+j}\tilde{c}_{N}\left[i\right]\tilde{c}_{N}\left[j\right]c_{i+j}\left[s\right]\ln^{s}\left(k_{0}\xi\right).\label{eq:outer_int}
\end{equation}
Here, we finally see why Eq.~\eqref{eq:I4} was needed (since $s$ can
vary up to $s=4$ for $N=3$).

}{
\begin{equation}
\intop_{0}^{\infty}\frac{dl}{l}\ln\left|\frac{k+l}{k-l}\right|\left(\ln\frac{l}{k_{0}}\right)^{n}=\frac{\pi^{2}}{2}\sum_{i=0}^{n}c_{n}\left[i\right]\intop_{0}^{\infty}\frac{d\zeta}{\zeta}m\left(k\zeta\right)\ln^{i}\left(k_{0}\zeta\right)=\frac{\pi^{2}}{2}\sum_{i=0}^{n}\sum_{j=0}^{i}c_{n}\left[i\right]c_{i}\left[j\right]\left(\ln\frac{k}{k_{0}}\right)^{j}\label{eq:logabscn2}
\end{equation}
and, substituting this into Eq.~\eqref{eq:integral2_expand} and changing
the order of summation, we are left with
\begin{equation}
\intop_{0}^{\infty}\frac{dl}{l}\ln\left|\frac{k+l}{k-l}\right|\frac{\mathcal{P}^{\prime}\left(\ln l\right)}{\mathcal{P}\left(\ln l\right)}=\sum_{j=0}^{N-1}\tilde{c}_{N}\left[j\right]\left(\ln\frac{k}{k_{0}}\right)^{j},\label{eq:inner_int}
\end{equation}
where we have defined
\begin{equation}
\tilde{c}_{N}\left[j\right]\equiv\sum_{n=j}^{N-1}\sum_{i=j}^{n}\frac{\pi^{2}}{2}\frac{\beta_{n+1}}{n!}c_{n}\left[i\right]c_{i}\left[j\right].\label{eq:cNtil_def}
\end{equation}

At last, we can tackle the full double integral, writing
\begin{equation}
-\frac{1}{2\pi^{2}}\intop_{0}^{\infty}\frac{dk}{k}m\left(k\xi\right)\left[\intop_{0}^{\infty}\frac{dl}{l}\ln\left|\frac{k+l}{k-l}\right|\frac{\mathcal{P}^{\prime}\left(\ln l\right)}{\mathcal{P}\left(\ln l\right)}\right]^{2}=\sum_{i=0}^{N-1}\sum_{j=0}^{N-1}-\frac{1}{2\pi^{2}}\tilde{c}_{N}\left[i\right]\tilde{c}_{N}\left[j\right]\intop_{0}^{\infty}\frac{dk}{k}m\left(k\xi\right)\left(\ln\frac{k}{k_{0}}\right)^{i+j}\label{eq:outer_int_expand}
\end{equation}
which, using Eq.~\eqref{eq:cn_def} once more, simplifies to
\begin{equation}
-\frac{1}{2\pi^{2}}\intop_{0}^{\infty}\frac{dk}{k}m\left(k\xi\right)\left[\intop_{0}^{\infty}\frac{dl}{l}\ln\left|\frac{k+l}{k-l}\right|\frac{\mathcal{P}^{\prime}\left(\ln l\right)}{\mathcal{P}\left(\ln l\right)}\right]^{2}=\sum_{i=0}^{N-1}\sum_{j=0}^{N-1}\sum_{s=0}^{i+j}-\frac{1}{2\pi^{2}}\tilde{c}_{N}\left[i\right]\tilde{c}_{N}\left[j\right]c_{i+j}\left[s\right]\ln^{s}\left(k_{0}\xi\right),\label{eq:outer_int}
\end{equation}
where we finally see why Eq.~\eqref{eq:I4} was needed (since $s$ can
vary up to $s=4$ for $N=3$).
}

\subsection{
$
\mathcal{I}_3(\xi)$ $ \equiv$
$\frac{\pi}{2}$$\intop_{0}^{\infty}\frac{dl}{l} \frac{\mathcal{P}'(\ln l)}{\mathcal{P}(\ln l)}$$ \intop_{0}^{\infty}\frac{dq}{q}
\frac{\mathcal{P}'(\ln q)}{\mathcal{P}(\ln q)}$
$\intop_{0}^{\infty}\frac{d\zeta}{\zeta}m\left(l\zeta\right)$$\intop_{0}^{\infty}\frac{dk}{k^{2}}m\left(k\xi\right)m\left(k\zeta\right)$ $\intop_{\zeta}^{\infty}\frac{d\chi}{\chi^{2}}m\left(q\chi\right)
$
}

Assuming Eq.~\eqref{eq:P_ansatz}, this term can be written as
\newform{
}{
\begin{equation}
\sum_{n=0}^{N-1}\sum_{s=0}^{N-1}\frac{\beta_{n+1}}{n!}\frac{\beta_{s+1}}{s!}\intop_{0}^{\infty}\frac{dl}{l}\intop_{0}^{\infty}\frac{dq}{q}M\left(l\xi,q\xi\right)\left(\ln\frac{l}{k_{0}}\right)^{n}\left(\ln\frac{q}{k_{0}}\right)^{s}\label{eq:integral3_expand}
\end{equation}
and we can use the known relation \cite{Stewart_inverse2}
\begin{equation}
\intop_{0}^{\infty}\frac{d\zeta}{\zeta}m\left(l\zeta\right)\intop_{0}^{\infty}\frac{dk}{k^{2}}m\left(k\xi\right)m\left(k\zeta\right)\intop_{\zeta}^{\infty}\frac{d\chi}{\chi^{2}}m\left(q\chi\right)=\frac{2}{\pi}M\left(l\xi,q\xi\right)\label{eq:M_integral}
\end{equation}
to express it as
}
\begin{multline}
\mathcal{I}_3(\xi) =
\frac{\pi}{2}\sum_{n=0}^{N-1}\sum_{s=0}^{N-1}\frac{\beta_{n+1}}{n!}\frac{\beta_{s+1}}{s!}
\times \\ \intop_{0}^{\infty}\frac{dl}{l}\intop_{0}^{\infty}\frac{dq}{q}\intop_{0}^{\infty}\frac{d\zeta}{\zeta}m\left(l\zeta\right)\intop_{0}^{\infty}\frac{dk}{k^{2}}m\left(k\xi\right)m\left(k\zeta\right)\intop_{\zeta}^{\infty}\frac{d\chi}{\chi^{2}}m\left(q\chi\right)\left(\ln\frac{l}{k_{0}}\right)^{n}\left(\ln\frac{q}{k_{0}}\right)^{s},\label{eq:2sum_5int}
\end{multline}
which we now can solve using a similar method to the previous subsections. For example, using Eq.~\eqref{eq:cn_def}
and integrating with respect to $l$ and $q$ we are left with
\begin{equation}
\sum_{n=0}^{N-1}\sum_{s=0}^{N-1}\sum_{i=0}^{n}\sum_{j=0}^{s}\frac{\pi}{2}\frac{\beta_{n+1}}{n!}\frac{\beta_{s+1}}{s!}c_{n}\left[i\right]c_{s}\left[j\right]\intop_{0}^{\infty}\frac{d\zeta}{\zeta}\ln^{i}\left(k_{0}\zeta\right)\intop_{0}^{\infty}\frac{dk}{k^{2}}m\left(k\xi\right)m\left(k\zeta\right)\intop_{\zeta}^{\infty}\frac{d\chi}{\chi^{2}}\ln^{j}\left(k_{0}\chi\right).\label{eq:2sum_5int-1}
\end{equation}
Focussing on the integral with respect to $\chi$, a simple
change of variables gives
\begin{equation}
\intop_{\zeta}^{\infty}\frac{d\chi}{\chi^{2}}\ln^{j}\left(k_{0}\chi\right)=k_{0}\intop_{k_{0}\zeta}^{\infty}\frac{dx}{x^{2}}\ln^{j}\left(x\right)=k_{0}\intop_{\ln\left(k_{0}\zeta\right)}^{\infty}t^{j}e^{-t}dt\equiv k_{0}\Gamma\left(j+1,\ln\left(k_{0}\zeta\right)\right)\label{eq:incomplete_gamma}
\end{equation}
where $\Gamma$ is the (upper) incomplete gamma function. Since $j$
is an integer, this can be explicitly written as the type of polynomial
we are interested in by using the known relation
\begin{equation}
\Gamma\left(j+1,\ln\left(k_{0}\zeta\right)\right)=\frac{j!}{k_{0}\zeta}\sum_{\sigma=0}^{j}\frac{\ln^{\sigma}\left(k_{0}\zeta\right)}{\sigma!}.\label{eq:integer_inc_gamma}
\end{equation}
The full integral thus becomes
\begin{equation}
\sum_{n=0}^{N-1}\sum_{s=0}^{N-1}\sum_{i=0}^{n}\sum_{j=0}^{s}\sum_{\sigma=0}^{j}\frac{\beta_{n+1}}{n!}\frac{\beta_{s+1}}{s!}\frac{\pi}{2}\frac{j!}{\sigma!}c_{n}\left[i\right]c_{s}\left[j\right]\intop_{0}^{\infty}\frac{d\zeta}{\zeta^{2}}\intop_{0}^{\infty}\frac{dk}{k^{2}}m\left(k\xi\right)m\left(k\zeta\right)\ln^{i+\sigma}\left(k_{0}\zeta\right),\label{eq:2sum_5int-1-1}
\end{equation}
which can be tackled by noticing that
\begin{multline}
\intop_{0}^{\infty}\frac{d\zeta}{\zeta^{2}}m\left(k\zeta\right)\ln^{i+\sigma}\left(k_{0}\zeta\right)=k\intop_{0}^{\infty}\frac{dx}{x^{2}}m\left(x\right)\ln^{i+\sigma}\left(\frac{k_{0}}{k}x\right)\\
=\sum_{\rho=0}^{i+\sigma}\binom{i+\sigma}{\rho}
k\left(\ln\frac{k_{0}}{k}\right)^{\rho}\intop_{0}^{\infty}\frac{dx}{x^{2}}m\left(x\right)\left(\ln x\right)^{i+\sigma-\rho}\equiv\sum_{\rho=0}^{i+\sigma}
\binom{i+\sigma}{\rho}
k\left(\ln\frac{k_{0}}{k}\right)^{\rho}\tilde{I}_{i+\sigma-\rho}\left(-1\right),\label{eq:Itil-1}
\end{multline}
where in the last step we used the definition of $\tilde{I}$
from Eq.~\eqref{eq:Itil_def}. The full integral is therefore reduced
to a sextuple sum of single integrals,
\begin{equation}
\sum_{n=0}^{N-1}\sum_{s=0}^{N-1}\sum_{i=0}^{n}\sum_{j=0}^{s}\sum_{\sigma=0}^{j}\sum_{\rho=0}^{i+\sigma}\frac{\beta_{n+1}}{n!}\frac{\beta_{s+1}}{s!}\frac{\pi}{2}
\binom{i+\sigma}{\rho}
\frac{j!}{\sigma!}\tilde{I}_{i+\sigma-\rho}\left(-1\right)c_{n}\left[i\right]c_{s}\left[j\right]\intop_{0}^{\infty}\frac{dk}{k}m\left(k\xi\right)\left(-\ln\frac{k}{k_{0}}\right)^{\rho}.\label{eq:12go}
\end{equation}
The remaining integral is simply $\left(-1\right)^{\rho}I_{\rho}\left(k_{0}\xi\right)$,
allowing us to write the final result as the following septuple sum
of known and given terms (keeping in mind that the method for finding
out any $\tilde{I}$ was shown in subsection \ref{sub:1st_int})
\begin{equation}
\sum_{n=0}^{N-1}\sum_{s=0}^{N-1}\sum_{i=0}^{n}\sum_{j=0}^{s}\sum_{\sigma=0}^{j}\sum_{\rho=0}^{i+\sigma}\sum_{\delta=0}^{\rho}\frac{\beta_{n+1}}{n!}\frac{\beta_{s+1}}{s!}\frac{\pi}{2}
\binom{i+\sigma}{\rho}\frac{j!}{\sigma!}\left(-1\right)^{\rho}\tilde{I}_{i+\sigma-\rho}\left(-1\right)c_{n}\left[i\right]c_{s}\left[j\right]c_{\rho}\left[\delta\right]\ln^{\delta}\left(k_{0}\xi\right).\label{eq:int3_final}
\end{equation}

\end{document}